\documentclass[11pt]{article}
\usepackage{pifont}
\usepackage{amsfonts}
\usepackage{subfigure}

\usepackage{amssymb}
\usepackage{amsmath}
\usepackage{epsfig}

\begin{document}
\newtheorem{theorem}{Theorem}
\newtheorem{corollary}{Corollary}
\newtheorem{conjecture}{Conjecture}
\newtheorem{definition}{Definition}
\newtheorem{lemma}{Lemma}

\newcommand{\define}{\stackrel{\triangle}{=}}

\pagestyle{empty}

\def\sDoF{\overline{\mbox{\normalfont \scriptsize DoF}}}

\def\QED{\mbox{\rule[0pt]{1.5ex}{1.5ex}}}
\def\proof{\noindent{\it Proof: }}

\date{}

\title{\vspace{-1cm}Subspace Alignment Chains and the Degrees of Freedom of the \\Three-User MIMO Interference Channel}
\author{\normalsize  Chenwei Wang, Tiangao Gou, Syed A. Jafar\\
%        {\small Department of Electrical Engineering and Computer Science}\\
%        {\small University of California, Irvine, Irvine, CA 92697}\\
      {\small \it E-mail~:~\{chenweiw, tgou, syed\}@uci.edu}\\
       }
%% Notes
\maketitle

\thispagestyle{empty}
\begin{abstract}
We show that the 3 user $M_T\times M_R$ MIMO interference channel
where each transmitter is equipped with $M_T$ antennas and each
receiver is equipped with $M_R$ antennas  has $d(M,N) \define
\min\left(\frac{M}{2-1/\kappa}, \frac{N}{2+1/\kappa}\right)$ degrees
of freedom (DoF) normalized by time, frequency, \emph{and space}
dimensions, where  $M\define\min(M_T,M_R), N\define\max(M_T,M_R),
\kappa \define \lceil\frac{M}{N-M}\rceil$.  While the DoF outer
bound of $d(M,N)$ is established for every $M_T, M_R$ value, the
achievability of $d(M,N)$ DoF is established \emph{in general}
subject to a normalization with respect to spatial-extensions, i.e.,
the scaling of the number of antennas at all nodes. Specifically, we
show that $q d(M,N)$ DoF are achievable for the $qM_T\times qM_R$
MIMO 3-user interference channel, for some positive integer $q$
which may be seen as a spatial-extension factor. $q$ is the scaling
factor needed to make the value $qd(M,N)$ an integer. Given
spatial-extensions, the achievability relies only on linear
beamforming based interference alignment schemes and requires
neither channel extensions nor channel variations in time or
frequency. In the absence of spatial extensions, it is shown through
examples how essentially the same interference alignment scheme may
be applied over time-extensions over either constant or time-varying
channels. The central new insight to emerge from this work is the
notion of subspace alignment chains as DoF bottlenecks. The subspace
alignment chains are instrumental both in identifying the extra
dimensions to be provided by a genie to a receiver for the DoF outer
bound, as well as in the construction of the optimal interference
alignment schemes.

The DoF value $d(M,N)$ is a piecewise linear function of $M,N$, with
either $M$ or $N$ being the bottleneck within each linear segment
while the other value contains some redundancy, i.e., it can be
reduced without reducing the DoF. The corner points of these
piecewise linear segments correspond to two sets,
$\mathcal{A}\define\{1/2, 2/3, 3/4, \cdots\}$ and
$\mathcal{B}\define\{1/3,3/5,5/7,\cdots\}$. The set $A$ contains all
those values of $M/N$ and only those values of $M/N$ for which there
is redundancy in \emph{both} $M$ and $N$, i.e., either can be
reduced without reducing DoF. The set $\mathcal{B}$ contains all
those values of $M/N$ and only those values of $M/N$ for which there
is \emph{no redundancy} in {either} $M$ or $N$, i.e., neither can be
reduced without reducing DoF. Because $\mathcal{A}$ and
$\mathcal{B}$ represent settings with maximum and minimum
redundancy, essentially they are the basis for the DoF outer bounds
and inner bounds, respectively.

Our results settle the question of feasibility of linear
interference alignment, introduced previously by Cenk et al.,  for
the $3$ user $M_T\times M_R$ MIMO interference channel, completely
for all values of $M_T,M_R$. Specifically, we show that the linear
interference alignment problem $(M_T\times M_R, d)^3$ (as defined in
previous work by Cenk et al.) is feasible if and only if $d\leq
\lfloor d(M,N)\rfloor$. With the exception of the values
$M/N\in\mathcal{B}$, and only with that exception, we show that for
every $M/N$ value there are proper systems (as defined  by Cenk et
al.) that are not feasible.   Evidently the redundancy contained in
all other values of $M/N$ manifests itself as superfluous variables
that are not discounted in the definition of proper systems, thus
creating a discrepancy between proper and feasible systems.

Our results show that  $M/N\in\mathcal{A}$ are the only values  for which
there is no DoF benefit of joint processing among co-located
antennas at the transmitters or receivers. This may also be seen as
a consequence of the maximum redundancy in the $M/N\in\mathcal{A}$
settings.
\end{abstract}

\newpage

\section{Introduction}

The number of degrees of freedom (DoF) of a communication network is  a
metric of great significance as it provides a  lens into the most
essential aspects of the communication problem. DoF investigations
have  motivated many fundamental ideas such as interference
alignment \cite{MMK, Cadambe_Jafar_int}, deterministic channel
models \cite{Avestimehr_Diggavi_Tse}, rational dimensions
\cite{Etkin_Ordentlich_rational, Motahari_Gharan_Khandani_real},
information dimensions \cite{Wu_Shamai_Verdu},  aligned interference
neutralization \cite{Gou_Jafar_Seong_Chung} and manageable interference
\cite{Shomorony_Avestimehr}. The DoF metric is especially valuable as a first order
capacity approximation. Since even the smallest gap between the best
available DoF inner and outer bounds translates into an unbounded
gap in the best available capacity approximations, communication
networks where DoF values are not known are some of the least understood
problems, and hence, also  the most promising research avenues for
significant and fundamental discoveries.

DoF characterizations have recently been obtained for a wide variety
of wireless networks. Since in this work we are interested primarily
in interference channels, it is notable that the DoF of interference
channels are known when all nodes are equipped with the same number
of antennas, for almost all channel realizations, and regardless of
whether the channels are time-varying, frequency-selective or
constant. However, if each node may have an arbitrary number of
antennas, then a general DoF characterization is not available
beyond the 2 user interference channel. One obvious reason is the
explosion of the number of parameters in considering arbitrary
antenna configurations. However, as we show in this work, the
problem involves fundamental challenges even when the number of
parameters is restricted by symmetry. In particular, the difficulty
of this problem has to do with the new notion of  ``depth" of
overlap  between  vector subspaces  that comes into play on the one
hand, and of translating this notion into information theoretic
bounds on the other. Specifically, in this work we explore what is
perhaps the simplest setting for MIMO interference channels where
the DoF remain unknown, and obtain its DoF characterization, while
highlighting the challenging nature of the general problem in the
process. The  assumptions that define our primary focus in this work
are:
\begin{enumerate}
\item {\bf 3-user symmetric MIMO interference channel:} {\it The number of users is set to $K=3$, the number of antennas at
every transmitter is set to the same value, $M_T$, and the number of
antennas at every receiver is set to the same value, $M_R$. Thus,
the problem space is parameterized by only two variables $M_T,
M_R$.}

\item{\bf Spatially-normalized DoF for almost-all channel realizations:} {\it The DoF characterizations that we seek are intended in the
``almost-surely" sense, with channel coefficients drawn from
continuous distributions. Further, the DoF are normalized not only
with respect to time and frequency dimensions, but also with respect
to the spatial dimension, i.e., we allow channel extensions not only
in time and frequency dimensions, but also in the spatial dimension
(i.e., through a scaling of antennas).}
\end{enumerate}
As the smallest, and therefore the most elementary interference
channel setting where interference alignment becomes relevant, the 3
user interference channel has special significance. The  assumption
of global channel knowledge, as well as the implicit assumption of
comparable signal strengths from all transmitters at all receivers
that follows from the definition of the DoF metric (as opposed to
Generalized DoF), is most relevant to small clusters of, e.g., no
more than 3,  mutually interfering users.
%TBD:  A discussion of non-symmetric settings will be presented at the end.

The rationale for the remaining assumptions is our interest
primarily in generic insights rather than the peculiarities and
caveats associated with special structures. While the restriction to
almost-all channel realizations is by now a standard assumption for
DoF studies, the normalization with respect to spatial dimension is
less common. Spatial extension, i.e., proportional scaling of the
number of antennas at each node and a corresponding normalization of
DoF by the same factor, is appealing in that it allows us to deal
with generic channels, thereby revealing generic insights into the
geometry of alignments and relative signal space dimensions without
having to deal with the added complexity of diagonal or block
diagonal channel structures that would result from channel
extensions over constant or time-varying channels, or the rational
dimension arguments that are often invoked in the absence of
sufficient channel diversity. As a further justification for the
normalization with respect to spatial dimension, we note that so
far, \emph{for every wireless network, with or without multiple
antennas, where the DoF characterizations are available for
almost-all channel realizations, the  DoF characterizations are
unaffected by spatial normalization}. Indeed, we conjecture that this 
should be the case in general, i.e., much like time, and frequency
dimensions, the DoF of a network (for almost all channel
realizations and with global channel knowledge as is assumed here)
should also scale with the proportional scaling of the spatial
dimension. Notably, the general question of the scaling of DoF with
spatial dimension appears as an open problem in \cite{Jafar_IA}.
Also notable is the use of spatial normalization to characterize the
DoF region for the general MIMO 2 user $X$ channel setting in one of
the earliest works on interference alignment \cite{Jafar_Shamai}.
Finally, as explained towards the end of this paper through several
examples, the interference alignment solutions developed in this
work under spatial extensions may be directly applied to time and/or
frequency extensions as well to obtain the same DoF
characterizations but without relying on spatial extensions.

\section{Background}\label{sec:background}
We are interested in both the information theoretic DoF of MIMO
interference channels, as well as the feasibility of linear
interference alignment schemes. We start with a summary of relevant work on these topics.

\subsection{DoF of  MIMO Interference Channels}
The two user ($K=2$) MIMO interference channel, where User 1 has
$M_1$ transmit and $N_1$ receive antennas, and User 2 has $M_2$
transmit and $N_2$ receive antennas, is shown by Jafar and
Fakhereddin in \cite{Jafar_Fakhereddin} to have $\min(M_1+M_2,
N_1+N_2,\max(M_1,N_2),\max(M_2,N_1))$ DoF. For this result, the
achievability is based on linear zero forcing beamforming schemes,
and the converse is based on DoF outer bounds for the multiple
access channel obtained by providing enough antennas to a receiver
so that after decoding and subtracting its own signal, it is able to
also decode the interfering signal.

In \cite{Cadambe_Jafar_int}, Cadambe and Jafar introduced an
asymptotic interference alignment scheme, referred to as the [CJ08]
scheme (see \cite{Jafar_IA} for an intuitive description of the
[CJ08] scheme), leading to the result that in the $K$-user MIMO
interference channel, each user can access half-the-cake in terms of
DoF\footnote{The ``cake" refers to the maximum DoF accessible by a
user when all interfering users are absent.}, for a total DoF value
of $KM/2$ DoF, almost surely, when all nodes are equipped with the
same number of antennas $M_T=M_R=M$ and when channels are
time-varying or frequency-selective. In
\cite{Motahari_Gharan_Maddah-Ali_Khandani}, Motahari et al.
introduced the rational dimensions framework  based on diophantine
approximation theory wherein the [CJ08] scheme is again applied to
establish the same DoF result, but with constant channels and without
time/frequency extensions. For the 3-user MIMO interference channel
setting with $M_T=M_R=M>1$, Cadambe and Jafar also present a closed
form linear beamforming solution in \cite{Cadambe_Jafar_int} that
requires no time-extensions for even $M$ and 2 time-extensions for
odd $M$, to achieve the DoF outer bound value of $M/2$ per user,
without the need for channel variations in time/frequency. The DoF
outer bound in each of these cases is based on the pairwise outer
bounds for any two users, as established previously for the single
antenna setting, $M=1$, by Host-Madsen and Nosratinia in
\cite{Host-Madsen_Nosratinia} and for the multiple antenna setting,
$M>1$, by Jafar and Fakhereddin in \cite{Jafar_Fakhereddin}.

In \cite{Gou_Jafar_MIMO}, Gou and Jafar studied the DoF of the
$K$-user $M_T\times M_R$ MIMO interference channel under the
assumption that  $\eta=\max(M_T,M_R)/\min(M_T,M_R)$ is an integer
and showed that each user has a fraction $\frac{\eta}{\eta+1}$ of
the cake\footnote{Here, the ``cake" corresponds to $\min(M_T,M_R)$
DoF.}, for a total DoF value of $K\min(M_T,M_R)\frac{\eta}{\eta+1}$,
almost surely, when the number of users $K>\eta$.

{\it Example: The 3-user $1\times 2$ MIMO interference channel,
i.e., the interference channel with 3 users where each transmitter
has $1$ antenna and each receiver has $2$ antennas, has $2/3$ DoF
per user, as does the 3-user $2\times 1$ MIMO interference channel. The results holds for more than $3$ users as well, i.e., 
the $K$-user $1\times 2$ and $2\times 1$ interference channels have
$2/3$ DoF per user, for all $K\geq 3$.}

The results of \cite{Gou_Jafar_MIMO}, established originally over time-varying/frequency-selective
channels, are extended to constant channels without the need for channel extensions,  by  Motahari et al. in \cite{Motahari_Gharan_Khandani_real}, by employing the
rational dimensions framework. 
Further, Motahari et al. show that each user in the $K$-user
$M_T\times M_R$ MIMO interference channel, has a fraction
$\frac{\eta}{\eta+1}$ of the cake, even when
$\eta=\frac{\max(M_T,M_R)}{\min(M_T,M_R)}$ is not an integer,
provided the number of users $K\geq
\frac{M_T+M_R}{\mbox{gcd}(M_T,M_R)}$. Interestingly, the
achievability of $\frac{\eta}{\eta+1}\min(M_T,M_R)$ DoF per user, or
equivalently $\frac{M_TM_R}{M_T+M_R}$ DoF per user,  follows from
the application of the [CJ08] scheme for every $M_T,M_R$ value, and for any number of users $K$, and
\emph{requires no joint signal processing between the multiple receive
(transmit) antennas at any receiver (transmitter)}. However, the
optimality of these achievable DoF has been shown only when either
$\eta$ is an integer and $K\geq \eta$ or when $K\geq
\frac{M_T+M_R}{\mbox{gcd}(M_T,M_R)}$ for any $\eta$. The outer
bounds in each of these cases are based on allowing full cooperation
among groups of transmitters/receivers and applying the DoF outer
bound for the resulting $2$-user MIMO interference channel
previously derived by Jafar and Fakhereddin in
\cite{Jafar_Fakhereddin}.

{\it Example: Consider the 5-user $2\times 3$ MIMO interference
channel. Allowing full cooperation between users 1, 2, 3 and
allowing full cooperation between users 4,5, we have a resulting two
user MIMO interference channel where the effective User 1 has 6
transmit and $9$ receive antennas, and the effective User 2 has $4$
transmit and $6$ receive antennas. According to the DoF result for
2-user MIMO interference channel shown by Jafar and Fakhereddin in
\cite{Jafar_Fakhereddin}, this channel has 6 DoF. Since cooperation
cannot reduce the total DoF, it follows that the original 5 user
$2\times 3$ interference channel has no more than $6/5$ DoF per
user. Since $M_TM_R/(M_T+M_R)$ DoF are always achievable per user,
$6/5$ is the optimal value of DoF for this channel. Further, since
DoF per user cannot increase with the number of users, $6/5$ is the
optimal value of DoF per user in the $K$ user $2\times 3$ MIMO
interference channel for all $K\geq 5$. The same conclusion applies
in the reciprocal $3\times 2$ MIMO setting as well.}

Note  that the DoF value of the $2\times 3$ or the $3\times 2$ MIMO
interference channel is not known if the number of users, $K$, is 3
or 4. As a special case of our results in this work, we will show
that $6/5$ is the optimal DoF value per user in the $2\times 3$ or
$3\times 2$ MIMO interference channel for all $K>2$, thereby
resolving the DoF value for all $K$ in the $2\times 3$ and $3\times
2$ settings. Since outer bounds based on full cooperation are not
enough, the challenge in this case will be to identify the genie
signals that will lead us to this conclusion.

\subsection{Feasibility of Linear Interference Alignment}
While the DoF of MIMO interference channels are of fundamental
interest, the achievable schemes are often built upon theoretical
constructs such as the rational dimensions framework or Renyi
information dimension, whose physical significance and robustness is not yet clear.
On the other hand, linear beamforming schemes are well understood
based on the abundance of MIMO literature. As a consequence, there
is much  interest in the DoF achievable through linear
beamforming schemes, i.e., through linear interference alignment
schemes. A central question in this research avenue is the
feasibility (almost surely) of linear interference alignment based on only spatial
beamforming, i.e., without the need for channel extensions or
variations in time/frequency/space. The feasibility problem is introduced
by Gomadam et al. in \cite{Gomadam_Cadambe_Jafar}, where iterative
algorithms were proposed to test the feasibility of desired
alignments. Recognizing the feasibility problem as equivalent to the
solvability of a system of polynomial equations, Cenk et al. in
\cite{Cenk_Gou_Jafar_feasibility} draw upon classical results in
algebraic geometry about the solvability of \emph{generic}
polynomial equations, to classify an alignment problem as
\emph{proper} if and only if the number of independent variables in
every set of equations is at least as large as the number of
equations in that set. While the polynomial equations involved in
the feasibility of interference alignment are not strictly generic,
Cenk et al. appeal to the intuition that proper systems are likely
to be feasible and improper systems to be infeasible. For a $K$ user
$M_T\times M_R$ MIMO interference channel where each user desires
$d$ DoF, denoted as the $(M_T\times M_R, d)^K$ setting, Cenk et al. identified the system as proper if and only if
\begin{eqnarray}
d&\leq&\frac{M_T+M_R}{K+1}\label{eqn:proper}
\end{eqnarray}

The conjectured correspondence between proper/improper systems and feasible/infeasible systems is settled completely in one direction, and partially in the other direction, in recent works by Bresler et al. in \cite{Bresler_Cavendish_Tse} and
Razaviyayn et al. in \cite{Razaviyayn_MIMO}, who show that:
\begin{enumerate}
\item Improper systems are infeasible \cite{Bresler_Cavendish_Tse, Razaviyayn_MIMO}.
\item If $M_T, M_R$ are divisible by $d$ then  proper systems are feasible \cite{Razaviyayn_MIMO}.
\item For square channels, $M_T=M_R$, proper systems are feasible \cite{Bresler_Cavendish_Tse}.
\end{enumerate}
While the properness of a system seems to work fairly well as an indicator of the feasibility  of linear interference alignment in most cases studied so far, it is also remarkable that based on existing results it is possible to find  examples of proper systems that are not feasible. For example. consider the 3 user interference channel where $(M, N)=(4,8)$ and where each user desires $d=3$ DoF. According to (\ref{eqn:proper})  this is a \emph{proper} system, and according to \cite{Gou_Jafar_MIMO} it is \emph{infeasible} because the information theoretic DoF outer bound value for this channel is only $8/3$ per user. The DoF outer bound is easily found by allowing two of the users to cooperate fully, so that the resulting 2 user MIMO interference channel with $(M_1,N_1,M_2,N_2)=(8,16,4,8)$ has a total DoF value of $8$ according to \cite{Jafar_Fakhereddin}. Since cooperation does not hurt, and linear schemes (or any other scheme for that matter) cannot beat an information-theoretic outer bound, it is clear that the $(4\times 8, d)^3$ linear interference alignment problem is infeasible for $d\geq 8/3$, and in particular for $d=3$.

The observation that some proper systems are not feasible is also
made by Cenk et al. in \cite{Cenk_Gou_Jafar_feasibility} who suggest
including known information theoretic DoF outer bounds to further
expand the set of infeasible systems. Interestingly, so far,
\emph{all known DoF outer bounds for $K$ user MIMO interference
channels come directly from the DoF result for the 2 user MIMO
interference channel } \cite{Jafar_Fakhereddin}, applied after
allowing various subsets of users to cooperate, while eliminating
other users to create a 2 user interference channel (as also
illustrated by the preceding example). As we show in this work,
these DoF outer bounds do not suffice, even for the symmetric 3 user MIMO
interference channel for all $M_T, M_R$ values. Thus, the
feasibility of proper systems remains an open problem in general,
even if restricted to the 3 user setting.

In this work, for the 3 user $M_T\times M_R$ MIMO
interference channel, we  settle the issue of
feasibility of linear interference alignment. Somewhat surprisingly
within this setting, especially considering systems near the
threshold of proper/improper distinction,  we  show that  \emph{most
proper systems are infeasible}.

\section{System Model and Metrics}\label{sec:model}

\begin{figure}[!h] \vspace{-0.1in}\centering
\includegraphics[width=3.0in]{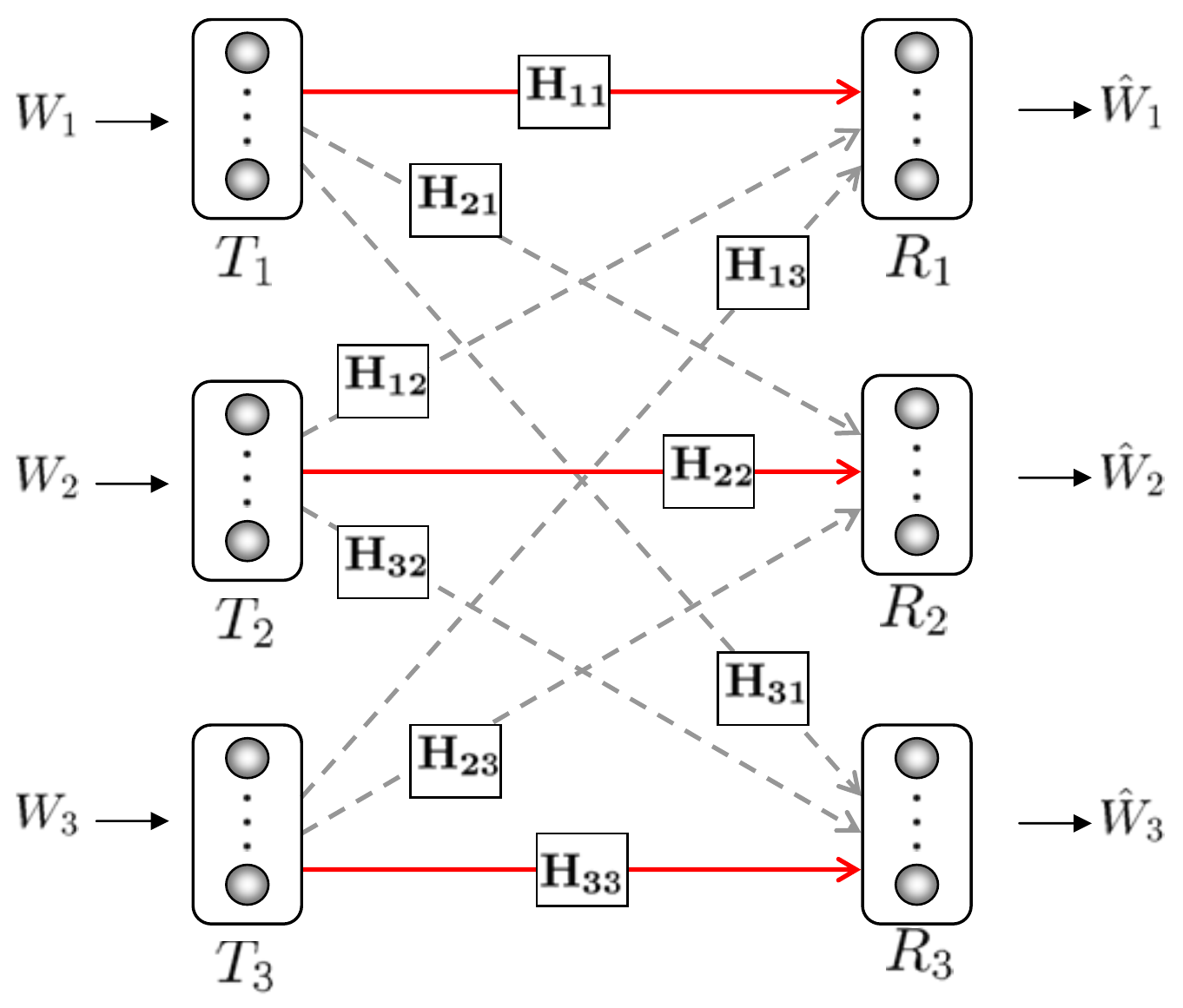}\vspace{-0.05in}
\caption{Three-User Interference Channel} \vspace{-0.1in}
\label{fig:system_model}
\end{figure}

We consider a fully connected three-user MIMO interference channel
where there are $M_T$ and $M_R$ antennas at each transmitter and
each receiver, respectively. As shown in Fig.
\ref{fig:system_model}, each transmitter sends one independent
message to its own desired receiver. Denote by ${\bf H}_{ji}$  the
$M_R\times M_T$ channel matrix from transmitter $i$ to receiver $j$
where $i,j\in\{1,2,3\}$. We assume that the channel coefficients are
independently drawn from continuous distributions. While our results
are valid regardless of whether the channel coefficients are
constant or varying in time/frequency, we will assume the channels
are constant through most of the paper. Global channel knowledge is assumed
to be available at all nodes.

At time index $t\in \mathbb{Z}^+$, Transmitter $i$ sends a
complex-valued $M_T\times 1$ signal vector $\bar{X}_i(t)$, which
satisfies an average power constraint
$\frac{1}{T}\sum_{t=1}^T\mathbb{E}[\|\bar{X}_i(t)\|^2]\leq \rho$ for
$T$ channel uses. At the receiver side, User $j$ receives an
$M_R\times 1$ signal vector $\bar{Y}_j(t)$ at  time index $t$, which
is given by:
\begin{eqnarray*}
\bar{Y}_j(t)=\sum_{i=1}^3{\bf H}_{ji}\bar{X}_i(t)+\bar{Z}_j(t)
\end{eqnarray*}
where $\bar{Z}_j(t)$ an $M_R\times 1$ column vector representing the
i.i.d. circularly symmetric complex additive white Gaussian noise
(AWGN) at  Receiver $j$,  each entry of which is an i.i.d. Gaussian
random variable with zero-mean and unit-variance.

Let $R_k({\rho})$ denote the achievable rate of User $k$ where
$\rho$ is also referred to as the Signal-to-Noise Ratio (SNR). The
capacity region $\mathcal {C}(\rho)$ of this network is the set of
achievable rate tuples $R(\rho)=(R_1(\rho),R_2(\rho),R_3(\rho))$,
such that each user can simultaneously decode its desired message
with arbitrarily small error probability. The maximum sum rate of
this channel is defined as $R_{\Sigma}(\rho)=\max_{R(\rho)\in
\mathcal {C}(\rho)}\sum_{k=1}^3 R_k(\rho)$, and
$R(\rho)=R_{\Sigma}(\rho)/3$ denotes the maximum rate normalized per
user. The sum DoF are defined as $d_\Sigma(M_T,
M_R)=\lim_{\rho\rightarrow \infty}R_{\Sigma}(\rho)/\log(\rho)$, and
DoF$(M_T,M_R)=d_{\Sigma}(M_T,M_R)/3$ stands for the normalized DoF
per user. Throughout this paper, we also write DoF$(M_T,M_R)$ as
$d(M_T,M_R)$ for simplicity. Moreover, we use $d_k(M_T,M_R)$ to
denote the number of DoF associated with User $k$, and the user
index $k$ is interpreted modulo $3$ so that, e.g., User 0 is the
same as User 3. Furthermore, we define the degrees of freedom
normalized by the spatial dimension, $\sDoF_\Sigma(M_T,M_R)$ as
\begin{eqnarray*}
\sDoF_\Sigma(M_T,M_R)&=&
\max_{q\in\mathbb{Z}^+}\frac{d_\Sigma(qM_T,qM_R)}{q}
\end{eqnarray*}
and $\sDoF(M_T,M_R) = \sDoF_\Sigma(M_T,M_R)/3$ is similarly defined.
The dependence on $M_T, M_R$ may be dropped for compact notation
when no ambiguity would be caused, i.e., instead of $\sDoF(M_T,M_R)$
we may write just $\sDoF$.

% Without loss of generality we assume that $M\leq N$, and thus we
%will consider three-user $M\times N$ interference channel and its
%reciprocal $N\times M$ interference channel, respectively in this
%paper.

{\it Notation:} For the matrix $\mathbf{A}$, $\mathbf{A}(i,:)$ and
$\mathbf{A}(:,j)$ denote its $i^{th}$ row and $j^{th}$ column
vector, respectively; $\mathbf{A}(i,m:n)$ denotes a $1\times
(n-m+1)$ row vector obtained from the $i^{th}$ row and the $m^{th}$
to $n^{th}$ columns of the matrix $\mathbf{A}$; $\mathbf{A}(m:n,:)$
denotes a matrix obtained from the $m^{th}$ to $n^{th}$ rows of the
matrix $\mathbf{A}$. We use the notation $o(x)$ to represent any
function $f(x)$ such that $\lim_{x\rightarrow \infty}f(x)/x = 0$.
Further, we define $M=\min(M_T, M_R), N=\max(M_T, M_R)$.

\section{Main Results}\label{sec:results}

In order to present the DoF results for the 3-user $M_T\times M_R$
MIMO interference channel in a compact form, let us define the
quantity DoF$^\star$ as follows.
\begin{definition}
\begin{eqnarray}
\mbox{\normalfont DoF}^\star&\define&\min\left(\frac{M}{2-1/\kappa},
\frac{N}{2+1/\kappa}\right)
\end{eqnarray}
where $M=\min(M_T,M_R), N=\max(M_T,M_R), \kappa=\lceil
\frac{M}{N-M}\rceil$.
\end{definition}
The quantity $\kappa$ denotes the length of the \emph{subspace
alignment chain}, to be described in the next section. Clearly, as
$N$ and $M$ become approximately equal, i.e., $\kappa$ becomes
large,  DoF$^\star$  converges to the ``half the cake" value, $M/2$.

To arrive at the main result of this paper, we proceed through two
intermediate results, presented here as lemmas.
\begin{lemma}[Outer Bound]\label{lemma:out}
For the 3-user $M_T\times M_R$ MIMO interference channel, the DoF
value per user is bounded above as:
\begin{eqnarray}\label{eqn:dofout}
\mbox{\normalfont DoF}\leq\mbox{\normalfont DoF}^\star
\end{eqnarray}
\end{lemma}
\noindent The proof of Lemma \ref{lemma:out} is presented in Section
\ref{sec:outerbound} and Section \ref{sec:outerbound_eg}.

{{\it Remark:} Note that the outer bound holds for {\em arbitrary}
values of $M_T,M_R$ {\em without} any spatial normalization.
However, also note that the outer bound does scale with spatial
dimension, e.g., if we double the number of antennas at each node
the value DoF$^\star$ would be doubled as well. In other words, the
outer bound holds both with and without spatial normalization.

{\it Remark:} Unlike all previously used DoF outer bounds for MIMO interference
channels, this outer bound does not follow from allowing cooperation among groups of users. Instead, what is needed is a careful choice of genie signals that provide a receiver access to  parts of the signal space originating at interfering transmitters. Precisely which parts of a signal space can be provided as side information to produce the tight DoF outer bounds, is perhaps the most challenging technical aspect that we deal with in this paper. As such, the outer bounds represent the most significant contribution and a majority of this paper is devoted to their derivation.

\begin{lemma}[Inner Bound]\label{lemma:in}
For the 3-user $M_T\times M_R$ MIMO interference channel, the DoF
per user value $\lfloor\mbox{\normalfont DoF}^\star\rfloor$  is
achievable with linear beamforming over constant channels without
the need for symbol extensions in time, frequency or space.
%value per user is bounded below as:
%\begin{eqnarray}\label{eqn:dofin}
%\mbox{\normalfont DoF}\geq \mbox{\normalfont DoF}^\star
%\end{eqnarray}
%whenever the quantity $\mbox{\normalfont DoF}^\star$ takes an integer value. Moreover, in all these cases, $\mbox{\normalfont DoF}^\star$ is achievable with linear beamforming  without the need for time extensions, frequency extensions, or spatial extensions.
\end{lemma}
The proof of Lemma \ref{lemma:in} is presented in Section
\ref{sec:innerbound}.

{\it Remark:}  Note that the achievability result stated in Lemma
\ref{lemma:in} is limited to integer values of DoF. Typically, the
issue of achievability for non-integer values of DoF is resolved by
using symbol extensions over time or frequency dimensions, often
with the need for time-varying/frequency-selective channels to
create sufficient diversity. Time/frequency extensions can be used
here as well, as will be discussed in Section \ref{subsec:symbolex}.
However, since our primary interest is in generic channels rather
than the structured (block-diagonal) channels that result from
time/frequency extensions, we will focus on spatial extensions
instead.

\noindent Lemma \ref{lemma:out} and Lemma \ref{lemma:in} lead us
immediately to our main results, stated as theorems.
\begin{theorem}[Spatially-Normalized DoF]\label{theorem:dof}
For the 3-user $M_T\times M_R$ MIMO interference channel, the
spatially-normalized degrees of freedom value  per user is given by:
\begin{eqnarray}\label{eqn:dof}
\sDoF&=&\mbox{\normalfont DoF}^\star
\end{eqnarray}
\end{theorem}

\proof The proof of Theorem \ref{theorem:dof} follows directly from
Lemma \ref{lemma:out} and Lemma \ref{lemma:in}. Clearly, whenever
DoF$^\star$ is an integer, we have an exact DoF characterization,
DoF = DoF$^\star$, without the need for any spatial extensions. For
the cases where DoF$^\star$ is not an integer, let us express it in
its rational form $p/q$. Then, scaling the number of antennas by
$q$, we have a 3 user $qM_T\times qM_R$ MIMO interference channel,
for which the value DoF$^\star = p$ is  both achievable and an outer
bound, i.e., it is optimal. Since the normalized DoF outer bound is
not affected by spatial scaling, no other spatial extension can
improve the spatially-normalized DoF, and we have the result of
Theorem \ref{theorem:dof}.\hfill\QED

To understand the result of Theorem \ref{theorem:dof}, the following
equivalent, but more explicit representation of $\sDoF$ will be
useful:
\begin{eqnarray}\label{eqn:dof}
\sDoF&=&\left\{\begin{array}{ccc}
M,&~& 0\leq\frac{M}{N}\leq 1/3\\
N/3,&~&1/3\leq\frac{M}{N}\leq 1/2\\
2M/3,&~&1/2\leq\frac{M}{N}\leq 3/5\\
2N/5,&~&3/5\leq\frac{M}{N}\leq 2/3\\
3M/5,&~&2/3\leq\frac{M}{N}\leq 5/7\\
3N/7,&~&5/7\leq\frac{M}{N}\leq 3/4\\
4M/7,&~&3/4\leq\frac{M}{N}\leq 7/9\\
4N/9,&~&7/9\leq\frac{M}{N}\leq 4/5\\
\vdots &~& \vdots\leq\frac{M}{N}\leq\vdots
\end{array} \right.
\end{eqnarray}
or in  compact form:
\begin{eqnarray}\label{eqn:compactdof}
\sDoF &=&\left\{\begin{array}{ccc}\frac{p}{2p-1}M,&~& \frac{p-1}{p}\leq\frac{M}{N}\leq \frac{2p-1}{2p+1}\\
\frac{p}{2p+1}N,&~& \frac{2p-1}{2p+1}\leq\frac{M}{N}\leq
\frac{p}{p+1}\end{array} \right. \quad p\in \mathbb{Z}^+.
\end{eqnarray}

While Theorem \ref{theorem:dof} allows spatial extensions to avoid dealing with structured channels resulting from symbol extensions in time/frequency, our results are not limited to settings with spatial extensions. As we illustrate  through numerous examples, the achievable schemes translate directly from spatial extensions to channel extensions over time/frequency dimensions instead. With few exceptions, these channel extensions do not require channel variations in time/frequency. A sample of our DoF results, \emph{without spatial extensions} appears in Fig. \ref{fig:map_table}.

\begin{figure}[!t]
\centering
\includegraphics[width=5.5in]{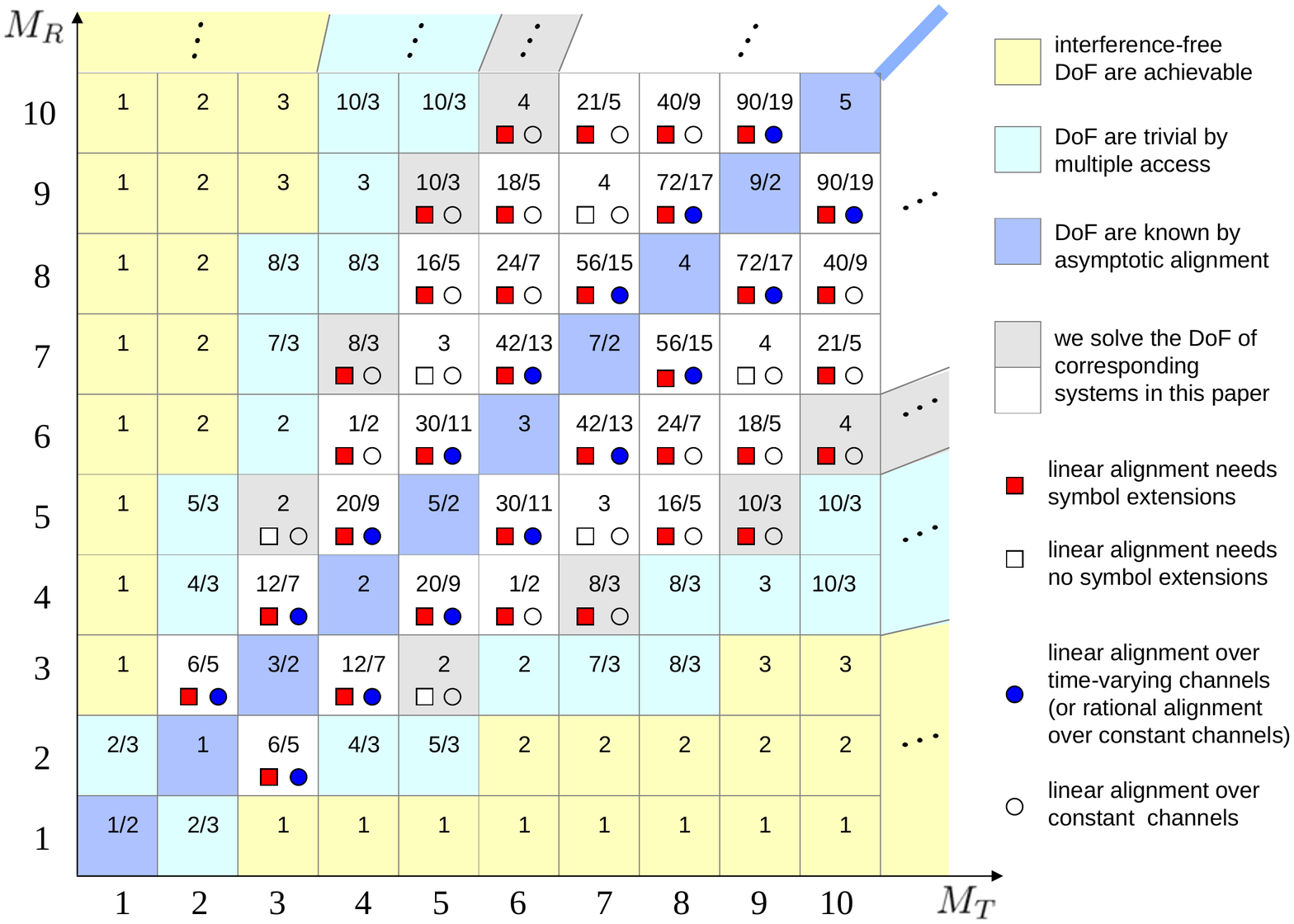}\vspace{-0.1in}
\caption{DoF per User (without space-extensions) of the Three-User $M_T\times M_R$ MIMO
Interference Channel} \label{fig:map_table}
\end{figure}

%We make the following observations based on the results stated in Lemma \ref{lemma:out}, Lemma \ref{lemma:in} and Theorem \ref{theorem:dof}. Based on (\ref{eqn:compactdof}) we observe:
%\begin{eqnarray}
%\mbox{DoF}((2p -1)m, (2p -1)n) = p, &~&\frac{p-1}{p}\leq\frac{m}{n}\leq \frac{2p-1}{2p+1}\\
%\mbox{DoF}((2p +1)m, (2p +1)n) = p, &~&\frac{2p-1}{2p+1}\leq\frac{m}{n}\leq \frac{p}{p+1}
%\end{eqnarray}

Lastly, we consider the most restricted setting where only linear alignment schemes are allowed, and no channel extensions are allowed in time/frequency/space. The next theorem settles the issue of feasibility of linear interference alignment for the 3 user $M_T\times M_R$ MIMO
interference channel.

\begin{theorem}[Feasibility of Linear Interference Alignment]\label{theorem:feasible}
For the 3-user $M_T\times M_R$ MIMO interference channel, the DoF
demand per user, $d$, is feasible with linear interference alignment
if and only if $d\leq \mbox{\normalfont DoF}^\star$.
\end{theorem}

{\it Remark: } Note that the feasibility of linear interference
alignment is intended here in the same sense as studied previously
by Cenk et al. in \cite{Cenk_Gou_Jafar_feasibility}, Bresler et al.
in \cite{Bresler_Cavendish_Tse} and Rezaviyayn et al. in
\cite{Razaviyayn_MIMO}, i.e., without symbol extensions in time/frequency/space.

\proof The proof of Theorem \ref{theorem:feasible} also follows
directly from Lemma \ref{lemma:out} and Lemma \ref{lemma:in}.  Since
the feasibility of linear interference alignment only concerns
integer values of DoF per user, where the inner and outer bounds are
tight, and the inner bound is achievable with linear interference
alignment without the need for symbol extensions in time, frequency
or space, the result of Theorem \ref{theorem:feasible} follows
immediately, and will be presented in Section \ref{sec:feasibility}
in detail.\hfill\QED

\section{Understanding the Result}
In this section we provide an intuitive understanding of the main
results based on linear dimension counting arguments, specifically
by introducing the notion of subspace alignment chains, and
highlight the key observations that follow from the main results.

\subsection{The Simple Cases: $M/N\leq 1/2$ and $M=N$}\label{subsec:achieveless1/2}

%For explaining linear dimension counting arguments we will  focus on
%the setting that $M_R\geq M_T$, so that $M=M_T, N=M_R$, i.e., the
%receiver has at least as many antennas as the transmitter. Since
%linear dimension counting arguments are the same for reciprocal
%networks, there is no loss of generality in this assumption.

If $M=N$, then the DoF are already known, corresponding to the
half-the-cake value reported in \cite{Cadambe_Jafar_int}.

%Let us consider Figure \ref{fig:map_table}. Given any $(M_T,M_R)$
%pair system, we show the value of DoF per user in the corresponding
%square. Among these squares, we use light yellow, green and blue
%shades to denote the settings where the DoF are trivially available,
%while the white and light grey shade denote the settings where the
%DoF remain open.

If $0< M/N\leq 1/2$, i.e., $2M\leq N$, then interference alignment
is not required. Consider first the setting $0<M/N\leq 1/3$, i.e.,
$3M\leq N$. Here, if $3M_T\leq  M_R$ then each receiver has enough
antennas to zero force all interference at no cost to desired
signals, and if $3M_R\leq M_T$ then each transmitter has enough
antennas to zero force all unintended receivers. Therefore, in this
case every user achieves all of the cake, i.e., his
interference-free DoF, $M$.

Next, consider the case $1/3\leq M/N\leq 1/2$, i.e., $2M\leq N\leq
3M$. Allowing any two users to cooperate, and using the DoF result
for the resulting 2 user MIMO interference channel, we find the
total DoF outer bound value of $N$. This value is easily seen to be
achievable with only zero forcing at the  receivers ($M_R>M_T$) or
at the transmitters ($M_T>M_R$), e.g., with users 1 and 2 achieving
$M$ DoF each and the third user achieving $N-2M$ DoF. Note that the
DoF assigned to each user can always be made equal by equal
time-sharing between all permutations of the users for a given DoF
allocation.

For the remaining cases, i.e., $1/2<M/N<1$, it turns out we need
interference alignment. The challenge lies not only in constructing
interference alignment schemes for this setting, but also, and to a
greater extent, in finding the required DoF outer bounds. A new
notion that emerges in this setting and that plays a central role in
limiting the DoF values, is the notion of  subspace alignment
chains, which is introduced next.

\subsection{Subspace Alignment Chains}\label{subsec:subspace_chain}
The main DoF outer bound, presented in Lemma \ref{lemma:out},
consists of two outer bounds.
\begin{eqnarray}
\mbox{\normalfont DoF}&\leq&\frac{N}{2+1/\kappa}\label{eqn:Nout}\\
\mbox{\normalfont DoF}&\leq&\frac{M}{2-1/\kappa}\label{eqn:Mout}
\end{eqnarray}
We will refer to these bounds as the $N$-bound and the $M$-bound,
respectively. The parameter $\kappa$ that appears in both bounds, is
the length of the \emph{subspace alignment chain}, a notion to be
introduced in this section. Note that the first outer bound value is
limited by $N$ and is an increasing function of $\kappa$ while the
second outer bound value is limited by $M$ and is a decreasing
function of $\kappa$. The significance of this will become clear in
the following description.

\subsubsection{The $N$-bound: DoF $\leq \frac{\kappa}{2\kappa+1}N$}

Consider the first outer bound, DoF $\leq
\frac{\kappa}{2\kappa+1}N$. Since linear dimension counting
arguments are identical for reciprocal networks
\cite{Gomadam_Cadambe_Jafar}, without loss of generality let us
focus on the setting $M_T<M_R$, so that $M=M_T, N=M_R$.  Now, since
each receiver has at least as many antennas as any transmitter,
zero-forcing of signals by the transmitters is  not possible. Since
interference cannot be eliminated, the next best thing is to align
interference. Ideally, since each transmitter causes interference at
two receivers, it should align with another interference vector at
each of those undesired receivers. For example, let us consider a
vector ${\bf V}_{1(1)}$ sent by Transmitter 1 that causes
interference at Receiver 2. This vector should align with a vector
${\bf V}_{3(1)}$ sent by Transmitter 3, which is also undesired at
Receiver 2. Now, the vector ${\bf V}_{3(1)}$  also causes
interference at Receiver 1, so it should align there with a vector
${\bf V}_{2(1)}$ sent by Transmitter 2. The vector ${\bf V}_{2(1)}$
in turn also causes interference to Receiver 3, so it should align
with a vector ${\bf V}_{1(2)}$ sent from Transmitter 1. Continuing
like this, we create a chain of desired alignments:
\begin{eqnarray*}
{\bf V}_{1(1)}\stackrel{\mbox{\tiny Rx 2}}{\longleftrightarrow} {\bf
V}_{3(1)}\stackrel{\mbox{\tiny Rx 1}}{\longleftrightarrow} {\bf
V}_{2(1)}\stackrel{\mbox{\tiny Rx 3}}{\longleftrightarrow} {\bf
V}_{1(2)}\stackrel{\mbox{\tiny Rx 2}}{\longleftrightarrow} {\bf
V}_{3(2)}\stackrel{\mbox{\tiny Rx 1}}{\longleftrightarrow} {\bf
V}_{2(2)}\stackrel{\mbox{\tiny Rx 3}}{\longleftrightarrow} \cdots
\end{eqnarray*}
The main question is whether this ideal scenario is possible, i.e.,
can we extend this subspace alignment chain indefinitely?  Consider,
for example the setting $M=N$ that is previously solved by Cadambe
and Jafar in \cite{Cadambe_Jafar_int}. Cadambe and Jafar create this
infinite chain of alignments  using the asymptotic alignment scheme
for $M=N=1$ and implicitly create an infinite alignment chain in the
non-asymptotic solution for, e.g.,  $M=N=2$, as the chain  closes
upon itself to form a loop, i.e., ${\bf V}_{1(1)}={\bf V}_{1(2)}$.
The chain closes upon itself mainly because in this setting (as well
as all cases where $M=N>1$) the optimal signal vectors are
eigenvectors of the cumulative channel encountered in traversing the
alignment chain starting from any transmitter and continuing until
we return to the same transmitter. Thus, as shown by Cadambe and
Jafar \cite{Cadambe_Jafar_int}, the ideal solution of perfect
alignment, achieved by an infinite (or closed-loop) alignment chain,
is possible when $M=N$.

When $M\neq N$, it turns out that the subspace alignment
chain can neither be continued indefinitely, nor be made to close
upon itself. The length of the subspace alignment chain is therefore
limited to a finite value $\kappa$ that is a function of $M$ and $N$.  The
limited length of the subspace alignment chain creates the
bottleneck on the extent to which interference can be aligned, and
is ultimately the main factor limiting the DoF value.

\begin{figure}[!ht]
\centering
\subfigure[$1/2<M/N\leq 2/3$, Chain length $\kappa=2$]{
\includegraphics[scale=0.45]{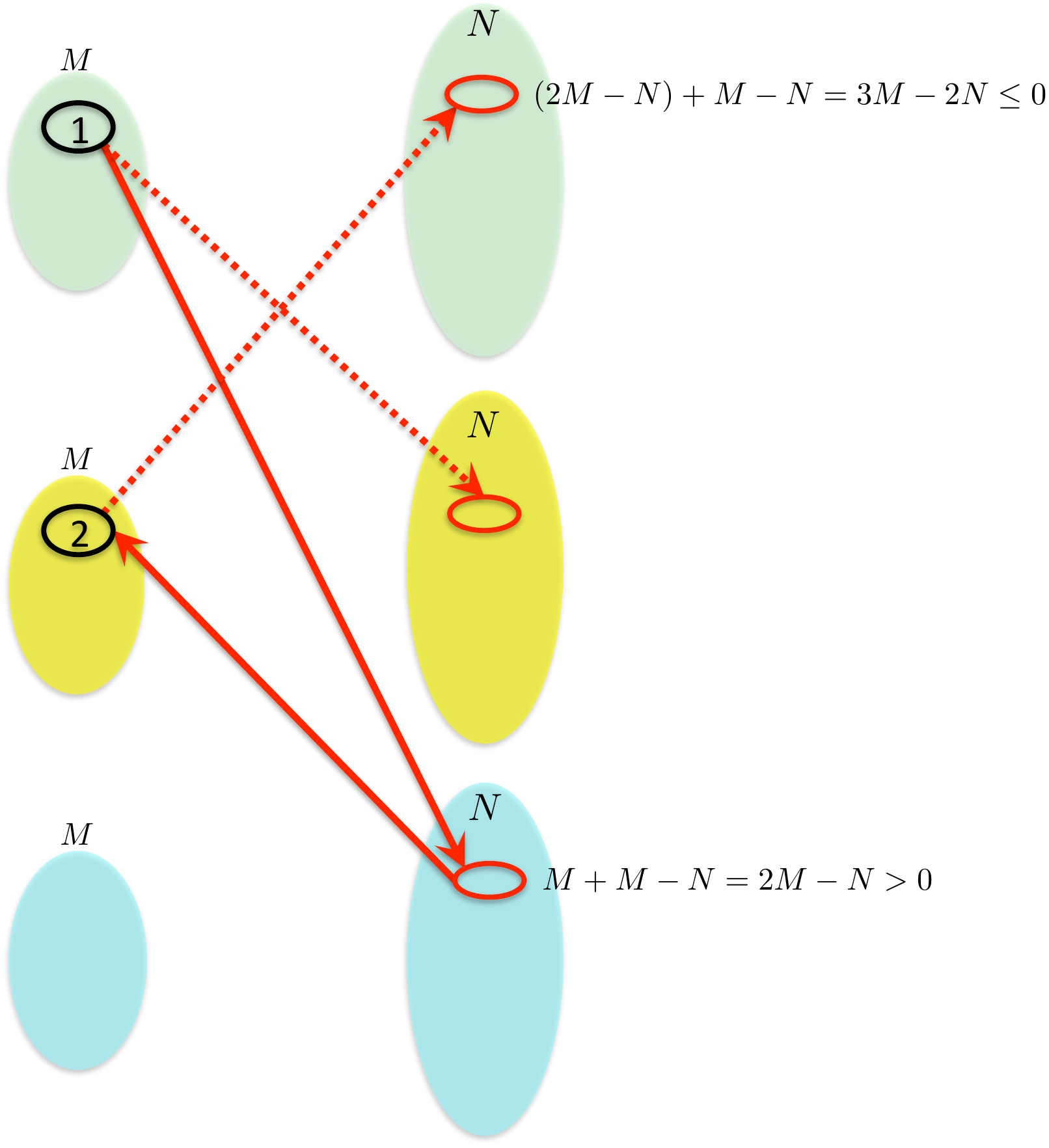}
\label{fig:Nsac2}
}
\subfigure[$3/4<M/N\leq 4/5$, Chain length $\kappa=4$]{
\includegraphics[scale=0.45]{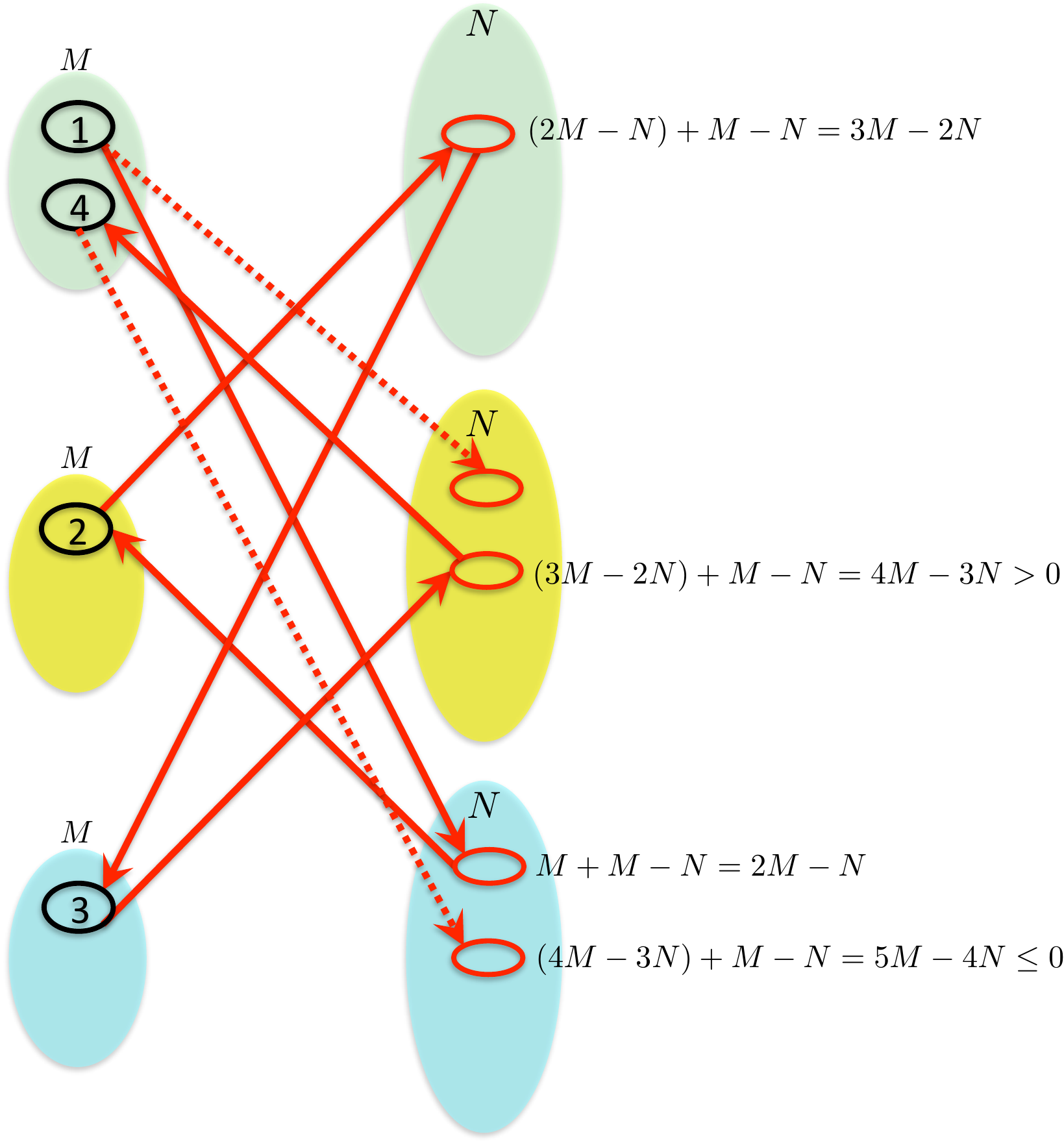}
\label{fig:Nsac4}
}
\caption{Subspace alignment chains leading to the $N$-bound, DoF $\leq \frac{\kappa}{2\kappa+1}N$ per user. Dashed arrows indicate unaligned interference. Numbered signal spaces shown in black ovals indicate the position of the signal space in the alignment chain. The directions of the solid arrows indicate the progression of the alignment chain.}\label{fig:chain_length}
\end{figure}

%
%\begin{figure}[!h]
%\centering
%\includegraphics[width=3in]{Nsac2}\vspace{-0.15in}
%\caption{Subspace Alignment Chain for $1/2<M/N\leq 2/3$} \label{fig:Nsac2}
%\end{figure}

Consider the setting $1/2<M/N\leq 2/3$, as shown in Fig.
\ref{fig:chain_length}\subref{fig:Nsac2}, from a linear dimension
counting perspective.  Users 1, 2, and 3 are shown in green, yellow
and blue in the figure. Let us start our subspace alignment chain
with a signal space at Transmitter 1, shown in the figure as a black
oval with the number 1. Consider the alignment that can occur at
Receiver 3  with a corresponding signal space (marked with the
number 2 to indicate the second signal in the alignment chain)
originating at Transmitter 2. Note that interference can be aligned
at each receiver only within that subspace which is accessible from
both interfering transmitters.  Since each transmitter can access
only a $M$ dimensional subspace of the $N$ dimensional signal space
available to Receiver 3, and because generic subspaces do not
overlap any more than they have to, it follows that the size of the
signal space accessible from both Transmitters 1 and 2, where
interference alignment can take place, is no more than $M+M-N=2M-N$
dimensional. Since $1/2<M/N$, we note that $2M-N$ is a positive
number, i.e., such a space exists. However, the accounting for
aligned dimensions does not stop here. Let us continue the alignment
chain to see if further alignment is possible. The $2M-N$
dimensional signal space 2 originating at Transmitter 2, can align
with a corresponding signal space from Transmitter 3 at Receiver 1,
in no more than $(2M-N) + M - N = 3M-2N$ dimensions. However, since
$M/N\leq 2/3$, we note that $3M-2N$ is not a positive number, i.e.,
such a space does not exist. Thus, the alignment chain must stop
here, and the maximum length of the alignment chain is $\kappa=2$,
representing the number of transmitted signal spaces that
participate in the alignment chain. Now, to find corresponding DoF
outer bound, let us account for the signal space dimensions. Fig.
\ref{fig:chain_length}\subref{fig:Nsac2} may be helpful here.
Assuming each transmitted signal space is $d_0$ dimensional, the
total number of dimensions transmitted is $d_0+d_0 = 2d_0$, and the
number of dimensions occupied by interference at all three receivers
is $d_0+d_0+d_0=3d_0$. Since the desired signal spaces must remain
distinct from interference, the sum of desired signal dimensions at
all receivers is $2d_0$. Thus, the total number of receive
dimensions needed to accommodate both the desired signals and the
interference is at least $5d_0$. Since the total number of receive
dimensions available is $3N$ we must have $5d_0\leq 3N$, i.e.,
$d_0\leq 3N/5$. Since the total number of transmitted dimensions
\emph{per user} is $2d_0/3$, we have the corresponding outer bound
value, DoF $\leq 2N/5$ per user, which corresponds to
$\frac{\kappa}{2\kappa+1}N$, as expected.

As the next example, consider the setting $3/4<M/N\leq 4/5$,  shown
in Fig. \ref{fig:chain_length}\subref{fig:Nsac4}. Again, we start
our alignment chain with the subspace numbered $1$ at Transmitter 1.
As seen from Receiver 3, this space can align with a corresponding
signal space (numbered 2) originating at Transmitter 2, in no more
than $2M-N>0$ dimensions. Continuing the chain, the next alignment
must occur at Receiver 1, where the $2M-N$ dimensional space
(numbered 2) originating at Transmitter 2, can align with a
corresponding signal space (numbered 3) originating at Transmitter
3, in no more than $(2M-N) + M - N=3M-2N$ dimensions, which is also
a positive number since $M/N>3/4>2/3$, i.e., such a space exists. At
this point the alignment chain has length 3. Continuing it further
we note that the next alignment must occur at Receiver 2, where the
$3M-2N$ dimensional space (numbered 3) originating at Transmitter 3
can align with a corresponding signal space (numbered 4) originating
at Transmitter 1, in no more than $(3M-2N)+M-N=4M-3N$ dimensions,
which is also a positive number since $M/N>3/4$, i.e., such a space
exists. Now the length of the alignment chain is 4. In order to
continue the alignment chain further, we next consider Receiver $3$,
where the $4M-3N$ dimensional space (numbered 4) originating at
Transmitter 1 can align with a corresponding space originating at
Transmitter 2 in no more than $(4M-3N)+M-N=5M-4N$ dimensions, which
is not a positive number since $M/N\leq 4/5$, i.e., such a space
does not exist. Thus, the alignment chain ends here, and the maximum
length of the alignment chain is $\kappa=4$ for this example. Now,
lets compute the implied DoF outer bound. Assuming each transmitted
signal space is $d_0$ dimensional, the total number of dimensions
transmitted is $4d_0$, and the number of dimensions occupied by
interference at all three receivers is $5d_0$. Including the desired
signal dimensions which must remain distinct from interference, the
total number of receive dimensions needed is at least
$4d_0+5d_0=9d_0$. Since the total number of receive dimensions
available is $3N$ we must have $9d_0\leq 3N$, i.e., $d_0\leq N/3$.
Since the total number of transmitted dimensions \emph{per user} is
$4d_0/3$, we have the corresponding outer bound, DoF $\leq 4N/9$ per
user, which again corresponds to $\frac{\kappa}{2\kappa+1}N$, as
expected.

These examples can be generalized in a straightforward manner, to
verify that for the setting $\frac{p-1}{p}<\frac{M}{N}\leq
\frac{p}{p+1}$, the length of the subspace alignment chain $\kappa =
p$, which can also be expressed as $\kappa =
\lceil\frac{M}{N-M}\rceil$. The corresponding outer bound value is
calculated as follows. A total of $\kappa d_0$ dimensions are
transmitted, creating interference spaces of  total dimension
$(\kappa+1)d_0$. The desired signal and interference together need a
total of $\kappa d_0 + (\kappa+1)d_0=(2\kappa+1)d_0$ dimensions.
Since only $3N$ dimensions are available, we must have $d_0\leq
\frac{3N}{2\kappa+1}$. Now, since the number of dimensions
transmitted per user is $\kappa d_0/3$, we have the $N$-bound, DoF
$\leq \frac{\kappa}{2\kappa+1}N$.

We make an interesting comparison between the subspace alignment
chains introduced above and the work of Bresler and Tse in
\cite{Bresler_Tse_diversity}. Bresler and Tse considered linear
interference alignment problem for a three-user SISO Gaussian
interference channel with time-varying/frequency-selective channel
coefficients when the maximum diversity order that the channel can
provide, e.g., the number of sub-carriers, is limited to $L$. They
show that the maximum DoF achievable per user through linear
interference alignment is a strictly increasing function of $L$ that
monotonically converges to the information theoretic outer bound
value of 1/2 as $L$ approaches infinity. Finite length chains of
aligned vectors appear in the derivation of Bresler and Tse's result
as well. However, in spite of this superficial similarity, the
alignment chains used by Bresler and Tse are  fundamentally
different  from the subspace alignment chains in this work. First,
the alignment chains of Bresler and Tse identify the limitations of
DoF achievable through \emph{linear} interference alignment schemes,
but it is known that these limitations are surpassed by rational
alignment schemes \cite{Motahari_Gharan_Khandani_real} that can
achieve the information theoretic outer bound value of $1/2$ without
the need for any channel variations. On the other hand, the subspace
alignment chains presented in this work are much more fundamental in
that they identify \emph{information theoretic} outer bounds, i.e.,
these bounds cannot be beaten by linear alignment, rational
alignment or any other  scheme to be invented in the future. Second,
the alignment chains in Bresler and Tse's work  reach their maximum
length when it becomes impossible to separate the desired signal
from interference.  Alignment of interference, per se, is not a
challenge in their setting. On the other hand, the subspace
alignment chains discussed above reach their maximum length when it
becomes impossible to align interference any further. Keeping the
desired signal separate from interference is not the main concern in
our setting.

\subsubsection{The $M$-bound: DoF $\leq \frac{\kappa}{2\kappa-1}M$}
Here we explain the second outer bound, DoF $\leq
\frac{\kappa}{2\kappa-1}M$, from linear dimension counting
arguments. Since linear dimension counting arguments are identical
for reciprocal networks \cite{Gomadam_Cadambe_Jafar}, without loss
of generality let us focus on the setting $M_T>M_R$, so that $M=M_R,
N=M_T$.  Now, since each transmitter has more antennas than any
receiver, zero-forcing of signals by the transmitters is  possible.
Ideally, we would like to zero force all interference. However,
since $M_T<2M_R$, it is not possible for any transmitter to
simultaneously zero-force its transmitted signal to both unintended
receivers. The next best thing is to zero-force interference to the
extent possible, and then align the remaining interference that
cannot be zero-forced. This observation gives rise to a slightly
different type of subspace alignment chains. The two ends of the
alignment chain correspond to transmitted signals that are zero
forced at one unintended receiver and cause interference at the
other unintended receiver. These two non-zero-forceable interference
terms are connected through a subspace alignment chain to alleviate
the impact of the non-zero-forceable interference. Since
zero-forcing is preferable to interference alignment, smaller
interference alignment chains are preferable. Therefore, in this
case, we will be limited by how quickly the semi-zero-forced signals
can be connected through a subspace alignment chain. Therefore, this
perspective gives rise to an outer bound $\frac{\kappa}{2\kappa-1}M$
which is a \emph{decreasing} function of the length of the subspace
alignment chain.

\begin{figure}[!ht]
\centering
\subfigure[$1/2\leq M/N<  2/3$, Chain length $\kappa=2$]{
\includegraphics[scale=0.45]{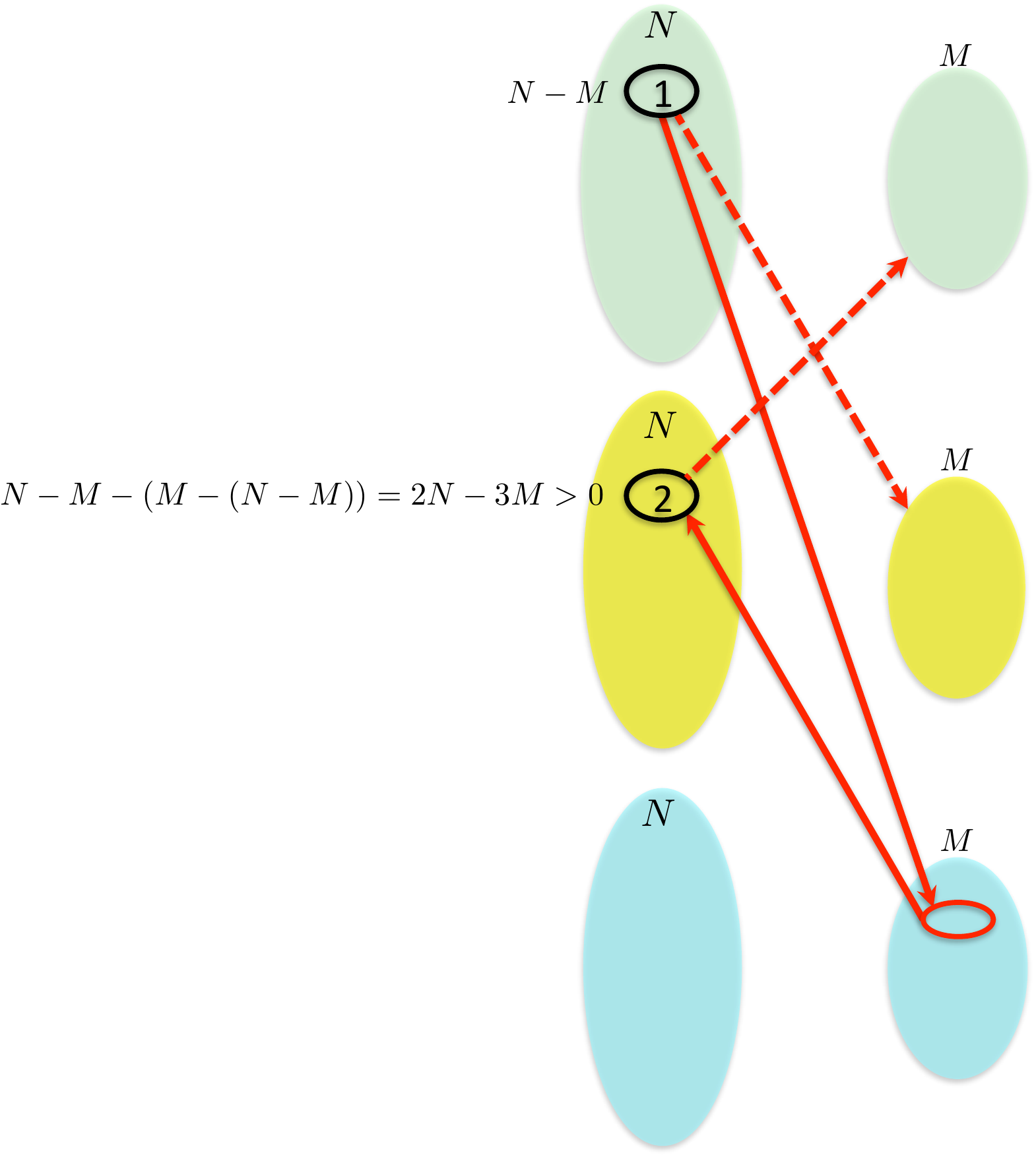}
\label{fig:Msac2}
}
\subfigure[$2/3\leq M/N< 3/4$, Chain length $\kappa=3$]{
\includegraphics[scale=0.45]{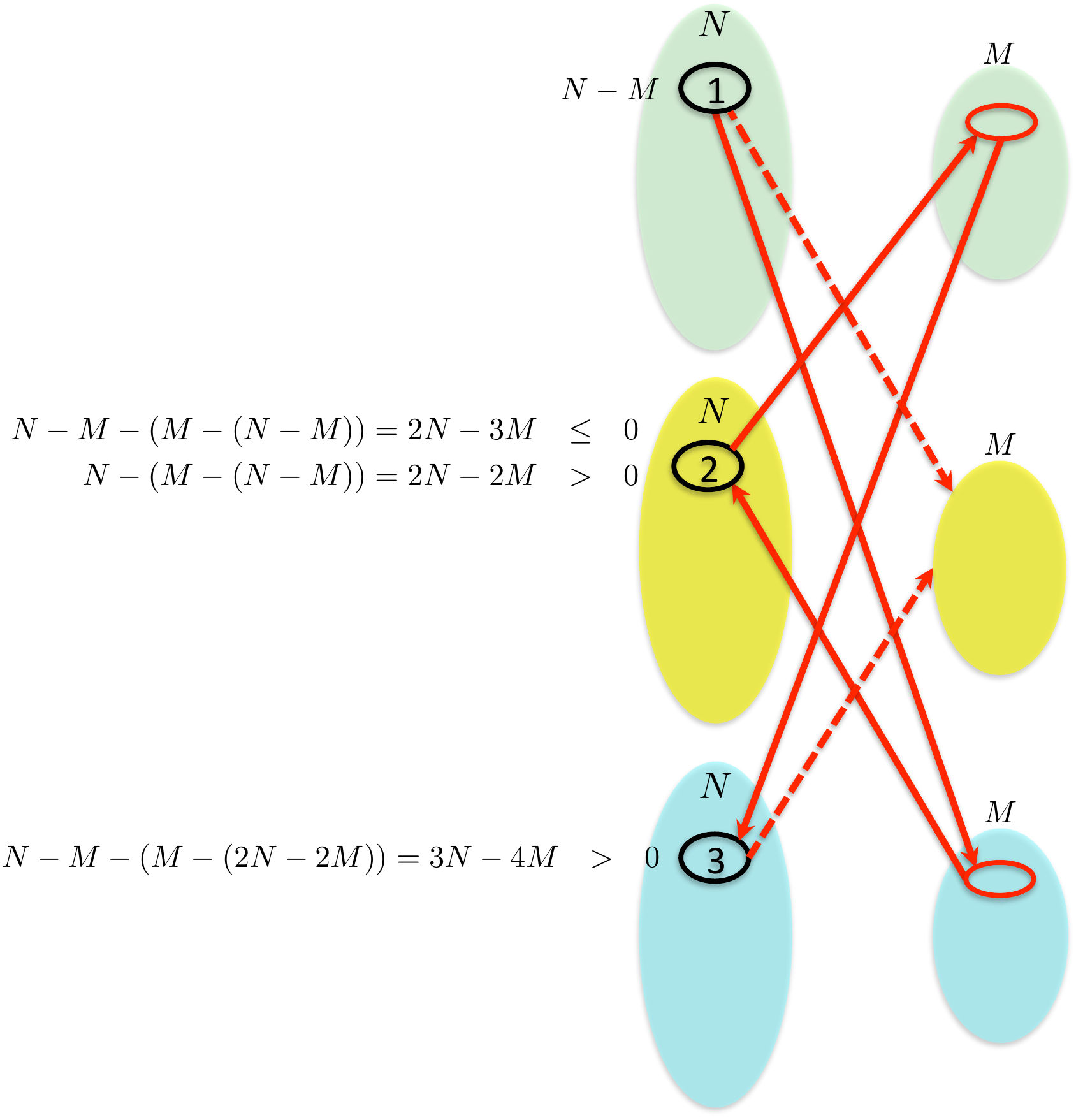}
\label{fig:Msac3}
}
\caption{Subspace alignment chains leading to the $M$-bound, DoF $\leq \frac{\kappa}{2\kappa-1}M$ per user. Dashed arrows indicate zero-forced interference. Numbered signal spaces shown in black ovals indicate the position of the signal space in the alignment chain. The directions of the solid arrows indicate the progression of the alignment chain.}\label{fig:chain_length2}
\end{figure}

Consider the setting $1/2\leq M/N<2/3$, illustrated in Fig.
\ref{fig:chain_length2}\subref{fig:Msac2}. We start the alignment
chain at Transmitter 1, where the number of transmitted dimensions
that can be zero-forced at Receiver 2 is no more than $N-M$. Since
$N-M\leq M$, these dimensions cannot be simultaneously zero-forced
at Receiver 3. This creates the non-zero-forceable interference at
Receiver 3 which initiates the subspace alignment chain. Recall that
we would like the alignment chain to be as short as possible. This
means that ideally, we would like the next signal in the chain
(marked as number 2 in the figure), which must originate at
Transmitter 2, to simultaneously accomplish the following
objectives:
\begin{enumerate}
\item Signal space 2 should align with pre-existing interference at Receiver 3.
\item Signal space 2 should be zero-forced at Receiver 1.
\end{enumerate}
Note that if it is possible to accomplish these objectives, signal
$2$ would not increase the interference space at any undesired
receiver, which is why this is the preferred goal in choosing signal
space 2. Let us see if this is possible. At Receiver 3, interference
already spans $N-M$ dimensions, leaving an interference free space
of $M-(N-M)$ dimensions. For signal space 2 to align with
pre-existing interference at Receiver 3, Transmitter 2 must
zero-force these $M-(N-M)$ interference-free dimensions for Receiver
3. In addition, if signal space 2 is to be zero-forced at Receiver
1, then another $M$ dimensions must be zero forced. With generic
signal spaces, the total number of dimensions to be zero forced is
$M+(M-(N-M))$. Since Transmitter 2 has only $N$ antennas, it leaves
$N-[M+(M-(N-M))]=2N-3M$ dimensions within which both objectives can
be accomplished. Since $M/N<2/3$, we  note that $2N-3M>0$, i.e.,
such a space exists and it is possible to terminate the alignment
chain. The resulting alignment chain length is $\kappa=2$. Now, let
us compute the implied DoF bound. With $d_0$ dimensions assigned to
each transmitted signal space, the total number of transmitted
dimensions is $2d_0$ and the total number of interference dimensions
is $d_0$. Adding up the desired signal dimensions and the
interference dimensions (because the two must not overlap) we need a
total of $3d_0$ dimensions at the three receivers. The total number
of dimensions at the three receivers is $3M$, which gives us the
outer bound $3d_0\leq 3M$, or $d_0\leq M$. Since the number of
transmitted dimensions \emph{per user} is $2d_0/3$ we have the outer
bound, DoF$\leq 2M/3$ per user, which coincides with
$\frac{\kappa}{2\kappa-1}M$, as expected.

Next, let us consider the setting $2/3\leq M/N<3/4$, illustrated in
Fig. \ref{fig:chain_length2}\subref{fig:Msac3}. Once again, we start
the alignment chain at Transmitter 1 and continue up to signal space
2. However, note that this time, because $2/3\leq M/N$, there does
not exist a signal space 2 which can be simultaneously aligned with
interference at Receiver 3 and be zero-forced at Receiver 1. Thus,
the alignment chain cannot be terminated at length 2. The next best
thing is to only achieve interference alignment with signal space 2
and extend the subspace alignment chain. Recall that in order to
align interference at Receiver 3, Transmitter 2 must zero force the
$M-(N-M)$ dimensional null-space of interference at Receiver 3,
leaving exactly $N-[M-(N-M)]=2N-2M$ dimensions at Transmitter 2 that
satisfy the interference alignment requirement. Since $2N-2M\leq M$,
note that the aligned dimensions cannot be zero forced at Receiver
1. Thus, the corresponding interference space at Receiver 1 is
$2N-2M$ dimensional.  Now, we extend the subspace alignment chain to
signal space 3, which originates at Transmitter 3  (see Fig.
\ref{fig:chain_length2}\subref{fig:Msac3}), and again we check if
the alignment chain can be terminated. For the alignment chain to
end with signal space 3, it must simultaneously satisfy the
following two objectives:
\begin{enumerate}
\item Signal space 3 should align with pre-existing interference at Receiver 1.
\item Signal space 3 should be zero-forced at Receiver 2.
\end{enumerate}
Let us see if this is possible. At Receiver 1, the space accessible
by interference produced so far occupies no more than  $2N-2M$
dimensions. Equivalently, an interference-free space of at least
$M-(2N-2M)$ dimensions must be maintained at Receiver 1. For signal
space 3 to align with pre-existing interference at Receiver 1,
Transmitter 3 must avoid the interference-free dimensions available
to Receiver 1, i.e., it needs to zero-force these $M-(2N-2M)$
interference-free dimensions. In addition, if signal space 3 is to
be zero-forced at Receiver 2, then another $M$ dimensions must be
zero forced. With generic signal spaces, the total number of
dimensions to be zero forced is $M+(M-(2N-2M))$. Since Transmitter 3
has only $N$ antennas, it leaves $N-[M+(M-(2N-2M))]=3N-4M$
dimensions within which both objectives can be accomplished. Since
$M/N<3/4$, $3N-4M>0$, i.e., such a space exists and it is possible
to terminate the alignment chain. The resulting alignment chain
length is $\kappa=3$. Now, let us compute the implied DoF bound.
With $d_0$ dimensions assigned to each transmitted signal space, the
total number of transmitted dimensions is $3d_0$ and the total
number of interference dimensions is $2d_0$. Adding up the desired
signal dimensions and the interference dimensions (because the two
must not overlap) we need a total of $5d_0$ dimensions at the three
receivers. The total number of dimensions at the three receivers is
$3M$, which gives us the outer bound $5d_0\leq 3M$, or $d_0\leq
3M/5$. Since the number of transmitted dimensions \emph{per user} is
$d_0$ we have the outer bound, DoF$\leq 3M/5$ per user, which
coincides with $\frac{\kappa}{2\kappa-1}M$, as expected.

Continuing along the same lines, these examples can also be
generalized in a straightforward manner to verify that for the
setting $\frac{p-1}{p}\leq\frac{M}{N}<\frac{p}{p+1}$, the length of
the subspace alignment chain $\kappa = p$, which can also be
expressed as $\kappa = \lceil\frac{M}{N-M}\rceil$. The corresponding
outer bound value is calculated as follows. A total of $\kappa d_0$
dimensions are transmitted, creating interference spaces of  total
dimension $(\kappa-1)d_0$. The desired signal and interference
together need a total of $\kappa d_0 + (\kappa+1)d_0=(2\kappa-1)d_0$
dimensions at the receivers. Since only $3M$ dimensions are
available at the receivers, we must have $d_0\leq
\frac{3M}{2\kappa-1}$. Now, since the number of dimensions
transmitted per user is $\kappa d_0/3$, we have the outer bound, DoF
$\leq \frac{\kappa}{2\kappa-1}M$.

We conclude this discussion with two  plots of the DoF
characterization presented in Theorem \ref{theorem:dof}, shown in
Fig. \ref{fig:result} and Fig. \ref{fig:resultM}. Both figures are
plotted as a function of  the ratio $\gamma\define M/N$. Clearly, as
$M, N$ become less disparate, $\gamma$ increases, and the length of
the subspace alignment chain, $\kappa$ increases as well,
approaching infinity as $\gamma\rightarrow 1$. While the two figures
are equivalent, the normalization with respect to $N$ in Fig.
\ref{fig:result} and the normalization with respect to $M$ in Fig.
\ref{fig:resultM} highlight the role of  interference-alignment and
zero-forcing, respectively. As $\gamma$ increases from 0 to 1,
subspace alignment chains become longer, a desirable outcome for
interference alignment, which is reflected in Fig. \ref{fig:result}.
On the other hand, as $\gamma$ increases from 0 to 1, alignment
chains become longer, an undesirable outcome for zero forcing, which
is reflected in Fig. \ref{fig:resultM}. The interplay between the
two bounds is evident in the piecewise analytical nature of the DoF
function, with either $M$ or $N$ being the bottleneck within each
analytical segment. A number of  interesting observations can be
made from the DoF result. These observations are the topic of the
remainder of this section.

\begin{figure}[!h]
\centering
\includegraphics[width=5.0in]{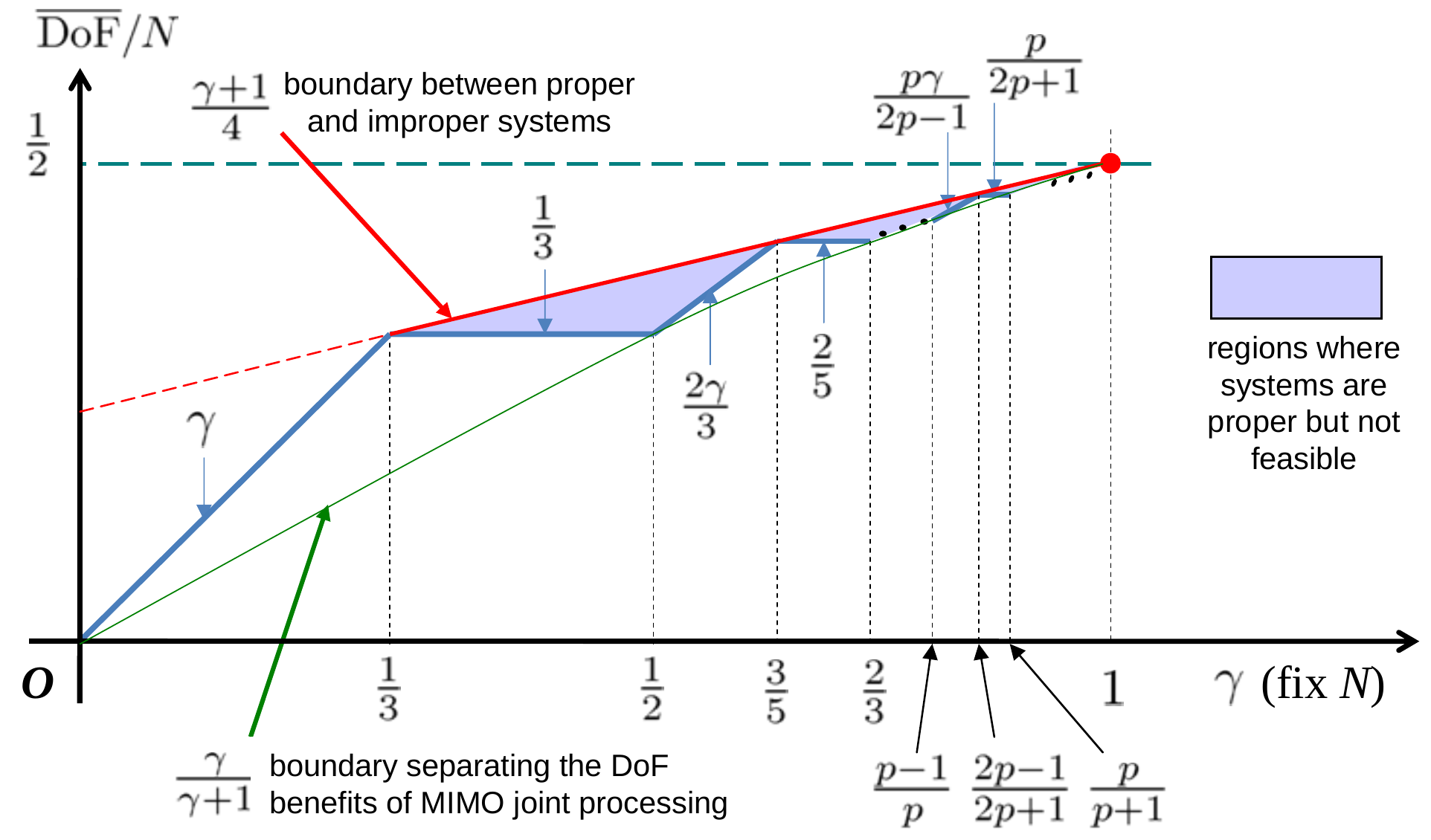}\vspace{-0.15in}
\caption{$\sDoF/N$ as a function of $\gamma=\frac{M}{N}$}
\label{fig:result}
\end{figure}

\begin{figure}[!h]
\centering
\includegraphics[width=5.0in]{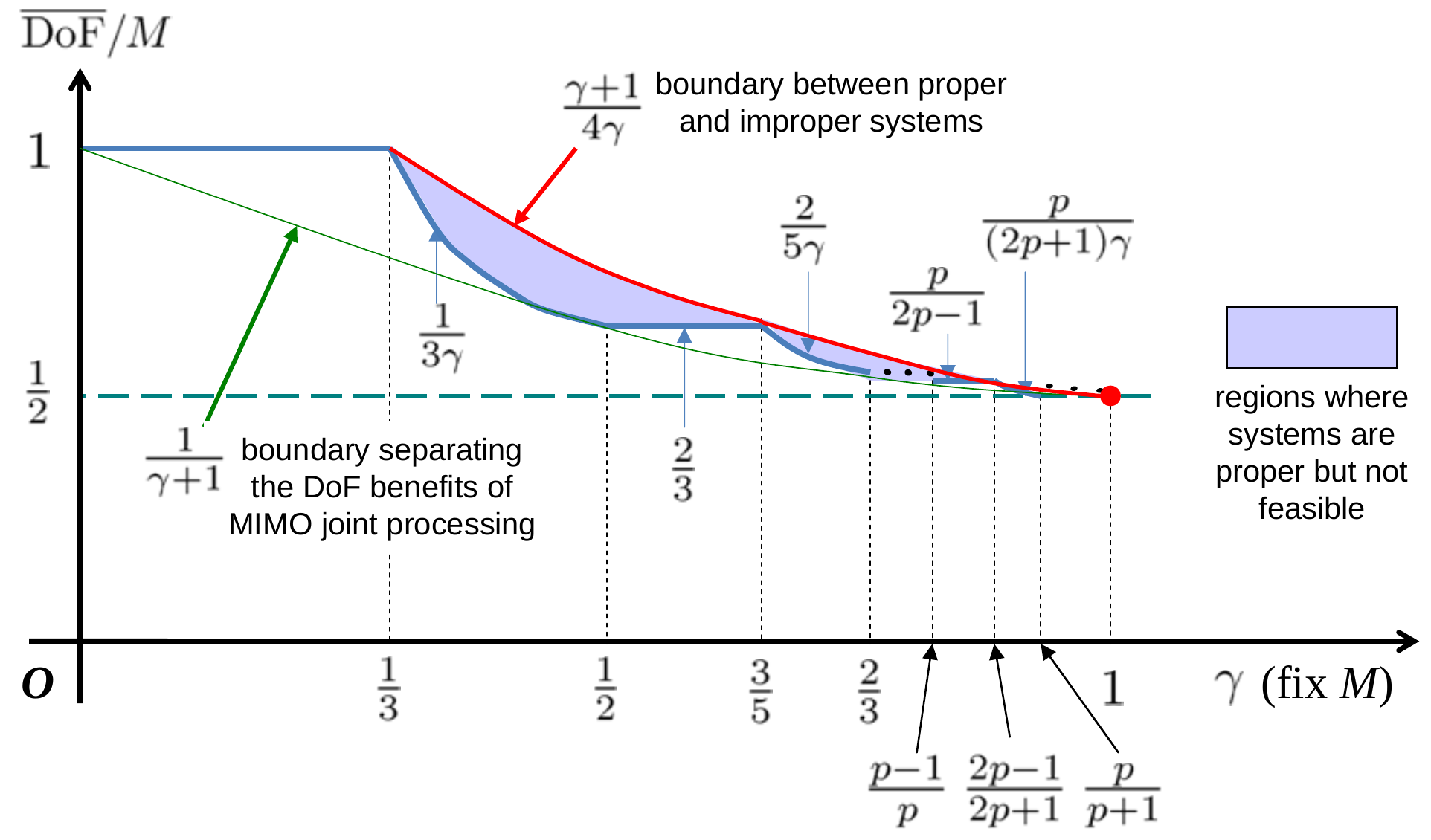}\vspace{-0.15in}
\caption{$\sDoF/M$ as a function of $\gamma=\frac{M}{N}$}
\label{fig:resultM}
\end{figure}

\subsection{Other Key Observations}
For these observations, we will refer to the $\sDoF$
characterization in Theorem \ref{theorem:dof} and its depiction as
the piecewise linear curve in Fig. \ref{fig:result}.

\subsubsection{Redundant Dimensions}\label{subsec:redundantdim}

A particularly interesting observation from (\ref{eqn:compactdof}) and Fig. \ref{fig:result}
is that within each piecewise linear interval, the  $\sDoF$ value
depends only on either $M$ or $N$. This makes the other parameter
somewhat redundant. In other words, within each piecewise linear
section of the $\sDoF$ curve, either the number of dimensions at the
transmitter or receiver can be increased/decreased without changing
the $\sDoF$  value. For example, consider the $3$ user $7\times 10$
MIMO interference channel. This channel lies in the range $2/3\leq
M/N\leq 5/7$, where the  $\sDoF$ value $3M/5$ depends only on $M$.
Therefore, there is some redundancy in $N$. Since the redundant
dimensions are fractional, they are much more explicitly seen in a
larger space. So let us consider the $3$ user $35\times 50$ MIMO
interference channel, which is simply a spatially scaled version of
the original $7\times 10$ setting. Now, we know that the $35\times
50$ setting has exactly $21$ DoF (no spatial scaling required, since
this is an integer value). However, note that the $35\times 49$
setting also has only $21$ DoF.  Incidentally, the $31\times 49$
setting  achieves the $21$ DoF with only linear beamforming based
interference alignment, i.e., without the need for symbol extensions
in time/frequency/space. Therefore, clearly, the $50^{th}$ receive
antenna is redundant from a DoF perspective.

The previous example illustrates the situation for $M/N$ values that
fall strictly inside the piece-wise linear intervals. Now let us
consider $M/N$ values that fall at the boundary points of the
piecewise linear intervals. For example, consider the  $3$ user
$10\times 15$ MIMO interference channel, which corresponds to the
$M/N$ value $2/3$, i.e., a corner point, and has DoF value $6$.
However, note that the $3$ user $9\times 15$ MIMO interference
channel also has $6$ DoF. Thus, the $10^{th}$ transmit antenna is
redundant from a DoF perspective. Alternatively, consider the $3$
user $10\times 14$ MIMO interference channel, which also has $6$
DoF. Thus, evidently, the $15^{th}$ receive antenna is redundant in
the $10\times 15$ setting. Thus, one can lose either a transmit
antenna or a receive antenna (but not both) without losing DoF in
the $3$ user $10\times 15$ MIMO interference channel. This is
because the $10\times 15$ setting, which corresponds to $M/N=2/3$,
sits at the boundary of two piece-wise linear segments where the
redundancies in $M$ and the redundancies in $N$ meet. Therefore it
contains both redundancies. The same observation is true for corner
points $M/N = 2/3, 3/4, 4/5, \cdots$.

Now consider the other set of corner values $M/N=1/3, 3/5, \cdots$.
As it turns out, these are the only values where neither the
transmitter, nor the receiver has any redundant dimensions.  We
summarize the observations below:
\begin{enumerate}
\item For $M/N\in (0, 1/3), (1/2, 3/5), (2/3, 5/7), \cdots$, the value of $M$ is the bottleneck, but the value of $N$ includes redundant dimensions that can be sacrificed without losing DoF.
\item For $M/N\in (1/3, 1/2), (3/5, 2/3), (5/7, 3/4), \cdots$, the value of $N$ is the bottleneck, but the value of $M$ includes redundant dimensions that can be sacrificed without losing DoF.
\item For $M/N\in \mathcal{A}\define\{1/2, 2/3, 3/4, 4/5, \cdots$\}, and only for these values, both $M$ and $N$ include redundant dimensions, either of which can be sacrificed without losing DoF.
\item For $M/N\in\mathcal{B}\define\{1/3, 3/5, 5/7, 7/9, \cdots$\}, and only for these values,  neither $M$ nor $N$ contains any redundant dimensions, i.e., reducing either will lead to loss of DoF.
\end{enumerate}
Thus, the sets $\mathcal{A}$ and $\mathcal{B}$  represent maximally and minimally redundant settings. The redundancy, and the lack thereof, has interesting implications. For instance, the set $\mathcal{A}$ corresponds to maximal redundancy and is also the precise set of $M/N$ values for which no joint processing is needed among the co-located antennas at any transmitter or receiver. On the other hand the set $\mathcal{B}$ corresponds to no-redundancy, and is also the precise set of $M/N$ values for which all proper systems are feasible from a linear interference alignment perspective. Next we elaborate upon these observations.

\subsubsection{The DoF Benefit of MIMO Processing}
In the $K$-user MIMO interference channel setting with $M=N$, i.e.,
equal number of antennas at every node, Cadambe and Jafar have shown
\cite{Cadambe_Jafar_int} that there is no DoF benefit of joint
processing among multiple antennas, because the network has $KM/2$
DoF even if each user is split into $M$ users, each with a single
transmit and single receive antenna and with only independent
messages originating at each transmitter. For the $K$ user
$M_T\times M_R$ MIMO  interference channel setting, Ghasemi et al.
have shown  in \cite{Ghasemi_Motahari_Khandani_MIMO} that
$\frac{M_TM_R}{M_T+M_R}=\frac{MN}{M+N}$ DoF are achievable even if
transmitter is split into $M_T$ single-antenna transmitters, each
receiver is split into $M_R$ single antenna receivers, and there are
no common messages. Note that while in general this achievable DoF
value is not optimal, Ghasemi et al. have shown that
$\frac{M_TM_R}{M_T+M_R}$ is the optimal DoF value per user for the
$K$ user $M_T\times M_R$ MIMO interference channel when the number
of users $K\geq\frac{M_T+M_R}{\mbox{gcd}(M_T,M_R)}$. Here we make
related observations for our setting.

\begin{enumerate}
\item The 3 user $M_T\times M_R$ MIMO interference channel has DoF$(M_T,
M_R) =  \frac{M_TM_R}{M_T+M_R}$ whenever $M/N\in\mathcal{A}$, i.e.,
$\frac{M}{N}=\frac{p}{p+1}$.  Note that this is a statement about
regular DoF, i.e., without requiring spatial extensions. This is
because the outer bound from Lemma \ref{lemma:out} matches the
achievability result of Ghasemi et al. in
\cite{Ghasemi_Motahari_Khandani_MIMO} for these settings.

\item The $K>2$ user $M_T\times M_R$ MIMO interference channel has
DoF$(M_T, M_R) =  \frac{M_TM_R}{M_T+M_R}$ whenever
$\frac{M}{N}=\frac{p}{p+1}$, for some $p\in\mathbb{Z}^+$. Note that
this statement is for \emph{any number of users} (greater than 2), and not just for 3 users.
Thus, we have a DoF characterization for any number of users
whenever $\frac{M}{N}=\frac{p}{p+1}$. The outer bound holds for any
number of users because the DoF per user cannot increase with the
number of users. The achievability is already established for any
number of users. Since the two match, we know the DoF for any number
of users. The result significantly strengthens the previous DoF
characterization by Ghasemi et al. in
\cite{Ghasemi_Motahari_Khandani_MIMO}. As an example, consider the
$(M,N)=(2,3)$ setting. Ghasemi et al. show that the this setting has
$6/5$ DoF per user if the number of users is  $5$ or higher.
However, our result shows that this setting has $6/5$ DoF if the
number of users is $3$ or higher. The improvement is even more stark
for larger values of $M,N$. For example, Ghasemi et al. show that
the setting $(M,N)=(9,10)$ has exactly $90/19$ DoF per user if the
number of users is 19 or higher, whereas our result shows that this
setting has  exactly $90/19$ DoF if the number of users is $3$ or
higher.

\item There are no DoF benefits of MIMO processing whenever
$\frac{M}{N}=\frac{p}{p+1}$, for $p\in\mathbb{Z}^+$. This can be
seen as follows.  For our result illustrated in Figure
\ref{fig:result}, the piecewise linear $\sDoF/N$ curve is bounded
below by the smooth curve shown with a light solid line plotting the
value $\frac{\gamma}{\gamma+1}$. Since $\gamma=M/N$, note that
$N\frac{\gamma}{\gamma+1} = \frac{MN}{M+N}$ corresponds to
$MN/(M+N)$ DoF, i.e., the  achievable DoF without any joint
processing across multiple antennas, or the  DoF achievable with
independent coding/decoding at each transmit/receive antenna. Since
the $\sDoF/N$ curve touches the $\frac{\gamma}{\gamma+1}$ curve
whenever $\gamma=\frac{p}{p+1}$,  there is no DoF benefit of MIMO
processing in these settings.

\item Conversely, MIMO processing has DoF benefits whenever
$\frac{M}{N}\neq\frac{p}{p+1}$, for some $p\in\mathbb{Z}^+$. This is
because for all these $(M,N)$ values, the $\sDoF$ curve is strictly
above the $MN/(M+N)$ value.

\item MIMO processing can enable linear achievable schemes when otherwise
asymptotic alignment would be needed. We illustrate this with an
example. Consider the 3 user $10\times 15$ interference channel
(where $M/N=2/3$),  which we now know, has $6$ DoF per user. We also
know from the work of Ghasemi et al. that this DoF value can be
achieved without any MIMO processing, but it requires the [CJ08]
asymptotic interference alignment scheme \cite{Jafar_FnT} either in
the rational dimensions framework or over time-varying channels. On
the other hand, as we will show in this work, the $6$ DoF per user
can be achieved purely through linear interference alignment based
on beamforming without requiring any symbol extensions in space,
time or frequency, by exploiting the MIMO benefit of joint
processing among  antennas located on the same node.
\end{enumerate}

\subsection{Infeasibility of Proper Systems}

Our DoF results settle the issue of  feasibility/infeasibility of
linear interference alignment for the $3$ user $M_T\times M_R$ MIMO
interference channel. Prior work on this topic is summarized in
Section \ref{sec:background}. As stated earlier, it is known that
improper systems are feasible, and also under certain conditions
proper systems are feasible. The feasibility of proper systems is
believed to be much more widely true. Somewhat surprisingly, our
results show that for the $3$ user MIMO interference channel, the
relationship between proper systems and feasibility of linear
interference alignment is very weak, in the sense that \emph{most}
proper systems are infeasible.

Specializing the characterization of proper systems by Cenk et al.
\cite{Cenk_Gou_Jafar_feasibility}  to the $3$ user $M_T\times M_R$
MIMO interference channel, the system is proper if and only if the
desired DoF per user
\begin{eqnarray}
d \leq \frac{M_T+M_R}{4} \label{eqn:strictproper}
\end{eqnarray}
We will say that a system is strictly proper if we have equality in
(\ref{eqn:strictproper}). In Figure \ref{fig:result}, the $\sDoF/N$
curve is bounded above by a solid red line that plots the value
$\frac{\gamma + 1}{4}$. Note that this line only starts from
$\gamma=1/3$ because for $\gamma\in(0,1/3)$ (denoted as the dashed
red line) the DoF per user is bounded above by the single user
bound. This solid red line differentiates proper systems from
improper systems. All systems above the curve are improper while
those below the curve are proper. However, the DoF outer bound lies
strictly below the curve except when $\gamma =
\frac{M}{N}=\frac{2p-1}{2p+1}$ for $p\in\mathbb{Z}^+$. Therefore,
\emph{many proper systems and most strictly proper systems are
infeasible.} Remarkably, these systems are not only infeasible in
terms of the restricted goal of achieving the desired DoF values
through interference alignment with linear beamforming and without
relying on channel extensions in time/frequency, but also they are
infeasible in terms of the much more relaxed goal of achieving the
desired DoF through any possible achievable scheme, linear or
non-linear,  using interference-alignment or otherwise, using
time-varying or constant channels, and using scalar or vector
coding. This is because the solid blue curve is the
information-theoretic DoF outer bound.

Let us make this observation explicit with a few examples. Consider
the $(M,N)=(8,12)$ setting (where $\gamma = M/N=2/3$), with desired
DoF value $d=5$ per user. This is a strictly proper system because
$d=\frac{M+N}{4}$. But from Lemma \ref{lemma:out} we know that the
information theoretic DoF per user  for this channel is bounded
above by $24/5$. Therefore $d=5$ is clearly infeasible. Similarly,
for any value of $\gamma\notin\mathcal{B}$, i.e., that is not of the form
$\frac{2p-1}{2p+1}$, one can create an infeasible proper system. For
example, suppose $\gamma = 0.61 $.  Equivalently,
$\frac{M}{N}=61/100$. Choosing $(M,N) = (244, 400)$, we arrive at
the desired DoF value $d= \frac{644}{4}=161$ per user which would
make the system strictly proper. However, we know from Lemma
\ref{lemma:out} that the information theoretic DoF (per user) outer
bound for $(M,N)=(244,400)$ is $800/5 = 160$. Thus, once again the
strictly proper system is infeasible. As yet another example consider
the choice $\gamma=0.74$, i.e., $\frac{M}{N}=\frac{74}{100}$. Choose
the setting $(M,N) = (148, 200)$ for which the information-theoretic
DoF  per user is bounded above by $\frac{600}{7} = 85.7...$.
Therefore, if the desired DoF value per user is $87$, the system is
strictly proper and infeasible, if the desired DoF value per user is
86, the system is proper and infeasible. Proceeding similarly, we
arrive at the following conclusions.

\begin{enumerate}
\item For every value of $\gamma$ except $\gamma\in\mathcal{B}$,
we can find proper systems that are infeasible, not only in terms of
linear interference alignment, but also information theoretically
infeasible.

\item For the values $\gamma\in\mathcal{B}$, proper systems are
always feasible, i.e., the DoF demand $d=(M+N)/4$ per user, which is
also the information theoretic outer bound, is actually achievable
with only linear interference alignment, without the need for symbol
extensions in time/frequency/space.
\end{enumerate}
These observations, especially the infeasibility of proper systems
for $M/N = 1/3, 3/5, 5/7, \cdots$, can be understood in terms of the
redundant dimensions explained in the previous section. Recall that
Cenk et al. \cite{Cenk_Gou_Jafar_feasibility} make the distinction
between proper/improper systems based on the number of variables
involved in the system of polynomial equations. From the
observations on redundant dimensions, we know that except for $M/N =
1/3, 3/5, 5/7, \cdots$, every other setting contains redundant
dimensions either in $M$ or $N$. Evidently these redundant
dimensions contribute superfluous variables which inflate the
variable count, thereby qualifying a system as proper even when it
is not feasible. We suspect that this observation may have  significant
implications in algebraic geometry where the solvability of systems
of polynomial equations remains an unsolved problem.

\subsection{Degrees of Freedom without Spatial Normalization}

From our results it is clear that for all $(M,N)$ settings where
DoF$^\star$ takes an integer value, we have a precise
characterization of DoF (without the need for spatial
normalizations). Precise DoF characterizations are also available
for $(M,N)$ settings where $\gamma=\frac{M}{N}=\frac{p}{p+1}$ for
some $p\in\mathbb{Z}^+$, because in all these settings the outer
bound of coincides with the achievability result of Ghasemi et al.
\cite{Ghasemi_Motahari_Khandani_MIMO}, and the DoF value is
$\frac{MN}{M+N}$. For the remaining cases, we propose a linear
beamforming construction for achieving the DoF outer bound without
relying on spatial extensions. Because the DoF outer bound is a
fractional value, symbol extensions over time/frequency are needed
to make the DoF value a whole number over the extended channel.
Interestingly, the construction is always non-asymptotic, as in, the
number of symbol extensions needed is only enough to make the DoF
value an integer.  We show analytically that while in many cases
symbol extensions in time over constant channels are sufficient to
achieve the information theoretic DoF outer bound, there are also
cases where time-variations/frequency-selectivity of the channel is
needed to achieve the DoF outer bound with linear interference
alignment schemes. The feasibility of interference alignment can be
settled in every case through a simple numerical test. We carry out
this test to establish the DoF values for all $(M,N)$ values upto
$M, N \leq 10$. In general, we end with the conjecture that in all
cases, the DoF outer bound value is tight, even with constant
channels, and may be achieved with non-linear interference alignment
schemes, e.g., exploiting the rational dimensions framework or the
Renyi information dimensions framework.

\section{DoF Outer Bounds: Preliminaries}\label{sec:outerbound}

Recall from the discussion on redundant dimensions,  that the
settings corresponding to $M/N=1/2, 2/3, 3/4, 4/5, \cdots,$ are the
ones that contain the most redundant dimensions, i.e., for these
settings, and only for these settings, it is possible to reduce
\emph{either} $M$ or $N$ without losing DoF. It follows then, that
in order to prove the strongest DoF outer bounds, it is these
settings that must be considered. Indeed, as we will see in the
information-theoretic derivations of the DoF outer bounds,
essentially the outer bounds must be shown for cases corresponding
to $M/N=1/2, 2/3, 3/4, \cdots,$ and then with only a little
additional effort, these outer bounds can be extended to all $M, N$
values.

A key step for the information-theoretic DoF outer bound proof is to
first perform a change of basis operation, corresponding to
invertible linear transformations at both the transmitters and
receivers. While invertible linear transformations at the
transmitters and/or receivers do not affect the DoF, the change of
basis identifies the subspace alignments which in turn helps
identify the side information to be provided by a genie for the DoF
outer bound. Next, we will present the invertible linear
transformations for the case of $(M_T,M_R)=(p,p+1)$. Since the
change of basis operations are linear transformations, they are also
directly applicable to the reciprocal channel according to the dual
nature of reciprocal channels \cite{Gomadam_Cadambe_Jafar}.

\subsection{Change of Basis for $(M_T, M_R)=(2,3)$}
We begin with the simplest case, i.e., the $2\times 3$ MIMO
interference channel. The linear transformations are carried out by
multiplying an invertible square matrix at each transmitter and each
receiver. The procedure of designing these transformation matrices
is illustrated as follows. For brevity, we use $\mathbf{T}_k$ to
denote the $2\times 2$ invertible square matrix at Transmitter $k$,
and $\mathbf{R}_k$ to denote the $3\times 3$ transformation matrix
at Receiver $k$.
\begin{figure}[!t]
\centering
\includegraphics[width=3.5in]{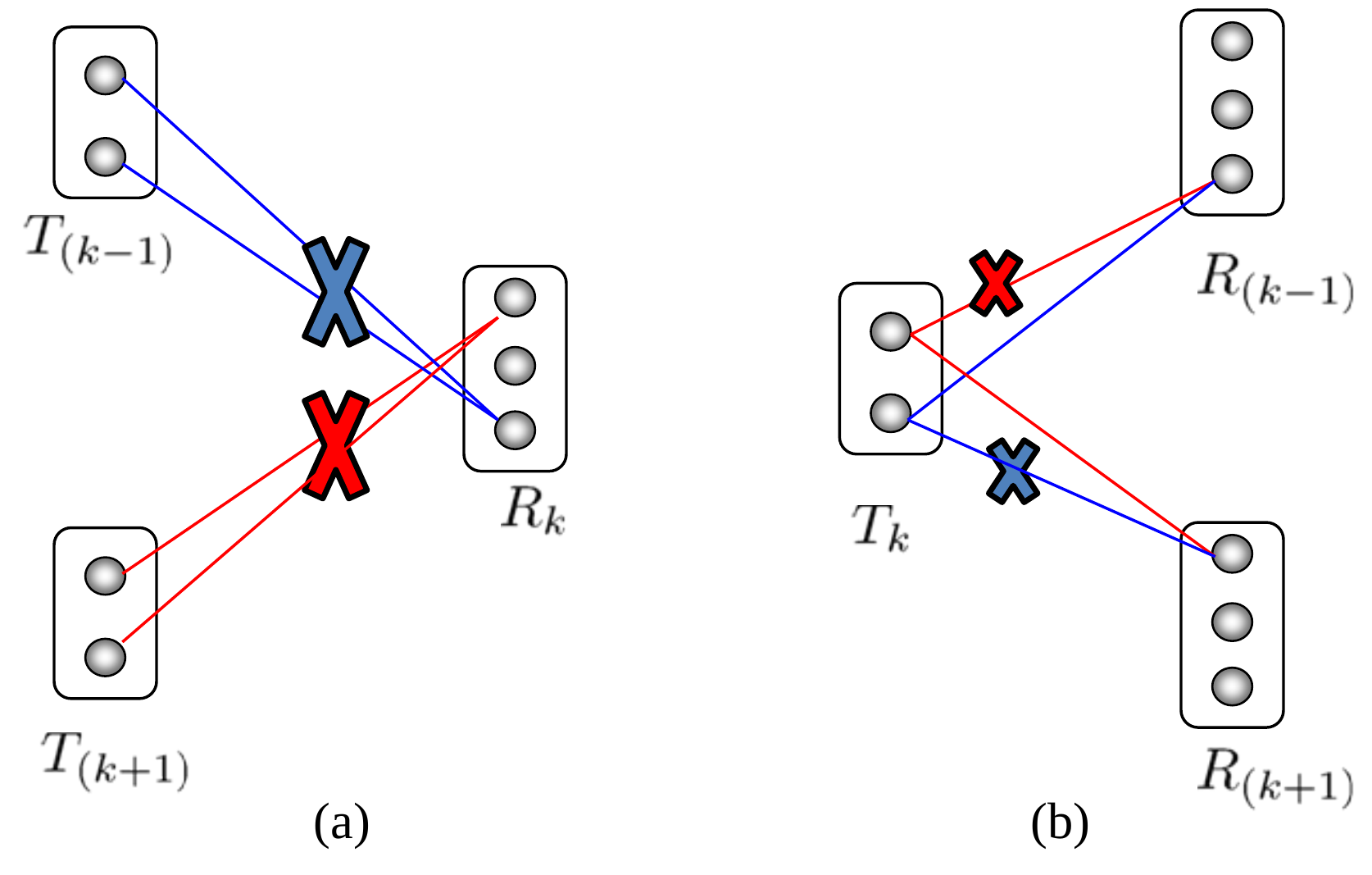}
\caption{Invertible Linear Transformations for the $2\times 3$ Case,
(a) Step 1, (b) Step 2} \label{fig:LT2by3}
\end{figure}

\emph{Step 1:} Consider the $3\times 3$ square matrix $\mathbf{R}_k$
which has three rows. First, we determine its first row and the last
row, corresponding to the first and the third antennas at Receiver
$k$. The linear transformation is designed in such a manner that the
first antenna of Receiver $k$ does not hear Transmitter $k+1$ and
the third antenna of Receiver $k$ does not hear Transmitter $k-1$.
This is illustrated in Fig.\ref{fig:LT2by3}(a). This operation is
guaranteed by the fact that $\mathbf{H}_{k(k+1)}$ and
$\mathbf{H}_{k(k-1)}$ are both $3\times 2$ matrices, and the left
null space of each of them has one dimension. Therefore, the first
row of $\mathbf{R}_k$ lies in the left null space of
$\mathbf{H}_{k(k+1)}$, and the last row lies in the left null space
of $\mathbf{H}_{k(k-1)}$. That is,
\begin{eqnarray}
\mathbf{R}_{k}(1,:)\mathbf{H}_{k(k+1)}&=&{\bf 0}\\
\mathbf{R}_{k}(3,:)\mathbf{H}_{k(k-1)}&=&{\bf 0}.
\end{eqnarray}
If we choose the first entry of $\mathbf{R}_{k}(1,:)$ and the last
entry of $\mathbf{R}_{k}(3,:)$ to be one, then their remaining
entries can be solved through following equations.
\begin{eqnarray}
\mathbf{R}_{k}(1,2:3)&=&-\left(\mathbf{H}_{k(k+1)}(2:3,:)\right)^{-1}\mathbf{H}_{k(k+1)}(1,:)\label{eqn:lineartran1}\\
\mathbf{R}_{k}(3,1:2)&=&-\left(\mathbf{H}_{k(k-1)}(1:2,:)\right)^{-1}\mathbf{H}_{k(k-1)}(3,:).\label{eqn:lineartran2}
\end{eqnarray}
%\begin{figure}[!t]
%\centering
%\includegraphics[width=1.5in]{LT2by3s2}
%\caption{Linear transformation for $2\times 3$ -- Step 2}
%\label{fig:LT2by3s2}
%\end{figure}

\emph{Step 2:} After the first and the third rows of $\mathbf{R}_k$
are specified, we switch to the transmitter side and determine the
two columns of $\mathbf{T}_{k}$, which correspond to the two
antennas at Transmitter $k$. Our goal is to ensure that the first
antenna of Transmitter $k$ is not heard by the last antenna of
Receiver $k-1$ but is heard by first antenna of Receiver $k+1$,
while the last antenna of Transmitter $k$ is not heard by first
antenna of Receiver $k+1$ but is heard by the last antenna of
Receiver $k-1$. This is illustrated in Fig. \ref{fig:LT2by3}(b). As
a result, the first column of $\mathbf{T}_k$ is chosen in the null
space of the channel associated with the last antenna of Receiver
$k-1$ (because the channels are generic, this is \emph{not} in the
null space of the first antenna of Receiver $k+1$). Similarly, the
last column of $\mathbf{T}_k$ is chosen in the null space of the
channel associated with the first antenna of Receiver $k+1$.
Mathematically,
\begin{eqnarray}
\mathbf{R}_{k-1}(3,:)\mathbf{H}_{(k-1)k}\mathbf{T}_{k}(:,1)&=&0\\
\mathbf{R}_{k+1}(1,:)\mathbf{H}_{(k+1)k}\mathbf{T}_{k}(:,2)&=&0.
\end{eqnarray}
Since $\mathbf{R}_{k-1}(3,:)\mathbf{H}_{(k-1)k}$ and
$\mathbf{R}_{k+1}(1,:)\mathbf{H}_{(k+1)k}$ are $1\times 2$ vectors,
the null space of each of them has one dimension. Because the
channel coefficients are generic, they are linearly independent
almost surely. Therefore, if we choose the first entry of
$\mathbf{T}_{k}(:,1)$ and last entry of $\mathbf{T}_{k}(:,2)$ to be
one, then their remaining entries can be solved from following
equations:
\begin{eqnarray}
\mathbf{T}_{k}(2,1)&=&-\mathbf{R}_{k-1}(:,1)\mathbf{H}_{(k-1)k}/\mathbf{R}_{k-1}(:,2)\mathbf{H}_{(k-1)k}\label{eqn:lineartran3}\\
\mathbf{T}_{k}(1,2)&=&-\mathbf{R}_{k+1}(:,2)\mathbf{H}_{(k+1)k}/\mathbf{R}_{k+1}(:,1)\mathbf{H}_{(k+1)k}.\label{eqn:lineartran4}
\end{eqnarray}
After this operation, the matrix ${\bf T}_k$ is uniquely determined.
%Since there are only two antennas at the transmitter, we have
%completely determined the linear transformation corresponding to the
%change of basis at the transmitter.

\emph{Step 3:} Now let us switch back to the receiver side again, to
determine the remaining row (the second row) of the ${\bf R}_k$ at
Receiver $k$. %In this case, there is only one row left for each
%matrix at the receiver side.
In order for the transformation to be invertible, we need to choose
this row such that it is linearly independent with the other rows.
For simplicity, we set the first and last entries to be zero and the
second to be one\footnote{Note that the manner we choose the second
row of ${\bf R}_k$ is not unique. In Section \ref{sec:innerbound} we
will consider another way to design this row for a simpler patten of
channel connectivity of the resulting network.}. Therefore, the
linear transformation matrices at both transmitter and receiver
sides have been determined. Note that the invertibility of these
transformations is guaranteed by the generic nature of channel
coefficients.

The resulting network connectivity after the change of basis
operations is shown in Fig. \ref{fig:2_by_3_mimo}.
\begin{figure}[!h] \vspace{-0.1in}\centering
\includegraphics[width=3.0in]{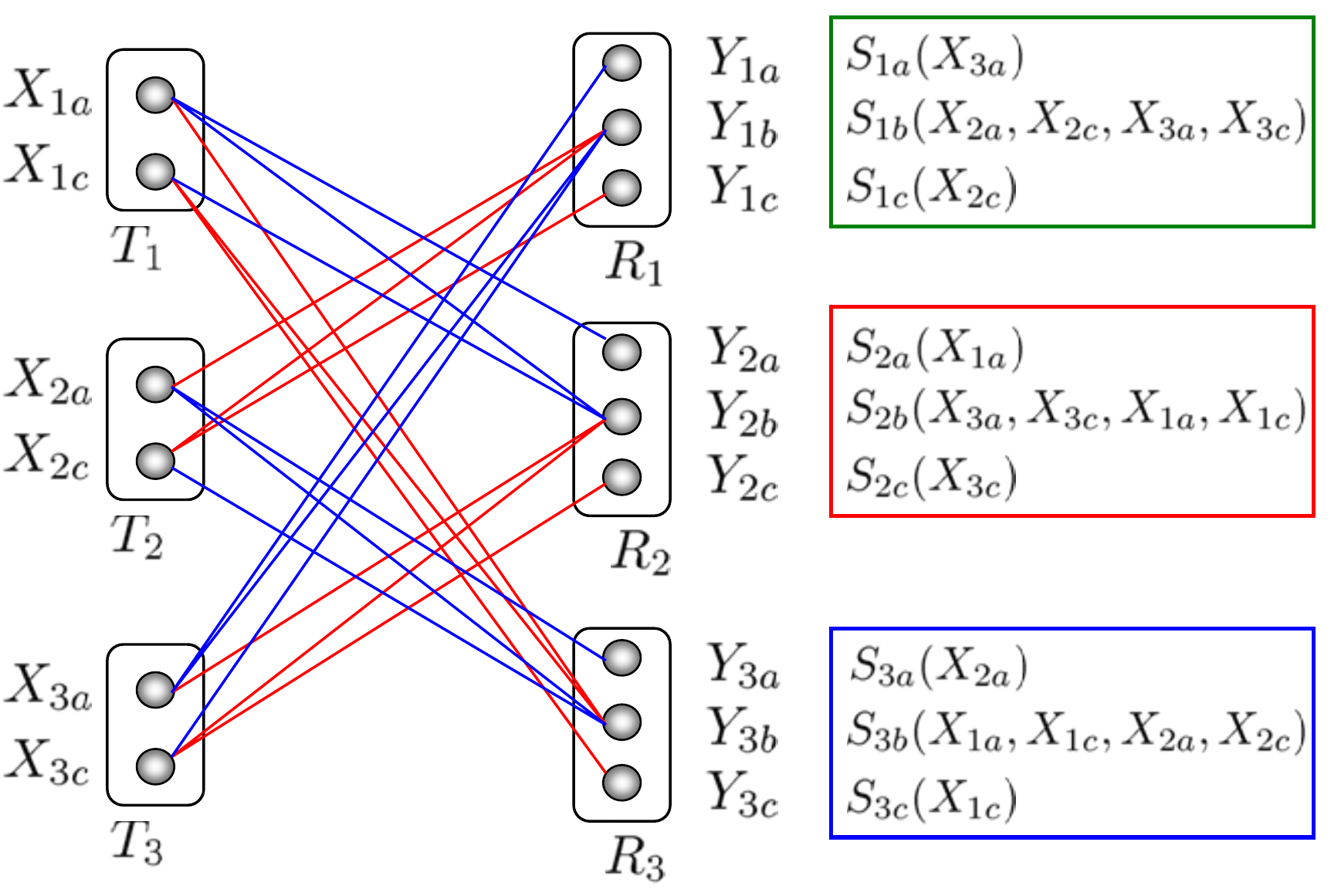}\vspace{-0.05in}
\caption{Three-User $2\times 3$ MIMO Interference Channel after the
Change of Basis Operation} \vspace{-0.1in} \label{fig:2_by_3_mimo}
\end{figure}

Since the channel coefficients of the original network are generic,
the coefficients of the connected links in Fig.
\ref{fig:2_by_3_mimo} are non-zero almost surely. For brevity, we
use ``$a,c$" at the transmitters and ``$a,b,c$" at the receivers to
denote the corresponding antennas.

Next, we show that these linear transformation matrices have full
rank almost surely and hereby invertible. To do that, we need to
prove the determinant of each linear transformation matrix is
non-zero almost surely. Notice that from \eqref{eqn:lineartran1},
\eqref{eqn:lineartran2}, \eqref{eqn:lineartran3} and
\eqref{eqn:lineartran4}, except some entries are chosen as specific
values, all the remaining entries of the transformation matrices are
in the form of the ratio of two polynomials of the channel
coefficients. Therefore, the determinant of the linear
transformation matrix can be also expressed as a ratio of two
polynomials of channel coefficients. Further, multiplication with
the denominator changes it to a polynomial. To prove that the
determinant is non-zero almost surely, it suffices to show that the
polynomial is not the zero polynomial. This can be easily verified
through constructing specific channel coefficients for this
resulting channel. Such a specific construction is as follows.
\begin{eqnarray}
\mathbf{H}_{(k-1)k}=\left[\begin{array}{cc}0&0\\1&0\\0&1\end{array}\right]~~~~~~~\mathbf{H}_{(k+1)k}=\left[\begin{array}{cc}1&0\\0&1\\0&0\end{array}\right]
\end{eqnarray}
It is easily seen that the linear transformation for such channel is
the identity matrix whose determinant is non-zero. Therefore, the
polynomial is not zero polynomial and thus it is non-zero almost
surely across all realizations of the channels.

%Since the linear transformation does not affect the DoF region of the network, each receiver can decode its own message in reliable communications.

In the new network obtained after the change of basis, each
receiver, after decoding its own message and then reconstructing its
signal vector sent from its own transmitter, can subtract it from
its received signal vector, leaving only the interference from the
other undesired transmitters. In Fig. \ref{fig:2_by_3_mimo}, we use
$S_{(\cdot)}$ to denote the noisy interference terms at the receiver
antenna $(\cdot)$, thus obtaining the interference signal vector
$\bar{S}_k=[S_{ka}~S_{kb}~S_{kc}]^T$ where subscripts are associated
with corresponding receive antennas. The transmit symbols contained
within each remaining interference term are indicated explicitly,
reflecting the connectivity of the network after the change of basis
operation.

%With the resulting channel connectivity, we can write out the specific combination of corresponding interference symbols at each receiver antenna.

\subsection{Change of Basis for $(M_T,M_R)=(3,4)$}
Consider the linear transformation for the $3 \times 4$ setting as
another example. We label antennas from the top to the bottom of
Transmitter $k$ as $ka_1,kb_1,kc_1$, respectively. The corresponding
columns in the transformation matrix are labeled in the same manner.
Also, we label the antennas from the top to the bottom of Receiver
$k$ as $ka_1,ka_0,kc_0,kc_1$, respectively, and the corresponding
rows of the transformation matrix are labeled in the same manner as
well.

\emph{Step 1:} We start with the linear transformations at the
receiver. First, we ensure that $ka_1$ does not hear Transmitter
$k+1$ and antenna $kc_1$ does not hear Transmitter $k-1$. This
operation is guaranteed because we can choose the columns $ka_1$ and
$kc_1$ in the left null space of the channel matrix
$\mathbf{H}_{k(k+1)}$ and $\mathbf{H}_{k(k-1)}$, respectively. Since
channel matrices are $4\times 3$, these two columns are uniquely
determined as the corresponding one dimensional left null spaces.

\emph{Step 2:} After determining rows $ka_1$ and $kc_1$ of ${\bf
R}_k$, we can design columns $ka_1$ and $kc_1$ of ${\bf T}_k$ at
Transmitter $k$. The goal is to ensure antenna $ka_1$ of Transmitter
$k$ is \emph{not} heard by antenna $(k-1)c_1$ of Receiver $k-1$.
Similarly, antenna $kc_1$ of Transmitter $k$ is \emph{not} heard by
antenna $(k+1)a_1$ of Receiver $k+1$. Note that in the
three-dimensional signal space at Transmitter $k$, there is a
two-dimensional subspace orthogonal to antenna $(k-1)c_1$ and
another two-dimensional subspace orthogonal to antenna $(k+1)a_1$.
These two two-dimensional subspaces have one-dimensional
intersection within the three-dimensional space seen from the
transmitter. Therefore, we choose columns $ka_1$ and $kc_1$ as the
directions within their corresponding two-dimensional subspaces but
\emph{not} in the one-dimensional intersection subspace. Moreover,
we will restrict remaining columns at Transmitter $k$ to lie in the
one-dimensional intersection. In other words, the remaining antennas
of Transmitter $k$ can see neither the antenna $(k+1)a_1$ nor
$(k-1)c_1$. In this case, since only one antenna $kb_1$ remains, the
column $kb_1$ is fixed as the one-dimensional intersection of those
two 2-dimensional subspaces.

\emph{Step 3:} Now we can switch to receivers again to determine
rows $ka_0$ and $kc_0$ of ${\bf R}_k$. First, in order for rows
$ka_0$ and $kc_0$ to be linearly independent with rows $ka_1$ and
$kc_1$ and hereby the transformations can be invertible, the
subspace spanned by rows $ka_0$ and $kc_0$ should have only null
intersection with that spanned by rows $ka_1$ and $kc_1$. For
simplicity, we set the first and last columns of these two rows to
zero. Then our goal is to ensure that $ka_0$ at the receiver does
not hear $(k+1)b_1$, and $kc_1$ at the receiver does not hear
$(k-1)b_1$. This can be easily achieved by choosing the rows of
$ka_1$ and $kc_1$ to be orthogonal to the channel vectors associated
with the antenna $(k+1)b_1$ and $(k-1)b_1$, respectively.

Similar to the $2\times 3$ setting, the invertibility of channel
transformation matrices is guaranteed almost surely owing to the
generic property of channel coefficients. The resulting channel
connectivity after the change of basis operations, as well as the
interference as the function of transmit signals are shown in
Fig.\ref{fig:3_by_4_mimo}.
\begin{figure}[!h] \vspace{-0.1in}\centering
\includegraphics[width=3.5in]{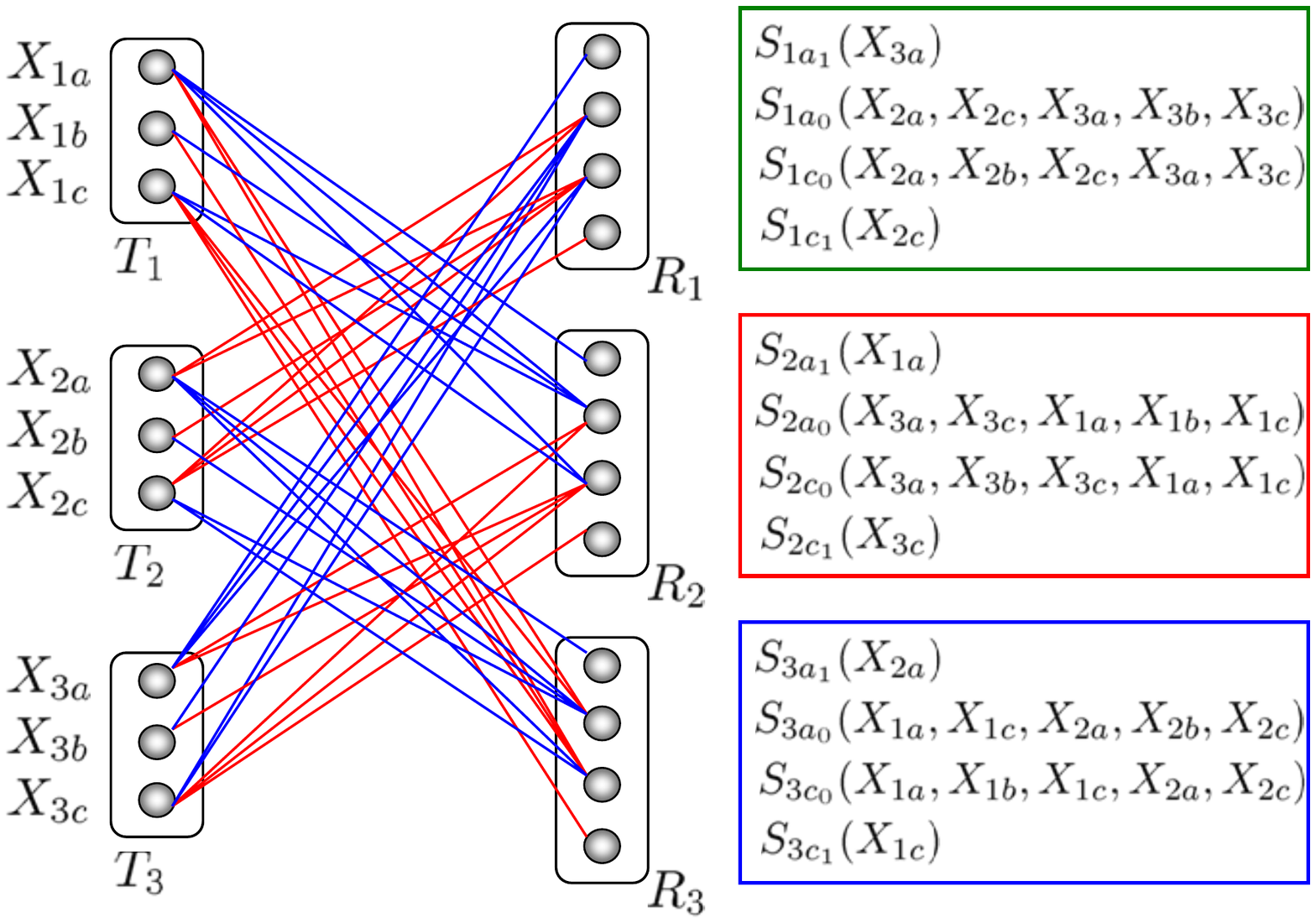}\vspace{-0.05in}
\caption{Three-User $3\times 4$ MIMO Interference Channel after the
Change of Basis Operation} \vspace{-0.1in} \label{fig:3_by_4_mimo}
\end{figure}

\subsection{Change of Basis for $(M_T,M_R)=(4,5)$}

The third example is the $4\times 5$ MIMO interference channel. As
we will see shortly, this case has a recursive relation with the
$2\times 3$ setting. As before, we label the antennas of Transmitter
$k$ from the top to the bottom as $ka_2, ka_1, kc_1, kc_2$ and the
antennas at Receiver $k$ from top to bottom as $ka_2, ka_1, kb_1,
kc_1, kc_2$. The corresponding columns and rows at transmitters and
receivers are labeled in the same manner.

\emph{Step 1:} As before, we start from the first and last antennas
at the receiver side. In the transformed basis, the first antenna at
Receiver $k$, $ka_2$, is placed into the
 null space of Transmitter $k+1$ such that it does not hear
Transmitter $k+1$. The last antenna at Receiver $k$, $kc_2$, is in
the null space of Transmitter $k-1$ such that it does not hear
Transmitter $k-1$.

\emph{Step 2:} After the first and last rows of the transformation
matrix at each receiver are designed, we now turn to the first and
last columns of the transformation matrix at each transmitter. The
goal is to ensure that antenna $ka_2$ and $kc_2$ at Transmitter $k$
cannot see receiver antenna $(k-1)c_2$ and $(k+1)a_2$. Since there
are 4 antennas at each transmitter, there are two 3-dimensional
subspaces orthogonal to $(k-1)c_2$ and $(k+1)a_2$, respectively.
However, these two subspaces have a 2-dimensional intersecting
subspace. To achieve the goal, we choose the rows $ka_2$ and $kc_2$
as the directions which are linearly independent. Moreover, we will
restrict the space spanned by remaining two columns ($ka_1$ and
$kc_1$) to be the 2-dimensional intersection subspace. As a result,
antennas $ka_1$ and $kc_1$ will not be heard by either $(k+1)a_2$ or
$(k-1)c_2$.

After Step 1 and Step 2, we have specified the first and last
columns and rows of the linear transformation matrices at each
transmitter and receiver, respectively. To determine the remaining
three rows of transformation matrices at each receiver, we first
need to ensure that they are linearly independent with the first and
last rows. One way to achieve this is to set the first and last
columns of these three rows to  zero. Therefore, although the
receiver has five antennas, the remaining three antennas are
restricted to access only three dimensional subspaces within this
five-dimensional signal space. Now ignoring the first and last
antennas, the receiver is equivalent to a receiver with 3 antennas.
On the other hand, since the remaining two antennas at each
transmitter are restricted to access only a two-dimensional subspace
in the four-dimensional transmitted signal space, it is equivalent
to a transmitter with 2 antennas. Therefore, we obtain a core of
$2\times 3$ interference channel, and thus the linear
transformations designed for the $2\times 3$ case previously can be
applied here to determine the remaining columns and rows at each
transmitter and receiver.

Again, it is easy to show that the linear transformation matrices at
each user have full rank almost surely due to the generic property
of the channel coefficients. The channel connectivity after the
change of basis operations and the interference as the function of
transmit signals plus noise are shown in Fig.\ref{fig:4_by_5_mimo}.
\begin{figure}[!t] \vspace{-0.15in}\centering
\includegraphics[width=4.5in]{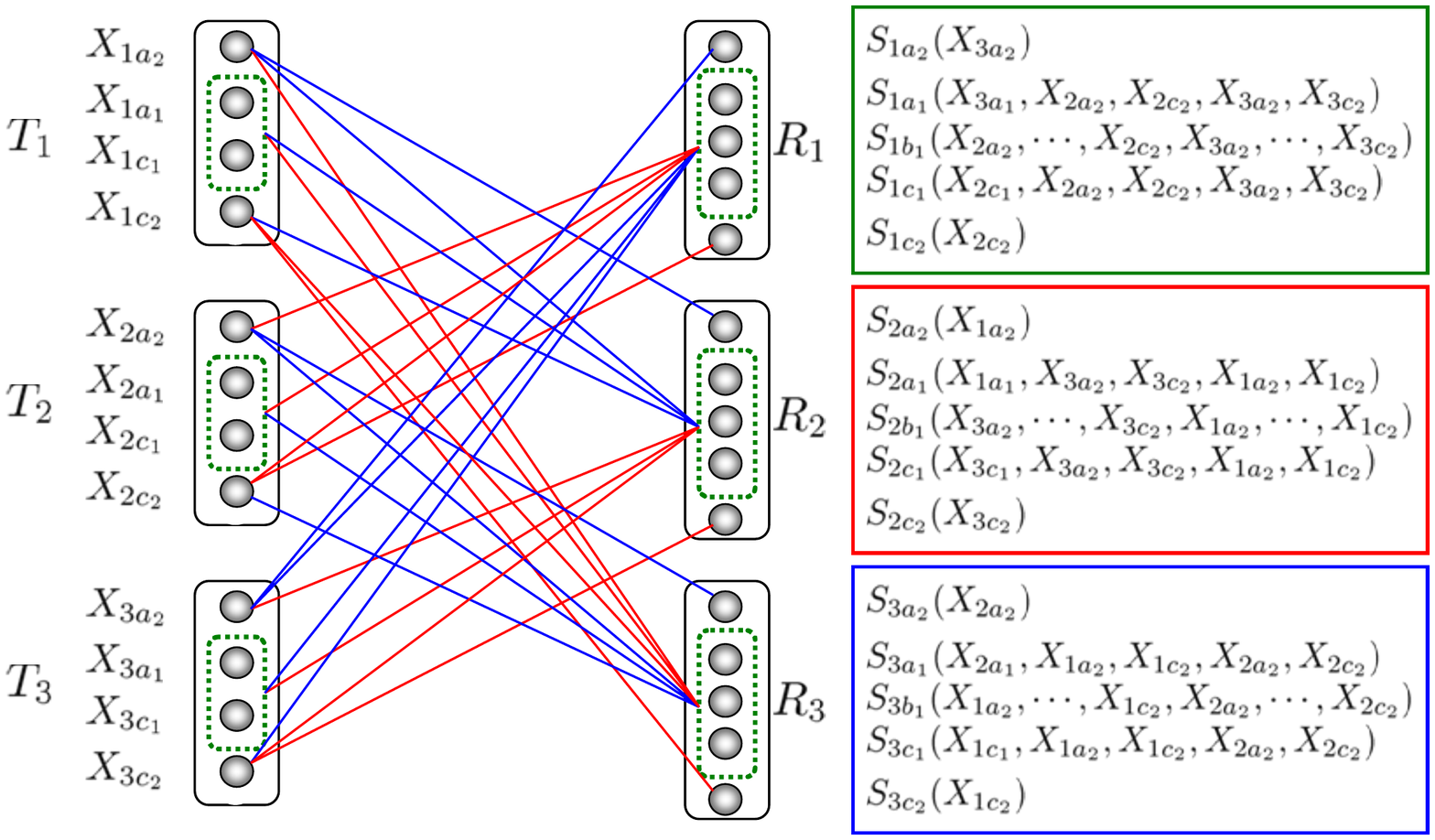}\vspace{-0.05in}
\caption{Three-User $4\times 5$ MIMO Interference Channel after the
Change of Basis Operation} \vspace{-0.1in} \label{fig:4_by_5_mimo}
\end{figure}

\subsection{Change of Basis for $(M_T,M_R)=(p,p+1)$}

After showing three specific examples, we present the change of
basis for the general $p\times (p+1)$ setting in this subsection. As
in the $4\times 5$ setting, the linear transformations for the
general setting can be determined in a recursive manner.
Specifically, for the $p\times(p+1)$ setting, we first design the
first and last columns and rows for the linear transformation
matrices at each transmitter and receiver, respectively. Once
designed, there remain $(p-2)$ and $(p-1)$ columns and rows in the
transformation matrices to be designed at each transmitter and each
receiver, respectively. Essentially, this is the $(p-2)\times (p-1)$
setting if we ignore the two antennas in the outer shell, and hereby
the design for the $(p-2)\times (p-1)$ case can be applied. As a
result, we only need to specify the design of the first and the last
columns and rows of transformation matrices at the transmitter and
receiver sides, respectively, for the general $p\times (p+1)$ case.
Since the $p \times (p+1)$ case can be reduced to the $(p-2)\times
(p-1)$ case, we further need to consider two subcases. One is the
$(2L,2L+1)$ setting which can be recursively boiled down to the
$2\times 3$ case discussed earlier, while the other is $(2L+1,2L+2)$
and can be recursively boiled down to the $3\times 4$ case discussed
earlier, $\forall L\in\mathbb{Z}^+$. The algorithm for these two
groups are essentially identical. In the following, we first
consider the $(2L,2L+1)$ setting.
\begin{figure}[!t]
\centering
\includegraphics[width=4.5in]{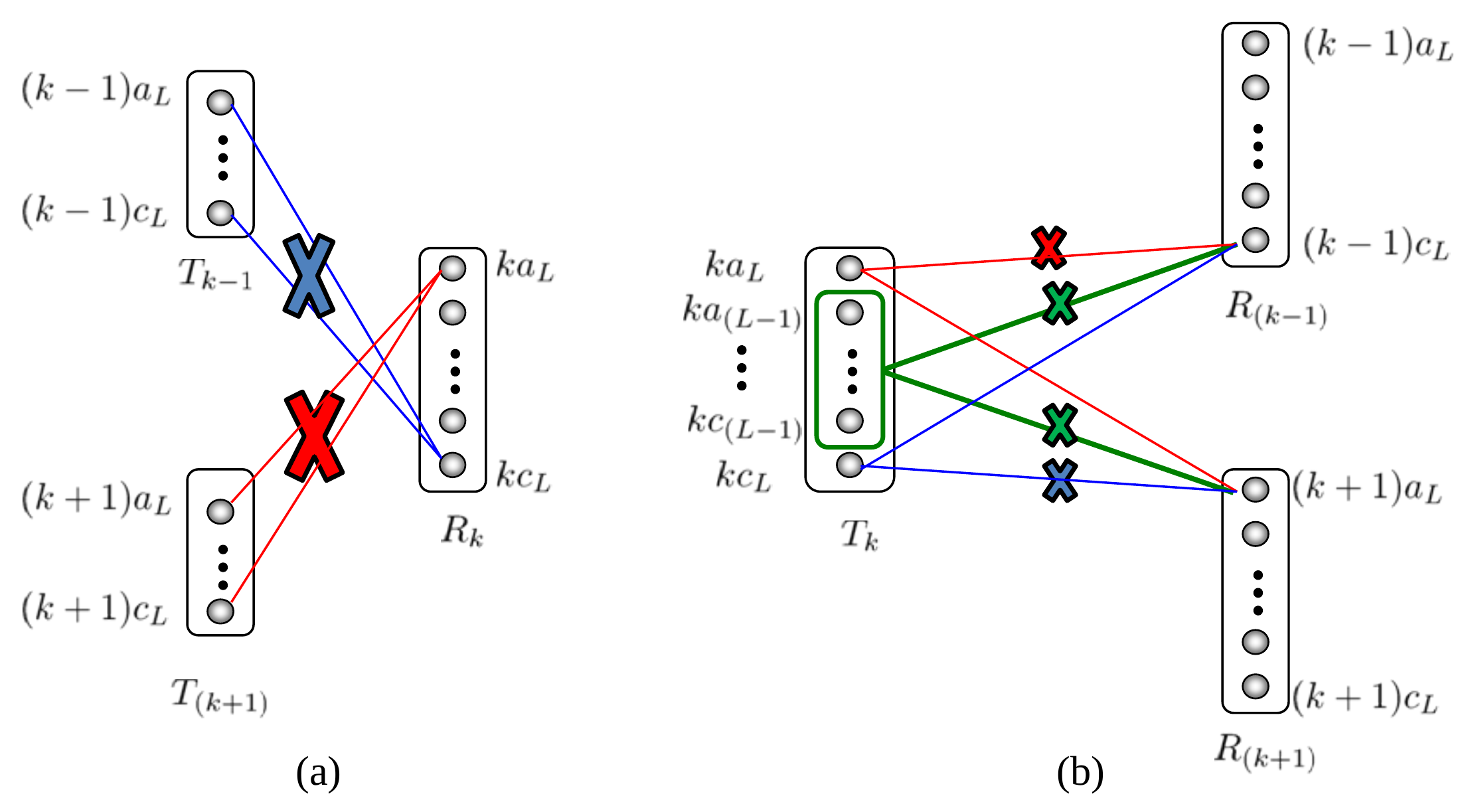}
\caption{Invertible Linear Transformations for the $2L \times
(2L+1)$ setting, (a) Step 1, (b) Step 2} \label{fig:LT}
\end{figure}

%\subsubsection{$(M,N)=(2L,2L+1)$}

\subsubsection{Change of Basis for $(M,N)=(2L,2L+1)$}

We label the antennas from the top to the bottom, as $ka_L$,
$ka_{L-1}$, $\cdots$, $ka_1$, $kc_1$, $\cdots, kc_L$ at Transmitter
$k$, and as $ka_L, ka_{L-1}$, $\cdots$, $ka_1$, $kb_1$, $kc_1$,
$\cdots, kc_L$ at Receiver $k$. Each antenna corresponds to one
column at each transmitter and one row at each receiver,
respectively. Therefore, the column labeling is from the left to the
right of the transformation matrices at each transmitter, and the
row labeling is from the top to the bottom of that at each receiver.
We will always first design the first and last rows of the
transformation matrix at each receiver, and then based on that
design determine the first and last columns of the transformation
matrix at each transmitter.

\emph{Step 1:} Consider Receiver $k$. The goal is to ensure that the
first antenna $ka_L$ and the last antenna $kc_L$ do not hear
transmitters $k-1$ and $k+1$, respectively, as illustrated in Fig.
\ref{fig:LT}(a). This can be done by choosing row $ka_L$ in the left
null space of $\mathbf{H}_{k(k-1)}$ and row $kc_L$ in the left null
space of $\mathbf{H}_{k(k+1)}$. Since these two matrices are
$(2L+1)\times 2L$, the left null space has one dimension and thus
two rows are uniquely determined. After that, the remaining $2L-1$
rows are restricted to span the $(2L-1)$-dimensional subspace that
does not overlap with the subspace spanned by the first and last
rows such that they are linearly independent. One way to achieve
this is to set the first and last columns, i.e., columns $ka_L$ and
$kc_L$, of all the remaining rows to be zero.
%\begin{figure}[!t]
%\centering
%\includegraphics[width=2.6in]{LTs2}
%\caption{Linear transformation for $2L \times (2L+1)$ -- Step 2}
%\label{fig:LTs2}
%\end{figure}

\emph{Step 2:} Once the rows $ka_L$ and $kc_L$ of ${\bf R}_k$ are
determined, we switch to determine the first and last columns of the
transformation matrix ${\bf T}_k$. As illustrated in
Fig.\ref{fig:LT}(b), the goal is to ensure that the antenna $ka_L$
of Transmitter $k$ cannot see antenna $(k-1)c_L$ of Receiver $k-1$,
while the antenna $kc_L$ cannot see the antenna $(k+1)a_n$ of
Receiver $k+1$. Since there are $2L$ antennas at each transmitter,
there are two $(2L-1)$-dimensional subspaces orthogonal to
$(k-1)c_L$ and $(k+1)a_L$, respectively. However, these two
subspaces have an intersection of $2L-2$ dimensions at the
transmitter in general, implying that there is only one dimension in
each subspace that does not overlap with the other. Therefore, we
choose rows $ka_L$ and $kc_L$ in that one dimension in each
subspace. Moreover, we restrict the space spanned by remaining
$2L-2$ columns to be the $(2L-2)$-dimensional intersection subspace.
As a result, antennas $ka_{L-1},\cdots, ka_1, kc_1, \cdots,
kc_{L-1}$ will not see either $(k+1)a_L$ or $(k-1)c_L$, as shown in
Figure \ref{fig:LT}(b).
%This is done by losing the degrees of freedom
%to arbitrarily pick entries in rows $ka_L$ and $kc_L$. In other
%words, these two rows for the remaining $2L-2$ columns will be
%determined uniquely after the rest rows are designed such that all
%these columns are orthogonal to antenna $(k+1)a_L$ or $(k-1)c_L$.

After Step 1 and Step 2, we essentially still need to design the
$(2L-2)\times (2L-1)$ setting. We can repeat these two steps until
we reach the $2\times 3$ core for which the linear transformations
have been illustrated before. Also, it is not difficult to see these
linear transformations are invertible, i.e., ${\bf T}_k$ and ${\bf
R}_k$ have full rank, almost surely.

%\subsubsection{$(M,N)=(2L+1,2L+2)$}

\subsubsection{Change of Basis for $(M,N)=(2L+1,2L+2)$}

The linear transformation for $(2L+1)\times(2L+2)$ is essentially
the same as the case $2L\times (2L+1)$. By designing the first and
last rows of the transformation matrix at each receiver first and
then designing the first and last columns at each transmitter in the
same manner as the $2L\times (2L+1)$ case, we end up with the
$(2L-1)\times 2L$ case. Repeating such a procedure until we arrive
at the $3\times4$ setting for which the linear transformations have
been illustrated. Moreover, it is not difficult to see these linear
transformations are invertible as well.

%Next, we will show that for generic channel coefficients, the linear
%transformation proposed is guaranteed to be invertible almost
%surely. Again, since the determinant of the transformation matrix is
%a ratio of two polynomials of channel coefficients, it suffices to
%show that the numerator is not identical zero polynomial. This is
%done by constructing a specific channel such that the linear
%transformation matrix obtained by the proposed method is full rank.
%The constructed channels for the $M\times (M+1)$ case are as
%follows:
%\begin{eqnarray}
%\mathbf{H}_{(k-1)k}=\left[\begin{array}{c}\mathbf{0}_{1\times M}\\ \mathbf{I}_{M\times M}\end{array}\right]\\
%\mathbf{H}_{(k+1)k}=\left[\begin{array}{c}\mathbf{I}_{M\times M}\\
%\mathbf{0}_{1\times M}\end{array}\right]
%\end{eqnarray}
%where $\mathbf{0}_{1\times M}$ is a $1\times M$ zero vector and
%$\mathbf{I}_{M\times M}$ is the $M \times M$ identity matrix. It can
%be easily seen that the linear transformation matrix at each
%transmitter and receiver is the identity matrix and thus is full
%rank. Therefore, we established that the linear transformation is
%invertible almost surely for generic channel coefficients.

\begin{figure}[!t] \vspace{-0.1in}\centering
\includegraphics[width=5.5in]{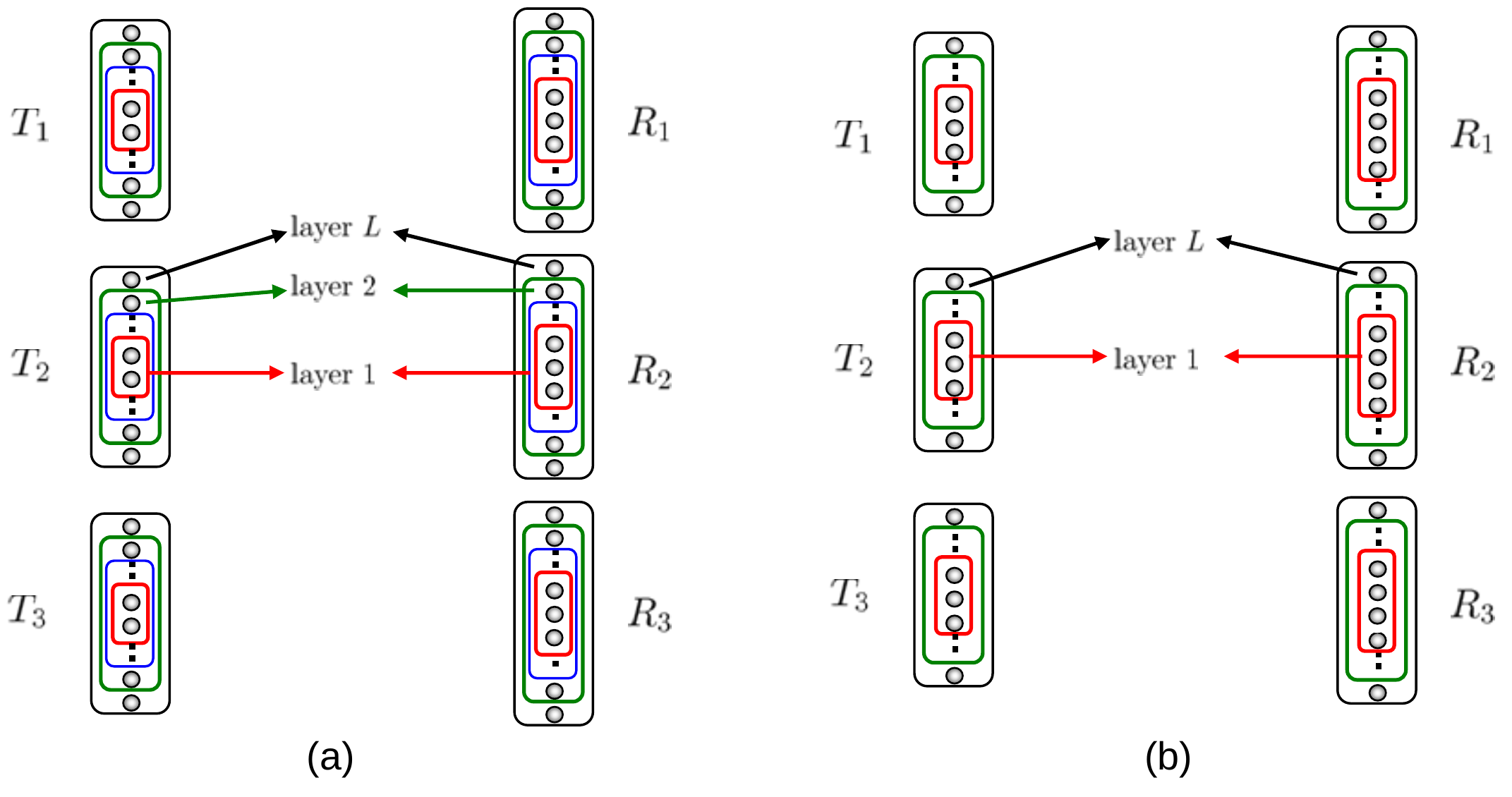}\vspace{-0.05in}
\caption{Intuition of Onion Peeling, (a) $M$ is even, (b) $M$ is
odd} \vspace{-0.1in} \label{fig:onion_peeling}
\end{figure}

{\it Remark:} Note that this transformation process looks like
layer-by-layer onion peeling, as shown in Figure
\ref{fig:onion_peeling}. Specifically, for the $p\times(p+1)$
setting, by peeling out the outer shell at each user, i.e., the
first and the last antennas, the remaining part is a
$(p-2)\times(p-1)$ interference channel. Repeating this procedure,
we eventually obtain the core of the $2\times 3$ interference
network if $p$ is even as shown in Figure
\ref{fig:onion_peeling}(a), or the core of the $3\times 4$
interference network if $p$ is odd as in Figure
\ref{fig:onion_peeling}(b). The onion peeling intuition is important
for the change of basis, as well as help us identify the genie
signals provided to each receiver for deriving the information
theoretic DoF outer bounds in Section \ref{sec:outerbound_eg}.

\section{Information Theoretic DoF Outer Bound}\label{sec:outerbound_eg}

%\section{Information Theoretic DoF Outer Bound for $M_T<M_R$ }\label{sec:outerbound_MsmallerN}

In this section, we will derive the information theoretic DoF outer
bound claimed in Lemma \ref{lemma:out}. We organize this section
as follows. We begin with the case when $M/N\leq 3/5$ where
cooperation outer bound is sufficient. For the remaining $M/N\leq
3/5$ case, we first consider the $M_T<M_R$ setting, i.e., networks
that have more antennas at the receiver nodes than at the
transmitters. We only show some specific examples to convey the
central idea to derive the information theoretic DoF outer bounds,
and defer the general proof to Appendix
\ref{sec:outerbound_MsmallerN}. Since the duality property does not
hold for the information theoretic statements, we also consider the
$M_T>M_R$ setting. Since the essential idea still follows the
onion peeling intuition, we defer all the rigorous proof to
Appendix \ref{sec:outerbound_MbiggerN}. We start with the $M/N\leq
3/5$ setting as follows.

{\bf DoF Outer Bound for $M/N\leq 3/5$:} Consider Figure
\ref{fig:map_table} where the squares are shaded with light yellow,
green and grey colors:
\begin{itemize}
\item[(A)] $M/N\in (0, 1/3] \Rightarrow d\leq M$
\item[(B)] $M/N\in (1/3, 1/2] \Rightarrow d\leq N/3$
\item[(C)] $M/N\in (1/2, 3/5] \Rightarrow d\leq 2M/3$
\end{itemize}
While we have shown in Section \ref{subsec:achieveless1/2} that the
single user or cooperation DoF outer bounds are sufficient for case
(A) and (B), we still need to consider the case (C). Fortunately, it
turns out that for this case, the cooperation DoF bound is still
sufficient. Specifically, the resulting two user interference
channel by allowing any two of three users to cooperate where User 1
has $M_T$ and $M_R$ antennas and User 2 has $2M_T$ and $2M_R$
antennas has been proved to have
$\min(3M_T,3M_R,\max(M_T,2M_R),\max(2M_T,M_R))$ DoF
\cite{Jafar_Fakhereddin}, which is equal to $2\min(M_T,M_R)=2M$.
Since cooperation among antennas cannot reduce the channel capacity,
$2M/3$ is the DoF per user bound for the original channel in case
(C).

%Since the outer bounds for these three cases are trivial, we
%directly show the results followed by simple arguments. The values
%of the DoF bounds can be obtained from Theorem \ref{theorem:dof} by
%letting $p=1,~2$. In fact, these bounds shown in three cases above
%are straight-froward, obtained from  the cooperation outer bound
%that was shown in \cite{Jafar_Fakhereddin, Gou_Jafar_MIMO}.
%Specifically, for case (A), $M$ DoF is the single user DoF bound.
%For case (B) and (C), by allowing any two of three users to
%cooperate, we obtain a two user interference channel in which one
%user has $M_T$ and $M_R$ antennas, and the other has $2M_T$ and
%$2M_R$ antennas at the transmitter and receiver, respectively. For
%this two-user interference channel, the sum DoF are bounded above by
%$\min(3M_T,3M_R,\max(M_T,2M_R),\max(2M_T,M_R))$
%\cite{Jafar_Fakhereddin}, which is equal to $\max(M_T,M_R)=N$ for
%case (B) and $2\min(M_T,M_R)=2M$ per user for case (C). Since
%cooperation cannot reduce the channel capacity region, the DoF of
%the original channel are bounded above by $N/3$ per user for case
%(B) and $2M/3$ per user for case (C) as well.

{\bf DoF Outer Bound for $M/N> 3/5$:} While applying cooperation
outer bounds for $M/N\leq 3/5$ is sufficient to establish the sum
DoF results (the achievability will be shown in Section
\ref{sec:innerbound}), the DoF remain open for $M/N> 3/5$ cases.

The nature of the DoF outer bound derivation is recursive. To set up the recursion we will first solve the core cases  and then construct a recursive argument for the general proof built upon the reduction to these core cases. The complexity of the problem is such that some partitioning into disjoint groups, each of which must be studied separately, is necessary. First, because no claim to duality can be made a-priori for information theoretic DoF results, we need to deal with the cases $M_T>M_R$ and $M_R>M_T$ separately. Indeed the outer bound proof for the two settings require different approaches. Second, within each of these cases, e.g., with $M_R>M_T$ there are also two groups of channels, $(2L, 2L+1)$ and $(2L+1, 2L+2)$, each of which must be considered separately through its own recursive reduction. As such we will need to establish the core results for each of these groups, which will  make the overall proof of Lemma \ref{lemma:out} quite lengthy, rather unavoidably so.

In the following, we will first show the information theoretic DoF
outer bounds for three specific examples of $(M,N)=(p,p+1)$ where
$p=2,3,4$. Then based on the observation on the proofs for these
examples, we generalize the information theoretic DoF outer bound
proof to $(M,N)=(pq,(p+1)q)$ where $p\in\mathbb{Z}^+,p>4$ and
$q\in\mathbb{Z}^+$ in Appendix \ref{sec:outerbound_MsmallerN}. Since
results for $q>1$ follow from $q=1$ due to a simple spatial expansion,
we only need to consider $q=1$, i.e., $(M,N)=(p,p+1)$. Before we
show the specific examples, we first introduce the following lemma
that will be often used in the converse argument in the remaining of
this section.

\begin{lemma}\label{lemma:genie}
If a genie provides a subset of signals, denoted as $\mathcal{G}$, to
Receiver $k$, such that it can decode all three messages from the
observation $(\bar{Y}_k^n,\mathcal{G})$, then we can always outer
bound the mutual information term
$I(W_1,W_2,W_3;\bar{Y}^n_k,\mathcal{G})$ as follows:
\begin{eqnarray}
I(W_1,W_2,W_3;\bar{Y}^n_k,\mathcal{G})
&=& I(W_1,W_2,W_3;\bar{Y}^n_k)+I(W_1,W_2,W_3;\mathcal{G}|\bar{Y}^n_k)\label{eqn:lemma_info_chain_rule}\\
&\leq & M_R~n\log\rho+I(W_1,W_2,W_3;\mathcal{G}|\bar{Y}^n_k)+n~o(\log\rho)\label{eqn:lemma_Rx}\\
&\leq& M_R~n\log\rho+h(\mathcal{G}|\bar{Y}^n_k)-h(\mathcal{G}|W_1,W_2,W_3,\bar{Y}^n_k)+n~o(\log\rho)\label{eqn:lemma_entropy_chain_rule}\\
&=&
M_R~n\log\rho+h(\mathcal{G}|\bar{Y}^n_k)+n~o(\log\rho)+o(n).\label{eqn:lemma_recover}
\end{eqnarray}
\end{lemma}
{\it Proof:} In the derivations above,
(\ref{eqn:lemma_info_chain_rule}) follows from the mutual
information chain rule. (\ref{eqn:lemma_Rx}) is obtained because
Receiver $k$ has only $M_R$ antennas.
(\ref{eqn:lemma_entropy_chain_rule}) follows from the entropy chain
rule, and (\ref{eqn:lemma_recover}) is obtained since given all the
three messages we can reconstruct the genie signals $\mathcal{G}$
subject to the noise distortion.\hfill\QED

%Thus, in the remaining part of this paper, when we derive the
%inequality, we directly start from the the term of
%$Nn\log\rho+I(\mathcal{G}|\bar{Y}^n_k)+n~o(\log\rho)+o(n)$ on the
%right-hand side.

\subsection{Case: $(M,N)=(2,3)~\Rightarrow$ DoF $\leq \frac{6}{5}$}

After the change of basis operation introduced in Section
\ref{sec:outerbound}, we obtain the network with connectivity in
Fig. \ref{fig:2_by_3_mimo}, shown again  in Fig.
\ref{fig:2_by_3_mimo2} for the reader's convenience.

%
%Note that because the
%channel coefficients for the original channel are generic, the
%channel coefficients in Fig.\ref{fig:2_by_3_mimo2} are non-zero
%almost surely. However, we need to emphasize that for each user $k$
%as long as the coefficients of $X_{(k-1)a}$ in $S_{ka}$,
%$X_{(k+1)c}$ in $S_{kc}$, $X_{(k-1)c}$ and $X_{(k+1)a}$ in $S_{kb}$
%are non-zero, the DoF outer bound that will be shown always hold.

\begin{figure}[!h] \vspace{-0.1in}\centering
\includegraphics[width=3.0in]{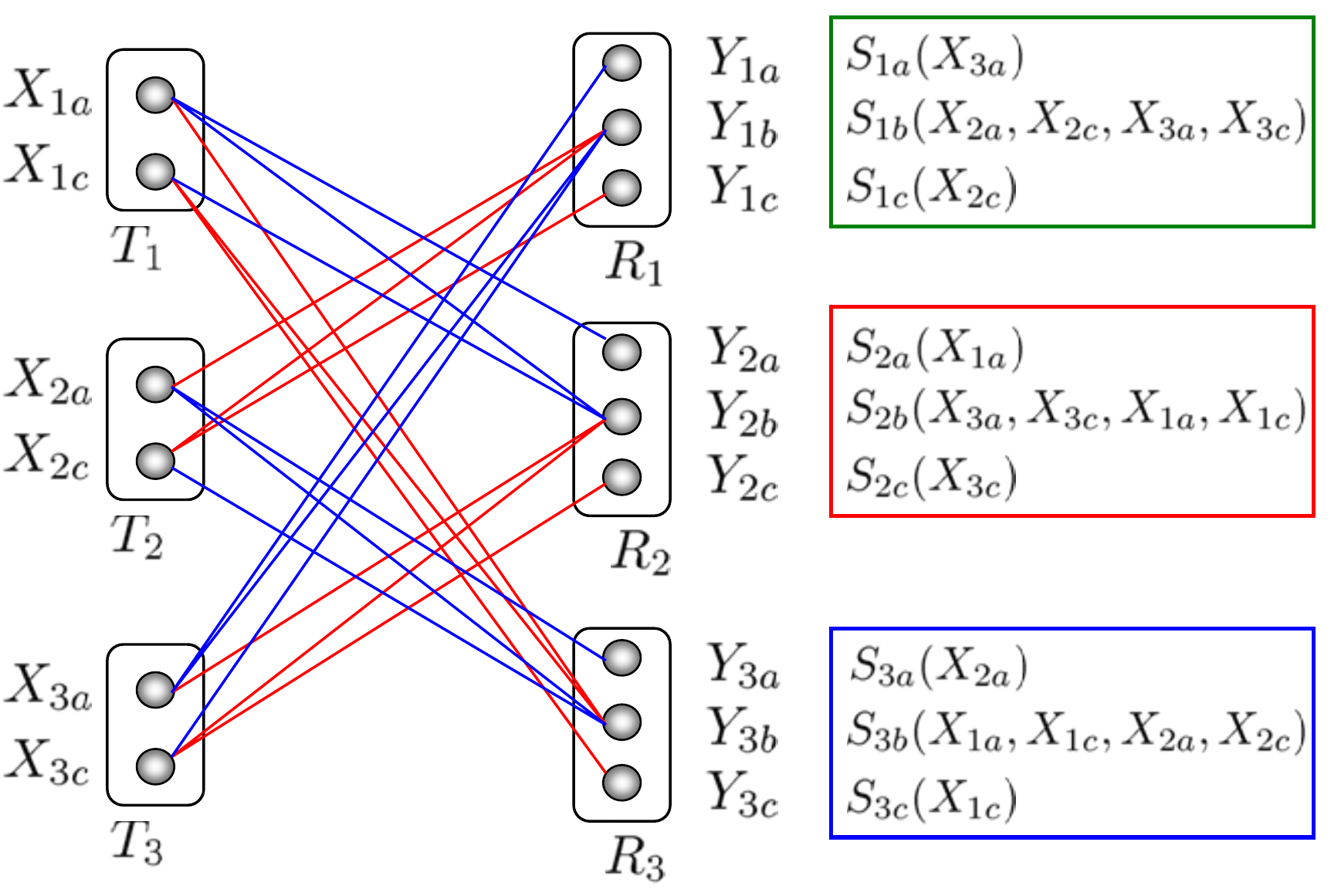}\vspace{-0.05in}
\caption{Three-User $2\times 3$ MIMO Interference Channel}
\vspace{-0.1in}\label{fig:2_by_3_mimo2}
\end{figure}

First, a genie provides the signal
$\mathcal{G}_1=\{X_{2a}^n+\tilde{Z}_{2a}^n\}$ to Receiver 1, where
$\tilde{Z}_{2a}^n\sim \mathcal{CN}(0,1)$ is an artificial i.i.d.
Gaussian noise. Let us consider Receiver 1. Since we are dealing
with a converse argument, it follows by assumption that the receiver
is able to decode and subtract out its desired signal. It therefore
has $S_{1c}^n(X_{2c}^n)=X_{2c}^n+Z_{2c}^n$
%(plus noise which will no longer be explicitly mentioned)
at the antenna ``$1c$". Thus, with the provided genie signal,
Receiver 1 can decode $W_2$ from the observation
$(X_{2a}^n+\tilde{Z}_{2a}^n,S_{1c}^n(X_{2c}^n))$ subject to the
noise distortion\footnote{We will often use the phrase ``subject to
noise distortion" in the remaining of this paper to indicate the
widely used (see e.g., \cite{Jafar_Fakhereddin, Jafar_Shamai}) DoF
outer bound argument whereby reducing noise at a node by an amount
that is SNR independent (and therefore inconsequential for DoF)
allows it to decode a message.}. After decoding $W_2$ we can
reconstruct the transmitted signals $(X_{2a}^n,X_{2c}^n)$ and
subtract them from $S^n_{1b}$ to obtain $S'^n_{1b}$ which is a
linear combination of transmitted signals $(X_{3a}^n,X_{3c}^n)$ plus
the noise. Now by the two linearly independent observations of
$(S_{1a}^n(X_{3a}^n),S'^n_{1b}(X_{3a}^n,X_{3c}^n))$, Receiver 1 can
resolve $(X_{3a}^n,X_{3c}^n)$ and thus can decode $W_3$ as well,
subject to the noise distortion. Since the genie information
$\mathcal{G}_1$ provided to Receiver 1 allows it to decode all three
messages subject to  noise distortion, we have:
\begin{eqnarray}
n(R_1+R_2+R_3)&\!\!\!\!\leq\!\!\!\!& I(W_1,W_2,W_3;\bar{Y}^n_1,\mathcal{G}_1)+n~o(\log\rho)+o(n)\label{eqn:2by3_ob1_fano}\\
&\!\!\!\!\leq\!\!\!\!& Nn\log\rho+h(X^n_{2a}+\tilde{Z}_{2a}^n|\bar{Y}^n_1)+n~o(\log\rho)+o(n)\label{eqn:2by3_ob1_reconstr}\\
&\!\!\!\!\leq\!\!\!\!& Nn\log\rho+h(X^n_{2a}+\tilde{Z}_{2a}^n|X^n_{2c}+Z_{2c}^n)+n~o(\log\rho)+o(n)\label{eqn:2by3_ob1_drop}\\
&\!\!\!\!\leq\!\!\!\!&
Nn\log\rho+nR_2-h(X^n_{2c}+Z_{2c}^n)+n~o(\log\rho)+o(n)\label{eqn:2by3_ob1_lemma1}
\end{eqnarray}
where (\ref{eqn:2by3_ob1_fano}) follows from Fano's inequality.
(\ref{eqn:2by3_ob1_reconstr}) follows from Lemma \ref{lemma:genie}.
(\ref{eqn:2by3_ob1_drop}) follows from the fact that dropping the
condition terms cannot decrease the differential entropy. Thus, we
only keep $S_{1c}^n$ as the condition term which is
$X_{2c}^n+Z_{2c}^n$. (\ref{eqn:2by3_ob1_lemma1}) is obtained because
from the observations of
$(X_{2a}^n+\tilde{Z}_{2a}^n,X_{2c}^n+Z_{2c}^n)$ we can decode $W_2$
subject to the noise distortion. By advancing the user indices, we
therefore obtain the following three inequalities:
\begin{subequations}\label{eqn:2by3_ob1_set}
\begin{eqnarray}
nR_{\Sigma}&\!\!\!\!\leq\!\!\!\!& Nn\log\rho+nR_2-h(X_{2c}^n+Z_{2c}^n)+n~o(\log\rho)+o(n)\\
nR_{\Sigma}&\!\!\!\!\leq\!\!\!\!& Nn\log\rho+nR_3-h(X_{3c}^n+Z_{3c}^n)+n~o(\log\rho)+o(n)\\
nR_{\Sigma}&\!\!\!\!\leq\!\!\!\!&
Nn\log\rho+nR_1-h(X_{1c}^n+Z_{1c}^n)+n~o(\log\rho)+o(n).
\end{eqnarray}
\end{subequations}
Since we always use ``advance the user indices" in a circularly
symmetric way, from now on we will use compact notations
\begin{eqnarray*}
R&=&R_{\Sigma}/3 \\
h(X_{(\cdot)})&=&[h(X_{1(\cdot)})+h(X_{2(\cdot)})+h(X_{3(\cdot)})]/3
\end{eqnarray*}
where quantities without user index represent the average of all
rotated indices. Thus (\ref{eqn:2by3_ob1_set}) can be rewritten as
\begin{eqnarray}\label{eqn:2by3_ob1}
3nR\leq Nn\log\rho+nR-h(X_{c}^n+Z_{c}^n)+n~o(\log\rho)+o(n).
\end{eqnarray}

Next, let a genie provide the signal
$\mathcal{G}_2=\{X_{3c}^n+\tilde{Z}_{3c}^n\}$ to Receiver 1 where
$\tilde{Z}_{2a}^n\sim \mathcal{CN}(0,1)$ is an artificial i.i.d.
Gaussian noise. Similarly, by providing $\mathcal{G}_2$ to Receiver
1, it can first decode $W_3$ subject to the noise distortion from
the observation $(S_{1a}^n(X_{3a}^n),X_{3c}^n+\tilde{Z}_{3c}^n)$.
After decoding $W_3$, Receiver 1 can reconstruct transmitted signals
$(X_{3a}^n,X_{3c}^n)$ and then subtract them from $S_{1b}^n$ to
obtain a noisy linear combination of $(X_{2a}^n,X_{2c}^n)$, from
which together with $S_{1c}^n(X_{2c}^n)$ Receiver 1 can decode $W_2$
as well subject to the noise distortion. Since Receiver 1 again can
decode all three messages, we have:
\begin{eqnarray}
nR_{\Sigma}&\!\!\!\!\leq\!\!\!\!& I(W_1,W_2,W_3;\bar{Y}^n_1,X_{3c}^n+\tilde{Z}_{3c}^n)+n~o(\log\rho)+o(n)\\
&\!\!\!\!\leq\!\!\!\!& Nn\log\rho+h(X_{3c}^n+\tilde{Z}_{3c}^n|\bar{Y}^n_1)+n~o(\log\rho)+o(n)\\
&\!\!\!\!\leq\!\!\!\!&
Nn\log\rho+h(X_{3c}^n+\tilde{Z}_{3c}^n)+n~o(\log\rho)+o(n).
\end{eqnarray}
The derivation above is similar to that for the previous bound. By
advancing the user indices, we obtain the second inequality as
follows.
\begin{eqnarray}\label{eqn:2by3_ob2}
3nR\leq Nn\log\rho+h(X_{c}^n+\tilde{Z}_{c}^n)+n~o(\log\rho)+o(n).
\end{eqnarray}

Adding up the two inequalities of (\ref{eqn:2by3_ob1}) and
(\ref{eqn:2by3_ob2}), each obtained by averaging over user indices,
subject to the noise distortion we have:
\begin{eqnarray}
6nR\leq 2Nn\log\rho+nR+n~o(\log\rho)+o(n).\label{eqn:2by3_ob3}
\end{eqnarray}
where
$h(X_{c}^n+\tilde{Z}_{c}^n)-h(X_{c}^n+Z_{c}^n)=n~o(\log\rho)+o(n)$.
By arranging terms of (\ref{eqn:2by3_ob3}) we have:
\begin{eqnarray}
5nR\leq 2Nn\log\rho+n~o(\log\rho)+o(n).\label{eqn:2by3_ob4}
\end{eqnarray}
By dividing $\log\rho$ and $n$ on both sides of
(\ref{eqn:2by3_ob4}), and letting $\rho\rightarrow \infty$ and
$n\rightarrow \infty$, we obtain:
\begin{eqnarray}
d\leq\frac{2N}{5}=\frac{6}{5}.
\end{eqnarray}
%\hfill\QED

Note that the genie signals provided to receivers contain Gaussian
noise. Since we are interested in the DoF characterization and due
to the ``subject to the noise distortion" statement, we will not
explicitly mention the noise terms in the genie signal and received
signals in the remaining of this exposition for simplicity.

{\it Remark:} In order to obtain the inequalities
(\ref{eqn:2by3_ob1}) and (\ref{eqn:2by3_ob2}), a genie provides the
signal set $\mathcal{G}_1$ and $\mathcal{G}_2$ to Receiver 1,
respectively (also to other receivers by advancing user indices). In
other words, with the genie signal provided to Receiver 1, it can
always decode all the three messages. Moreover, in the remaining of
this paper, unless otherwise specified, it is not difficult to see
that by providing the genie signal set that will be specified later
to Receiver 1, it can always decode all three messages subject to
the noise distortion.

\subsection{Case: $(M,N)=(3,4)~\Rightarrow$ DoF $\leq \frac{12}{7}$}

The second example is the $(M,N)=(3,4)$ setting. After taking the
linear transformation introduced in Section \ref{sec:outerbound}, we
obtain the network with connectivity shown in
Fig.\ref{fig:3_by_4_mimo2}, and the resulting equivalent channel
coefficients are non-zero.

\begin{figure}[!h] \vspace{-0.1in}\centering
\includegraphics[width=3.5in]{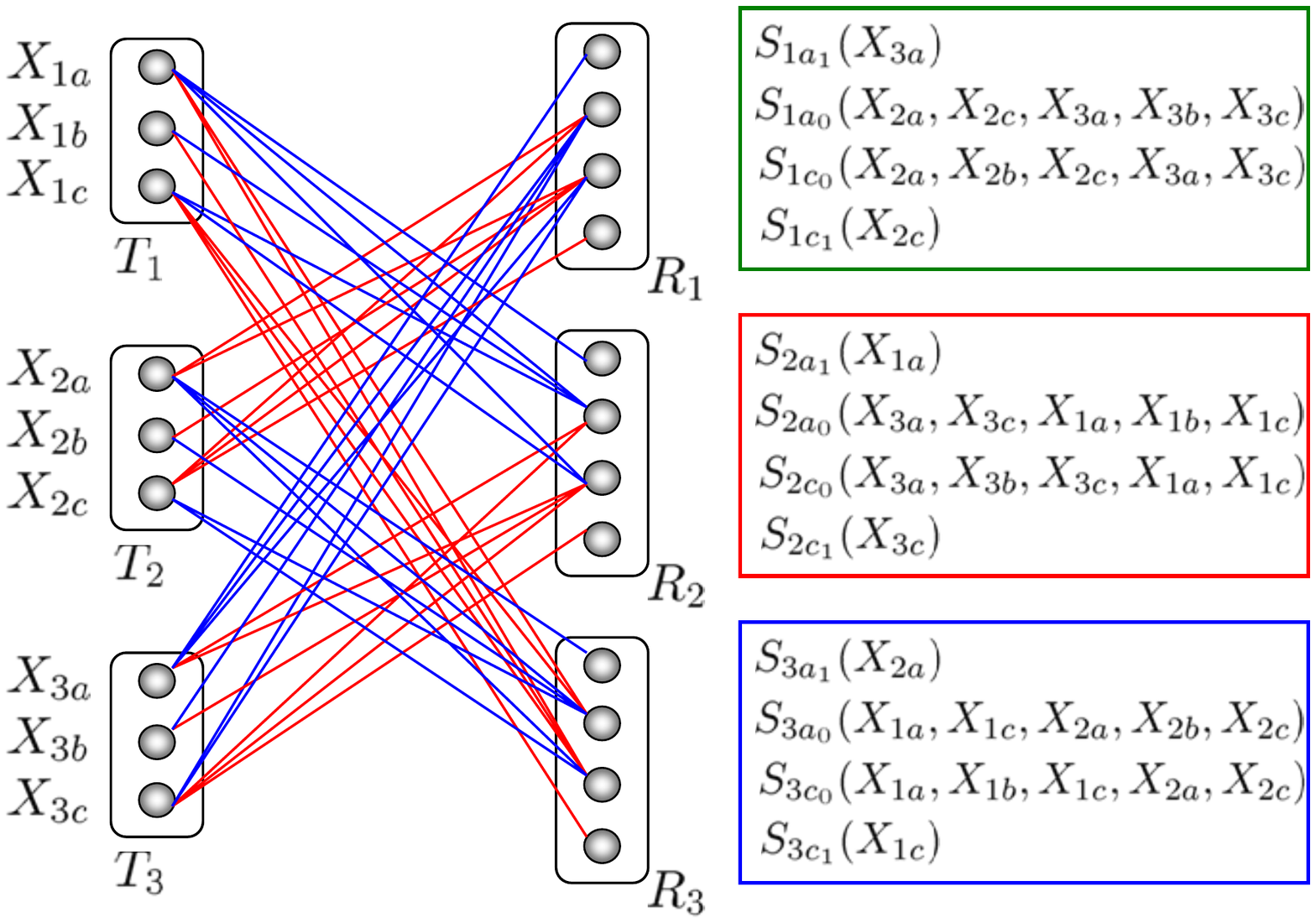}\vspace{-0.05in}
\caption{Three-User $3\times 4$ MIMO Interference
Channel}\label{fig:3_by_4_mimo2} \vspace{-0.1in}
\end{figure}

First, a genie provides the signal set
$\mathcal{G}_1=\{X_{2a}^n,X_{3c}^n\}$ to Receiver 1. As in the
$(M,N)=(2,3)$ case, it is easy to see by providing this set of genie
signals to Receiver 1, it can decode all three messages subject to
the noise distortion. Consider the sum rate of three messages:
\begin{eqnarray}
nR_{\Sigma}&\!\!\!\!\leq\!\!\!\!& I(W_1,W_2,W_3;\bar{Y}^n_1,X_{2a}^n,X_{3c}^n)+n~o(\log\rho)+o(n)\label{eqn:3by4_ob1_fano}\\
&\!\!\!\!\leq\!\!\!\!& Nn\log\rho+h(X_{2a}^n,X_{3c}^n|\bar{Y}^n_1)+n~o(\log\rho)+o(n)\label{eqn:3by4_ob1_lemma}\\
&\!\!\!\!\leq\!\!\!\!&
Nn\log\rho+h(X_{2a}^n)+h(X_{3c}^n)+n~o(\log\rho)+o(n)\label{eqn:3by4_ob1_drop}
\end{eqnarray}
where (\ref{eqn:3by4_ob1_fano}) follows from Fano's inequality, and
(\ref{eqn:3by4_ob1_lemma}) is because of Lemma \ref{lemma:genie}.
(\ref{eqn:3by4_ob1_drop}) follows from the chain rule and dropping
the condition terms cannot decrease the differential entropy. By
averaging over the user indices, we obtain the first inequality as
follows.
\begin{eqnarray}\label{eqn:3by4_ob1}
3nR\leq Nn\log\rho+h(X_{a}^n)+h(X_{c}^n)+n~o(\log\rho)+o(n).
\end{eqnarray}

Next, a genie provides the signal set
$\mathcal{G}_2=\{X_{2a}^n,X_{2b}^n\}$ to Receiver 1, then the sum
rate of three messages is bounded above by:
\begin{eqnarray}
nR_{\Sigma}&\!\!\!\!\leq\!\!\!\!& I(W_1,W_2,W_3;\bar{Y}^n_1,X_{2a}^n,X_{2b}^n)+n~o(\log\rho)+o(n)\\
&\!\!\!\!\leq\!\!\!\!& Nn\log\rho+h(X_{2a}^n,X_{2b}^n|\bar{Y}^n_1)+n~o(\log\rho)+o(n)\\
&\!\!\!\!=\!\!\!\!& Nn\log\rho+h(X_{2a}^n|\bar{Y}^n_1)+h(X_{2b}^n|X_{2a}^n,\bar{Y}^n_1)+n~o(\log\rho)+o(n)\\
&\!\!\!\!\leq\!\!\!\!& Nn\log\rho+h(X_{2a}^n|X_{2c}^n)+h(X_{2b}^n|X_{2a}^n,X_{2c}^n)+n~o(\log\rho)+o(n)\\
&\!\!\!\!\leq\!\!\!\!&
Nn\log\rho+nR_2-h(X_{2c}^n)+n~o(\log\rho)+o(n).
\end{eqnarray}
The derivation above is similar to that for the previous bound. Note
that the last inequality also follows that subject to the noise
distortion, we can decode the message $W_2$ from the observation of
$(X_{2a}^n,X_{2b}^n,X_{2c}^n)$. By averaging over the user indices,
we obtain the second inequality as follows.
\begin{eqnarray}\label{eqn:3by4_ob2}
3nR\leq Nn\log\rho+nR-h(X_{c}^n)+n~o(\log\rho)+o(n).
\end{eqnarray}

Similarly, if a genie provides the signal set
$\mathcal{G}_3=\{X_{3b}^n,X_{3c}^n\}$ to Receiver 1, then we have:
\begin{eqnarray}
nR_{\Sigma}&\!\!\!\!\leq\!\!\!\!& I(W_1,W_2,W_3;\bar{Y}^n_1,X_{3b}^n,X_{3c}^n)+n~o(\log\rho)+o(n)\\
&\!\!\!\!\leq\!\!\!\!& Nn\log\rho+h(X_{3b}^n,X_{3c}^n|\bar{Y}^n_1)+n~o(\log\rho)+o(n)\\
&\!\!\!\!=\!\!\!\!& Nn\log\rho+h(X_{3c}^n|\bar{Y}^n_1)+h(X_{3b}^n|X_{3c}^n,\bar{Y}^n_1)+n~o(\log\rho)+o(n)\\
&\!\!\!\!\leq\!\!\!\!& Nn\log\rho+h(X_{3c}^n|X_{3a}^n)+h(X_{3b}^n|X_{3c}^n,X_{3a}^n)+n~o(\log\rho)+o(n)\\
&\!\!\!\!\leq\!\!\!\!&
Nn\log\rho+nR_3-h(X_{3a}^n)+n~o(\log\rho)+o(n).
\end{eqnarray}
where the last inequality follows that subject to the noise
distortion we can decode $W_3$ from the observation of
$(X_{3a}^n,X_{3b}^n,X_{3c}^n)$. By averaging over the user indices,
we obtain the third inequality:
\begin{eqnarray}\label{eqn:3by4_ob3}
3nR\leq Nn\log\rho+nR-h(X_{a}^n)+n~o(\log\rho)+o(n).
\end{eqnarray}

Adding up the three inequalities (\ref{eqn:3by4_ob1}),
(\ref{eqn:3by4_ob2}) and (\ref{eqn:3by4_ob3}), we have:
\begin{eqnarray}
9nR\leq 3Nn\log\rho+2nR+n~o(\log\rho)+o(n).\label{eqn:3by4_ob4}
\end{eqnarray}
By arranging terms of (\ref{eqn:3by4_ob4}) we have:
\begin{eqnarray}
7nR\leq 3Nn\log\rho+n~o(\log\rho)+o(n).\label{eqn:3by4_ob5}
\end{eqnarray}
By dividing $\log\rho$ and $n$ on both sides of
(\ref{eqn:3by4_ob5}), and letting $\rho\rightarrow \infty$ and
$n\rightarrow \infty$, we obtain:
\begin{eqnarray}
d\leq \frac{3N}{7}=\frac{12}{7}.
\end{eqnarray}

\subsection{Case: $(M,N)=(4,5)~\Rightarrow$ DoF $\leq \frac{20}{9}$}

We show the third example is this subsection. Let us consider the
$(M,N)=(4,5)$ setting. Similar to the two examples we have shown, we
again take linear transformations introduced in Section
\ref{sec:outerbound}, thus obtaining the network with resulting
connectivity shown in Fig.\ref{fig:4_by_5_mimo2}, and the equivalent
channel coefficients are non-zero.

\begin{figure}[!h] \vspace{-0.1in}\centering
\includegraphics[width=4.5in]{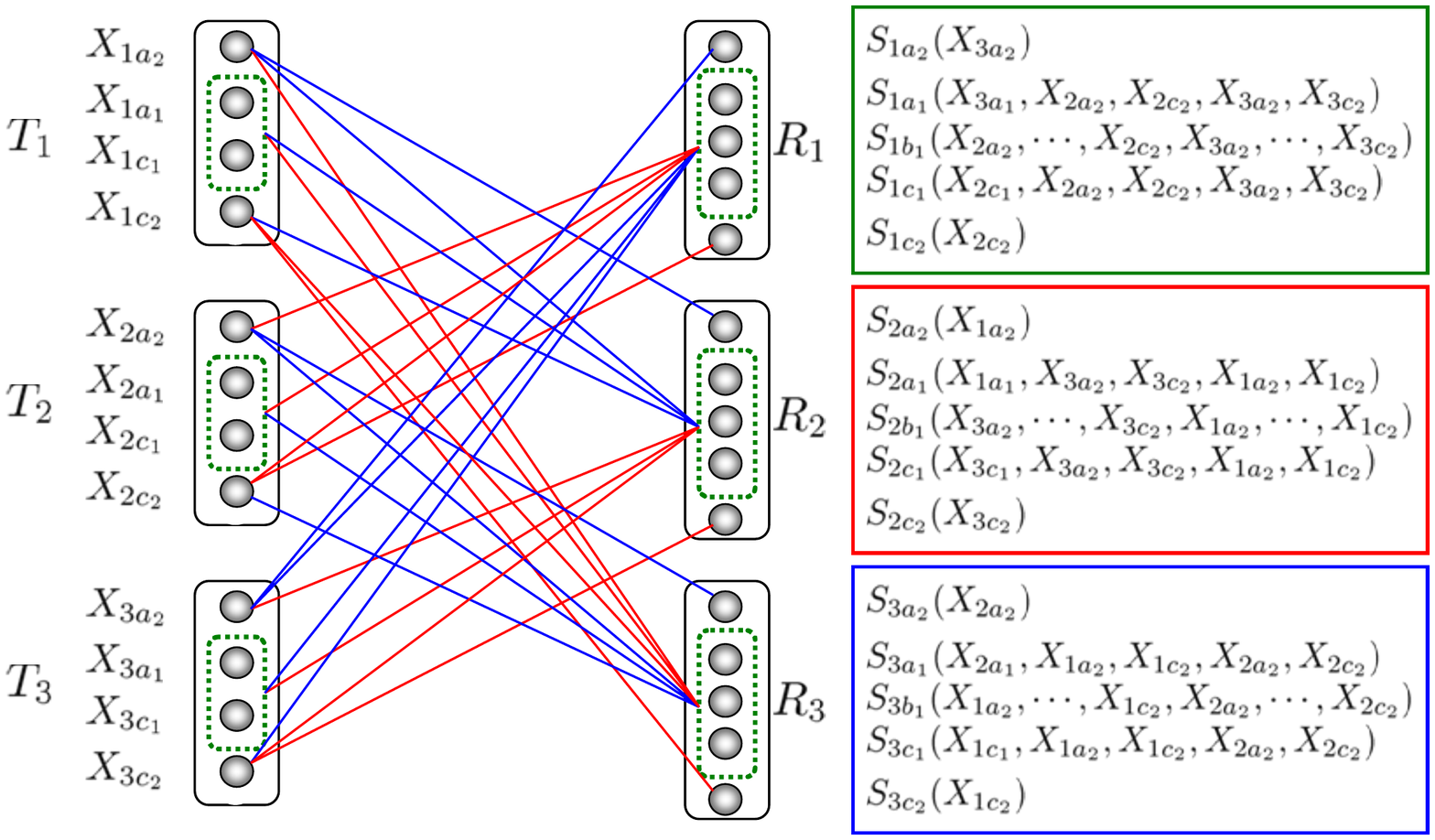}\vspace{-0.05in}
\caption{Three-User $4\times 5$ MIMO Interference
Channel}\label{fig:4_by_5_mimo2} \vspace{-0.1in}
\end{figure}

As what we have shown in Section \ref{sec:outerbound} that the
linear transformations of the $4\times 5$ setting follow from a
recursive manner (onion peeling insights), in this subsection, the
readers can easily observe that the genie signals in this network
also have a recursive relationship with those for the $(M,N)=(2,3)$
setting. We will emphasize this observation in the next subsection.

First, a genie provides the signal set
$\mathcal{G}_1=\{X_{2a_2}^n,X_{3c_2}^n,X_{2a_1}^n\}$ to Receiver 1.
Let us consider the sum rate of three messages:
\begin{eqnarray}
nR_{\Sigma}&\!\!\!\!\leq\!\!\!\!& I(W_1,W_2,W_3;\bar{Y}^n_1,X_{2a_2}^n,X_{3c_2}^n,X_{2a_1}^n)+n~o(\log\rho)+o(n)\label{eqn:4by5_ob1_fano}\\
&\!\!\!\!\leq\!\!\!\!& Nn\log\rho+h(X_{2a_2}^n,X_{3c_2}^n,X_{2a_1}^n|\bar{Y}^n_1)+n~o(\log\rho)+o(n)\label{eqn:4by5_ob1_lemma}\\
&\!\!\!\!\leq\!\!\!\!& Nn\log\rho+h(X_{2a_2}^n|\bar{Y}^n_1)+h(X_{3c_2}^n|\bar{Y}^n_1)+h(X_{2a_1}^n|\bar{Y}^n_1,X_{2a_2}^n,X_{3c_2}^n)+n~o(\log\rho)+o(n)\\
&\!\!\!\!\leq\!\!\!\!& Nn\log\rho+h(X_{2a_2}^n|X_{2c_2}^n)+h(X_{3c_2}^n)+h(X_{2a_1}^n|X_{2a_2}^n,X_{2c_2}^n,X_{2c_1}^n)+n~o(\log\rho)+o(n)\label{eqn:4by5_ob1_drop}\\
&\!\!\!\!=\!\!\!\!&
Nn\log\rho+h(X_{3c_2}^n)+nR_2-h(X_{2c_2}^n)-h(X_{2c_1}^n|X_{2a_2}^n,X_{2c_2}^n)+n~o(\log\rho)+o(n)\label{eqn:4by5_ob1_rate}
\end{eqnarray}
where (\ref{eqn:4by5_ob1_fano}) follows from Fano's inequality, and
(\ref{eqn:4by5_ob1_lemma}) is due to Lemma \ref{lemma:genie}.
(\ref{eqn:4by5_ob1_drop}) is obtained because dropping conditioning
terms does not decrease the differential entropy. By averaging over
the user indices, we therefore obtain the first inequality as
follows.
\begin{eqnarray}
3nR&\!\!\!\!\leq\!\!\!\!& Nn\log\rho+h(X_{c_2}^n)+nR-h(X_{c_2}^n)-h(X_{c_1}^n|X_{a_2}^n,X_{c_2}^n)+n~o(\log\rho)+o(n)\notag\\
&\!\!\!\!\!\!\!\!&Nn\log\rho+nR-h(X_{c_1}^n|X_{a_2}^n,X_{c_2}^n)+n~o(\log\rho)+o(n).\label{eqn:4by5_ob1}
\end{eqnarray}

Second, a genie provides the signal set
$\mathcal{G}_2=\{X_{2a_2}^n,X_{3c_2}^n,X_{3c_1}^n\}$ to Receiver 1.
With the similar analysis we have the sum rate bound:
\begin{eqnarray}
nR_{\Sigma}&\!\!\!\!\leq\!\!\!\!& I(W_1,W_2,W_3;\bar{Y}^n_1,X_{2a_2}^n,X_{3c_2}^n,X_{3c_1}^n)+n~o(\log\rho)+o(n)\\
&\!\!\!\!\leq\!\!\!\!& Nn\log\rho+h(X_{2a_2}^n,X_{3c_2}^n,X_{3c_1}^n|\bar{Y}^n_1)+n~o(\log\rho)+o(n)\\
&\!\!\!\!\leq\!\!\!\!& Nn\log\rho+h(X_{2a_2}^n)+h(X_{3c_2}^n)+h(X_{3c_1}^n|\bar{Y}^n_1,X_{2a_2}^n,X_{3c_2}^n)+n~o(\log\rho)+o(n)\\
&\!\!\!\!\leq\!\!\!\!&
Nn\log\rho+h(X_{2a_2}^n)+h(X_{3c_2}^n)+h(X_{3c_1}^n|X_{3a_2}^n,X_{3c_2}^n)+n~o(\log\rho)+o(n)
\end{eqnarray}
where the last inequality is obtained because knowing $\bar{Y}_1^n$
we can decode $W_1$, and thus we obtain $S_{1a_2}^n$ which is a
noisy version of $X_{3a_2}^n$. Then we use the fact that dropping
the condition terms cannot decrease the differential entropy. By
averaging over the user indices, we have the second inequality:
\begin{eqnarray}\label{eqn:4by5_ob2}
3nR\leq
Nn\log\rho+h(X_{a_2}^n)+h(X_{c_2}^n)+h(X_{c_1}^n|X_{a_2}^n,X_{c_2}^n)+n~o(\log\rho)+o(n).
\end{eqnarray}

Third, a genie provides signals
$\mathcal{G}_3=\{X_{2a_2}^n,X_{2a_1}^n,X_{2c_1}^n\}$ to Receiver 1.
Thus, we have:
\begin{eqnarray}
nR_{\Sigma}&\!\!\!\!\leq\!\!\!\!& I(W_1,W_2,W_3;\bar{Y}^n_1,X_{2a_2}^n,X_{2a_1}^n,X_{2c_1}^n)+n~o(\log\rho)+o(n)\\
&\!\!\!\!\leq\!\!\!\!& Nn\log\rho+h(X_{2a_2}^n,X_{2a_1}^n,X_{2c_1}^n|\bar{Y}^n_1)+n~o(\log\rho)+o(n)\\
&\!\!\!\!\leq\!\!\!\!& Nn\log\rho+h(X_{2a_2}^n,X_{2a_1}^n,X_{2c_1}^n|X_{2c_2}^n)+n~o(\log\rho)+o(n)\\
&\!\!\!\!=\!\!\!\!& Nn\log\rho+nR_2-h(X_{2c_2}^n)+n~o(\log\rho)+o(n)
\end{eqnarray}
and by averaging over the user indices, we have the third
inequality:
\begin{eqnarray}\label{eqn:4by5_ob3}
3nR\leq Nn\log\rho+nR-h(X_{c_2}^n)+n~o(\log\rho)+o(n).
\end{eqnarray}

At last, a genie provides signals
$\mathcal{G}_4=\{X_{3a_1}^n,X_{3c_1}^n,X_{3c_2}^n\}$ to Receiver 1,
and we have:
\begin{eqnarray}
nR_{\Sigma}&\!\!\!\!\leq\!\!\!\!& I(W_1,W_2,W_3;\bar{Y}^n_1,X_{3a_1}^n,X_{3c_1}^n,X_{3c_2}^n)+n~o(\log\rho)+o(n)\\
&\!\!\!\!\leq\!\!\!\!& Nn\log\rho+h(X_{3a_1}^n,X_{3c_1}^n,X_{3c_2}^n|\bar{Y}^n_1)+n~o(\log\rho)+o(n)\\
&\!\!\!\!\leq\!\!\!\!& Nn\log\rho+h(X_{3a_1}^n,X_{3c_1}^n,X_{3c_2}^n|X_{3a_2}^n)+n~o(\log\rho)+o(n)\\
&\!\!\!\!=\!\!\!\!& Nn\log\rho+nR_3-h(X_{3a_2}^n)+n~o(\log\rho)+o(n)
\end{eqnarray}
and we obtain the last inequality after averaging over the user
indices:
\begin{eqnarray}\label{eqn:4by5_ob4}
3nR\leq Nn\log\rho+nR-h(X_{a_2}^n)+n~o(\log\rho)+o(n).
\end{eqnarray}

Now add up  all the sum rate inequalities in (\ref{eqn:4by5_ob1}), (\ref{eqn:4by5_ob2}), (\ref{eqn:4by5_ob3}) and
(\ref{eqn:4by5_ob4}) we have:
\begin{eqnarray}
12nR\leq 4Nn\log\rho+3nR+n~o(\log\rho)+o(n).\label{eqn:4by5_ob5}
\end{eqnarray}
By arranging terms of (\ref{eqn:4by5_ob5}) we have:
\begin{eqnarray}
9nR_{\Sigma}\leq 4Nn\log\rho+n~o(\log\rho)+o(n).\label{eqn:4by5_ob6}
\end{eqnarray}
By dividing $\log\rho$ and $n$ on both sides of
(\ref{eqn:4by5_ob6}), and letting $\rho\rightarrow \infty$ and
$n\rightarrow \infty$, we obtain:
\begin{eqnarray}
d\leq\frac{4N}{9}=\frac{20}{9}.
\end{eqnarray}

%{\it Remark:} By providing the genie signal $\mathcal{G}$ to
%Receiver 1, it can decode all three messages from the observation
%$(\bar{Y}_1^n,\mathcal{G})$. Thus, we can always start the
%derivation of the inequalities by expanding the mutual information
%term $I(W_1,W_2,W_3;\bar{Y}^n_1,\mathcal{G})$ as follows
%\begin{eqnarray*}
%I(W_1,W_2,W_3;\bar{Y}^n_1,\mathcal{G})
%&=& I(W_1,W_2,W_3;\bar{Y}^n_1)+I(W_1,W_2,W_3;\mathcal{G}|\bar{Y}^n_1)\\
%&\leq& Nn\log\rho+h(\mathcal{G}|\bar{Y}^n_1)-h(\mathcal{G}|W_1,W_2,W_3,\bar{Y}^n_1)\\
%&=& Nn\log\rho+h(\mathcal{G}|\bar{Y}^n_1)+n~o(\log\rho)+o(n).
%\end{eqnarray*}
%Thus, in the remaining part of this paper, when we derive the inequality,
%we directly start from the the term of
%$Nn\log\rho+I(\mathcal{G}|\bar{Y}^n_1)+n~o(\log\rho)+o(n)$ on the
%right-hand side.

\subsection{Onion Peeling: The Intuition of the DoF Outer Bound Proof}

While we have only shown the information theoretic DoF outer bounds for $(M,N)=(p,p+1),~p=2,3,4$ settings, it is not difficult to see that they share some similarities in both genie signals and sum rate bounding techniques. In this subsection, we will specify these similarities and build the intuitions behind the outer bound derivations so as to solve general $(M,N)=(p,p+1)$ cases where $p\in\mathbb{Z}$. There are two questions that we will answer:

\begin{enumerate}
\item How should we design the genie signal sets with minimum
cardinality such that each receiver with each genie signal set can
decode all three messages?
\item In order to finalize the converse arguments, how many sum rate bounds do we need totally?
\end{enumerate}

{\bf Answer to Question 1:} In the three examples that we have
shown, we always provide genie signals to Receiver 1 so as to upper
bound the sum rate of three users, then we obtain the other two sum
rate inequalities by advancing user indices and eventually have one
inequality by averaging over three users for compact notations. That
is to say, each receiver needs enough extra signaling dimensions
provided by the genie to distinguish the signal vectors sent from
the other two users as well. For an $M\times N$ interference
channel, since each receiver is able to decode its own message, it
can subtract the signal carrying its desired message and obtains $N$
linearly independent combinations of the interference signals from
the $2M$ transmit antennas (dimensions) of the two interferers.
Thus, in order for the receiver to decode the other two undesired
messages, a genie needs to provide a signal set containing at least
$\max(0,2M\!\!-\!\!N)$ linearly independent dimensional signals to
that receiver such that it can recover the signal vectors sent from
the two interferers directly or by an invertible linear
transformation subject to the noise distortion. Therefore, as in
$(M,N)=(2,3),~(3,4),(4,5)$ cases, each genie signal set has one,
two, three dimensions, respectively.

After knowing each genie signal set has $\max(0,2M\!\!-\!\!N)$
dimensions, we still need to specify which $\max(0,2M\!\!-\!\!N)$
dimensions that a genie provides. Let us first recall $2\times 3$
and $4\times 5$ cases. Note that we only consider providing genie
signals to Receiver 1 because by advancing user indices we can
circularly symmetrically obtain that for Receiver 2 and 3. For the
$2\times 3$ setting, a genie provides $X_{2a}^n$ and $X_{3c}^n$,
respectively, one for each time. For the $4\times 5$ setting, if we
peel out the two antennas in the outer shell, i.e., by providing the
corresponding transmitted signals $(X_{2a_2}^n,X_{3c_2}^n)$ in the
information theoretic statements, the remaining part is an embedded
$(M,N)=(2,3)$ setting. For the $2\times 3$ setting, the two genie
signals are $X_{2a_2}^n$ and $X_{3c_2}^n$, respectively, and thus
for the $4\times 5$ setting, at least we should have two sets of
genie signals, which include $(X_{2a_2}^n,X_{3c_2}^n)$ and one of
$X_{2a_2}^n$ and $X_{3c_2}^n$, respectively. After doing this, the
remaining task is to bound the sum rate of the signals sent from the
peeled out layer. For example, in the $(M,N)=(4,5)$ case, we have
four sets of genie signals, two more than that in the $(M,N)=(2,3)$
case. In fact, for the general $(M,N)=(p,p+1)$ case, as in the
proofs that we show for general cases, the genie provides a total of
$p$ sets of signals to Receiver 1. In the $p-2$ sets among them, the
signals include that designed for the $(p-2,p-1)$ setting plus the
two signals sent from the first antenna of Transmitter 2 and the
last antenna sent from Transmitter 3.

%\begin{figure}[!h] \vspace{-0.1in}\centering
%\includegraphics[width=5.5in]{onion_peeling}\vspace{-0.05in}
%\caption{Intuition of Onion Peeling, (a) $M$ is even, (b) $M$ is
%odd} \vspace{-0.1in} \label{fig:onion_peeling}
%\end{figure}

%If we extend the analysis above to more general $(M,N)=(p,p+1)$
%interference network where $p$ is an even number, we can take
%invertible linear transformations iteratively, and eventually obtain
%the core $2\times 3$ interference channel. As shown in
%Fig.\ref{fig:onion_peeling}, given all the signals from transmit
%antennas in outer layers, the remaining is an interference channel
%with less antennas at each user.

{\bf Answer to Question 2:} The number of sum rate inequalities
depends on how many sets of signals that genie provides to Receiver 1
(also for Receiver 2 and 3 by advancing user indices). In the cases
of $(M,N)=(2,3)$ and $(3,4)$, the number of genie signal sets is 2
and 3, respectively, i.e., identical to the value of
$M$\footnote{The number of genie signal sets provided to each
receiver is associated with the length of subspace alignment chain
$\kappa$. In the special $(M,N)=(p,p+1)$ case where
$p\in\mathbb{Z}^+$, this number is equal to $M$.}.
%Adding one out layer (i.e., two antennas) at each user, we also add
%two more sets of genie signals. For each set, out of a total of
%$2M-(M+1)=M-1$ dimensions of the genie signals, one is
%$(X_{2a_L}^n,\cdots,X_{2c_{L\!-\!1}}^n)$, and the other is
%$(X_{3a_{L\!-\!1}}^n,\cdots,X_{3c_L}^n)$.
Essentially, one set of genie signals produces one sum
rate inequality, and thus we have a total of $M$ sum rate
inequalities. From the three examples, it can be seen that in each
inequality, on the right-hand-side, there is an $Nn\log\rho$ term
plus extra terms. These extra terms, if added through the total of
$M$ inequalities, produce the term $(M-1)nR$ subject to the noise
distortion, i.e.,
\begin{eqnarray}
\frac{1}{3}\sum_{k=1}^3\sum_{m=1}^Mh(\mathcal{G}_{km}^M|\bar{Y}_k)=(M-1)nR+n~o(\log\rho)+o(n)
\end{eqnarray}
where $\frac{1}{3}\sum_{k=1}^3$ indicates averaging over user
indices, and $\mathcal{G}_{km}^M$ denotes the $m^{th}$ set of genie
signals provided to Receiver $k$ for the $(p,p+1)$ case. In this
paper we also write $\mathcal{G}_{1m}^M$ as $\mathcal{G}_{m}^M$ or
$\mathcal{G}_{m}$ with respect to Receiver 1 for simplicity if there
is no ambiguity, and $\mathcal{G}_{2m},\mathcal{G}_{3m}$ can be
obtained by advancing user indices of $\mathcal{G}_{m}$. Therefore,
we have the following inequality:
\begin{eqnarray}
3MnR\leq MNn\log\rho+(M-1)nR+n~o(\log\rho)+o(n)
\end{eqnarray}
which implies the DoF per user bound
\begin{eqnarray}
d\leq \frac{MN}{2M+1}=\frac{MN}{M+N}.
\end{eqnarray}

The arguments we show above based on the onion peeling intuition are
summarized in Table \ref{tab:genie_table} where $\mathcal{G}_m^M$
denotes the $m^{th}$ genie signals set provided to Receiver 1 for
the $(M,N)=(p,p+1)$ case. From this table, it can be easily seen
that for the $(M,M+1)$ case, the first $M-2$ genie signal sets,
i.e., $\mathcal{G}_m^M,~m=1,2,\cdots,M-2$, are always equal to
$\mathcal{G}_m^{M-2},~m=1,2,\cdots,M-2$ plus
$(X_{2a_L}^n,X_{3c_L}^n)$. This observation builds the foundation of
the recursive proof that will be shown in detail in Appendix
\ref{sec:outerbound_MsmallerN}. \vspace{-0.1in}
\begin{table}[htb]
\caption{Examples of Genie Signal Sets Provided to Receiver 1 for
the $(M,M+1)$ Setting where $M=2,3,\cdots,7$ (Note that in the
previous analysis, for $L=1$ cases we omit the layer indices
associated with the antennas for
simplicity.)}\label{tab:genie_table}
\centerline{\begin{tabular}{|c|l|l|l|l}\hline Layers & $L=1$ & $L=2$
& $L=3$  & $\cdots$
\\\hline
            & $\mathcal{G}_1^2=\{X_{2a_1}^n\}$ & $\mathcal{G}_1^4=\{\mathcal{G}_1^2,X_{2a_2}^n,X_{3c_2}^n\}$ & $\mathcal{G}_1^6=\{\mathcal{G}_1^4,X_{2a_3}^n,X_{3c_3}^n\}$ & \\
            & $\mathcal{G}_2^2=\{X_{3c_1}^n\}$ & $\mathcal{G}_2^4=\{\mathcal{G}_2^2,X_{2a_2}^n,X_{3c_2}^n\}$ & $\mathcal{G}_2^6=\{\mathcal{G}_2^4,X_{2a_3}^n,X_{3c_3}^n\}$ & \\
$M$ is even &  & $\mathcal{G}_3^4$ & $\mathcal{G}_3^6=\{\mathcal{G}_3^4,X_{2a_3}^n,X_{3c_3}^n\}$ & $\cdots$ \\
$M=2L$      &  & $\mathcal{G}_4^4$ & $\mathcal{G}_4^6=\{\mathcal{G}_4^4,X_{2a_3}^n,X_{3c_3}^n\}$ & \\
            &  &  & $\mathcal{G}_5^6$ & \\
            &  &  & $\mathcal{G}_6^6$ & \\
\hline
            & $\mathcal{G}_1^3=\{X_{2a_1}^n,X_{3c_1}^n\}$ & $\mathcal{G}_1^5=\{\mathcal{G}_1^3,X_{2a_2}^n,X_{3c_2}^n\}$ & $\mathcal{G}_1^7=\{\mathcal{G}_1^5,X_{2a_3}^n,X_{3c_3}^n\}$ & \\
            & $\mathcal{G}_2^3=\{X_{2a_1}^n,X_{2b_1}^n\}$ & $\mathcal{G}_2^5=\{\mathcal{G}_2^3,X_{2a_2}^n,X_{3c_2}^n\}$ & $\mathcal{G}_2^7=\{\mathcal{G}_2^5,X_{2a_3}^n,X_{3c_3}^n\}$ & \\
$M$ is odd  & $\mathcal{G}_3^3=\{X_{3b_1}^n,X_{3c_1}^n\}$ & $\mathcal{G}_3^5=\{\mathcal{G}_3^3,X_{2a_2}^n,X_{3c_2}^n\}$ & $\mathcal{G}_3^7=\{\mathcal{G}_3^5,X_{2a_3}^n,X_{3c_3}^n\}$ & \\
$M=2L+1$    &  & $\mathcal{G}_4^5$ & $\mathcal{G}_4^7=\{\mathcal{G}_4^5,X_{2a_3}^n,X_{3c_3}^n\}$ & $\cdots$ \\
            &  & $\mathcal{G}_5^5$ & $\mathcal{G}_5^7=\{\mathcal{G}_5^5,X_{2a_3}^n,X_{3c_3}^n\}$ & \\
            &  &  & $\mathcal{G}_6^7$ & \\
            &  &  & $\mathcal{G}_7^7$ & \\\hline
\end{tabular}}
\end{table}

For general $(M,N)=(p,p+1)$ where $p>4$ as well as $M/N\neq p/(p+1)$
cases, since the proofs are much more cumbersome but still follow
from the intuition shown above, we defer all of them to Appendix
\ref{sec:outerbound_MsmallerN}.

For the setting $M_T>M_R$, the proofs also follow from the onion
peeling intuition introduced in Section \ref{sec:outerbound}, and
 we defer them to Appendix \ref{sec:outerbound_MbiggerN}.

\section{DoF Achievability}\label{sec:innerbound}

The achievable schemes are based on linear beamforming at the
transmitters and zero forcing at the receivers. Due to the
reciprocity of the linear scheme, without loss of generality, we
only provide achievable schemes for the case when $M_T<M_R$,
so that $M_T=M$ and $M_R=N$.

As we explained in Section \ref{subsec:achieveless1/2}, when
$M/N\leq 1/2$, interference alignment is not needed and zero-forcing
at receivers is sufficient to achieve the optimal DoF. When
$M/N>1/2$, interference is needed. In this regime, as mentioned in
Section \ref{subsec:redundantdim}, only for $M/N
\in\mathcal{B}=\{1/3,3/5,5/7,7/9,\cdots\}$, neither $M$ nor $N$
contains any redundant dimensions. For all other cases, either $M$
or $N$ can be reduced  without losing DoF. Therefore, we can reduce
$M$ or $N$ depending on their ratio such that the ratio becomes one
of the ratios in $\mathcal{B}$ to achieve the same DoF for the
original $M$ and $N$. As a consequence, we only need to provide
achievable schemes for the case when $M/N \in\mathcal{B}$ and
$M/N>1/2$. The achievable schemes for all other cases follow
directly from these schemes.

%
%\subsection{Case: $0<M/N\leq 1/2$ (Zero-forcing)}
%
%Consider $0<M/N\leq 1/2$ in Figure \ref{fig:map_table} where the
%squares are shaded with light yellow and green colors. In fact
%simple zero-forcing is sufficient to achieve the optimal DoF.
%Specifically, let us first consider $0<M/N\leq 1/3$. In this case,
%each user can achieve $M$ DoF by sending $M$ symbols using randomly
%generated beamforming vectors at each transmitter. At each receiver,
%since the number of antennas is no less than the total number of
%antennas of all three transmitters, $N\geq 3M$, each receiver can
%separate all symbols and decode the desired symbols by zero-forcing
%the interference. For the other case where $1/3< M/N\leq 1/2$, it is
%trivial that $N$ DoF per user are achievable. Again this is achieved
%by sending a total of $N$ symbols across all users using randomly
%generated beamforming vectors. Since the receiver has $N$ antennas,
%it can separate all symbols by inverting the channel to decode its
%desired symbols.

%As introduced in Lemma \ref{lemma:in} in Section \ref{sec:results},
%we

\subsection{Case: $M/N>1/2$ and $M/N=(2p-1)/(2p+1)$}\label{subsec:achievek}

The cases $M/N>1/2$ and $M/N \in\mathcal{B}$ correspond to the
settings where each transmitter has $(2p-1)q$ antennas and each
receiver has $(2p+1)q$ antennas, $\forall p, q\in \mathbb{Z}^+,
p\geq 2$. We will show that each user can achieve $pq$ DoF, for a
total of $3pq$ DoF. Since integer DoF are achieved, symbol
extensions or spatial extensions are not needed.

The achievable scheme can be understood intuitively as follows. As
explain in Section \ref{subsec:subspace_chain}, the length of the
subspace alignment chain is $\kappa=\lceil\frac{M}{N-M}\rceil=p$.
Each subspace participated in the chain is $q$ dimensional and thus
a total of $pq$ DoF are sent for a chain, occupying a total of
$(2p+1)p$ dimensions at all receivers. Since there are $(2p+1)p$
dimensions at each receiver, a total of three subspace alignment
chains, each originating from one of three transmitters, can be
packed for this channel, thus achieving $3pq$ DoF. Next, we will
show how to design these three subspace alignment chains. Before we
present the general achievable schemes, let us first consider two
simple cases, i.e., $p=2,3$.

\subsubsection{Case: $(M,N)=(3q,5q)~~ (p=2)$}\label{subsec:achievek=1}

For the $3q\times 5q$ case, we will show each user can achieve $2q$
DoF. Specifically, Transmitter $i$ sends $2q$ independent symbols
using a $3q\times 2q$ beamforming matrix $\mathbf{V}_i$. First note
that the length of the alignment chain is 2 in this case, so there
are two signal spaces that participate in each alignment chain. Due
to symmetry, there are also two signal spaces at each transmitter
that participate in the other two alignment chains, one for each
chain. In addition, since there are three alignment chains, we will
use a superscript to denote the chain index to which the subspaces
belong. Specifically, we use $\mathbf{V}^k_{i(s)}$ where
$k,i\in\{1,2,3\}$ to denote the $s^{th}$ $q$-dimensional subspace at
Transmitter $i$ that participates in the $k^{th}$ chain, and in this
case $s=1$.

Let us start with the first alignment chain which originates from Transmitter 1:
\begin{eqnarray}
{\bf V}^1_{1(1)}\stackrel{\mbox{\tiny Rx 2}}{\longleftrightarrow} {\bf V}^1_{3(1)}.
\end{eqnarray}
Mathematically, the alignment equation is
\begin{eqnarray}
&&\mathbf{H}_{21}\mathbf{V}^1_{1(1)}=\mathbf{H}_{23}\mathbf{V}^1_{3(1)}\label{eqn:align3by5_1}\\
&\Rightarrow&
\underbrace{\left[\mathbf{H}_{21}~~-\mathbf{H}_{23}\right]_{5q
\times
6q}}_{\mathbf{A}_{p}}\underbrace{\left[\begin{array}{c}\mathbf{V}^1_{1(1)}\\
\mathbf{V}^1_{3(1)}\end{array}\right]}_{\mathbf{a}_{p}}=\mathbf{0}.\label{eqn:alignm3by5_1}
\end{eqnarray}
Since $\mathbf{A}_{p}$ is a $5q \times 6q$ matrix, $\mathbf{a}_{p}$
can be obtained as $q$ basis vectors of the null space of
$\mathbf{A}_{p}$. From $\mathbf{a}_{p}$, we can obtain
$\mathbf{V}^1_{1(1)}$ and $\mathbf{V}^1_{3(1)}$.

Then the second alignment chain which originates from Transmitter 2 is
\begin{eqnarray}
{\bf V}^2_{2(1)}\stackrel{\mbox{\tiny Rx 3}}{\longleftrightarrow} {\bf V}^2_{1(1)}.
\end{eqnarray}
Mathematically, the second alignment equation can be written as
\begin{eqnarray}
&&\mathbf{H}_{32}\mathbf{V}^2_{2(1)}=\mathbf{H}_{31}\mathbf{V}^2_{1(1)}\label{eqn:align3by5_2}\\
&\Rightarrow&
\underbrace{\left[\mathbf{H}_{32}~~-\mathbf{H}_{31}\right]}_{\mathbf{B}_{p}}\underbrace{\left[\begin{array}{c}\mathbf{V}^2_{2(1)}\\
\mathbf{V}^2_{1(1)}\end{array}\right]}_{\mathbf{b}_{p}}=\mathbf{0}.\label{eqn:alignm3by5_2}
\end{eqnarray}
Again, since $\mathbf{B}_{p}$ is a $5q\times 6q$,  $\mathbf{b}_{p}$
is chosen as $q$ basis vectors of the null space of
$\mathbf{B}_{p}$, from which we can obtain can obtain
$\mathbf{V}^2_{2(1)}$ and $\mathbf{V}^2_{1(1)}$.

Finally, the third alignment chain which originates from Transmitter 3 is
\begin{eqnarray}
{\bf V}^3_{3(1)}\stackrel{\mbox{\tiny Rx 1}}{\longleftrightarrow}
{\bf V}^3_{2(1)}
\end{eqnarray}
which produces the third alignment equation:
\begin{eqnarray}
&&\mathbf{H}_{13}\mathbf{V}^3_{3(1)}=\mathbf{H}_{12}\mathbf{V}^3_{2(1)}\label{eqn:align3by5_3}\\
&\Rightarrow&
\underbrace{\left[\mathbf{H}_{13}~~-\mathbf{H}_{12}\right]}_{\mathbf{C}_{p}}\underbrace{\left[\begin{array}{c}\mathbf{V}^3_{3(1)}\\
\mathbf{V}^3_{2(1)}\end{array}\right]}_{\mathbf{c}_{p}}=\mathbf{0}.\label{eqn:alignm3by5_3}
\end{eqnarray}
$\mathbf{c}_{p}$ is chosen as $q$ basis vectors of the null
space of $\mathbf{C}_{p}$, from which we can obtain
$\mathbf{V}^3_{3(1)}$ and $\mathbf{V}^3_{2(1)}$.

After aligning interference, we ensure that the dimensions occupied
by the interference is small enough. To decode the desired signals,
the desired signal space cannot overlap with the space of
interference. This is guaranteed by an independent linear
transformation of the desired signals from the transmitter to its
corresponding receiver. Notice that the direct channels are not used
to align interference and they are generic channel matrices without
special structure. Therefore, after going through an independent
linear transformation, the desired signals do not overlap with the
interference almost surely.

\subsubsection{Case: $(M,N)=(5q, 7q)~~(p=3)$}\label{subsec:achievek=2}

Now let us consider the case $5q \times 7q$, for which every user
can achieve $3q$ DoF. Again, we need to design three alignment
chains each with length 3, each originating from one of three
transmitters.

The first alignment chain which originates from Transmitter 1 is
\begin{eqnarray*}
{\bf V}^1_{1(1)}\stackrel{\mbox{\tiny Rx 2}}{\longleftrightarrow} {\bf
V}^1_{3(1)}\stackrel{\mbox{\tiny Rx 1}}{\longleftrightarrow} {\bf
V}^1_{2(1)}.
\end{eqnarray*}
Mathematically, the alignment equations are
\begin{eqnarray}
&&\left\{\begin{array}{ccc}\mathbf{H}_{21}\mathbf{V}^1_{1(1)}&=&\mathbf{H}_{23}\mathbf{V}^1_{3(1)}\\
\mathbf{H}_{13}\mathbf{V}^1_{3(1)}&=&\mathbf{H}_{12}\mathbf{V}^1_{2(1)}\end{array} \right.\\
&\Rightarrow&
\underbrace{\left[\begin{array}{ccc}\mathbf{H}_{21}&-\mathbf{H}_{23}&\mathbf{0}\\
\mathbf{0}&\mathbf{H}_{13}
&-\mathbf{H}_{12}\end{array}\right]_{14q\times
15q}}_{\mathbf{A}_{p}}
\underbrace{\left[\begin{array}{c}\mathbf{V}^1_{1(1)}\\
\mathbf{V}^1_{3(1)} \\
\mathbf{V}^1_{2(1)}\end{array}\right]}_{\mathbf{a}_{p}}=\mathbf{0}.\label{eqn:alignsol5by7_1}
\end{eqnarray}
Note that $\mathbf{A}_{p}$ is a $14q\times 15q$ matrix. Therefore,
$\mathbf{a}_{p}$ is chosen as the $q$ basis of the null space of
$\mathbf{A}_{p}$, from which we can obtain can obtain
$\mathbf{V}^1_{1(1)}, \mathbf{V}^1_{2(1)}, \mathbf{V}^1_{3(1)}$.

Then, the second alignment chain which originates from Transmitter 2 is
\begin{eqnarray*}
{\bf V}^2_{2(1)}\stackrel{\mbox{\tiny Rx 3}}{\longleftrightarrow} {\bf
V}^2_{1(1)}\stackrel{\mbox{\tiny Rx 2}}{\longleftrightarrow} {\bf
V}^2_{3(1)}.
\end{eqnarray*}
Mathematically, we have the alignment equations
\begin{eqnarray}
&&\left\{\begin{array}{ccc}\mathbf{H}_{32}\mathbf{V}^2_{2(1)}&=&\mathbf{H}_{31}\mathbf{V}^2_{1(1)}\\
\mathbf{H}_{21}\mathbf{V}^2_{1(1)}&=&\mathbf{H}_{23}\mathbf{V}^2_{3(1)}\end{array} \right.\\
&\Rightarrow&
\underbrace{\left[\begin{array}{ccc}\mathbf{H}_{32}&-\mathbf{H}_{31}&\mathbf{0}\\
\mathbf{0}&\mathbf{H}_{21}
&-\mathbf{H}_{23}\end{array}\right]_{14q\times
15q}}_{\mathbf{B}_{p}}
\underbrace{\left[\begin{array}{c}\mathbf{V}^2_{2(1)}\\
\mathbf{V}^2_{1(1)} \\
\mathbf{V}^2_{3(1)}\end{array}\right]}_{\mathbf{b}_p}=\mathbf{0}.\label{eqn:alignsol5by7_1}
\end{eqnarray}
Similarly, $\mathbf{b}_p$ is chosen as the $q$ basis of the null space of
$\mathbf{B}_p$, from which we can obtain can obtain $\mathbf{V}^2_{2(1)}, \mathbf{V}^2_{1(1)}, \mathbf{V}^2_{3(1)}$.

Finally, the third alignment chain which originates from Transmitter 3 is
\begin{eqnarray*}
{\bf V}^3_{3(1)}\stackrel{\mbox{\tiny Rx 1}}{\longleftrightarrow} {\bf
V}^3_{2(1)}\stackrel{\mbox{\tiny Rx 3}}{\longleftrightarrow} {\bf
V}^3_{1(1)}.
\end{eqnarray*}
This subspace alignment chain produces the following alignment
equations:
\begin{eqnarray}
&&\left\{\begin{array}{ccc}\mathbf{H}_{13}\mathbf{V}^3_{3(1)}&=&\mathbf{H}_{12}\mathbf{V}^3_{2(1)}\\
\mathbf{H}_{32}\mathbf{V}^2_{2(1)}&=&\mathbf{H}_{31}\mathbf{V}^3_{1(1)}\end{array} \right.\\
&\Rightarrow&
\underbrace{\left[\begin{array}{ccc}\mathbf{H}_{13}&-\mathbf{H}_{12}&\mathbf{0}\\
\mathbf{0}&\mathbf{H}_{32}
&-\mathbf{H}_{31}\end{array}\right]_{14q\times
15q}}_{\mathbf{C}_{p}}
\underbrace{\left[\begin{array}{c}\mathbf{V}^3_{3(1)}\\
\mathbf{V}^3_{2(1)} \\
\mathbf{V}^3_{1(1)}\end{array}\right]}_{\mathbf{c}_p}=\mathbf{0}.\label{eqn:alignsol5by7_1}
\end{eqnarray}
$\mathbf{c}_p$ is chosen as the $q$ basis of the null space of
$\mathbf{C}_p$, from which we can obtain can obtain
$\mathbf{V}^3_{3(1)}$, $\mathbf{V}^3_{2(1)}$ and
$\mathbf{V}^3_{1(1)}$.

\subsubsection{Case: $(M,N)=\left((2p-1)q,(2p+1)q\right)$}\label{subsec:integer_dof}

Now, consider the general $(2p-1)q\times (2p+1)q$ case. Again, we
will design three $q$-dimensional subspace alignment chains with
length $p$, each originating from one of three transmitters. Let
$\mathbf{V}^k_{i(s)}$, $k,i\in\{1,2,3\},
s\in\{1,\cdots,\lceil{p/3}\rceil\}$  denote the $s^{th}$
$q$-dimensional subspace at Transmitter $i$ that participates in the
$k^{th}$ subspace alignment chain. In addition, let
$\underline{p}=(p\mod 3)$ and $\overline{p}=\lceil{p/3}\rceil$.
Next, we can write three subspace alignment chains.

We start with the first alignment chain which originates from Transmitter 1. Let $a(1)=1, a(2)=3, a(0)=2$.
Then the first alignment chain is
\begin{eqnarray}\label{eqn:chain1}
{\bf V}^1_{1(1)}\stackrel{\mbox{\tiny Rx 2}}{\longleftrightarrow}
{\bf V}^1_{3(1)}\stackrel{\mbox{\tiny Rx 1}}{\longleftrightarrow}
{\bf V}^1_{2(1)}\stackrel{\mbox{\tiny Rx 3}}{\longleftrightarrow}
{\bf V}^1_{1(2)}\stackrel{\mbox{\tiny Rx 2}}{\longleftrightarrow}
{\bf V}^1_{3(2)}\stackrel{\mbox{\tiny Rx 1}}{\longleftrightarrow}
{\bf V}^1_{2(2)}\cdots {\bf
V}^1_{a(\underline{p-1})(\overline{p-1})} \stackrel{\mbox{\tiny Rx
$a(\underline{p+1})$}}{\longleftrightarrow} {\bf
V}^1_{a(\underline{p})(\overline{p})}.
\end{eqnarray}
Mathematically, we have the following alignment equation:
\begin{eqnarray}\label{eqn:align_eqn1}
\mathbf{A}_p\mathbf{a}_p=\mathbf{0}
\end{eqnarray}
where
\begin{eqnarray}
\mathbf{A}_p=\left[\begin{array}{cccccc}\mathbf{H}_{21}&-\mathbf{H}_{23}&\mathbf{0}&\mathbf{0}&\mathbf{0}&\mathbf{0}\\
\mathbf{0}&\mathbf{H}_{13}&-\mathbf{H}_{12}&\mathbf{0}&\mathbf{0}&\mathbf{0}\\
\mathbf{0}&\mathbf{0} &\mathbf{H}_{32} &-\mathbf{H}_{31}&\mathbf{0}&\mathbf{0}\\
\mathbf{0}&\mathbf{0}&\ddots&\ddots&\ddots&\mathbf{0}\\
\mathbf{0}&\mathbf{0}&\cdots&\mathbf{0}&\mathbf{H}_{a(\underline{p+1})a(\underline{p-1})}&-\mathbf{H}_{a(\underline{p+1})a(\underline{p})}\end{array}\right],\quad
\mathbf{a}_p=\left[\begin{array}{c}{\bf V}^1_{1(1)}\\ {\bf
V}^1_{3(1)}\\ \vdots\\ {\bf
V}^1_{a(\underline{p})(\overline{p})}\end{array} \right].
\end{eqnarray}
Since $\mathbf{A}_p$ is a $(p-1)(2p+1)q \times p(2p-1)q$ matrix, $q$
columns of $\mathbf{a}_p$ are chosen as the $q$ basis vectors of the
null space of $\mathbf{A}_p$, from which we can obtain
$\mathbf{V}^1_{1(1)}, \cdots, {\bf
V}^1_{a(\underline{p})(\overline{p})}$.

Similarly, let $b(1)=2, b(2)=1, b(0)=3$. Then the second alignment chain which originates from Transmitter 2 is
\begin{eqnarray}\label{eqn:chain2}
{\bf V}^2_{2(1)}\stackrel{\mbox{\tiny Rx 3}}{\longleftrightarrow}
{\bf V}^2_{1(1)}\stackrel{\mbox{\tiny Rx 2}}{\longleftrightarrow}
{\bf V}^2_{3(1)}\stackrel{\mbox{\tiny Rx 1}}{\longleftrightarrow}
{\bf V}^2_{2(2)}\stackrel{\mbox{\tiny Rx 3}}{\longleftrightarrow}
{\bf V}^2_{1(2)}\stackrel{\mbox{\tiny Rx 2}}{\longleftrightarrow}
{\bf V}^2_{3(2)} \cdots {\bf
V}^2_{b(\underline{p-1})(\overline{p-1})} \stackrel{\mbox{\tiny Rx
$b(\underline{p+1})$}}{\longleftrightarrow} {\bf
V}^2_{b(\underline{p})(\overline{p})}.
\end{eqnarray}
Mathematically, we have the second alignment equation:
\begin{eqnarray}\label{eqn:align_eqn2}
\mathbf{B}_p\mathbf{b}_p=\mathbf{0}
\end{eqnarray}
where
\begin{eqnarray}
\mathbf{B}_p=\left[\begin{array}{cccccc}\mathbf{H}_{32}&-\mathbf{H}_{31}&\mathbf{0}&\mathbf{0}&\mathbf{0}&\mathbf{0}\\
\mathbf{0}&\mathbf{H}_{21}&-\mathbf{H}_{23}&\mathbf{0}&\mathbf{0}&\mathbf{0}\\
\mathbf{0}&\mathbf{0} &\mathbf{H}_{13} &-\mathbf{H}_{12}&\mathbf{0}&\mathbf{0}\\
\mathbf{0}&\mathbf{0}&\ddots&\ddots&\ddots&\mathbf{0}\\
\mathbf{0}&\mathbf{0}&\cdots&\mathbf{0}&\mathbf{H}_{b(\underline{p+1})b(\underline{p-1})}&-\mathbf{H}_{b(\underline{p+1})b(\underline{p})}\end{array}\right],\quad \mathbf{b}_p=\left[\begin{array}{c}{\bf V}^2_{2(1)}\\ {\bf
V}^2_{1(1)}\\ \vdots\\ {\bf V}^2_{b(\underline{p})(\overline{p})}\end{array} \right].
\end{eqnarray}
Again, since $\mathbf{B}_p$ is a $(p-1)(2p+1)q \times p(2p-1)q$ matrix, $q$ columns of $\mathbf{b}_p$ are chosen as the $q$ basis vectors of the null space of $\mathbf{B}_p$, from which we can obtain $\mathbf{V}^2_{2(1)}, \cdots, {\bf V}^2_{b(\underline{p})(\overline{p})}$.

Finally, let $c(1)=3, c(2)=2, c(0)=1$. Then the third alignment chain which originates from Transmitter 3 is
\begin{eqnarray}\label{eqn:chain3}
{\bf V}^3_{3(1)}\stackrel{\mbox{\tiny Rx 1}}{\longleftrightarrow}
{\bf V}^3_{2(1)}\stackrel{\mbox{\tiny Rx 3}}{\longleftrightarrow}
{\bf V}^3_{1(1)}\stackrel{\mbox{\tiny Rx 2}}{\longleftrightarrow}
{\bf V}^3_{3(2)}\stackrel{\mbox{\tiny Rx 1}}{\longleftrightarrow}
{\bf V}^3_{2(2)}\stackrel{\mbox{\tiny Rx 3}}{\longleftrightarrow}
{\bf V}^3_{1(2)} \cdots {\bf
V}^3_{c(\underline{p-1})(\overline{p-1})} \stackrel{\mbox{\tiny Rx
$c(\underline{p+1})$}}{\longleftrightarrow} {\bf
V}^3_{c(\underline{p})(\overline{p})}.
\end{eqnarray}
Mathematically, this subspace alignment chain implies the following
alignment equation:
\begin{eqnarray}\label{eqn:align_eqn3}
\mathbf{C}_p\mathbf{c}_p=\mathbf{0}
\end{eqnarray}
where
\begin{eqnarray}
\mathbf{C}_p=\left[\begin{array}{cccccc}\mathbf{H}_{13}&-\mathbf{H}_{12}&\mathbf{0}&\mathbf{0}&\mathbf{0}&\mathbf{0}\\
\mathbf{0}&\mathbf{H}_{32}&-\mathbf{H}_{31}&\mathbf{0}&\mathbf{0}&\mathbf{0}\\
\mathbf{0}&\mathbf{0} &\mathbf{H}_{21} &-\mathbf{H}_{23}&\mathbf{0}&\mathbf{0}\\
\mathbf{0}&\mathbf{0}&\ddots&\ddots&\ddots&\mathbf{0}\\
\mathbf{0}&\mathbf{0}&\cdots&\mathbf{0}&\mathbf{H}_{c(\underline{p+1})c(\underline{p-1})}&-\mathbf{H}_{c(\underline{p+1})c(\underline{p})}\end{array}\right],\quad \mathbf{c}_p=\left[\begin{array}{c}{\bf V}^3_{3(1)}\\ {\bf
V}^3_{2(1)}\\ \vdots\\ {\bf V}^3_{b(\underline{p})(\overline{p})}\end{array} \right].
\end{eqnarray}
$q$ columns of $\mathbf{c}_p$ are chosen as the $q$ basis vectors of the null space of $\mathbf{C}_p$, from which we can obtain $\mathbf{V}^3_{3(1)}, \cdots, {\bf V}^3_{b(\underline{p})(\overline{p})}$.

After the interference is aligned, we ensure the dimension of the
space spanned by the interference is small enough. To make sure each
user can achieve $pq$ DoF, it remains to show 1) the space spanned
by the desired signal and that spanned by the interference do not
overlap, and 2) the constructed beamforming vectors for each user
using the proposed alignment scheme are linearly independent. For
the first one, since the direct channels do not appear in the
alignment equations and are generic without any special structure,
the desired signals and interference do not overlap each other
almost surely. Next we prove the second one.

First note that all entries of the beamforming vectors are ratios of
polynomial functions of the channel coefficients. To prove all the
$(2p-1)q\times pq$ beamforming matrix $\mathbf{V}_i$ at
$\mathbf{V}_i$ at transmitter $i$ is full rank almost surely, it is
sufficient to show the determinant of the following square matrix is
not equal to zero almost surely:
\begin{eqnarray}
\mathbf{V}'_i=\left[\mathbf{V}_i~\mathbf{U}_i\right]
\end{eqnarray}
where $\mathbf{U}_i$ is a randomly generated $(2p-1)q\times (p-1)q$
matrix. Now the determinant of the above matrix is a polynomial
function of its entries. This polynomial is either a zero polynomial
or not equal to zero almost surely, since there are finite number of
solutions of the polynomial equation, which has measure zero.
Therefore, if the polynomial is not a zero polynomial, the
polynomial is not equal to zero almost surely for randomly generated
channel coefficients. Next, we will show the polynomial is not a
zero polynomial. To do that, we only need to find one specific set
of channel coefficients such that the polynomial is not equal to
zero. For brevity, we next construct the channels for the case when
$q=1$. The construction for arbitrary $q$ follows in a straightforward
manner.

Suppose the channel matrix $\mathbf{H}_{(i-1)i}$ is
\begin{eqnarray}
\mathbf{H}_{(i-1)i}= \left[\begin{array}{cc}\mathbf{I}_{(p-1)\times
(p-1)}& \mathbf{0}_{(p-1)\times
p}\\ \mathbf{0}_{1\times (p-1)}&\mathbf{0}_{1\times p}\\
\mathbf{0}_{p\times (p-1)}& \mathbf{I}_{p\times p}\\
\mathbf{0}_{1\times (p-1)}&\mathbf{0}_{1\times p}\end{array}\right],
\forall i\in\{1,2,3\}.
\end{eqnarray}
And the channel matrix $\mathbf{H}_{(i+1)i}$ is
\begin{eqnarray}
\mathbf{H}_{(i+1)i}= \left[\begin{array}{cc}\mathbf{0}_{1\times
(p-1)}&\mathbf{0}_{1\times p}\\
\mathbf{I}_{(p-1)\times (p-1)}&\mathbf{0}_{(p-1)\times p}\\
\mathbf{0}_{1\times (p-1)}&\mathbf{0}_{1\times p} \\
\mathbf{0}_{p\times (p-1)}& \mathbf{I}_{p\times p}
\end{array}\right], \forall i\in\{1,2,3\}.
\end{eqnarray}
It can be verified that the alignment schemes proposed before
produce the following beamforming matrix for each transmitter:
\begin{eqnarray}
\mathbf{V}_i=\left[\begin{array}{c}\mathbf{0}_{(p-1)\times p}\\
\mathbf{I}_{p\times p}\end{array}\right]
\end{eqnarray}
With this full rank beamforming matrix, it can be easily seen that
$\mathbf{V}'_i$ has full rank and its determinant is not equal to
zero almost surely. Therefore, the beamforming vectors of each user
designed by the alignment schemes before are linear independent
almost surely.

\subsection{Achievability for General Cases with only Space Extension}

With the understanding of the achievable schemes for the settings
$M/N=(2p-1)/(2p+1)$ for $p=2,3,\cdots$, we can consider the
achievable schemes for all other cases. Recall that the optimal DoF
given by Theorem \ref{theorem:dof} are as follows:
\begin{eqnarray}\label{eqn:dof}
\sDoF=\left\{\begin{array}{ccc}\frac{p}{2p-1}M,&~& \frac{p-1}{p}\leq\frac{M}{N}\leq \frac{2p-1}{2p+1}\\
\frac{p}{2p+1}N,&~& \frac{2p-1}{2p+1}\leq\frac{M}{N}\leq
\frac{p}{p+1}\end{array} \right. \quad p=2,3,\cdots.
\end{eqnarray}
We will prove that for a finite integer $q$ such that $qd$ is an
integer, the $q M\times q N$ interference channel achieves $qd$ DoF
per user. Therefore, the spatially-normalized DoF are $d$.

Let us first consider the case $\frac{p-1}{p}\leq\frac{M}{N}\leq
\frac{2p-1}{2p+1}$. In this regime, the DoF only depend on $M$,
i.e., $d=\frac{p}{2p-1}M$. In other words, the bottleneck of the DoF
in this regime is the value of $M$. If we fix $M$, that DoF value
does not change even if $N$ changes as long as the ratio remains in
that regime. If $M$ is an integer multiple of $2p-1$, we can reduce
the number of antennas $N$ such that the ratio becomes
$\frac{2p-1}{2p+1}$. And once the ratio is equal to
$\frac{2p-1}{2p+1}$, the achievable scheme proposed in last
subsection can be applied. On the other hand, if $M$ is \emph{not}
an integer multiple of $2p-1$, we scale $M$ and $N$ by a factor of
$2p-1$ to obtain a $(2p-1)M\times (2p-1)N$ interference channel
which is a spatially scaled version of the original channel. Again
for this channel, we reduce the number of antennas at the receiver
from $(2p-1)N$ to $(2p+1)M$ such that the ratio of the number of
transmit and receive antennas is equal to $\frac{2p-1}{2p+1}$. By
applying the achievable scheme in last subsection, we can achieve
$pM$ DoF per user. Therefore, the spatially-normalized DoF of
$\frac{p}{2p-1}M$ per user can be achieved for the original $M
\times N$ interference channel.

Now consider the other case $\frac{2p-1}{2p+1}\leq\frac{M}{N}\leq
\frac{p}{p+1}$, where the optimal DoF per user, $\frac{p}{2p+1}N$,
only depends on $N$. Similar to the previous case, if $N$ is an
integer multiple of $2p+1$, we can simply reduce the number of
transmit antennas (i.e., some transmit antennas are redundant), such
that the ratio becomes $\frac{2p-1}{2p+1}$. Therefore, the
achievable scheme proposed before can be applied here. If $N$ is not
an integer $2p+1$, we can scale $M$ and $N$ by a factor of $2p+1$.
By reducing the number of transmit antennas for the spatially scaled
version of the original channel, we can achieve $pN$ DoF per user,
thus spatially-normalized $\frac{p}{2p+1}N$ DoF for the original
channel.

\subsection{Achievability with Symbol Extensions in Time/Frequency Domains}\label{subsec:symbolex}

While we have shown the achievability for all cases (with spatial
normalization for the cases where fractional DoF are optimal), one
question that remains unresolved is whether the outer bound can be
achieved as well if we restrict to only symbol extensions over
time/frequency, i.e., without spatial normalization. The difficulty
is to deal with the added complexity of block diagonal channel
structure that would result from channel extensions. In other words,
the channels are not generic with symbol extensions. As a
consequence, although the alignment schemes presented before can
still be applied to align interference, it is not clear that
desired signals remain distinguishable from interference after going
through the direct channel due to the special channel structure. In
this subsection, we first explore some simple cases and show that
the outer bound is still achievable for these simple settings with
linear schemes using symbol extensions. Then we will construct a
linear scheme with symbol extensions for general cases. While in
many cases symbol extensions in time over constant channels are
sufficient to achieve the information theoretic DoF outer bound,
there are also cases where time-variations/frequency-selectivity of
the channel is needed with linear interference alignment schemes.
The feasibility of such schemes can be tested through simple
numerical tests. Let us start with some settings where symbol
extensions in time over constant channels are \emph{not} sufficient
to achieve the outer bound, i.e., $(M,N)=(p,p+1)$ where $p\geq 2$.

%In this subsection, we investigate the DoF of the $M_T\times M_R$
%setting using linear beamforming schemes with time extensions over
%{\em constant} channels in this subsection.

\subsubsection{Case: $(M,N)=(p,p+1)$}\label{subsec:ach_symbol_ext_pplus1}
We will first consider the $2\times 3$ setting and analytically
prove that over the {\em constant} channels, the linear scheme
introduced in this paper does not apply to this setting. Such
analysis can be generalized to show that for $(M,N)=(p,p+1)$ cases
where $p\geq 2$, linear schemes with symbol extensions over constant
channels are not sufficient to achieve the DoF outer bound. Then we
will introduce time-variation/frequency-selectivity to the channel
and show that in this case, linear schemes with symbol extensions
can achieve the DoF outer bound.

\begin{figure}[!h] \vspace{-0.1in}\centering
\includegraphics[width=2.5in]{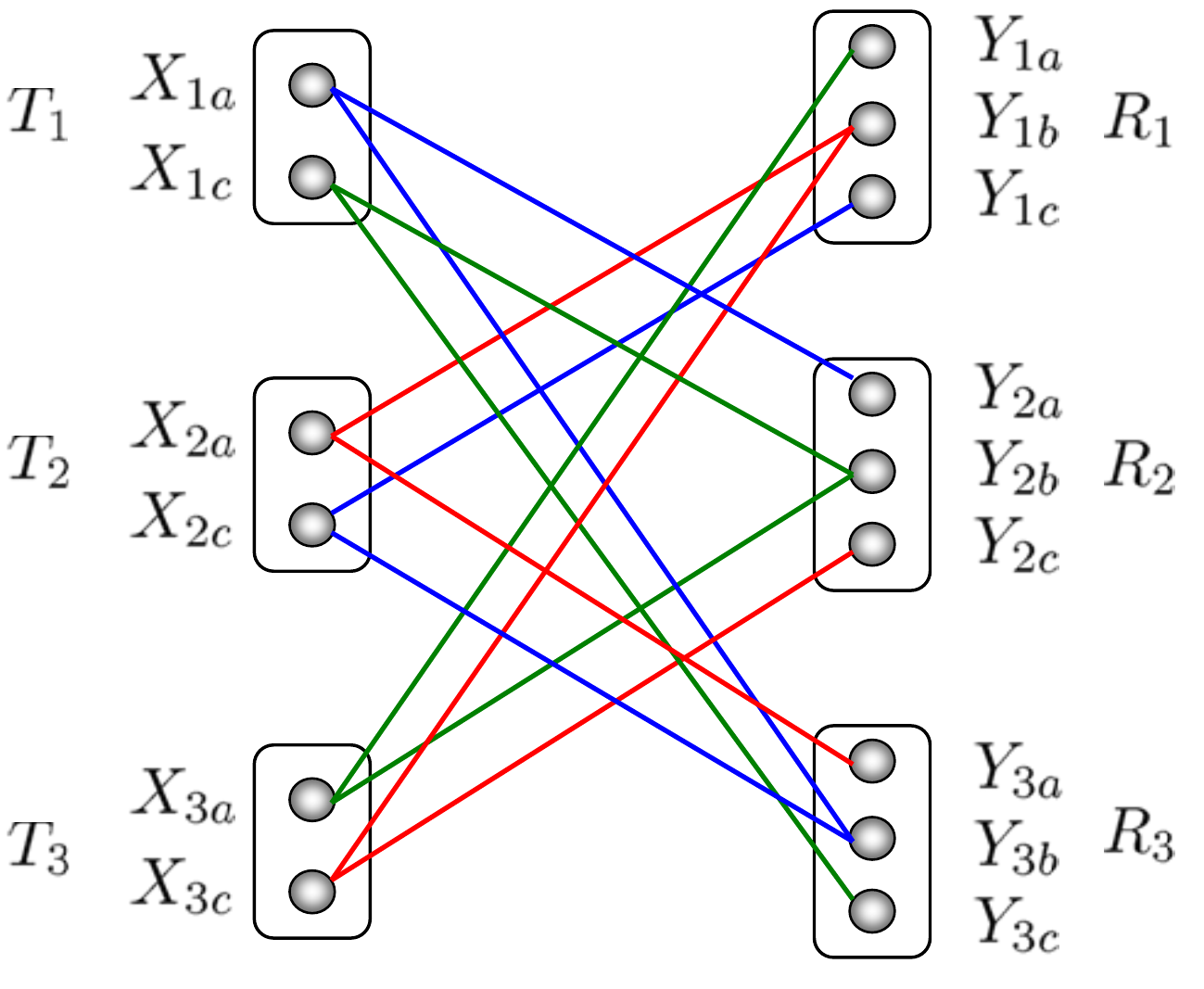}\vspace{-0.05in}
\caption{Normalizing the Interference-carrying Links of the $2\times
3$ Setting to Identity Matrices} \vspace{-0.1in}
\label{fig:2by3_infeasibility}
\end{figure}

Let us start with the $2\times 3$ setting. First let us recall the
invertible linear transformation introduced in Figure
\ref{fig:2_by_3_mimo} in Section \ref{sec:outerbound}. It can be
seen that at Receiver $k$ after subtracting the signal carrying its
desired message, the first antenna only sees $X_{(k-1)a}$ and the
third antenna only sees $X_{(k+1)a}$. Therefore, Receiver $k$ is
able to subtract $X_{(k-1)a},X_{(k+1)a}$ from $S_{kb}$, such that
the resulting channel connectivity becomes the one shown in Figure
\ref{fig:2by3_infeasibility}. In Figure \ref{fig:2by3_infeasibility}
there are three open chains, denoted as blue, green and red colors,
each implying a subspace alignment chain with length 2. Because they
are open loop, we can normalize the channel coefficient of each
segment to be one. As a result, we can write the cross channel
matrices as follows,
\begin{eqnarray}\label{eqn:2by3_infea_inf_links}
{\bf
H}_{k(k+1)}=\left[\begin{array}{cc}0&0\\1&0\\0&1\end{array}\right]~~~~~~{\bf
H}_{k(k-1)}=\left[\begin{array}{cc}1&0\\0&1\\0&0\end{array}\right].
\end{eqnarray}
For simplicity, we still use ${\bf H}_{kk}$ to denote the direct
channel matrix of User $k$ after the change of basis operations. Now
Suppose we can achieve $6/5$ DoF per user, which can be done by
achieving $6$ DoF per user over $5$ time slots. With $5$ symbol
extensions, the effective channel matrix becomes
\begin{eqnarray}
\bar{\mathbf{H}}_{ji}=\mathbf{I}_5\otimes\mathbf{H}_{ji}
\end{eqnarray}
where $\mathbf{I}_5$ is the $5\times 5$ identity matrix. To ensure
each user can separate  6 desired signal vectors from the
interference in its $15$-dimensional signal space, the dimension of
the space spanned by $12$ interference vectors cannot be more than
9. Therefore, at each receiver, we need to align 3 interference
vectors. Note that in the signal space at each receiver, there is a
5-dimensional subspace that can be accessed by two interferers. Then
we can randomly choose 3-dimensional subspace in this common
subspace as the subspace where the 3 interference vectors align.
Mapping these 3-dimensional subspace back to the interferers
determines the beamforming matrix at the transmitter. Since each
transmitter interferes with 2 receivers, each of two unintended
receivers will determine 3 beamforming vectors, for a total of 6
beamforming vectors.

With this intuitive understanding, on the symbol extended channel,
we write the $10\times 6$ beamforming matrix of user $i$,
$\bar{\mathbf{V}}_i$, as
$\bar{\mathbf{V}}_i=\left[\bar{\mathbf{V}}_{i,1}~\bar{\mathbf{V}}_{i,2}\right]$
where $\bar{\mathbf{V}}_{i,1}$ and $\bar{\mathbf{V}}_{i,2}$ are $10
\times 3$ matrices. Then at Receiver 1, we align the first $3$ beams
from Transmitter 2 with those from Transmitter 3. Mathematically, we
have
\begin{eqnarray}
&&\bar{\mathbf{H}}_{13}\bar{\mathbf{V}}_{3,1}=\bar{\mathbf{H}}_{12}\bar{\mathbf{V}}_{2,1}\\
&\Rightarrow&
\underbrace{\left[\bar{\mathbf{H}}_{13}~~-\bar{\mathbf{H}}_{12}\right]}_{\bar{\mathbf{A}}}\underbrace{\left[\begin{array}{c}\bar{\mathbf{V}}_{3,1}\\
\bar{\mathbf{V}}_{2,1}\end{array}\right]}_{\bar{\mathbf{a}}}=\mathbf{0}
\end{eqnarray}
Since $\bar{\mathbf{A}}$ is a $15 \times 20$ matrix,
$\bar{\mathbf{a}}$ can be obtained as 3 linearly independent vectors
in the $5$-dimensional null space of $\bar{\mathbf{A}}$. Notice that
the five basis vectors of the null space of $\bar{\mathbf{A}}$ are
columns of the matrix $\mathbf{I}_5\otimes\mathbf{V}_1 $ where
$\mathbf{V}_1$ is the unique vector in the null space of the
$3\times 4$ matrix $\left[\mathbf{H}_{13}~~-\mathbf{H}_{12}\right]$,
i.e.,
\begin{eqnarray}
\left[\mathbf{H}_{13}~~-\mathbf{H}_{12}\right] \mathbf{V}_1
=\mathbf{0}
\end{eqnarray}
where
\begin{eqnarray}
\left[\mathbf{H}_{13}~-\mathbf{H}_{12}\right]=\left[\begin{array}{rrrr}1&0&0&0\\0&1&-1&0\\0&0&0&-1\end{array}\right].
\end{eqnarray}
Therefore, the direction of ${\bf V}_1$ is uniquely determined and
can be written as:
\begin{eqnarray}
{\bf V}_1=[0~~1~~1~~0]^T.
\end{eqnarray}
Then, we obtain three columns of $\bar{\mathbf{a}}$ as
\begin{eqnarray}
\bar{\mathbf{a}}=(\mathbf{I}_5\otimes\mathbf{V}_1)\mathbf{a}
\end{eqnarray}
where $\mathbf{a}=(a_{ij})$ is a $5\times 3$ combining matrix with
i.i.d. randomly generated entries.

Similarly, at Receiver 2, we align the first 3 beams from
Transmitter 1 with the last 3 beams from Transmitter 3.
Mathematically,
\begin{eqnarray}
&&\bar{\mathbf{H}}_{21}\bar{\mathbf{V}}_{1,1}=\bar{\mathbf{H}}_{23}\bar{\mathbf{V}}_{3,2}\\
&\Rightarrow&
\left[\bar{\mathbf{H}}_{21}~~-\bar{\mathbf{H}}_{23}\right]\underbrace{\left[\begin{array}{c}\bar{\mathbf{V}}_{1,1}\\
\bar{\mathbf{V}}_{3,2}\end{array}\right]}_{\bar{\mathbf{b}}}=\mathbf{0}
\end{eqnarray}
Then we can choose $\bar{\mathbf{b}}$ as
\begin{eqnarray}
\bar{\mathbf{b}}=(\mathbf{I}_5\otimes\mathbf{V}_2)\mathbf{b}
\end{eqnarray}
where $\mathbf{V}_2$ is the unique vector in the null space of the
$3\times 4$ matrix $\left[\mathbf{H}_{21}~~-\mathbf{H}_{23}\right]$,
i.e., ${\bf V}_2={\bf V}_1$, and $\mathbf{b}=(b_{ij})$ is a $5\times
3$ matrix with i.i.d. randomly generated entries.

At Receiver 3, we align the last 3 beams from Transmitter 1 with
those from Transmitter 2, hereby we have the equation:
\begin{eqnarray}
&&\bar{\mathbf{H}}_{32}\bar{\mathbf{V}}_{2,2}=\bar{\mathbf{H}}_{31}\bar{\mathbf{V}}_{1,2}\\
&\Rightarrow&
\left[\bar{\mathbf{H}}_{32}~~-\bar{\mathbf{H}}_{31}\right]\underbrace{\left[\begin{array}{c}\bar{\mathbf{V}}_{2,2}\\
\bar{\mathbf{V}}_{1,2}\end{array}\right]}_{\bar{\mathbf{c}}}=\mathbf{0}
\end{eqnarray}
Then we can choose $\bar{\mathbf{c}}$ as
\begin{eqnarray}
\bar{\mathbf{c}}=(\mathbf{I}_5\otimes\mathbf{V}_3)\mathbf{c}
\end{eqnarray}
where $\mathbf{V}_3$ is the unique vector in the null space of the
$3\times 4$ matrix $\left[\mathbf{H}_{32}~~-\mathbf{H}_{31}\right]$,
i.e., ${\bf V}_3={\bf V}_1$, and $\mathbf{c}$ is a $5\times 3$
matrix with i.i.d. randomly generated entries.

After aligning interference, we ensure that the dimension of the
space spanned by interference is small enough. In order for each
receiver to decode the desired message, it remains to see if the
desired signals and interference do not overlap at each receiver.
Specifically, we need to see if the $15 \times 15$ matrix consisting
of 6 desired signal vectors and 9 effective interference vectors has
full rank. Due to the symmetry of the signaling, let us consider
Receiver 1. The $15\times 15$ matrix at Receiver 1 is given by:
\begin{eqnarray}
{\bf G}=\left[\bar{{\bf H}}_{11}\bar{{\bf V}}_{1,1}~~\bar{{\bf
H}}_{11}\bar{{\bf V}}_{1,2}~~\bar{{\bf H}}_{12}\bar{{\bf
V}}_{2,1}~~\bar{{\bf H}}_{12}\bar{{\bf V}}_{2,2}~~\bar{{\bf
H}}_{13}\bar{{\bf V}}_{3,2}\right]
\end{eqnarray}
where $\bar{{\bf H}}_{13}\bar{{\bf V}}_{3,1}$ does not appear
because it aligns with $\bar{{\bf H}}_{12}\bar{{\bf V}}_{2,1}$ at
Receiver 1. Now let us substitute the channels as well as
beamforming matrices into the equation above, and rearrange the rows
and columns, then we obtain:
\begin{eqnarray}\label{eqn:G_at_Rx1}
{\bf G}=\left[\begin{array}{ccccc}h_{21}{\bf a}&h_{11}{\bf b}&{\bf
0}&{\bf 0}&{\bf a}\\h_{22}{\bf a}&h_{12}{\bf b}&{\bf 0}&{\bf c}&{\bf
0}\\h_{23}{\bf a}&h_{13}{\bf b}&{\bf b}&{\bf 0}&{\bf
0}\end{array}\right]
\end{eqnarray}
where $h_{ij}=[\bar{{\bf H}}_{11}]_{ij}$ and ${\bf 0}$ is a $2\times
3$ zero matrix. Furthermore, through an invertible linear transformation, ${\bf G}$ becomes:
\begin{eqnarray}
{\bf G}'= \left[\begin{array}{ccccc}{\bf 0}&{\bf b}&{\bf 0}&{\bf
0}&{\bf a}\\{\bf a}&{\bf b}&{\bf 0}&{\bf c}&{\bf 0}\\{\bf a}&{\bf
0}&{\bf b}&{\bf 0}&{\bf 0}\end{array}\right]
\end{eqnarray}
In order to see if ${\bf G}'$ has full rank, we only need to see if
the following equation has non-zero solution:
\begin{eqnarray}
{\bf G}'\left[{\bf \lambda}_1^T~~{\bf \lambda}_2^T~~{\bf
\lambda}_3^T~~{\bf \lambda}_4^T~~{\bf \lambda}_5^T\right]^T={\bf 0}
\end{eqnarray}
where ${\bf \lambda}_l\in\mathbb{C}^{3\times 1},~l=1,2,\cdots,5$.
The equation above implies that
\begin{eqnarray}\label{eqn:non-zero-solution}
\left\{\begin{array}{r}{\bf a}{\bf \lambda}_5+{\bf b}{\bf \lambda}_2
={\bf 0}\\{\bf a}{\bf \lambda}_1+{\bf b}{\bf \lambda}_2+{\bf c}{\bf
\lambda}_4 ={\bf 0}\\{\bf a}{\bf \lambda}_1+{\bf b}{\bf \lambda}_3
={\bf 0}.\\\end{array}\right.
\end{eqnarray}
If we let ${\bf \lambda}_5={\bf \lambda}_1$, ${\bf \lambda}_3={\bf
\lambda}_2$ and ${\bf \lambda}_4={\bf 0}$, then (\ref{eqn:non-zero-solution}) can be equivalently simplified as:
\begin{eqnarray}
{\bf a}{\bf \lambda}_1+{\bf b}{\bf \lambda}_2 ={\bf 0}.
\end{eqnarray}
Now it can be easily seen that ${\bf G}'$ is rank deficient because
$[{\bf a}~~{\bf b}]$ is a $5\times 6$ matrix which has at least
one-dimensional null space, and thus ${\bf \lambda}_1,{\bf
\lambda}_2$ can be non-zero vectors. Therefore, we cannot achieve
$6/5$ DoF per user using the linear beamforming schemes with symbol
extensions over the constant channels.

{\it Remark:} While we only show the infeasibility of linear
beamforming schemes for the $2\times 3$ setting using time
extensions over the constant channels, in fact for all
$(M,N)=(p,p+1)$ cases where $p\geq 2$, we can take similar
invertible linear transformations at each transmitter and receiver,
such that the resulting interference-carrying links are identity
matrices. The proof of the infeasibility is similar to that for the
$2\times 3$ setting, and thus omitted here. On the other hand,
although linear schemes cannot be applied to $(M_T,M_R)=(p,p+1)$
cases where $p\geq 2$ over the constant channels, the DoF outer
bounds can still be achieved by rational alignment proposed in
\cite{Ghasemi_Motahari_Khandani_MIMO}.

While the linear scheme with time extensions over constant channels
fails to achieve the DoF outer bound for the setting $2\times3$, it
does not imply that linear scheme is not optimal for this setting in
general. In fact, the linear scheme is still sufficient to achieve
the optimal DoF if the channel is time-varying or
frequency-selective. To see this, let us consider Figure
\ref{fig:2by3_infeasibility} again. Note that we can always obtain
the resulting channel connectivity no matter if the channels are
constant or time-varying/ frequency-selective. Therefore, the
interference-carrying links are identical to that in
(\ref{eqn:2by3_infea_inf_links}), and ${\bf H}_{kk}(t)$ stands for
the direct channel matrices at the $t^{th}$ time slot. Again,
consider 5 symbol extension. Then the effective cross channel
matrices become
\begin{eqnarray}
\bar{\mathbf{H}}_{ji}=\mathbf{I}_5\otimes\mathbf{H}_{ji},~~~~~j\neq
i.
\end{eqnarray}
Note that the direct channel cannot be expressed in the same manner
since it varies over time or frequency although it is still a
block-diagonal matrix. Then we can express the matrix ${\bf G}$  in
(\ref{eqn:G_at_Rx1}) as
\begin{eqnarray}\label{eqn:Gnew_at_Rx1}
{\bf G}=\left[\begin{array}{ccccc}{\bf R}^h_{21}{\bf a}&{\bf
R}^h_{11}{\bf b}&{\bf 0}&{\bf 0}&{\bf a}\\{\bf R}^h_{22}{\bf a}&{\bf
R}^h_{12}{\bf b}&{\bf 0}&{\bf c}&{\bf 0}\\{\bf R}^h_{23}{\bf a}&{\bf
R}^h_{13}{\bf b}&{\bf b}&{\bf 0}&{\bf 0}\end{array}\right]
\end{eqnarray}
where ${\bf R}^h_{nm},~n\in\{1,2,3\},~m\in\{1,2\}$ is given by
\begin{eqnarray}
{\bf R}^h_{nm}=\textrm
{diag}\{[h_{nm}(1)~~h_{nm}(2)~~h_{nm}(3)~~h_{nm}(4)~~h_{nm}(5)]\}
\end{eqnarray}
and $h_{nm}(t)=[{\bf H}_{11}(t)]_{nm}$. Note that each entry of the
diagonal matrix ${\bf R}^h_{nm}$ is generic, and thus the rank
deficient argument for the matrix ${\bf G}$ used for constant channel no long holds here.
Moreover, through simple numerical simulations, we can easily verify
${\bf G}$ has full rank almost surely if we randomly pick the
entries of ${\bf a}$, ${\bf b}$ and ${\bf c}$. Thus, $6/5$
DoF per user can be achieved almost surely by using this
non-asymptotical linear scheme if the channel coefficients are
time-varying/ frequency-selective.

{\it Remark:} It should be noted that the rational alignment scheme
proposed for the constant channel in
\cite{Ghasemi_Motahari_Khandani_MIMO} can also be translated to the
linear scheme for time-varying/frequency-selective channels to
achieve the DoF outer bound. Such scheme does not explore antenna
cooperations and achieves the outer bound asymptotically by using
{\em infinite} symbol extensions. In contrast, the achievable
schemes proposed above which explore antenna cooperation achieve the
outer bound exactly with only a {\em finite} number of symbol
extensions.

\subsubsection{A Linear Scheme with  Symbol Extensions in Time/Frequency for General Cases}
In this subsection, we will construct a linear scheme with symbol
extensions for general cases. The feasibility of such scheme can be
settled in every case through a simple numerical test. Since the DoF
value depends on either $M$ or $N$, we consider them separately.

{\bf (1) $d(M,N)=\frac{p}{2p-1}M$ if
$\frac{p-1}{p}\leq\frac{M}{N}\leq \frac{2p-1}{2p+1}$}

Since the DoF  outer bound is a fractional value,  we take  $2p-1$
symbol extensions to make the DoF value an integer, i.e., $pM$. Then
over the extended channel, our goal is to achieve $pM$ DoF.  Note
that after the symbol extensions, each transmitter has $(2p-1)M$
dimensions and each receiver has $(2p-1)N$ dimensions. The DoF value
depending on $M$ implies that the receiver includes redundant
dimensions. Therefore, we randomly generate a $(2p+1)M\times
(2p-1)N$ matrix at each receiver independently and multiply it to
the channel seen at each {\em receiver}, such that each receiver
effectively has a $(2p+1)M$ dimensional space. As a consequence, we
have an effective $((2p-1)M,(2p+1)N)$ setting, for which
interference alignment scheme introduced in Section
\ref{subsec:integer_dof} can be applied here such that the dimension
of the space spanned by interference is small enough. What remains
to be shown is that the desired $pM$ symbols do not overlap with the
$(p+1)M$ interference dimensions. While we do not prove this
statement in general, we provide the recipe for completing the proof
subject to a numerical test. The independence of the desired signal
space from the interference space is equivalent to the condition
that the matrix containing the interference vectors and the desired
signal vectors has full rank, i.e., the determinant of the matrix is
non-zero. Since the determinant is a polynomial with the channel
coefficients and the elements of the random projection matrix as the
variables, there are only two possibilities. Either it is the zero
polynomial or it is not. To prove that it is not the zero
polynomial, it suffices to substitute any numerical values for all
variables and show that the polynomial evaluates to a non-zero
value. Then, since the polynomial is not the zero polynomial, it
must almost surely be non-zero for generic values of channel
coefficients and the projection matrix, thus completing the proof
(subject to the condition that the numerical evaluation produces a
non-zero value).

{\bf (2) $d(M,N)=\frac{p}{2p+1}N$ if
$\frac{2p-1}{2p+1}\leq\frac{M}{N}\leq \frac{p}{p+1}$}

Similar to the previous case, consider $2p+1$ symbol extensions. Our
goal is to achieve $pN$ DoF over the extended channel. After the
symbol extensions, each transmitter has $(2p+1)M$ dimensions and
each receiver has $(2p+1)N$ dimensions. Now we randomly generate a
$(2p+1)M\times (2p-1)N$ matrix at each {\em transmitter}, and
multiply it to the channel seen at each transmitter, such that each
transmitter effectively has a $(2p-1)N$-dimensional space. As a
result, we end up with an effective $((2p-1)N,(2p+1)N)$ setting, for
which interference alignment schemes introduced in Subsection
\ref{subsec:integer_dof} can be applied. Again we can apply numerical
tests to verify if the desired signal overlaps with the
interference.

While we do not prove that the numerical test will produce a
non-zero value for all $M,N$, we have carried out the numerical test
to complete the proof for all $(M,N)$ values upto $M, N \leq 10$ as
shown in Fig. \ref{fig:map_table}. As we can see, linear schemes
with symbol extensions over constant channels are sufficient to
achieve the information theoretic DoF outer bounds for all $(M,N)$
values upto $M, N \leq 10$ except for the $(q,q+1)$ cases where
time-variations/frequency-selectivity are needed as shown in Section
\ref{subsec:ach_symbol_ext_pplus1}.

%Notice that each entry of the matrix is a polynomial in the channel
%coefficients and the randomly generated matrices $\mathbf{a}$,
%$\mathbf{b}$ and $\mathbf{c}$. Therefore, the determinant of the
%$15\times 15$ matrix is also a polynomial in these variables.
%Moreover, the polynomial is either identically the zero polynomial,
%or it is non-zero almost surely for all realization of $\mathbf{H}$,
%$\mathbf{a}$, $\mathbf{b}$ and $\mathbf{c}$. To prove that it is
%non-zero almost surely, it suffices to show that it is not the zero
%polynomial. This can be verified by a numerical example. Since such
%examples are easy to find (almost all choices work fine) we will
%omit the explicit construction.

%Moreover, when $M/N\neq p/(p+1)$, the non asymptotic linear
%beamforming schemes can be applied using symbol extensions as well
%to achieve the optimal DoF results.
%
%
%
%Notice that each entry of the matrix is a polynomial in the channel
%coefficients and the randomly generated matrices $\mathbf{a}$,
%$\mathbf{b}$ and $\mathbf{c}$. Therefore, the determinant of the
%$15\times 15$
%
%
%matrix is also a polynomial in these variables. Moreover, the
%polynomial is either identically the zero polynomial, or it is
%non-zero almost surely for all realization of $\mathbf{H}$,
%$\mathbf{a}$, $\mathbf{b}$ and $\mathbf{c}$.
%
%To prove that it is non-zero almost surely, it suffices to show that
%it is not the zero polynomial. This can be verified by a numerical
%example. Since such examples are easy to find (almost all choices
%work fine) we will omit the explicit construction.

\subsection{Feasibility of Linear Interference Alignment without any Symbol Extensions in Time/Frequency/Space}\label{sec:feasibility}

In this section, we consider the feasibility of linear interference
alignment for the 3 user $M_T\times M_R$ MIMO interference channel.
As stated in Theorem \ref{theorem:feasible}, the DoF demand per
user, $d$, is feasible with linear interference alignment if and
only if $d\leq \lfloor\mbox{\normalfont DoF}^\star\rfloor$. The
outer bound follows directly from Lemma \ref{lemma:out}. Next, we
provide a proof of Lemma \ref{lemma:in} to show that
$d=\lfloor\mbox{\normalfont DoF}^\star\rfloor$ per user is
achievable.

Since $\mbox{\normalfont DoF}^\star$  is limited by either the
$M$-bound or the $N$-bound depending on $M/N$, we consider the two
cases separately. We begin with the setting when
$\frac{2p-1}{2p+1}\leq\frac{M}{N}\leq \frac{p}{p+1}$ and show that
$\lfloor\frac{p}{2p+1}N\rfloor$ DoF per user is achievable using
linear interference alignment \emph{without} the need for symbol
extensions in time/frequency/space.

\subsubsection{Case: $\frac{2p-1}{2p+1}\leq\frac{M}{N}\leq
\frac{p}{p+1}\Rightarrow d=\lfloor\frac{p}{2p+1}N\rfloor$:}

For this case, we will provide the achievable scheme for the case
$M_T=M$ and $M_R=N$. Due to reciprocity of linear schemes, such
schemes can be applied to the case $M_T=N$ and $M_R=M$ as well. Let
us first consider the case when $N$ is an integer multiple of
$2p+1$. In this case,
$d=\lfloor\frac{p}{2p+1}N\rfloor=\frac{p}{2p+1}N$ which is an
integer. As proved in Section \ref{subsec:achievek}, for this case,
no symbol extensions are needed to achieve the information theoretic
DoF outer bound. %Therefore, without symbol extensions
%$\lfloor\frac{p}{2p+1}N\rfloor$ can be achieved.

Now let us consider the remaining cases when $N$ is not an integer
multiple of $2p+1$. We first provide an intuition from the linear
dimension counting perspective, and then show the rigorous proof
through constructing specific channels as in Section
\ref{subsec:achievek}. As shown previously, for
$\frac{2p-1}{2p+1}\leq\frac{M}{N}\leq \frac{p}{p+1}$, the maximum
length of the alignment chain is $\kappa=\lceil \frac{M}{N-M}\rceil
=p$. Since the alignment chain with the maximum length corresponds
to the most efficient alignment scheme, we would like the dimensions
of the subspaces that participate in such chains to be as large as
possible. Recall that if the dimension of the subspace is $d_0$,
then the total number of dimensions transmitted from all
transmitters is $pd_0$ and the total number of dimensions occupied
at all receivers is $(2p+1)d_0$. Due to symmetric antenna
configurations, we can pack three such chains, each originating from
one of three transmitters, as shown in \eqref{eqn:chain1},
\eqref{eqn:chain2} and \eqref{eqn:chain3}. Then each transmitter
achieves $pd_0$ DoF, and $(2p+1)d_0$ dimensions are occupied at each
receiver. Since the number of dimensions at each receiver is $N$,
$(2p+1)d_0$ cannot exceed $N$, implying that
$d_0\leq\lfloor\frac{N}{2p+1}\rfloor$. Thus, $d_0$ is set to be
equal to $\lfloor\frac{N}{2p+1}\rfloor$. After this, each
transmitter still needs to achieve
$\lfloor\frac{p}{2p+1}N\rfloor-p\lfloor\frac{N}{2p+1}\rfloor$ DoF.
This can be done if we can pack three more subspace alignment chains
with length
$p'=\lfloor\frac{p}{2p+1}N\rfloor-p\lfloor\frac{N}{2p+1}\rfloor$,
each originating from one transmitter, and the dimension of the
subspaces that participate in each chain is one. These symbols will
occupy $2p'+1$ dimensions at each receiver. Note that this is
possible if the total number of dimensions occupied by these six
alignment chains is less than $N$. In order to see this is true, we
need to check that
$2p'+1=2(\lfloor\frac{p}{2p+1}N\rfloor-p\lfloor\frac{N}{2p+1}\rfloor)+1$
does not exceed the remaining dimensions at each receiver after
accommodating the first three alignment chains with length $p$,
i.e., $N-\lfloor\frac{N}{2p+1}\rfloor (2p+1)$.
\begin{eqnarray}
&&\left[2\left(\left\lfloor\frac{p}{2p+1}N\right\rfloor-\left\lfloor\frac{N}{2p+1}\right\rfloor
p\right)+1\right]-\left[N-\left\lfloor\frac{N}{2p+1}\right\rfloor (2p+1)\right]\notag\\
&=&2\left\lfloor\frac{p}{2p+1}N\right\rfloor+\left\lfloor\frac{N}{2p+1}\right\rfloor+1-N\\
&<&2\frac{p}{2p+1}N+\frac{N}{2p+1}+1-N \label{eqn:N_not_int_mul}\\
&=&1\label{eqn:less_than_one}
\end{eqnarray}
where (\ref{eqn:N_not_int_mul}) is obtained because  $N$ is not an
integer multiple of $2p+1$. Because $2p'+1$ and
$(N-\lfloor\frac{N}{2p+1}\rfloor (2p+1))$ are both integers,
(\ref{eqn:less_than_one}) implies that $2p'+1\leq
(N-\lfloor\frac{N}{2p+1}\rfloor (2p+1))$. As a result, each user can
send $\lfloor\frac{p}{2p+1}N\rfloor$ DoF and interference occupies
small enough dimensions. Intuitively, the
$\lfloor\frac{p}{2p+1}N\rfloor$ vectors of each user should be
linearly independent, almost surely, because the channels are
generic and these vectors do not align among themselves unless they
have to. Next we provide the proof to show this is true through
constructing specific channels.

\begin{figure}[!h] \vspace{-0.1in}\centering
\includegraphics[width=5.0in]{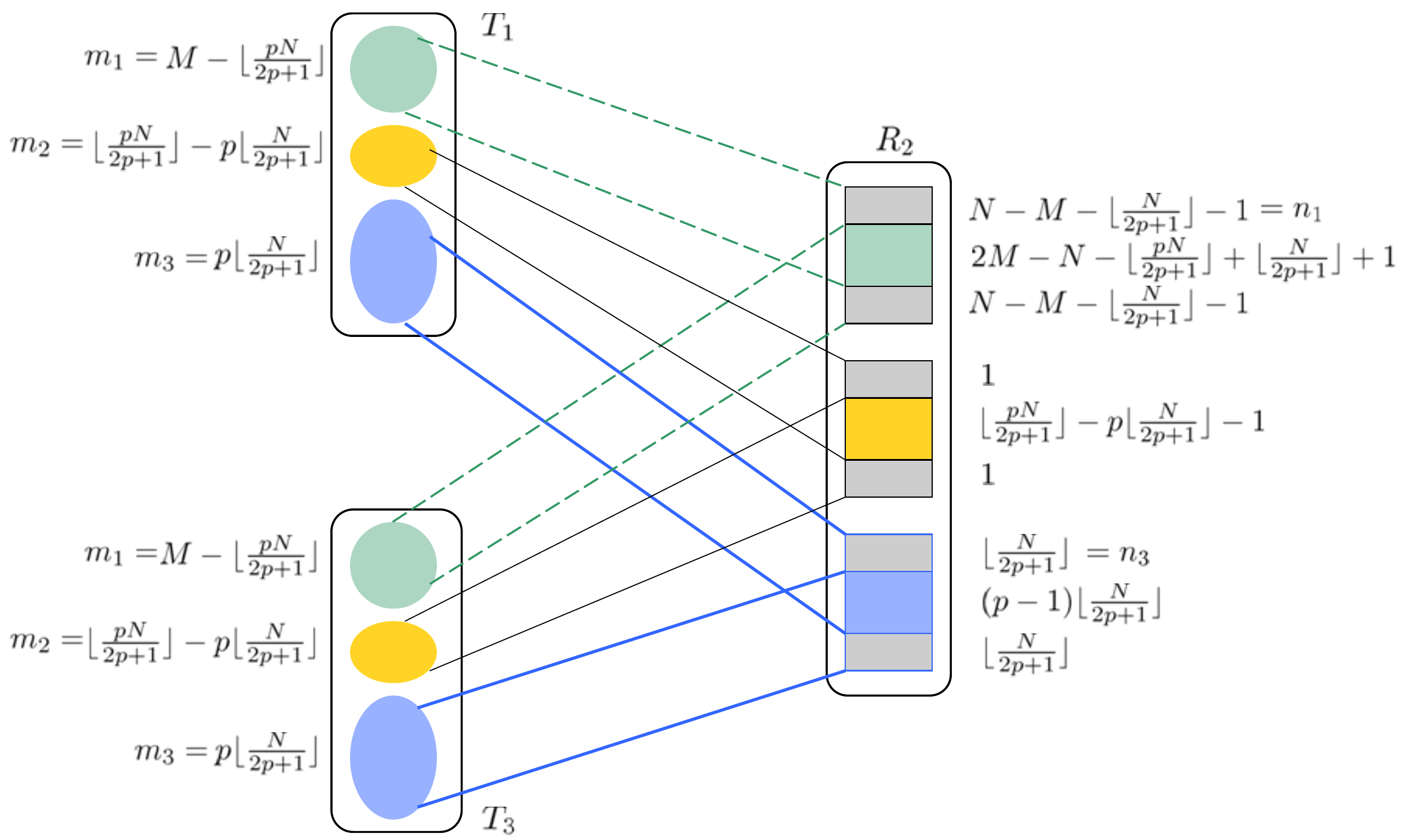}\vspace{-0.05in}
\caption{Linear Dimension Counting of Subspaces Participating in the
Interference Alignment for the $M\times N$ setting (the values
denote the dimensions of each corresponding subspace, and it is easy
to verify each value is non-negative.)} \vspace{-0.1in}
\label{fig:feasibility}
\end{figure}

The linear dimension counting argument shown above is characterized
in Figure \ref{fig:feasibility} where we consider the signal
subspaces at Transmitter 1, Transmitter 3 and Receiver 2.
Specifically, we need to design six subspace alignment chains. Three
of them are $p$-length $\lfloor\frac{N}{2p+1}\rfloor$ dimensional,
for a total of $p\lfloor\frac{N}{2p+1}\rfloor$ dimensions at each
transmitter, denoted as the blue color at Transmitter 1 and
Transmitter 3. Since three $p$-length $\lfloor\frac{N}{2p+1}\rfloor$
dimensional chains will occupy $(2p+1)\lfloor\frac{N}{2p+1}\rfloor$
dimensions at each receiver, we need to align
$(p-1)\lfloor\frac{N}{2p+1}\rfloor$ dimensions, i.e., the blue color
box at Receiver 2, and the non-aligned
$\lfloor\frac{N}{2p+1}\rfloor$ dimensional subspace coming from each
transmitter is denoted as the adjacent grey color region. The other
three chains are $p'$-length one dimensional, denoted as the yellow
color region at Transmitter 1 and Transmitter 3. Again, since these
three $p'$-length one dimensional chains occupy $2p'+1$ dimensions
at each receiver, we need to align $p'-1$ dimensions, i.e., the
yellow color box at Receiver 2, and the non-aligned one dimensional
subspace from each transmitter is denoted as the adjacent grey color
region. Last, since Transmitter 1 and Transmitter 3 have a total of
$2M-N$ common dimensions at Receiver 2 almost surely, the remaining
two $M-\lfloor\frac{pN}{2p+1}\rfloor$ dimensional subspaces (denoted
as the green color) from these two transmitters have to have the
remaining
$2M-N-\lfloor\frac{pN}{2p+1}\rfloor+\lfloor\frac{N}{2p+1}\rfloor+1$
dimensional intersection at Receiver 2 (denoted as the green color),
and the non-aligned $N-M-\lfloor\frac{N}{2p+1}\rfloor-1$ dimensional
subspace from each transmitter is denoted as the adjacent grey color
box.

Let us first design the three $p$-length subspace alignment chains,
which are shown in \eqref{eqn:chain1}, \eqref{eqn:chain2} and
\eqref{eqn:chain3}. These three chains yield the three alignment
equations shown in \eqref{eqn:align_eqn1}, \eqref{eqn:align_eqn2}
and \eqref{eqn:align_eqn3}, where each of three matrices ${\bf A}_p,
{\bf B}_p,{\bf C}_p$ is a $(p-1)N\times pM$ matrix. Since null
spaces of ${\bf A}_p, {\bf B}_p,{\bf C}_p$ all have $pM-(p-1)N$
dimensions, and also because $pM-(p-1)N\geq
\frac{N}{2p+1}>\lfloor\frac{N}{2p+1}\rfloor$, we can choose ${\bf
a}_p, {\bf b}_p, {\bf c}_p$ as the $\lfloor\frac{N}{2p+1}\rfloor$
basis vectors in the $pM-(p-1)N$ dimensional null spaces of ${\bf
A}_p, {\bf B}_p,{\bf C}_p$, respectively. That is, assuming ${\bf
Q}_{k,p}$ are three $(pM-(p-1)N)\times \lfloor\frac{N}{2p+1}\rfloor$
randomly generated matrices, then we have
\begin{eqnarray}
\bar{\bf A}_p\triangleq \mathcal{N}({\bf A}_p),~~\bar{\bf
B}_p\triangleq \mathcal{N}({\bf B}_p),~~\bar{\bf C}_p\triangleq
\mathcal{N}({\bf C}_p)\notag\\
{\bf a}_p=\bar{\bf A}_p{\bf Q}_{1,p},~~{\bf b}_p=\bar{\bf B}_p{\bf
Q}_{2,p},~~{\bf c}_p=\bar{\bf C}_p{\bf Q}_{3,p}
\end{eqnarray}
where $\mathcal{N}(\cdot)$ denotes the matrix whose columns span the
whole null space of matrix $(\cdot)$.

Next we design the three $p'$-length one-dimensional subspace
alignment chains, which are still identical to \eqref{eqn:chain1},
\eqref{eqn:chain2} and \eqref{eqn:chain3} by replacing $p$ with
$p'$. These three chains yield the three alignment equations shown
in \eqref{eqn:align_eqn1}, \eqref{eqn:align_eqn2} and
\eqref{eqn:align_eqn3} by replacing $p$ with $p'$, where each of
three matrices ${\bf A}_{p'}, {\bf B}_{p'},{\bf C}_{p'}$ is a
$(p'-1)N\times p'M$ matrix. Since null spaces of  ${\bf A}_{p'},
{\bf B}_{p'},{\bf C}_{p'}$ all have $p'M-(p'-1)N$ dimensions, which
is always no less than one, we can choose ${\bf a}_{p'}, {\bf
b}_{p'}, {\bf c}_{p'}$ as the basis vector in the $p'M-(p'-1)N$
dimensional null spaces of ${\bf A}_{p'}, {\bf B}_{p'},{\bf
C}_{p'}$, respectively. Assuming ${\bf Q}_{k,p'}$ are three
$(p'M-(p'-1)N)\times 1$ randomly generated vectors, then we have
\begin{eqnarray}
\bar{\bf A}_{p'}\triangleq \mathcal{N}({\bf A}_{p'}),~~\bar{\bf
B}_{p'}\triangleq \mathcal{N}({\bf B}_{p'}),~~\bar{\bf
C}_{p'}\triangleq
\mathcal{N}({\bf C}_{p'})\notag\\
{\bf a}_{p'}=\bar{\bf A}_{p'}{\bf Q}_{1,{p'}},~~{\bf
b}_{p'}=\bar{\bf B}_{p'}{\bf Q}_{2,{p'}},~~{\bf c}_{p'}=\bar{\bf
C}_{p'}{\bf Q}_{3,{p'}}.
\end{eqnarray}

Consider the square matrix of each receiver which consists of the
desired signal and interference vectors. Now its determinant is a
polynomial function of the channel coefficients and entries of ${\bf
Q}_{k,p},{\bf Q}_{k,p'},k=1,2,3$, and this polynomial is an either
zero or non-zero polynomial almost surely. As in Section
\ref{subsec:achievek}, we only need to construct specific channels
as well as ${\bf Q}_{k,p},{\bf Q}_{k,p'}$, and show the beamforming
vectors of each user are linearly independent. We choose the channel
matrices as follows:
\begin{eqnarray}
\mathbf{H}_{(i-1)i}= \left[\begin{array}{ccc}\mathbf{0}_{n_1\times
m_1}& \mathbf{0}_{n_1\times m_2}&\mathbf{0}_{n_1\times m_3}\\
\mathbf{I}_{m_1\times m_1}&\mathbf{0}_{m_1\times
m_2}&\mathbf{0}_{m_1\times m_3}\\ \mathbf{0}_{1\times m_1}&
\mathbf{0}_{1\times m_2} & \mathbf{0}_{1\times m_3}\\
\mathbf{0}_{m_2\times m_1}& \mathbf{I}_{m_2\times m_2} &
\mathbf{0}_{m_2\times m_3}\\ \mathbf{0}_{n_3\times m_1} &
\mathbf{0}_{n_3\times m_2} & \mathbf{0}_{n_3\times m_3} \\
\mathbf{0}_{m_3\times m_1} & \mathbf{0}_{m_3\times m_2} &
\mathbf{I}_{m_3\times m_3}\end{array}\right],~~~~
\mathbf{H}_{(i+1)i}= \left[\begin{array}{ccc}\mathbf{I}_{m_1\times
m_1}& \mathbf{0}_{m_1\times m_2}&\mathbf{0}_{m_1\times m_3}\\
\mathbf{0}_{n_1\times m_1}&\mathbf{0}_{n_1\times
m_2}&\mathbf{0}_{n_1\times m_3}\\ \mathbf{0}_{m_2\times m_1}&
\mathbf{I}_{m_2\times m_2} & \mathbf{0}_{m_2\times m_3}\\
\mathbf{0}_{1\times m_1}& \mathbf{0}_{1\times m_2} &
\mathbf{0}_{1\times m_3}\\ \mathbf{0}_{m_3\times m_1} &
\mathbf{0}_{m_3\times m_2} & \mathbf{I}_{m_3\times m_3} \\
\mathbf{0}_{n_3\times m_1} & \mathbf{0}_{n_3\times m_2} &
\mathbf{0}_{n_3\times m_3}\end{array}\right]
\end{eqnarray}
where $\forall i\in\{1,2,3\}$ and the values of $m_i,n_i$ are shown
in Figure \ref{fig:feasibility}. It can be easily verified that the
alignment schemes we describe above produce one solution of the
beamforming matrix for each transmitter as follows:
\begin{eqnarray}
\mathbf{V}_i=[\mathbf{V}_{i,p}~~\mathbf{V}_{i,p'}],~~~~~~\mathbf{V}_{i,p}=\left[\begin{array}{c}\mathbf{0}_{m_1\times
m_3}\\\mathbf{0}_{m_2\times m_3} \\ \mathbf{I}_{m_3\times
m_3}\end{array}\right],~~~~\mathbf{V}_{i,p'}=\left[\begin{array}{c}\mathbf{0}_{m_1\times
m_2}\\\mathbf{I}_{m_2\times m_2} \\ \mathbf{0}_{m_3\times
m_2}\end{array}\right]
\end{eqnarray}
where $\mathbf{V}_{i,p}$ and $\mathbf{V}_{i,p'}$ are obtained from
the $p$-length $\lfloor\frac{N}{2p+1}\rfloor$ dimensional chains and
$p'$-length one dimensional chains, respectively. Clearly, since the
column spaces of $\mathbf{V}_{i,p}$ and $\mathbf{V}_{i,p'}$ are
orthogonal to each other, the rank of $\mathbf{V}_i$ is
$m_2+m_3=\lfloor\frac{p}{2p+1}N\rfloor$. Therefore, the beamforming
vectors of each user are linearly independent among themselves,
almost surely. The linear independence established through this specific example proves that the polynomial (comprised of channel variables and {\bf Q} variables) representing the determinant of a square matrix containing the beamforming vectors originating at a transmitter, is not the zero-polynomial. Since it is not the zero-polynomial, it is almost surely not zero for random choices of channel coefficients and {\bf Q} matrices, i.e., the linear interference alignment solution exists almost surely.

\subsubsection{Case: $\frac{p-1}{p}\leq\frac{M}{N}\leq
\frac{2p-1}{2p+1}\Rightarrow d=\lfloor\frac{p}{2p-1}M\rfloor$:}

For this case, we provide the achievable schemes for the setting
$M_T=N$ and $M_R=M$. The proof for this case is similar as the
previous case with $2p+1$ and $N$ in the previous argument replaced
by $2p-1$ and $M$, respectively. The detailed proof followed is
omitted due to the similarity. \hfill\QED

\section{Conclusion}

In this paper we characterize both the spatially-normalized
information-theoretic DoF as well as the DoF achievable through
linear beamforming schemes without channel extensions in
time/frequency/ space, for the symmetric three-user $M_T\times M_R$
Gaussian MIMO interference channel where each transmitter has $M_T$
antennas and each receiver has $M_R$ receive antennas.  In order to
establish this result, we derive  information theoretic DoF outer
bounds for arbitrarily values of $M_T$ and $M_R$.  The  information
theoretic DoF outer bounds are facilitated by a change of basis
operation at both transmitter and receiver sides which helps
identify the projections of the transmitted signal space to be
provided as genie signals  for the outer bound arguments. One of the
main ideas involved in this problem is the notion of  {\em subspace
alignment chains}, the length of which indicates the limitations of
aligning interference symbols. Achievable schemes take advantage of
the linear-dimension counting arguments and generic property of
channels, both of which are captured by subspace alignment chains.

Several interesting observations follow as a byproduct of our
analysis. First we precisely identify settings with redundant
dimensions at the transmitters, receivers, both or neither. The
maximally redundant settings correspond to $M/N=1/2,2/3,3/4,\cdots$
and contain redundant dimensions at both the transmitters and
receivers. The DoF outer bounds essentially boil down to these
settings. These settings are also the only ones where there is no
DoF benefit of joint processing across the multiple antennas located
at any transmitter or receiver node. The minimally redundant
settings correspond to $M/N=3/5, 5/7, 7/9, \cdots$ and contain no
redundant dimensions at either the transmitters or receivers. The
achievability results essentially boil own to these settings. These
settings are also the only ones where proper systems are guaranteed
to be feasible.

We also note the recent work in \cite{Nafie_etal} where partial
achievability results  are independently obtained and shown to be
optimal in some cases with respect to linear interference alignment.
To the best of our understanding, all the results of
\cite{Nafie_etal} can also be recovered as special cases of the
results presented in this paper.

Finally, we conclude with a pointer to the open problem of
characterizing the DoF of asymmetric settings, where each
transmitter and receiver may have different number of antennas. DoF
results for certain asymmetric settings can be obtained directly
from our results whenever the asymmetric setting differs from the
symmetric case only in redundant dimensions. However, in general,
the complexity of the asymmetric problem may be understood in light
of the difficulty of packing subspace alignment chains within the
space constraints imposed by the number of antennas at each node,
which may be tight or redundant. Both outer bounds and inner bounds
are challenging for this general setting. Dimension counting
arguments, motivated by subspace alignment chains may be a good
starting point for this general setting.

%
%Several issues are worthy of further study. One issue is that while
%the rigorous proof of the DoF achievability relies on the space
%extension, we do not have a rigorous systematic proof of the DoF
%results for $(M_T,M_R)$ settings in general with symbols extensions
%on constant channels except for $M/N\leq 1/2$ and $M/N\in
%\{3/5,5/7,7/9,\cdots\}$. Although numerical simulations corroborate
%our results for several examples and we kindly believe it is correct
%for for any $M_T\times M_R$ settings, it still needs some efforts to
%completely the rigorous proof in general. Another interesting issue,
%as mentioned above, is to characterize the DoF region of three-user
%symmetric MIMO Gaussian interference channel. With the trivial
%single-user DoF bound, cooperation DoF bounds
%\cite{Jafar_Fakhereddin}, as well as 3-user DoF bounds that we
%derive in this paper, one question is that are these bounds
%sufficient to characterize the DoF region of the 3-user symmetric
%interference channel. For some simple networks, it is not difficult
%to see that these DoF bounds are sufficient, but in general the
%problem remains open.

\appendix{}
\section*{Appendix}

\section{Information Theoretic DoF Outer Bound for $M_T<M_R$ }\label{sec:outerbound_MsmallerN}

We will show the proofs in the following two subsections. In the
first, we investigate the cases of $M/N=p/(p+1)$ for which we only
need to consider $(M,N)=(p,p+1)$ as we mentioned before. In the
second subsection, we investigate the $M/N\neq p/(p+1)$ cases.

\subsection{Cases: $(M,N)=(p,p+1)$ $\Rightarrow$ DoF $\leq \frac{MN}{M+N}$}

\subsubsection{$M$ is an Even Number}

We consider $M$ is even in this subsection. We first take recursive
invertible linear transformations introduced in Section
\ref{sec:outerbound} at each user. The transmitted signals from each
antenna and what transmitted signals that each receiver antenna can
see are specified in Fig.\ref{fig:onion_even}. Note that again we
use $S_{k(\cdot)}$ to denote received signal at the antenna
$(\cdot)$ of Receiver $k$ subtracting the signal carrying its
desired message, which implies linearly independent combinations of
the interference symbols at the antenna $(\cdot)$ of Receiver $k$.

\begin{figure}[!h] \vspace{-0.15in}\centering
\includegraphics[width=6.0in]{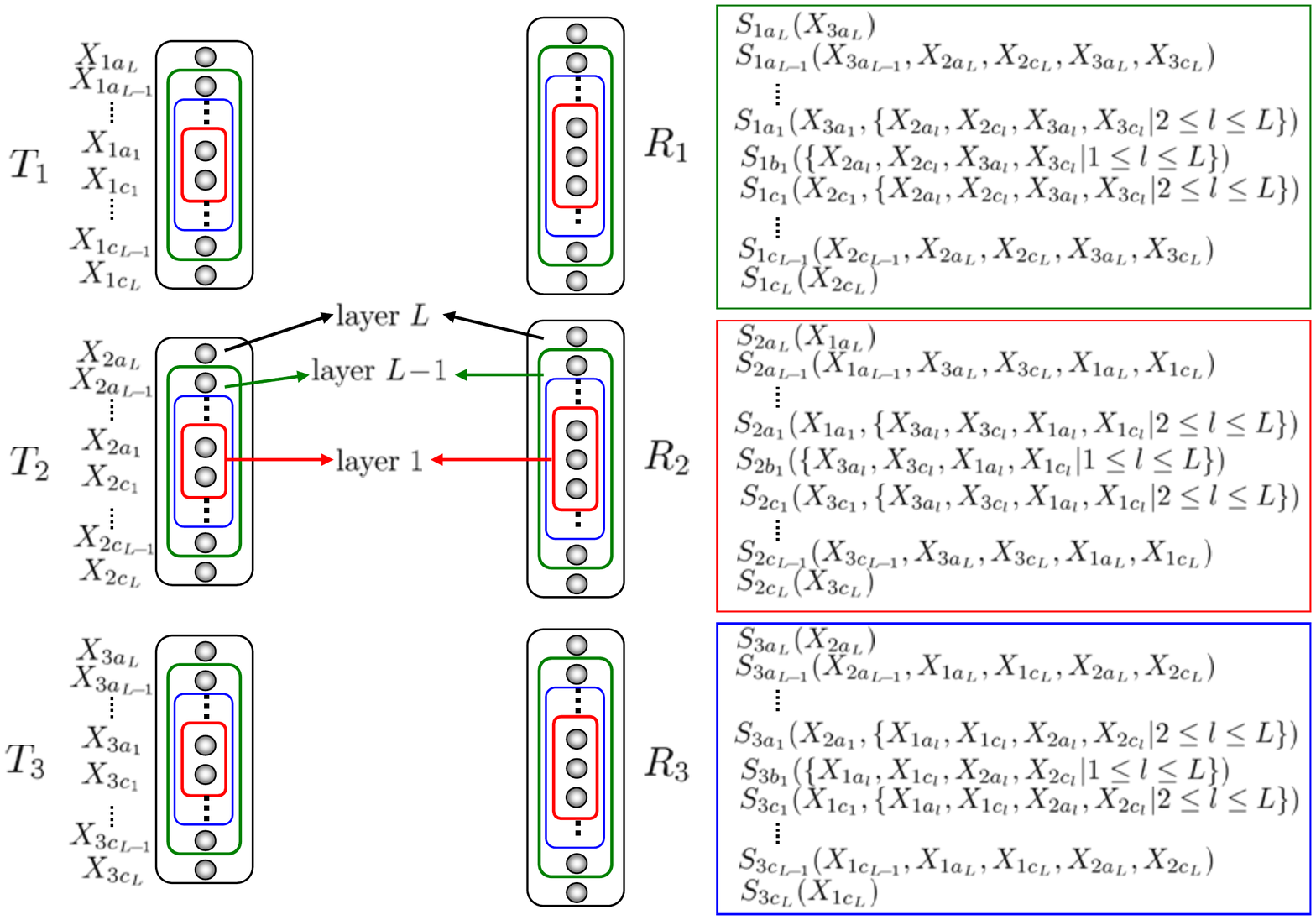}\vspace{-0.05in}
\caption{Intuition of Onion Peeling if $M$ is even} \vspace{-0.1in}
\label{fig:onion_even}
\end{figure}

For the general $(M,N)=(p,p+1)$ setting, as what have shown in
Section \ref{sec:outerbound_eg}, we need a total of $M$ sets of
genie signals and each set has $M-1$ dimensions. We denote $\mathcal
{G}_{km}^M,~k\in\{1,2,3\},~m\in\{1,\cdots,M\}$ as the $M$ sets of
genie signals provided to Receiver $k$ for the $(M,N)=(p,p+1)$
network. If $k=1$, $\mathcal {G}_{1m}^M$ or directly $\mathcal
{G}_m^M$ if without ambiguity is given by:
%for
%$m'=\{1,2,\cdots,L\}$ where $L=M/2$ is the total number of layers,
%\begin{eqnarray}
%\mathcal{G}_{2m'-1}^M&\!\!\!\!=\!\!\!\!&\{\{X_{2a_l}^n,X_{3c_l}^n|m'+1\leq l \leq L\},X_{2a_{m'}}^n,\{X_{2a_l}^n,X_{2c_l}^n|1\leq l \leq m'-1\}\}\notag\\
%\mathcal{G}_{2m'}^M&\!\!\!\!=\!\!\!\!&\{\{X_{2a_l}^n,X_{3c_l}^n|m'+1\leq
%l \leq
%L\},X_{3c_{m'}}^n,\{X_{3a_l}^n,X_{3c_l}^n|1\leq
%l \leq m'-1\}\}\label{eqn:general_genie_even}
%\end{eqnarray}
\begin{eqnarray}
{\rm for}~~M=2,~~~~~\mathcal{G}_1^M&\!\!\!\!=\!\!\!\!&\{X_{2a_1}^n\}\notag\\
\mathcal{G}_2^M&\!\!\!\!=\!\!\!\!&\{X_{3c_1}^n\}\notag\\
{\rm then~for}~~M>2,~~~~~\mathcal{G}_{m}^M&\!\!\!\!=\!\!\!\!&\{X_{2a_L}^n,X_{3c_L}^n,\mathcal{G}_m^{M-2}\}~~~~~m=\{1,2,\cdots,M-2\},\notag\\
{\rm and}~~~~~\mathcal{G}_{M-1}^M&\!\!\!\!=\!\!\!\!&\{X_{2a_L}^n,\{X_{2a_l}^n,X_{2c_l}^n|1\leq l \leq L-1\}\}\notag\\
\mathcal{G}_M^M&\!\!\!\!=\!\!\!\!&\{X_{3c_L}^n,\{X_{3a_l}^n,X_{3c_l}^n|1\leq
l \leq L-1\}\} \label{eqn:general_genie_even}
\end{eqnarray}
and $\mathcal {G}_{2m}^M,~\mathcal {G}_{3m}^M$ can be obtained by
advancing the user indices. It is easy to check that the cardinality
of each set is $M-1$. Also, note that if Receiver 1 knows the two
signals $(X_{2a_L}^n,X_{3c_L}^n)$ in the $L^{th}$ layer, then after
removing $(X_{2a_L}^n,X_{3c_L}^n)$ from $\mathcal
{G}_m^M,~m\in\{1,\cdots,M-2\}$ in (\ref{eqn:general_genie_even}), we
obtain $\mathcal {G}_m^{M-2},~m\in\{1,\cdots,M-2\}$. This is in fact
essential why we can design the genie signals in an iterative way,
which help us recursively derive the DoF outer bound based on
mathematical induction. Consider an $(M,M+1)$ interference network,
which has a total of $L=M/2$ layers, with two transmit signals
$(X_{ka_l}^n,X_{kc_l}^n)$ at the $l^{th}$ layer where
$l=\{1,2,\cdots,L\}$. The transmit signals in the $l^{th}$ layer
will be heard by all antennas in all $l'\leq l$ layers. Now let us
peel the $L^{th}$ layer at both sides, i.e., a genie provides the
two symbols $(X_{(k+1)a_L}^n,X_{(k-1)c_L}^n)$ to Receiver $k$. Since
the peeled $S_{ka_L}^n$ and $S_{kc_L}^n$ at Receiver $k$ are already
noised versions $X_{(k-1)a_L}^n$ and $X_{(k+1)c_L}^n$ respectively,
Receiver 1 can recover
$(X_{(k+1)a_L}^n,X_{(k+1)c_L}^n,X_{(k-1)a_L}^n,X_{(k-1)c_L}^n)$
subject to the noise distortion, and thus it can subtract them from
the received signals at the other $M-1$ antennas. Therefore, we
obtain an embedded $(M-2,M-1)$ interference network.

The proof is based on mathematical induction. Since we have already
finished the proof for $(M,M+1)=(2,3)$ and $(4,5)$ previously, now
let us assume it works for the $(M-2,M-1)$ case. That is to say, by
providing genie signals $\mathcal {G}_m^{M-2},~m\in\{1,\cdots,M-2\}$
to Receiver 1, we obtain a total of $M-2$ sum rate inequalities,
each obtained by averaging over user indices. If we add up all these
$(M-2)$-inequalities, at the left-hand side we have $(M-2)nR$; at
the right-hand side we have
$(M-2)N\log\rho+(M-3)nR+n~o(\log\rho)+o(n)$. In other words, we can
bound the sum of differential entropy of all genie signals provided
to three receivers by:
\begin{eqnarray}
\frac{1}{3}\sum_{m=1}^{M-2}\sum_{k=1}^3 h(\mathcal
{G}_{km}^{M-2}|\bar{Y}_k^n)\leq (M-3)nR_{\Sigma}+n~o(\log\rho)+o(n)
\end{eqnarray}
where $\frac{1}{3}\sum_{k=1}^3$ stands for averaging over user
indices and hereinafter.

Next let consider the $(M,M+1)$ case, i.e., adding one layer to the
previous network. Since by providing
$(X_{(k+1)a_L}^n,X_{(k-1)c_L}^n)$ to Receiver $k$, we can boil down
the $(M,M+1)$ setting to the $(M-2,M-1)$ setting. By our
assumptions, we can easily obtain:
\begin{eqnarray}
&&\frac{1}{3}\sum_{m=1}^{M-2}\sum_{k=1}^3 h(\mathcal
{G}_{km}^M\setminus\{X_{(k+1)a_L}^n,X_{(k-1)c_L}^n\}|\bar{Y}_k^n,X_{(k+1)a_L}^n,X_{(k-1)c_L}^n)\leq \notag\\
&&(M-3)\frac{1}{3}\sum_{k=1}^3 h(\{X_{ka_l}^n,X_{kc_l}^n|1\leq l
\leq L-1\}|X_{ka_L}^n,X_{kc_L}^n)+n~o(\log\rho)+o(n).
\end{eqnarray}
For the $(M,M+1)$ case, we consider the sum differential entropy of
all genie signals provided to all receivers in the first $M-2$ sets
conditioning on the observations of each receiver, respectively.
That is,
\begin{small}
\begin{eqnarray}
&\!\!\!\!\!\!\!\!&\sum_{m=1}^{M-2}\sum_{k=1}^3 h(\mathcal {G}_{km}^M|\bar{Y}_k^n)\notag\\
&\!\!\!\!=\!\!\!\!&\sum_{m=1}^{M-2}\sum_{k=1}^3 \left(h(X_{(k+1)a_L}^n,X_{(k-1)c_L}^n|\bar{Y}_k^n)+h(\mathcal {G}_{km}^M\setminus\{X_{(k+1)a_L}^n,X_{(k-1)c_L}^n\}|\bar{Y}_k^n,X_{(k+1)a_L}^n,X_{(k-1)c_L}^n)\right)\\
&\!\!\!\!=\!\!\!\!& (M\!-\!2)\!\sum_{k=1}^3h(X_{(k+1)a_L}^n,X_{(k-1)c_L}^n|\bar{Y}_k^n)+\!\sum_{m=1}^{M-2}\!\sum_{k=1}^3 h(\mathcal {G}_{km}^M\!\!\setminus\!\!\{X_{(k+1)a_L}^n,X_{(k-1)c_L}^n\!\}|\bar{Y}_k^n,X_{(k+1)a_L}^n,X_{(k-1)c_L}^n\!)\ \ \ \ \ \ \ \\
&\!\!\!\!\leq\!\!\!\!& \sum_{k=1}^3h(X_{(k+1)a_L}^n,X_{(k-1)c_L}^n)+(M-3)\sum_{k=1}^3\left(h(X_{(k-1)c_L}^n)+h(X_{(k+1)a_L}^n|X_{(k+1)c_L}^n)\right)\notag\\
&\!\!\!\!\!\!\!\!&+\sum_{m=1}^{M-2}\sum_{k=1}^3 h(\mathcal {G}_{km}^M\setminus\{X_{(k+1)a_L}^n,X_{(k-1)c_L}^n\}|\bar{Y}_k^n,X_{(k+1)a_L}^n,X_{(k-1)c_L}^n)\\
&\!\!\!\!=\!\!\!\!& \sum_{k=1}^3h(X_{(k+1)a_L}^n,X_{(k-1)c_L}^n)+(M-3)\sum_{k=1}^3\left(h(X_{(k-1)c_L}^n)-h(X_{(k+1)c_L}^n)+h(X_{(k+1)a_L}^n,X_{(k+1)c_L}^n)\right)\notag\\
&\!\!\!\!\!\!\!\!&+(M-3)n\sum_{k=1}^3 h(\{X_{ka_l}^n,X_{kc_l}^n|1\leq l \leq L-1\}|X_{ka_L}^n,X_{kc_L}^n)+n~o(\log\rho)+o(n)\\
&\!\!\!\!=\!\!\!\!& \sum_{k=1}^3h(X_{(k+1)a_L}^n,X_{(k-1)c_L}^n)+n~o(\log\rho)+o(n)\notag\\
&\!\!\!\!\!\!\!\!&+(M-3)\sum_{k=1}^3\left(h(X_{(k+1)a_L}^n,X_{(k+1)c_L}^n)+h(\{X_{ka_l}^n,X_{kc_l}^n|1\leq l \leq L-1\}|X_{ka_L}^n,X_{kc_L}^n)\right)\\
&\!\!\!\!\leq \!\!\!\!& \sum_{k=1}^3\left(h(X_{(k+1)a_L}^n)+h(X_{(k-1)c_L}^n)\right)+(M-3)\sum_{k=1}^3 h(\{X_{ka_l}^n,X_{kc_l}^n|1\leq l \leq L\})+n~o(\log\rho)+o(n)\\
&\!\!\!\!= \!\!\!\!&
\sum_{k=1}^3\left(h(X_{ka_L}^n)+h(X_{kc_L}^n)\right)+(M-3)nR_{\Sigma}+n~o(\log\rho)+o(n).
\end{eqnarray}
\end{small}
Therefore, the summation of the $M-2$ sum rate inequalities produce
the following inequality:
\begin{eqnarray*}
3(M-2)nR&\!\!\!\!\leq \!\!\!\!& (M-2)Nn\log\rho+(M-3)nR\notag\\
&\!\!\!\!\!\!\!\!&+\frac{1}{3}\sum_{k=1}^3\left(h(X_{ka_L}^n)+h(X_{kc_L}^n)\right)+n~o(\log\rho)+o(n).
\end{eqnarray*}
Or equivalently it can be rewritten as:
\begin{eqnarray}
3(M-2)nR&\!\!\!\!\leq \!\!\!\!&
(M-2)Nn\log\rho+(M-3)nR+\left(h(X_{a_L}^n)+h(X_{c_L}^n)\right)+n~o(\log\rho)+o(n).\
\ \ \label{eqn:MbyN_ob1}
\end{eqnarray}

For the remaining last two genie signal sets, if a genie provides
$\mathcal {G}_{M-1}^M=\{X_{2a_L}^n,\{X_{2a_l}^n,X_{2c_l}^n|1\leq l
\leq L-1\}\}$ to Receiver 1, we have:
\begin{eqnarray}
nR_{\Sigma}&\!\!\!\!\leq\!\!\!\!& I(W_1,W_2,W_3;\bar{Y}^n_1,\mathcal{G}_{M-1}^M)+n~o(\log\rho)+o(n)\\
&\!\!\!\!\leq\!\!\!\!& Nn\log\rho+h(\mathcal{G}_{M-1}^M|\bar{Y}^n_1)+n~o(\log\rho)+o(n)\\
&\!\!\!\!\leq\!\!\!\!& Nn\log\rho+h(\mathcal{G}_{M-1}^M|X^n_{2c_L})+n~o(\log\rho)+o(n)\\
&\!\!\!\!\leq\!\!\!\!&
Nn\log\rho+nR_2-h(X^n_{2c_L})+n~o(\log\rho)+o(n)
\end{eqnarray}
and thus by averaging over user indices, we have the $(M-1)^{th}$
inequality:
\begin{eqnarray}\label{eqn:MbyN_ob2}
3nR\leq Nn\log\rho+nR-h(X^n_{c_L})+n~o(\log\rho)+o(n).
\end{eqnarray}
With similar analysis if a genie provides $\mathcal
{G}_{M-1}^M=\{X_{3c_L}^n,\{X_{3a_l}^n,X_{3c_l}^n|1\leq l \leq
L-1\}\}$ to Receiver 1, we have the $M^{th}$ inequality:
\begin{eqnarray}\label{eqn:MbyN_ob3}
3nR\leq Nn\log\rho+nR-h(X^n_{a_L})+n~o(\log\rho)+o(n).
\end{eqnarray}

Add up the inequalities of (\ref{eqn:MbyN_ob1}),
(\ref{eqn:MbyN_ob2}) and (\ref{eqn:MbyN_ob3}), we eventually obtain
\begin{eqnarray}
3MnR\leq MN\log\rho+(M-1)nR+n~o(\log\rho)+o(n),
\end{eqnarray}
which implies that the DoF per user outer bound
\begin{eqnarray}
d\leq \frac{MN}{2M+1}\leq\frac{MN}{M+N}.
\end{eqnarray}

Therefore, we establish the outer bound results. \hfill\QED

\subsubsection{$M$ is an Odd Number}\label{subsubsec:M_odd}

For $(M,M+1)$ case where $M$ is odd, the proof for the DoF outer
bound can be carried out in a similar way as what we have shown if
$M$ is even. Specifically, the total number of layers is still equal
to $L$ where $M=2L+1$. By peeling out the layers one by one, we
eventually obtain the $3\times 4$ core. The interference signallings
comprised at the received signal at each antenna after the linear
transformations is shown in Fig.\ref{fig:onion_odd}.
\begin{figure}[!h] \vspace{-0.15in}\centering
\includegraphics[width=5.5in]{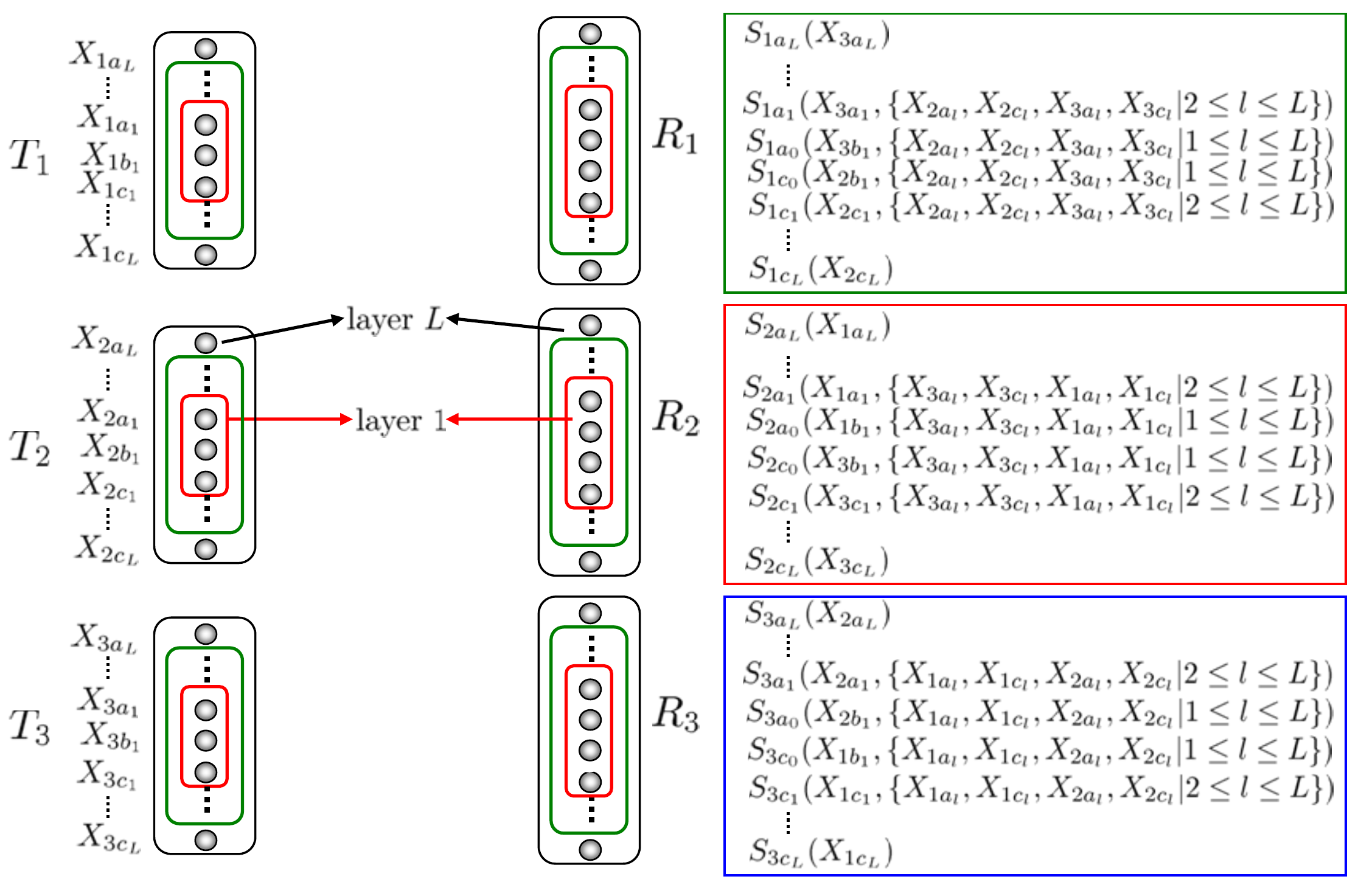}\vspace{-0.05in}
\caption{Intuition of Onion Peeling if $M$ is odd} \vspace{-0.1in}
\label{fig:onion_odd}
\end{figure}

For this case, we will provide $M$-sets of genie signals to each
receiver, each set with $M-1$ dimensional signals. Again, we show
the genie provided to receiver as follows, and the genie signals
provided to Receiver 2 and 3 can be obtained by advancing the user
indices.
\begin{eqnarray}
{\rm for}~~M=3,~~~~~\mathcal{G}_1^M&\!\!\!\!=\!\!\!\!&\{X_{2a_1}^n,X_{3c_1}^n\}\notag\\
\mathcal{G}_2^M&\!\!\!\!=\!\!\!\!&\{X_{2a_1}^n,X_{2b_1}^n\}\notag\\
\mathcal{G}_3^M&\!\!\!\!=\!\!\!\!&\{X_{3b_1}^n,X_{3c_1}^n\}\notag\\
{\rm then~for}~~M>3,~~~~~\mathcal{G}_{m}^M&\!\!\!\!=\!\!\!\!&\{X_{2a_L}^n,X_{3c_L}^n,\mathcal{G}_m^{M-2}\}~~~~~m=\{1,2,\cdots,M-2\},\notag\\
{\rm and}~~~~~\mathcal{G}_{M-1}^M&\!\!\!\!=\!\!\!\!&\{X_{2a_L}^n,\{X_{2a_l}^n,X_{2c_l}^n|1\leq l \leq L-1\},X_{2b_1}^n\}\notag\\
\mathcal{G}_M^M&\!\!\!\!=\!\!\!\!&\{X_{3c_L}^n,\{X_{3a_l}^n,X_{3c_l}^n|1\leq
l \leq L-1\},X_{3b_1}^n\} \label{eqn:general_genie_odd}
\end{eqnarray}
The remaining proof is the similar as that if $M$ is even by
replacing the genie signals $\mathcal{G}_m^M$ in
(\ref{eqn:general_genie_even}) with that shown in
(\ref{eqn:general_genie_odd}).

\subsection{DoF Outer Bound for $\frac{M}{N}\neq \frac{p}{p+1}$}

In this subsection, we only show the information-theoretic DoF outer
bound proofs for $\frac{M}{N}\neq \frac{p}{p+1}$. In fact, all the
derivations follow directly from what we have shown for $M/N=
p/(p+1)$. We will first show some examples and then extend the
analysis to the general cases.

\subsubsection{Case: Any $M/N\geq 2/3~\Rightarrow$ DoF $\leq \frac{3M}{5}$}
In this section, consider arbitrary $(M,N)$ values such that $M/N
\geq 2/3$. For all such values of $(M,N)$, the DoF outer bound that
holds is $d\leq \frac{3M}{5}$. In order to keep the presentation
clean and simple, from now on we will use tables instead of figures
to show the channel connectivity. Note that the linear
transformation still follows the similar manner as that introduced
in Section \ref{sec:outerbound}.

\begin{small}
\begin{eqnarray*}
\begin{array}{r|r|c|}
\hline |X_{1a}|=(N-M)&X_{1a}&\circ\\\hline
|X_{1b}|=3M-2N\geq0&X_{1b}&\circ\\\hline
|X_{1c}|=(N-M)&X_{1c}&\circ\\\hline
\end{array}
&&
\begin{array}{|c|l|l}
\hline \circ&S_{1a}(X_{3a})&|S_{1a}|=(N-M)\\\hline
\circ&S_{1b}(X_{2a},X_{2b},X_{2c},X_{3a},X_{3b},X_{3c})&|S_{1b}|=2M-N\geq(N-M)\\\hline
\circ&S_{1c}(X_{2c})&|S_{1c}|=(N-M)\\\hline
\end{array}
\\
&&\\
&&\\\begin{array}{r|r|c|} \hline |X_{2a}|=(N-M)&X_{2a}&\circ\\\hline
|X_{2b}|=3M-2N\geq0&X_{2b}&\circ\\\hline
|X_{2c}|=(N-M)&X_{2c}&\circ\\\hline
\end{array}
&&
\begin{array}{|c|l|l}
\hline \circ&S_{2a}(X_{1a})&|S_{2a}|=(N-M)\\\hline
\circ&S_{2b}(X_{3a},X_{3b},X_{3c},X_{1a},X_{1b},X_{1c})&|S_{2b}|=2M-N\geq(N-M)\\\hline
\circ&S_{2c}(X_{3c})&|S_{2c}|=(N-M)\\\hline
\end{array}
\\
&&\\
&&\\
\begin{array}{r|r|c|}
\hline |X_{3a}|=(N-M)&X_{3a}&\circ\\\hline
|X_{3b}|=3M-2N\geq0&X_{3b}&\circ\\\hline
|X_{3c}|=(N-M)&X_{3c}&\circ\\\hline
\end{array}
&&
\begin{array}{|c|l|l}
\hline \circ&S_{3a}(X_{2a})&|S_{3a}|=(N-M)\\\hline
\circ&S_{3b}(X_{1a},X_{1b},X_{1c},X_{2a},X_{2b},X_{2c})&|S_{3b}|=2M-N\geq(N-M)\\\hline
\circ&S_{3c}(X_{1c})&|S_{3c}|=(N-M)\\\hline
\end{array}
\end{eqnarray*}
\end{small}

In the table above, notice that one circle could represent more than
one antennas. For simplicity we still use $X$ to denote the signal
vector at corresponding antennas if without ambiguity and
hereinafter.

First a genie provides $\mathcal{G}_1=\{X_{2a}^n,X_{2b}^n\}$ to
receiver 1. Then the total number of dimensions available to
receiver 1 (including those provided by the genie) is equal to:
\begin{eqnarray*}
N+|\mathcal{G}_1|=N+|X_{2a}^n|+|X_{2b}^n|=N+(N-M)+(3M-2N)=2M
\end{eqnarray*}
With these (at least) $2M$ dimensions, it is easy to see that
receiver 1 is able to resolve both interfering signals and thus it
can decode all three messages. Therefore, we have:
\begin{eqnarray}
nR_{\Sigma}&\!\!\!\!\leq\!\!\!\!& Nn\log\rho+h(X_{2a}^n,X_{2b}^n|\bar{Y}_1^n)+n~o(\log\rho)+o(n)\\
&\!\!\!\!\leq\!\!\!\!& Nn\log\rho + h(X_{2a}^n,X_{2b}^n|X_{2c}^n)+n~o(\log\rho)+o(n)\\
&\!\!\!\!=\!\!\!\!& Nn\log\rho + nR_2-h(X_{2c}^n)+n~o(\log\rho)+o(n)
\end{eqnarray}
By advancing user indices, therefore we have:
%\begin{subequations}\label{eqn:big_2by3_ob1_set}
%\begin{eqnarray}
%nR_{\Sigma}&\!\!\!\!\leq\!\!\!\!&Nn\log\rho+nR_2-h(X_{2c}^n)+n~o(\log\rho)+o(n)\\
%nR_{\Sigma}&\!\!\!\!\leq\!\!\!\!&Nn\log\rho+nR_3-h(X_{3c}^n)+n~o(\log\rho)+o(n)\\
%nR_{\Sigma}&\!\!\!\!\leq\!\!\!\!&Nn\log\rho+nR_1-h(X_{1c}^n)+n~o(\log\rho)+o(n).
%\end{eqnarray}
%\end{subequations}
%Or equivalently (\ref{eqn:big_2by3_ob1_set}) can be rewritten as
\begin{eqnarray}
3nR\leq
Nn\log\rho+nR-h(X_{c}^n)+n~o(\log\rho)+o(n).\label{eqn:big_2by3_ob1}
\end{eqnarray}

Second a genie provides $\mathcal{G}_2=\{X_{3b}^n,X_{3c}^n\}$ to
receiver 1. Then the total number of dimensions available to
receiver 1 (including those provided by the genie) is equal to:
\begin{eqnarray*}
N+|\mathcal{G}_2|=N+|X_{3b}^n|+|X_{3c}^n|=N+(3M-2N)+(N-M)=2M
\end{eqnarray*}
Thus, receiver 1 is able to resolve both interfering signals.
Therefore, once again, we have:
\begin{eqnarray}
nR_{\Sigma}&\!\!\!\!\leq\!\!\!\!& Nn\log\rho+h(X_{3b}^n,X_{3c}^n|\bar{Y}_1^n)+n~o(\log\rho)+o(n)\\
&\!\!\!\!\leq\!\!\!\!& Nn\log\rho + h(X_{3b}^n)+h(X_{3c}^n)+n~o(\log\rho)+o(n)\\
&\!\!\!\!=\!\!\!\!& Nn\log\rho + (3M-2N)n\log\rho
+h(X_{3c}^n)+n~o(\log\rho)+o(n)
\end{eqnarray}
and therefore we have:
\begin{eqnarray}
3nR\leq Nn\log\rho+(3M-2N)n\log\rho
+h(X_{c}^n)+n~o(\log\rho)+o(n)\label{eqn:big_2by3_ob2}
\end{eqnarray}
Adding up the inequalities in (\ref{eqn:big_2by3_ob1}) and
(\ref{eqn:big_2by3_ob2}) we obtain:
\begin{eqnarray*}
6nR\leq 3Mn\log\rho+nR+n~o(\log\rho)+o(n)
\end{eqnarray*}
which implies that
\begin{eqnarray*}
d\leq \frac{3M}{5}.
\end{eqnarray*}

\subsubsection{Case: Any $M/N \in [1/2,2/3]\Rightarrow$ DoF $\leq \frac{2N}{5}$}
In this section, consider arbitrary $(M,N)$ values such that $M/N
\in [1/2,2/3]$. For all such values of $(M,N)$, the DoF outer bound
that holds is $d\leq \frac{2N}{5}$.

We still take the linear transformation introduced in Section 4
first at the receiver side. In addition, we take linear
transformation at the transmitter side by multiplying an $M\times M$
square matrix at transmitter $k$ which is the inverse of the channel
matrix from $X_k$ to $(Y_{(k-1)b},Y_{(k-1)c})$. This operation will
force $Y_{(k-1)c}$ only see the signals of $(X_2^n)_{[(2M-N+1):M]}$
where $(X)_{[m:n]}$ denotes the $m^{th}$ to $n^{th}$ entries of the
signal vector $X$ and hereinafter. The resulting network and
corresponding connectivity are present in the following table.

\begin{eqnarray*}
\begin{array}{r|r|c|}
\hline |X_1|=M&X_{1}&\circ\\\hline
\end{array}
&&
\begin{array}{|c|l|l}
\hline \circ&S_{1a}(X_{3})&|S_{1a}|=(N-M)\\\hline
\circ&S_{1b}(X_{2},X_{3})&|S_{1b}|=2M-N\leq (N-M)\\\hline
\circ&S_{1c}(X_{2})&|S_{1c}|=(N-M)\\\hline
\end{array}
\\
&&\\
&&\\
\begin{array}{r|r|c|}
\hline |X_2|=M&X_{2}&\circ\\\hline
\end{array}
&&
\begin{array}{|c|l|l}
\hline \circ&S_{2a}(X_{1})&|S_{2a}|=(N-M)\\\hline
\circ&S_{2b}(X_{3},X_{1})&|S_{2b}|=2M-N\leq (N-M)\\\hline
\circ&S_{2c}(X_{3})&|S_{2c}|=(N-M)\\\hline
\end{array}
\\
&&\\
&&\\
\begin{array}{r|r|c|}
\hline |X_3|=M&X_{3}&\circ\\\hline
\end{array}
&&
\begin{array}{|c|l|l}
\hline \circ&S_{3a}(X_{2})&|S_{3a}|=(N-M)\\\hline
\circ&S_{3b}(X_{1},X_{2})&|S_{3b}|=2M-N\leq (N-M)\\\hline
\circ&S_{3c}(X_{1})&|S_{3c}|=(N-M)\\\hline
\end{array}
\end{eqnarray*}

First a genie provides $\mathcal{G}_1=(X_2^n)_{[1:(N-M)]}$ to
receiver 1. Then the total number of dimensions available to
receiver 1 (including those provided by the genie) is equal to:
\begin{eqnarray*}
N+|\mathcal{G}_1|=N+(N-M)\geq 3M-M=2M
\end{eqnarray*}
With at least $2M$ dimensions among them, it is easy to see that
receiver 1 is able to resolve both interfering signals. Therefore,
we have:
\begin{eqnarray}
nR_{\Sigma}&\!\!\!\!\leq\!\!\!\!& Nn\log\rho+h((X_2^n)_{[1:(N-M)]}|\bar{Y}_1^n)+n~o(\log\rho)+o(n)\\
&\!\!\!\!\leq\!\!\!\!& Nn\log\rho + h((X_2^n)_{[1:(N-M)]}|(X_2^n)_{[(N-M+1):M]})+n~o(\log\rho)+o(n)\label{eqn:small_2by3_drop}\\
&\!\!\!\!=\!\!\!\!& Nn\log\rho +
nR_2-h((X_2^n)_{[(N-M+1):M]})+n~o(\log\rho)+o(n)
\end{eqnarray}
where (\ref{eqn:small_2by3_drop}) follows from dropping condition
terms does not decrease the capacity. Specifically, $Y_{(k-1)c}$
only see the signals of $(X_2^n)_{[(2M-N+1):M]}$. Since we are
considering $\frac{1}{2}\leq \frac{M}{N} \leq \frac{2}{3}$ which
implies that $3M\leq 2N$, we have $(2M-N+1)\leq (N-M+1)$. In other
words, we only keep the last $2M-N$ entries of $\bar{Y}_1$ in the
condition and drop the other terms. By advancing user indices,
therefore we have:
\begin{eqnarray}
3nR\leq
Nn\log\rho+nR-h((X^n)_{[(N-M+1):M]})+n~o(\log\rho)+o(n).\label{eqn:small_2by3_ob1_set}
\end{eqnarray}

Second a genie provides $\mathcal{G}_2=(X_3^n)_{[(N-M+1):M]}$ to
receiver 1. Then the total number of dimensions available to
receiver 1 including those provided by the genie is equal to:
\begin{eqnarray*}
N+|\mathcal{G}_2|=N+(M-(N-M+1)+1)=2M
\end{eqnarray*}
With these $2M$ dimensions, receiver 1 once again is able to resolve
both interfering signals. Therefore, we have:
\begin{eqnarray}
nR_{\Sigma}&\!\!\!\!\leq\!\!\!\!& Nn\log\rho+h((X_3^n)_{[(N-M+1):M]}|\bar{Y}_1^n)+n~o(\log\rho)+o(n)\\
&\!\!\!\!\leq\!\!\!\!& Nn\log\rho+
h((X_3^n)_{[(N-M+1):M]})+n~o(\log\rho)+o(n)
\end{eqnarray}
By advancing user indices, therefore we have:
\begin{eqnarray}
3nR\leq
Nn\log\rho+h((X^n)_{[(N-M+1):M]})+n~o(\log\rho)+o(n).\label{eqn:small_2by3_ob2_set}
\end{eqnarray}

Adding up the inequalities in (\ref{eqn:small_2by3_ob1_set}) and
(\ref{eqn:small_2by3_ob2_set}) we obtain:
\begin{eqnarray*}
6nR\leq 2Nn\log\rho+nR+n~o(\log\rho)+o(n)
\end{eqnarray*}
which implies that
\begin{eqnarray*}
d\leq \frac{2N}{5}.
\end{eqnarray*}

\subsubsection{Case: Any $M/N\geq 4/5~\Rightarrow$ DoF $\leq \frac{5M}{9}$}
In this section, consider arbitrary $(M,N)$ values such that $M/N
\geq 4/5$. For all such values of $(M,N)$, the DoF outer bound that
holds is $d\leq \frac{5M}{9}$.

After the linear transformation that we introduce in Section 4, the
resulting network and channel connectivity are shown in the
following table.

\begin{small}
\begin{eqnarray*}
\begin{array}{r|r|c|}
\hline |X_{1a_2}|=N-M&X_{1a_2}&\circ\\\hline
|X_{1a_1}|=N-M&X_{1a}&\circ\\\hline |X_{1b_1}|=
5M-4N&X_{1b_1}&\circ\\\hline |X_{1c_1}|=N-M&X_{1c_1}&\circ\\\hline
|X_{1c_2}|=N-M&X_{1c_2}&\circ\\\hline
\end{array}
&&
\begin{array}{|c|l|l}
\hline \circ&S_{1a_2}(X_{3a_2})&|S_{1a_1}|=N-M\\\hline
\circ&S_{1a_1}(X_{3a_1},X_{3a_2},X_{3c_2},X_{2a_2},X_{2c_2})&|S_{1a}|=N-M\\\hline
\circ&S_{1b_1}(X_{2},X_{3})&|S_{1b}|=4M-3N\geq (N-M)\\\hline
\circ&S_{1c_1}(X_{2c_1},X_{3a_2},X_{3c_2},X_{2a_2},X_{2c_2})&|S_{1c}|=N-M\\\hline
\circ&S_{1c_2}(X_{2c_2})&|S_{1c_1}|=N-M\\\hline
\end{array}
\\
&&\\
&&\\
\begin{array}{r|c|}
\hline X_{2a_2}&\circ\\\hline X_{2a_1}&\circ\\\hline
X_{2b_1}&\circ\\\hline X_{2c_1}&\circ\\\hline X_{2c_2}&\circ\\\hline
\end{array}
&&
\begin{array}{|c|l}
\hline \circ&S_{2a_2}(X_{1a_2})\\\hline
\circ&S_{2a_1}(X_{1a_1},X_{1a_2},X_{1c_2},X_{3a_2},X_{3c_2})\\\hline
\circ&S_{2b_1}(X_{3},X_{1})\\\hline
\circ&S_{2c_1}(X_{3c_1},X_{1a_2},X_{1c_2},X_{3a_2},X_{3c_2})\\\hline
\circ&S_{2c_2}(X_{3c_2})\\\hline\end{array}
\\
&&\\
&&\\
\begin{array}{r|c|}
\hline X_{3a_2}&\circ\\\hline X_{3a_1}&\circ\\\hline
X_{3b_1}&\circ\\\hline X_{3c_1}&\circ\\\hline X_{3c_2}&\circ\\\hline
\end{array}
&&
\begin{array}{|c|l}
\hline \circ&S_{3a_2}(X_{2a_2})\\\hline
\circ&S_{3a_1}(X_{2a_1},X_{2a_2},X_{2c_2},X_{1a_2},X_{1c_2})\\\hline
\circ&S_{3b_1}(X_{1},X_{2})\\\hline
\circ&S_{3c_1}(X_{1c_1},X_{2a_2},X_{2c_2},X_{1a_2},X_{1c_2})\\\hline
\circ&S_{3c_2}(X_{1c_2})\\\hline
\end{array}
\end{eqnarray*}
\end{small}

First a genie provides
$\mathcal{G}_1=\{X_{2a_2}^n,X_{3c_2}^n,X_{2a_1}^n,X_{2b_1}^n\}$ to
receiver 1. Then the total number of dimensions available to
receiver 1 (including those provided by the genie) is equal to:
\begin{eqnarray*}
N+|\mathcal{G}_1|&\!\!\!\!=\!\!\!\!& N+|X_{2a_2}^n|+|X_{3c_2}^n|+|X_{2a_1}^n|+|X_{2b_1}^n|\\
&\!\!\!\!=\!\!\!\!& N+(N-M)+(N-M)+(N-M)+(5M-4N)=2M
\end{eqnarray*}
With these $2M$ dimensions, receiver 1 is able to resolve both
interfering signals. Therefore, we have:
\begin{small}
\begin{eqnarray}
nR_{\Sigma}&\!\!\!\!\leq\!\!\!\!& Nn\log\rho+h(X_{2a_2}^n,X_{3c_2}^n,X_{2a_1}^n,X_{2b_1}^n|\bar{Y}_1^n)+n~o(\log\rho)+o(n)\\
&\!\!\!\!\leq\!\!\!\!& Nn\log\rho+h(X_{3c_2}^n|\bar{Y}_1^n)+h(X_{2a_2}^n|\bar{Y}_1^n)+h(X_{2a_1}^n,X_{2b_1}^n|\bar{Y}_1^n,X_{2a_2}^n,X_{3c_2}^n)+n~o(\log\rho)+o(n)\\
&\!\!\!\!\leq\!\!\!\!& Nn\log\rho+h(X_{3c_2}^n)+h(X_{2a_2}^n|X_{2c_2}^n)+h(X_{2a_1}^n,X_{2b_1}^n|X_{2c_1}^n,X_{2a_2}^n,X_{3c_2}^n)+n~o(\log\rho)+o(n)\\
&\!\!\!\!=\!\!\!\!& Nn\log\rho+h(X_{3c_2}^n)+
nR_2-h(X_{2c_2}^n)-h(X_{2c_1}^n|X_{2a_2}^n,X_{2c_2}^n)+n~o(\log\rho)+o(n)
\end{eqnarray}
\end{small}
and by advancing user indices, therefore we have:
\begin{eqnarray}
3nR\leq
Nn\log\rho+nR-h(X_{c_1}^n|X_{a_2}^n,X_{c_2}^n)+n~o(\log\rho)+o(n).\label{eqn:big_4by5_ob1_set}
\end{eqnarray}

Second a genie provides
$\mathcal{G}_2=\{X_{2a_2}^n,X_{3c_2}^n,X_{3b_1}^n,X_{3c_1}^n\}$ to
receiver 1. Then the total number of dimensions available to
receiver 1 (including those provided by the genie) is equal to:
\begin{eqnarray*}
N+|\mathcal{G}_1|&\!\!\!\!=\!\!\!\!& N+|X_{2a_2}^n|+|X_{3c_2}^n|+|X_{3b_1}^n|+|X_{3c_1}^n|\\
&\!\!\!\!=\!\!\!\!& N+(N-M)+(N-M)+(5M-4N)+(N-M)=2M
\end{eqnarray*}
With these $2M$ dimensions, receiver 1 is able to resolve both
interfering signals. Therefore, we have:
\begin{small}
\begin{eqnarray}
nR_{\Sigma}&\!\!\!\!\leq\!\!\!\!& Nn\log\rho+h(X_{2a_2}^n,X_{3c_2}^n,X_{3b_1}^n,X_{3c_1}^n|\bar{Y}_1^n)+n~o(\log\rho)+o(n)\\
&\!\!\!\!\leq\!\!\!\!& Nn\log\rho+h(X_{2a_2}^n)+h(X_{3c_2}^n)+h(X_{3b_1}^n)+h(X_{3c_1}^n|X_{3a_2}^n,X_{3c_2}^n)+n~o(\log\rho)+o(n)\\
&\!\!\!\!\leq\!\!\!\!&
Nn\log\rho+h(X_{2a_2}^n)+h(X_{3c_2}^n)+(5M-4N)n\log\rho+h(X_{3c_1}^n|X_{3a_2}^n,X_{3c_2}^n)+n~o(\log\rho)+o(n)
\end{eqnarray}
\end{small}
and by advancing user indices, therefore we have:
\begin{eqnarray}
3nR\leq
(5M-3N)n\log\rho+nR+h(X_{a_2}^n)+h(X_{c_2}^n)-h(X_{c_1}^n|X_{a_2}^n,X_{c_2}^n)+n~o(\log\rho)+o(n).\label{eqn:big_4by5_ob2_set}
\end{eqnarray}

Third a genie provides
$\mathcal{G}_3=\{X_{2a_2}^n,X_{2a_1}^n,X_{2b_1}^n,X_{2c_1}^n\}$ to
receiver 1. Then the total number of dimensions available to
receiver 1 (including those provided by the genie) is equal to:
\begin{eqnarray*}
N+|\mathcal{G}_1|=N+(N-M)+(N-M)+(5M-4N)+(N-M)=2M
\end{eqnarray*}
With these $2M$ dimensions, receiver 1 is able to resolve both
interfering signals. Therefore, we have:
\begin{eqnarray}
nR_{\Sigma}&\!\!\!\!\leq\!\!\!\!& Nn\log\rho+h(X_{2a_2}^n,X_{2a_1}^n,X_{2b_1}^n,X_{2c_1}^n|\bar{Y}_1^n)+n~o(\log\rho)+o(n)\\
&\!\!\!\!\leq\!\!\!\!& Nn\log\rho+h(X_{2a_2}^n,X_{2a_1}^n,X_{2b_1}^n,X_{2c_1}^n|X_{2c_2}^n)+n~o(\log\rho)+o(n)\\
&\!\!\!\!=\!\!\!\!&
Nn\log\rho+nR_2-h(X_{2c_2}^n)+n~o(\log\rho)+o(n).
\end{eqnarray}
By advancing user indices, therefore we have:
\begin{eqnarray}
3nR\leq
Nn\log\rho+nR-h(X_{c_2}^n)+n~o(\log\rho)+o(n).\label{eqn:big_4by5_ob3_set}
\end{eqnarray}

Last a genie provides
$\mathcal{G}_4=\{X_{3c_2}^n,X_{3a_1}^n,X_{3b_1}^n,X_{3c_1}^n\}$ to
receiver 1. With the similar analysis shown above, we obtain
\begin{eqnarray}
3nR\leq
Nn\log\rho+nR-h(X_{a_2}^n)+n~o(\log\rho)+o(n).\label{eqn:big_4by5_ob4_set}
\end{eqnarray}

Adding up the four inequalities in (\ref{eqn:big_4by5_ob1_set}),
(\ref{eqn:big_4by5_ob2_set}), (\ref{eqn:big_4by5_ob3_set}) and
(\ref{eqn:big_4by5_ob4_set}) we obtain:
\begin{eqnarray*}
12nR\leq 5Mn\log\rho+3nR+n~o(\log\rho)+o(n)
\end{eqnarray*}
which implies that
\begin{eqnarray*}
d\leq \frac{5M}{9}.
\end{eqnarray*}

\subsubsection{Case: Any $M/N \in [3/4,4/5]\Rightarrow$ DoF $\leq \frac{4N}{9}$}

In this section, consider arbitrary $(M,N)$ values such that $M/N
\in [3/4,4/5]$. For all such values of $(M,N)$, the DoF outer bound
that holds is $d\leq \frac{4N}{9}$.

With the linear transformation at both transmitter and receiver
sides, we have the following table representing the network and
corresponding connectivity.

\begin{small}
\begin{eqnarray*}
\begin{array}{r|r|c|}
\hline |X_{1a_2}|=N-M&X_{1a_2}&\circ\\\hline
|X_{1b}|=3M-2N>(N-M)&X_{1b}&\circ\\\hline |X_{1c_2}|\leq
N-M&X_{1c_2}&\circ\\\hline
\end{array}
&&
\begin{array}{|c|l|l}
\hline \circ&S_{1a_2}(X_{3a_2})&|S_{1a_2}|=N-M\\\hline
\circ&S_{1a_1}(X_{3b},X_{2a_2},X_{2c_2},X_{3a_2},X_{3c_2})&|S_{1a_1}|=N-M\\\hline
\circ&S_{1b_1}(X_{2},X_{3})&|S_{1b}|=4M-3N\\\hline
\circ&S_{1c_1}(X_{2b},X_{2a_2},X_{2c_2},X_{3a_2},X_{3c_2})&|S_{1c_1}|=N-M\\\hline
\circ&S_{1c_2}(X_{2c_2})&|S_{1c_2}|=N-M\\\hline
\end{array}
\\
&&\\
&&\\
\begin{array}{r|c|}
\hline X_{2a_2}&\circ\\\hline X_{2b}&\circ\\\hline
X_{2c_2}&\circ\\\hline
\end{array}
&&
\begin{array}{|c|l}
\hline \circ&S_{2a_2}(X_{1a_2})\\\hline
\circ&S_{2a_1}(X_{1b},X_{3a_2},X_{3c_2},X_{1a_2},X_{1c_2})\\\hline
\circ&S_{2b_1}(X_{3},X_{1})\\\hline
\circ&S_{2c_1}(X_{3b},X_{3a_2},X_{3c_2},X_{1a_2},X_{1c_2})\\\hline
\circ&S_{2a_2}(X_{3c_2})\\\hline
\end{array}
\\
&&\\
&&\\
\begin{array}{r|c|}
\hline X_{3a_2}&\circ\\\hline X_{3b}&\circ\\\hline
X_{3c_2}&\circ\\\hline
\end{array}
&&
\begin{array}{|c|l}
\hline \circ&S_{3a_2}(X_{2a_2})\\\hline
\circ&S_{3a_1}(X_{2b},X_{1a_2},X_{1c_2},X_{2a_2},X_{2c_2})\\\hline
\circ&S_{3b_1}(X_{1},X_{2})\\\hline
\circ&S_{3c_1}(X_{1b},X_{1a_2},X_{1c_2},X_{2a_2},X_{2c_2})\\\hline
\circ&S_{3c_2}(X_{1c_2})\\\hline
\end{array}
\end{eqnarray*}
\end{small}

Similar to $M/N \in [1/2,2/3]$ case, we at last multiply at the
transmitter $k$ a square matrix which is inverse of the channel
matrix from $X_{kb}$ to $(X_{(k-1)b_1},X_{(k-1)c_1})$ for
$k\in\{1,2,3\}$. This operation will help us only keep the necessary
condition terms when we bound the sum rate of three messages, as
what we have shown in $M/N \in [1/2,2/3]$ case.

First a genie provides
$\mathcal{G}_1=\{X_{2a_2}^n,X_{3c_2}^n,(X_{2b}^n)_{[1:(N-M)]}\}$ to
receiver 1. Then the total number of dimensions available to
receiver 1 is equal to:
\begin{eqnarray*}
N+|\mathcal{G}_1|=N+(N-M)+(N-M)+(N-M)=4N-3M\geq 2M
\end{eqnarray*}
With at least $2M$ dimensions among them, receiver 1 is able to
resolve both interfering signals. Therefore, we have:
\begin{small}
\begin{eqnarray}
nR_{\Sigma}&\!\!\!\!\leq\!\!\!\!& Nn\log\rho+h(X_{2a_2}^n,X_{3c_2}^n,(X_{2b}^n)_{[1:(N-M)]}|\bar{Y}_1^n)+n~o(\log\rho)+o(n)\\
&\!\!\!\!\leq\!\!\!\!& Nn\log\rho+h(X_{3c_2}^n)+h(X_{2a_2}^n|X_{2c_2}^n)+h((X_{2b}^n)_{[1:(N-M)]}|\bar{Y}_1^n,X_{2a_2}^n,X_{3c_2}^n)+n~o(\log\rho)+o(n)\\
&\!\!\!\!\leq\!\!\!\!& Nn\log\rho+h(X_{3c_2}^n)+h(X_{2a_2}^n|X_{2c_2}^n)+h((X_{2b}^n)_{[1:(N-M)]}|(X_{2b}^n)_{[(N-M+1):M]},X_{2a_2}^n,X_{2c_2}^n)\notag\\
&\!\!\!\!\!\!\!\!&+n~o(\log\rho)+o(n)\label{eqn:small_4by5_drop}\\
&\!\!\!\!=\!\!\!\!& Nn\log\rho+h(X_{3c_2}^n)+
nR_2-h(X_{2c_2}^n)-h((X_{2b}^n)_{[(N-M+1):(3M-2N)]}|X_{2a_2}^n,X_{2c_2}^n)\notag\\
&\!\!\!\!\!\!\!\!&+n~o(\log\rho)+o(n)
\end{eqnarray}
\end{small}
where (\ref{eqn:small_4by5_drop}) follows from dropping condition
terms does not decrease the channel capacity. Specifically, with
$X_{2a_2}^n,X_{3c_2}^n$ and
$S_{1a_2}(X_{3a_2}^n),S_{1c_2}(X_{2c_2}^n)$, we can subtract them
from $S_{1c_1}$ and thus only leave a clean $X_{2b}^n$. Since we
have invert the associated channel at the transmitter 2, we can only
keep the term $(X_{2b}^n)_{[(N-M+1):M]}$ in the condition. By
advancing user indices, therefore we have:
\begin{eqnarray}
3nR\leq
Nn\log\rho+nR-h((X_{b}^n)_{[(N-M+1):(3M-2N)]}|X_{a_2}^n,X_{c_2}^n)+n~o(\log\rho)+o(n).\label{eqn:small_4by5_ob1_set}
\end{eqnarray}

Second a genie provides
$\mathcal{G}_2=\{X_{2a_2}^n,X_{3c_2}^n,(X_{3b}^n)_{[(N-M+1):(3M-2N)]}\}$
to receiver 1. Then the total number of dimensions available to
receiver 1 is equal to:
\begin{eqnarray*}
N+|\mathcal{G}_2|=N+(N-M)+(N-M)+((3M-2N)-(N-M+1)+1)=2M
\end{eqnarray*}
With these $2M$ dimensions, receiver 1 is able to resolve both
interfering signals. Therefore, we have:
\begin{small}
\begin{eqnarray}
nR_{\Sigma}&\!\!\!\!\leq\!\!\!\!& Nn\log\rho+h(X_{2a_2}^n,X_{3c_2}^n,(X_{3b}^n)_{[(N-M+1):(3M-2N)]}|\bar{Y}_1^n)+n~o(\log\rho)+o(n)\\
&\!\!\!\!\leq\!\!\!\!& Nn\log\rho+h(X_{2a_2}^n,X_{3c_2}^n)+h((X_{3b}^n)_{[(N-M+1):(3M-2N)]}|\bar{Y}_1^n,X_{2a_2}^n,X_{3c_2}^n)+n~o(\log\rho)+o(n)\\
&\!\!\!\!\leq\!\!\!\!& Nn\log\rho+h(X_{2a_2}^n)+h(X_{3c_2}^n)+
h((X_{3b}^n)_{[(N-M+1):(3M-2N)]}|X_{3a_2}^n,X_{3c_2}^n)+n~o(\log\rho)+o(n)
\end{eqnarray}
\end{small}
and by advancing user indices, therefore we have:
\begin{eqnarray}
3nR\leq Nn\log\rho+h(X_{a_2}^n)+h(X_{c_2}^n)+
h(X_{b}^n)_{[(N-M+1):(3M-2N)]}|X_{a_2}^n,X_{c_2}^n)+n~o(\log\rho)+o(n).\label{eqn:small_4by5_ob2_set}
\end{eqnarray}

Third, if a genie provides $\mathcal{G}_3=\{X_{2a_2}^n,X_{2b}^n\}$
to receiver 1. Then the total number of dimensions available to
receiver 1 (including those provided by the genie) is equal to:
\begin{eqnarray*}
N+|\mathcal{G}_1|=N+(N-M)+(3M-2N)=2M
\end{eqnarray*}
With these $2M$ dimensions, receiver 1 is able to resolve both
interfering signals. Therefore, we have:
\begin{eqnarray}
nR_{\Sigma}&\!\!\!\!\leq\!\!\!\!& Nn\log\rho+h(X_{2a_2}^n,X_{2b}^n|\bar{Y}_1^n)+n~o(\log\rho)+o(n)\\
&\!\!\!\!\leq\!\!\!\!&
Nn\log\rho+nR_2-h(X_{2c_2}^n)+n~o(\log\rho)+o(n).
\end{eqnarray}
By advancing user indices, therefore we have:
\begin{eqnarray}
3nR\leq
Nn\log\rho+nR-h(X_{c_2}^n)+n~o(\log\rho)+o(n).\label{eqn:small_4by5_ob3_set}
\end{eqnarray}

Last a genie provides $\mathcal{G}_4=\{X_{3b}^n,X_{3c_2}^n\}$ to
receiver 1. With the similar analysis shown above, we obtain
\begin{eqnarray}
3nR\leq
Nn\log\rho+nR-h(X_{a_2}^n)+n~o(\log\rho)+o(n).\label{eqn:small_4by5_ob4_set}
\end{eqnarray}

Adding up the inequalities in (\ref{eqn:small_4by5_ob1_set}),
(\ref{eqn:small_4by5_ob2_set}), (\ref{eqn:small_4by5_ob3_set}) and
(\ref{eqn:small_4by5_ob4_set}) we obtain:
\begin{eqnarray*}
12nR\leq 4Nn\log\rho+3nR+n~o(\log\rho)+o(n)
\end{eqnarray*}
which implies that
\begin{eqnarray*}
d\leq \frac{4N}{9}.
\end{eqnarray*}

%\subsection{$M=2L$ is an Even Number}

\subsubsection{Case: Any $M/N\geq \frac{2L}{2L+1}~\Rightarrow$ DoF $\leq \frac{2L+1}{4L+1}M$}
For this case, the resulting network and channel connectivity after
the linear transformation are shown in the following table.

\begin{small}
\begin{eqnarray*}
\begin{array}{r|c|}
\hline X_{1a_L}&\circ\\\hline \vdots&\circ\\\hline
X_{1a_1}&\circ\\\hline X_{1b_1}&\circ\\\hline X_{1c_1}&\circ\\\hline
\vdots&\circ\\\hline X_{1c_L}&\circ\\\hline
\end{array}
&&
\begin{array}{|c|l}\hline
\circ&S_{1a_L}(X_{3a_L})\\\hline \circ&\vdots\\\hline
\circ&S_{1a_1}(X_{3a_1},\{X_{2a_l},X_{2c_l},X_{3a_l},X_{3c_l}|2\leq
l\leq L\})\\\hline \circ&S_{1b_1}(X_{2},X_{3})\\\hline
\circ&S_{1c_1}(X_{2c_1},\{X_{2a_l},X_{2c_l},X_{3a_l},X_{3c_l}|2\leq
l\leq L\})\\\hline \circ&\vdots\\\hline
\circ&S_{1c_L}(X_{2c_L})\\\hline
\end{array}
\\
&&\\
&&\\
\begin{array}{r|c|}
\hline X_{2a_L}&\circ\\\hline \vdots&\circ\\\hline
X_{2a_1}&\circ\\\hline X_{2b_1}&\circ\\\hline X_{2c_1}&\circ\\\hline
\vdots&\circ\\\hline X_{2c_L}&\circ\\\hline
\end{array}
&&
\begin{array}{|c|l}
\hline \circ&S_{2a_L}(X_{1a_L})\\\hline \circ&\vdots\\\hline
\circ&S_{2a_1}(X_{1a_1},\{X_{3a_l},X_{3c_l},X_{1a_l},X_{1c_l}|2\leq
l\leq L\})\\\hline \circ&S_{2b_1}(X_{3},X_{1})\\\hline
\circ&S_{2c_1}(X_{3c_1},\{X_{3a_l},X_{3c_l},X_{1a_l},X_{1c_l}|2\leq
l\leq L\})\\\hline \circ&\vdots\\\hline
\circ&S_{2c_L}(X_{3c_L})\\\hline

\end{array}
\\
&&\\
&&\\
\begin{array}{r|c|}
\hline X_{3a_L}&\circ\\\hline \vdots&\circ\\\hline
X_{3a_1}&\circ\\\hline X_{3b_1}&\circ\\\hline X_{3c_1}&\circ\\\hline
\vdots&\circ\\\hline X_{3c_L}&\circ\\\hline
\end{array}
&&
\begin{array}{|c|l}
\hline \circ&S_{3a_L}\\\hline \circ&\vdots\\\hline
\circ&S_{3a_1}\\\hline \circ&S_{3b_1}\\\hline \circ&S_{3c_1}\\\hline
\circ&\vdots\\\hline \circ&S_{3c_L}\\\hline\end{array}
\end{eqnarray*}
\end{small}

We denote $\mathcal {G}_m^{2L},~m\in\{1,\cdots,2L\}$ as the the $2L$
sets of the provided genie signals to receiver 1 for the cases we
consider, which can be written in an iterative form as follows: for
$m'=\{1,2,\cdots,L\}$ where $L$ is the total number of layers,
\begin{eqnarray}
\mathcal{G}_{2m'-1}^{2L}&\!\!\!\!=\!\!\!\!&\{\{X_{2a_l}^n,X_{3c_l}^n|m'+1\leq l \leq L\},X_{2a_{m'}}^n,\{X_{2a_l}^n,X_{2c_l}^n|1\leq l \leq m'-1\},X_{2b_1}^n\}\notag\\
\mathcal{G}_{2m'}^{2L}&\!\!\!\!=\!\!\!\!&\{\{X_{2a_l}^n,X_{3c_l}^n|m'+1\leq
l \leq L\},X_{3c_{m'}}^n,\{X_{3a_l}^n,X_{3c_l}^n|1\leq l \leq
m'-1\},X_{3b_1}^n\}\label{eqn:general_big_genie_even}
\end{eqnarray}
%\begin{eqnarray}
%{\rm for}~~L=1,~~~~~\mathcal{G}_1^{2L}&\!\!\!\!=\!\!\!\!&\{X_{2a_1}^n,X_{2b_1}^n\}\notag\\
%\mathcal{G}_2^{2L}&\!\!\!\!=\!\!\!\!&\{X_{3b_1}^n,X_{3c_1}^n\}\notag\\
%{\rm then~for}~~L>1,~~~~~\mathcal{G}_{m}^{2L}&\!\!\!\!=\!\!\!\!&\{X_{2a_L}^n,X_{3c_L}^n,\mathcal{G}_m^{M-2}\}~~~~~m=\{1,2,\cdots,2L-2\},\notag\\
%{\rm and}~~~~~\mathcal{G}_{2L-1}^{2L}&\!\!\!\!=\!\!\!\!&\{X_{2a_L}^n,\{X_{2a_l}^n,X_{2c_l}^n|1\leq l \leq L-1\},X_{2b_1}^n\}\notag\\
%\mathcal{G}_{2L}^{2L}&\!\!\!\!=\!\!\!\!&\{X_{3c_L}^n,\{X_{3a_l}^n,X_{3c_l}^n|1\leq
%l \leq L-1\},X_{3b_1}^n\} \label{eqn:general_big_genie_even}
%\end{eqnarray}
For each set shown above, the total number of dimensions available
to receiver 1 including the dimensions provided by the genie is
equal to:
\begin{eqnarray*}
N+|\mathcal{G}_m^{2L}|=N+(N-M)[2(L-1)+1]+[(2L+1)M-2LN]=2M.
\end{eqnarray*}
Similar to the proof shown in Appendix A.1, we still use
mathematical induction to prove the general result. Due to the
similarities between the proofs in this subsection and in Appendix
A.1, we will only highlight their difference.

With the intuition of onion peeling, we have a total of $L$ layers
at each user in this case. For the $l^{th}$ layer where $2\leq l\leq
L$, there are two groups of signals $(X_{ka_l}^n,X_{kc_l}^n)$ for
user $k$ with the cardinality $|X_{ka_l}^n|=|X_{kc_l}^n|=N-M$, while
in in the first layer, there are three groups of signals
$(X_{ka_1}^n,X_{kb_1}^n,X_{kc_1}^n)$ with the cardinality
$|X_{ka_1}^n|=|X_{kc_1}^n|=N-M$ and $|X_{kb_1}^n|=(2L+1)M-2LN$.
Since we have proved for $L=1,~2$ cases previously, now let us
assume it works for the $L-1$ case. That is to say, by providing
genie signals $\mathcal {G}_m^{2(L-1)},~m\in\{1,\cdots,L-1\}$ to
receiver 1 and by advancing user indices, we obtain a total of
$(2L-2)$-rate per user inequalities. If we add up all these
inequalities, at the left-hand side we have $3(2L-2)nR$; at the
right-hand side we have
$(2L-2)Nn\log\rho+[(2L+1)M-2LN]n\log\rho+(2L-3)nR+n~o(\log\rho)+o(n)$.
In other words, we can bound the sum differential entropy of all
provided genie signals to three users above by:
\begin{eqnarray}
\sum_{m=1}^{2L-2}\sum_{k=1}^3 h(\mathcal
{G}_{km}^{2L-2}|\bar{Y}_k^n)\leq
[(2L+1)M-2LN]n\log\rho+(2L-3)nR+n~o(\log\rho)+o(n).
\end{eqnarray}
Thus, if we add the $L^{th}$ layer, the resulting summation of rate
outer bounds provided by the first $2L-2$ inequalities, as shown in
(\ref{eqn:MbyN_ob1}), in this case becomes:
\begin{small}
\begin{eqnarray}
3(2L-2)nR&\!\!\!\!\leq \!\!\!\!& (2L-2)Nn\log\rho+[(2L+1)M-2LN]n\log\rho+(2L-3)nR\notag\\
&\!\!\!\!\!\!\!\!&+h(X_{a_L}^n)+h(X_{c_L}^n)+n~o(\log\rho)+o(n)\\
&\!\!\!\!= \!\!\!\!&
[(2L+1)M-2N]n\log\rho+(2L-3)nR+h(X_{ka_L}^n)+h(X_{kc_L}^n)+n~o(\log\rho)+o(n).\label{eqn:bigeven_MbyN_ob1_set}
\end{eqnarray}
\end{small}
Next, if the genie provides the last two sets of signals
$\mathcal{G}_{2L-1}^{2L}$ and $\mathcal{G}_{2L}^{2L}$ to receiver 1,
respectively, we can easily obtain the last two inequalities as:
\begin{eqnarray}
3nR&\!\!\!\!\leq \!\!\!\!& Nn\log\rho+nR-h(X_{c_L}^n)+n~o(\log\rho)+o(n)\label{eqn:bigeven_MbyN_ob2_set}\\
3nR&\!\!\!\!\leq \!\!\!\!&
Nn\log\rho+nR-h(X_{a_L}^n)+n~o(\log\rho)+o(n).\label{eqn:bigeven_MbyN_ob3_set}
\end{eqnarray}

Add up inequalities in (\ref{eqn:bigeven_MbyN_ob1_set}),
(\ref{eqn:bigeven_MbyN_ob2_set}) and
(\ref{eqn:bigeven_MbyN_ob3_set}), we eventually obtain
\begin{eqnarray}
6LnR\leq (2L+1)Mn\log\rho+(2L-1)nR+n~o(\log\rho)+o(n),
\end{eqnarray}
which implies that the sum DoF outer bound
\begin{eqnarray}
d\leq \frac{2L+1}{4L+1}M.
\end{eqnarray}

\subsubsection{Case: Any $M/N \in [\frac{2L-1}{2L},\frac{2L}{2L+1}]\Rightarrow$ DoF $\leq \frac{2L}{4L+1}N$}

\begin{small}
\begin{eqnarray*}
\begin{array}{r|c|}\hline
X_{1a_L}&\circ\\\hline \vdots&\circ\\\hline X_{1a_2}&\circ\\\hline
X_{1b}&\circ\\\hline X_{1c_2}&\circ\\\hline \vdots&\circ\\\hline
X_{1c_L}&\circ\\\hline
\end{array}
&&
\begin{array}{|c|l}
\hline \circ&S_{1a_L}(X_{3a_L})\\\hline \circ&\vdots\\\hline
\circ&S_{1a_2}(X_{3a_2},\{X_{2a_l},X_{2c_l},X_{3a_l},X_{3c_l}|3\leq
l \leq L\})\\\hline
\circ&S_{1a_1}(X_{3b},\{X_{2a_l},X_{2c_l},X_{3a_l},X_{3c_l}|2\leq l
\leq L\})\\\hline \circ&S_{1b_1}(X_{2},X_{3})\\\hline
\circ&S_{1c_1}(X_{2b},\{X_{2a_l},X_{2c_l},X_{3a_l},X_{3c_l}|2\leq l
\leq L\})\\\hline
\circ&S_{1c_2}(X_{2c_2},\{X_{2a_l},X_{2c_l},X_{3a_l},X_{3c_l}||3\leq
l \leq L\})\\\hline \circ&\vdots\\\hline
\circ&S_{1c_L}(X_{2c_L})\\\hline
\end{array}
\\
&&\\
\begin{array}{r|c|}\hline
X_{2a_L}&\circ\\\hline \vdots&\circ\\\hline X_{2a_2}&\circ\\\hline
X_{2b}&\circ\\\hline X_{2c_2}&\circ\\\hline \vdots&\circ\\\hline
X_{2c_2}&\circ\\\hline
\end{array}
&&
\begin{array}{|c|l}
\hline \circ&S_{2a_L}\\\hline \circ&\vdots\\\hline
\circ&S_{2a_2}\\\hline \circ&S_{2a_1}\\\hline \circ&S_{2b_1}\\\hline
\circ&S_{2c_1}\\\hline \circ&S_{2c_2}\\\hline \circ&\vdots\\\hline
\circ&S_{2a_L}\\\hline
\end{array}
\\
&&\\
\begin{array}{r|c|}\hline
X_{3a_L}&\circ\\\hline \vdots&\circ\\\hline X_{3a_2}&\circ\\\hline
X_{3b}&\circ\\\hline X_{3c_2}&\circ\\\hline \vdots&\circ\\\hline
X_{3c_2}&\circ\\\hline
\end{array}
&&
\begin{array}{|c|l}
\hline \circ&S_{3a_L}\\\hline \vdots&\circ\\\hline
\circ&S_{3a_2}\\\hline \circ&S_{3a_1}\\\hline \circ&S_{3b_1}\\\hline
\circ&S_{3c_1}\\\hline \circ&S_{3c_2}\\\hline \vdots&\circ\\\hline
\circ&S_{3c_L}\\\hline
\end{array}
\end{eqnarray*}
\end{small}

With the intuition of onion peeling, we have a total of $L$ layers
at each user in this case. For the $l^{th}$ layer where $2\leq l\leq
L$, there are two groups of signals $(X_{ka_l}^n,X_{kc_l}^n)$ for
user $k$ with the cardinality $|X_{ka_l}^n|=|X_{kc_l}^n|=N-M$, while
in the first layer, we only have $X_{kb_1}^n$ with the cardinality
$|X_{kb}^n|=(2L-1)M-(2L-2)N$.

We denote $\mathcal {G}_m^{2L},~m\in\{1,\cdots,2L\}$ as the the $2L$
sets of the provided genie signals to receiver 1, which can be
written in an iterative form as follows:
\begin{eqnarray}
{\rm for}~~~~m'&\!\!\!\!=\!\!\!\!&1,\notag\\
\mathcal{G}_{2m'-1}^{2L}&\!\!\!\!=\!\!\!\!&\{\{X_{2a_l}^n,X_{3c_l}^n|m'+1\leq l \leq L\},(X_{2b}^n)_{[1:(N-M)]}\}\notag\\
\mathcal{G}_{2m'}^{2L}&\!\!\!\!=\!\!\!\!&\{\{X_{2a_l}^n,X_{3c_l}^n|m'+1\leq l \leq L\},(X_{3b}^n)_{[(N-M+1):((2L-1)M-(2L-2)N)]}\}\notag\\
{\rm for}~~~~m'&\!\!\!\!=\!\!\!\!&\{2,3,\cdots,L\},\notag\\
\mathcal{G}_{2m'-1}^{2L}&\!\!\!\!=\!\!\!\!&\{\{X_{2a_l}^n,X_{2c_l}^n|m'+1\leq l \leq L\},\{X_{2a_l}^n,X_{2c_l}^n|2\leq l \leq m'-1\},X_{2a_{m'}}^n,X_{2b}^n\}\notag\\
\mathcal{G}_{2m'}^{2L}&\!\!\!\!=\!\!\!\!&\{\{X_{3a_l}^n,X_{3c_l}^n|m'+1\leq
l \leq L\},\{X_{3a_l}^n,X_{3c_l}^n|2\leq l \leq
m'-1\},X_{3c_{m'}}^n,X_{3b}^n\}\label{eqn:general_small_genie_even}
\end{eqnarray}
%\begin{eqnarray}
%{\rm for}~~L=1,~~~~~\mathcal{G}_1^{2L}&\!\!\!\!=\!\!\!\!&\{(X_{2b}^n)_{[1:(N-M)]}\}\notag\\
%\mathcal{G}_2^{2L}&\!\!\!\!=\!\!\!\!&\{(X_{3b}^n)_{[(N-M+1):((2L-1)M-(2L-2)N)]}\}\notag\\
%{\rm then~for}~~L>1,~~~~~\mathcal{G}_{m}^{2L}&\!\!\!\!=\!\!\!\!&\{X_{2a_L}^n,X_{3c_L}^n,\mathcal{G}_m^{M-2}\}~~~~~m=\{1,2,\cdots,2L-2\},\notag\\
%{\rm and}~~~~~\mathcal{G}_{2L-1}^{2L}&\!\!\!\!=\!\!\!\!&\{X_{2a_L}^n,\{X_{2a_l}^n,X_{2c_l}^n|2\leq l \leq L-1\},X_{2b}^n\}\notag\\
%\mathcal{G}_{2L}^{2L}&\!\!\!\!=\!\!\!\!&\{X_{3c_L}^n,\{X_{3a_l}^n,X_{3c_l}^n|2\leq
%l \leq L-1\},X_{3b}^n\} \label{eqn:general_small_genie_even}
%\end{eqnarray}
For each set shown above, the total number of dimensions available
to receiver 1 including the dimensions provided by the genie is
equal to:
\begin{eqnarray*}
N+|\mathcal{G}_{1}^{2L}|&\!\!\!\!=\!\!\!\!&N+(N-M)2(L-1)+(N-M)\geq 2M\\
N+|\mathcal{G}_{2}^{2L}|&\!\!\!\!=\!\!\!\!&N+(N-M)2(L-1)+[((2L-1)M-(2L-2)N)-(N-M+1)+1]=2M\\
N+|\mathcal{G}_{m}^{2L}|&\!\!\!\!=\!\!\!\!&N+(N-M)2(L-2)+(N-M)+[(2L-1)M-(2L-2)N]=2M~~~{\rm
for}~m\geq 3.
\end{eqnarray*}
The rigorous proof is still based on mathematical induction. Since
we have shown the proof for $L=1,~2$, the remaining part directly
follows from what we have shown in Appendix A.1 by replacing the
genie signals sets with that in
(\ref{eqn:general_small_genie_even}).

%\subsection{$M=2L+1$ is an Odd Number}

\subsubsection{Case: Any $M/N\geq \frac{2L+1}{2L+2}~\Rightarrow$ DoF $\leq \frac{2L+2}{4L+3}M$}
% Example: (M,N)=(7,9)

\begin{small}
\begin{eqnarray*}
\begin{array}{r|c|}\hline
X_{1a_L}&\circ\\\hline \vdots&\circ\\\hline X_{1a_1}&\circ\\\hline
X_{1b_1}&\circ\\\hline X_{1c_1}&\circ\\\hline \vdots&\circ\\\hline
X_{1c_2}&\circ\\\hline
\end{array}
&&
\begin{array}{|c|l}
\hline \circ&S_{1a_L}(X_{3a_L})\\\hline \circ&\vdots\\\hline
\circ&S_{1a_1}(X_{3a_1},\{X_{2a_l},X_{2c_l},X_{3a_l},X_{3c_l}|2\leq
l \leq L\})\\\hline \circ&S_{1b_1}(X_{2},X_{3})\\\hline
\circ&S_{1c_1}(X_{2c_1},\{X_{2a_l},X_{2c_l},X_{3a_l},X_{3c_l}|2\leq
l \leq L\})\\\hline \circ&\vdots\\\hline
\circ&S_{1c_L}(X_{2c_L})\\\hline
\end{array}
\\
&&\\
&&\\
\begin{array}{r|c|}\hline
X_{2a_L}&\circ\\\hline \vdots&\circ\\\hline X_{2a_1}&\circ\\\hline
X_{2b_1}&\circ\\\hline X_{2c_1}&\circ\\\hline \vdots&\circ\\\hline
X_{2c_L}&\circ\\\hline
\end{array}
&&
\begin{array}{|c|l}
\hline \circ&S_{2a_L}\\\hline \circ&\vdots\\\hline
\circ&S_{2a_1}\\\hline \circ&S_{2b_1}\\\hline \circ&S_{2c_1}\\\hline
\circ&\vdots\\\hline \circ&S_{2a_L}\\\hline
\end{array}
\\
&&\\
&&\\
\begin{array}{r|c|}\hline
X_{3a_L}&\circ\\\hline \vdots&\circ\\\hline X_{3a_2}&\circ\\\hline
X_{3b}&\circ\\\hline X_{3c_2}&\circ\\\hline \vdots&\circ\\\hline
X_{3c_2}&\circ\\\hline
\end{array}
&&
\begin{array}{|c|l}
\hline \circ&S_{3a_L}\\\hline \vdots&\circ\\\hline
\circ&S_{3a_1}\\\hline \circ&S_{3b_1}\\\hline \circ&S_{3c_1}\\\hline
\vdots&\circ\\\hline \circ&S_{3c_L}\\\hline
\end{array}
\end{eqnarray*}
\end{small}

The proof for this case still directly follows what we have shown in
Appendix A.2 and B.1.1. The only difference is the first set of
signals provided by the genie. Specifically, we write out the genie
signal sets as follows:
\begin{eqnarray}
\mathcal{G}_1^{2L+1}&\!\!\!\!=\!\!\!\!&\{\{X_{2a_l}^n,X_{3c_l}^n|1\leq l\leq L\},(X_{3b_1}^n)_{[1:((2L+2)M-(2L+1)N)]}\}\notag\\
{\rm and ~for}~~~~m'&\!\!\!\!=\!\!\!\!&\{2,3,\cdots,L\},\notag\\
\mathcal{G}_{2m'}^{2L+1}&\!\!\!\!=\!\!\!\!&\{\{X_{2a_l}^n,X_{2c_l}^n|m'+1\leq l \leq L\},\{X_{2a_l}^n,X_{2c_l}^n|2\leq l \leq m'-1\},X_{2a_{m'}}^n,X_{2b}^n\}\notag\\
\mathcal{G}_{2m'+1}^{2L+1}&\!\!\!\!=\!\!\!\!&\{\{X_{3a_l}^n,X_{3c_l}^n|m'+1\leq
l \leq L\},\{X_{3a_l}^n,X_{3c_l}^n|2\leq l \leq
m'-1\},X_{3c_{m'}}^n,X_{3b}^n\}\label{eqn:general_big_genie_odd}
\end{eqnarray}
Note that if we peel out all the $l^{th}$ layers where $l\geq 2$,
the aim of the first set of signals provided by the genie, e.g., the
example $(M,N)=(3,4)$ in Section 4.1.2 and Appendix A.2, is to
produce the following terms
\begin{eqnarray}
Nn\log\rho+h(X_{2a_1}^n|\{X_{2a_l}^n|2\leq l\leq
L\})+h(X_{3c_1}^n|\{X_{3c_l}^n|2\leq l\leq
L\})+n~o(\log\rho)+o(n)\label{eqn:eqn:general_big_genie_odd_ob1}
\end{eqnarray}
on the right-hand side of the inequality, where the intermediate
terms $h(X_{2a_1}^n|\{X_{2a_l}^n|2\leq l\leq L\})$ and
$h(X_{3c_1}^n|\{X_{3c_l}^n|2\leq l\leq L\})$ can be be canceled by
the summation of the remaining $2L$ inequalities (by providing
$\mathcal{G}_{m}^{2L+1},~2\leq m \leq 2L+1$) on the right-hand side.
In this case, however, because $\mathcal{G}_1^{2L+1}$ includes an
extra signal $(X_{3b_1}^n)_{[1:((2L+2)M-(2L+1)N)]}$ which will
produce an extra term $((2L+2)M-(2L+1)N)n\log\rho$ in
(\ref{eqn:eqn:general_big_genie_odd_ob1}). Therefore, following the
idea of previous proofs, by adding up the total of
$2L+1$-inequalities, we can obtain:
\begin{small}
\begin{eqnarray*}
3(2L+1)nR\leq (2L+1)Nn\log\rho+((2L+2)M-(2L+1)N)n\log\rho+
2LnR+n~o(\log\rho)+o(n)
\end{eqnarray*}
\end{small}
which implies that
\begin{eqnarray*}
d\leq \frac{2L+2}{4L+3}M.
\end{eqnarray*}

\subsubsection{Case: Any $M/N \in [\frac{2L}{2L+1},\frac{2L+1}{2L+2}]\Rightarrow$ DoF $\leq \frac{2L+1}{4L+3}N$}
% Example: (M,N)=(5,7) or (8,11)

The proof for this case is identical to that in Appendix
\ref{subsubsec:M_odd} by only noting $|X_{ka_l}|=|X_{kc_l}|=N-M$
where $l=1,2,\cdots,L$ and $|X_{kb_1}|=(2L+1)M-2LN$.

\section{Information Theoretic DoF Outer Bound for $M_T>M_R$}\label{sec:outerbound_MbiggerN}

In this section we consider the DoF outer bound for $M_T>M_R$, i.e.,
the $N\times M$ setting. Note that the linear transformation for the
$(N,M)$ case is still identical to that for the $(M,N)$ case that we
introduce in Section \ref{sec:outerbound} but in a reciprocal
manner. Therefore, we directly show the table representing each
resulting network and corresponding channel connectivity for each
case without specifying the linear transformation process in this
section.

\subsection{Outerbound Proofs around $M/N$ ratios $2/3, 4/5, 6/7,
\cdots$} Consider 3 user $N\times M$ MIMO interference channel,
i.e., with $N$ transmit antennas and $M$ receive antennas for each
user. We assume $M\leq N$ throughout.

\subsubsection{Case $(N,M)=(3,2) \Rightarrow $ DoF $\leq
\frac{6}{5}$}

\begin{eqnarray*}
\begin{array}{r|c|}
\hline X_{1a}&\circ\\\hline X_{1b}&\circ\\\hline
X_{1c}&\circ\\\hline
\end{array}
&&
\begin{array}{|c|l}
\hline \circ&S_{1a}(X_{2a},X_{2b},X_{3b})\\\hline
\circ&S_{1c}(X_{2b},X_{3b},X_{3c})\\\hline
\end{array}
\\
&&\\
&&\\
\begin{array}{r|c|}
\hline X_{2a}&\circ\\\hline X_{2b}&\circ\\\hline
X_{2c}&\circ\\\hline
\end{array}
&&
\begin{array}{|c|l}
\hline \circ&S_{2a}(X_{3a},X_{3b},X_{1b})\\\hline
\circ&S_{2c}(X_{3b},X_{1b},X_{1c})\\\hline
\end{array}
\\
&&\\
&&\\
\begin{array}{r|c|}
\hline X_{3a}&\circ\\\hline X_{3b}&\circ\\\hline
X_{3c}&\circ\\\hline
\end{array}
&&
\begin{array}{|c|l}
\hline \circ&S_{3a}(X_{1a},X_{1b},X_{2b})\\\hline
\circ&S_{3c}(X_{1b},X_{2b},X_{2c})\\\hline
\end{array}
\end{eqnarray*}

First, a genie provides $\mathcal{G}_1=(X_{3a}^n, X_{3b}^n,
X_{2b}^n, X_{2c}^n)$ to receiver 1:
\begin{eqnarray}
nR_{\Sigma}&\leq& I(W_1,W_2,W_3;\bar{Y}_1^n,\mathcal{G}_1)+n~o(\log\rho)+o(n)\\
&=& I(W_1,W_2,W_3;\bar{Y}_1^n)+I(W_1,W_2,W_3;\mathcal{G}_1|\bar{Y}_1^n)+n~o(\log\rho)+o(n)\\
&\leq& Mn\log\rho + h(X_{3a}^n, X_{3b}^n)+h(X_{2b}^n,X_{2c}^n)+n~o(\log\rho)+o(n)\\
\hookrightarrow 3nR&\leq& Mn\log\rho +
h(X_a^n,X_b^n)+h(X_b^n,X_c^n)+n~o(\log\rho)+o(n)\label{eq:3by2first}
\end{eqnarray}
The symbol $\hookrightarrow$ stands for ``follows from symmetry"
that we have applied in previous section.

Second, a genie provides $\mathcal{G}_2=(W_2, X_{3a}^n)\equiv
(X_{2a}^n, X_{2b}^n, X_{2c}^n, X_{3a}^n)$ to receiver 1:
\begin{eqnarray}
nR_{\Sigma} &\leq& I(W_1,W_2,W_3;\bar{Y}_1^n,\mathcal{G}_2)+n~o(\log\rho)+o(n)\\
&\leq& Mn\log\rho+h(\mathcal{G}_2|\bar{Y}_1^n)+n~o(\log\rho)+o(n)\\
&\leq& Mn\log\rho+ nR_2 + h(X_{3a}^n| X_{3b}^n,X_{3c}^n)+n~o(\log\rho)+o(n)\\
\hookrightarrow 3nR&\leq& Mn\log\rho +nR+
h(X_a^n|X_b^n,X_c^n)+n~o(\log\rho)+o(n)\label{eq:3by2second}
\end{eqnarray}

Similarly if a genie provides $\mathcal{G}_3=(W_3, X_{2c}^n)\equiv
(X_{3a}^n, X_{3b}^n, X_{3c}^n, X_{2c}^n)$ to receiver 1, we obtain
\begin{eqnarray}
3nR&\leq& Mn\log\rho +nR+
h(X_c^n|X_b^n,X_a^n)+n~o(\log\rho)+o(n)\label{eq:3by2third}
\end{eqnarray}

Adding (\ref{eq:3by2first}), (\ref{eq:3by2second}),
(\ref{eq:3by2third}) we have:
\begin{eqnarray}
9nR&\leq& 3Mn\log\rho +4nR+n~o(\log\rho)+o(n)
\end{eqnarray}
which implies that
\begin{eqnarray}
d&\leq&\frac{3M}{5}=\frac{6}{5}
\end{eqnarray}

\subsubsection{Case: Any $M/N\geq2/3 \Rightarrow $ DoF $\leq
\frac{3M}{5}$} In this section, consider arbitrary $M, N$ values
such that $M/N\geq 2/3$. For all such values of $M,N$, the DoF outer
bound that holds is DoF $\leq 3M/5$.
\begin{small}
\begin{eqnarray*}
\begin{array}{r|r|c|}
\hline |X_{1a}|=(N-M)&X_{1a}&\circ\\\hline
|X_{1b}|=N-2(N-M)\geq(N-M)&X_{1b}&\circ\\\hline
|X_{1c}|=(N-M)&X_{1c}&\circ\\\hline
\end{array}
&&
\begin{array}{|c|l|l}
\hline \circ&S_{1a}(X_{2a},X_{2b},X_{3b})&|S_{1a}|=(N-M)\\\hline
\circ&S_{1b}(X_{2b},X_{3b})&|S_{1b}|=M-2(N-M)\geq0\\\hline
\circ&S_{1c}(X_{2b},X_{3b},X_{3c})&|S_{1c}|=(N-M)\\\hline
\end{array}
\\
&&\\
&&\\\begin{array}{r|r|c|} \hline |X_{2a}|=(N-M)&X_{2a}&\circ\\\hline
|X_{2b}|=N-2(N-M)\geq(N-M)&X_{2b}&\circ\\\hline
|X_{2c}|=(N-M)&X_{2c}&\circ\\\hline
\end{array}
&&
\begin{array}{|c|l|l}
\hline \circ&S_{2a}(X_{3a},X_{3b},X_{1b})&|S_{2a}|=(N-M)\\\hline
\circ&S_{2b}(X_{3b},X_{1b})&|S_{2b}|=M-2(N-M)\geq0\\\hline
\circ&S_{2c}(X_{3b},X_{1b},X_{1c})&|S_{2c}|=(N-M)\\\hline
\end{array}
\\
&&\\
&&\\
\begin{array}{r|r|c|}
\hline |X_{3a}|=(N-M)&X_{3a}&\circ\\\hline
|X_{3b}|=N-2(N-M)\geq(N-M)&X_{3b}&\circ\\\hline
|X_{3c}|=(N-M)&X_{3c}&\circ\\\hline
\end{array}
&&
\begin{array}{|c|l|l}
\hline \circ&S_{3a}(X_{1a},X_{1b},X_{2b})&|S_{3a}|=(N-M)\\\hline
\circ&S_{3b}(X_{1b},X_{2b})&|S_{3b}|=M-2(N-M)\geq0\\\hline
\circ&S_{3c}(X_{1b},X_{2b},X_{2c})&|S_{3c}|=(N-M)\\\hline
\end{array}
\end{eqnarray*}
\end{small}

\noindent Note that the size of $|X_b|\geq N-M$ and the size of
$|S_{b}|\geq 0$ as indicated above, only because $M/N\geq 2/3$.

First, a genie provides $\mathcal{G}_1=(X_{3a}, X_{3b}, X_{2b},
X_{2c})$ to receiver 1. Thus, the total number of dimensions
available to the receiver 1 (including those provided by the genie)
is:

$$M+|G|=M+|X_{a}|+2|X_b|+|X_c|=M+N+|X_b|\geq M+N+(N-M)=2N$$

\noindent With these (at least) $2N$ dimensions, the receiver is
able to resolve both interfering signals. Therefore, once again, we
have
\begin{eqnarray}
nR_{\Sigma} &\leq& I(W_1,W_2,W_3;\bar{Y}_1^n,\mathcal{G}_1)+n~o(\log\rho)+o(n)\\
&\leq& Mn\log\rho+h(\mathcal{G}_1|\bar{Y}_1^n)+n~o(\log\rho)+o(n)\\
&\leq& Mn\log\rho + h(X_{3a}^n, X_{3b}^n)+h(X_{2b}^n,X_{2c}^n)+n~o(\log\rho)+o(n)\\
\hookrightarrow 3nR&\leq& Mn\log\rho +
h(X_a^n,X_b^n)+h(X_b^n,X_c^n)+n~o(\log\rho)+o(n)\label{eq:g3by2first}
\end{eqnarray}

Second, a genie provides $\mathcal{G}_2=(W_2, X_{3a}^n)\equiv
(X_{2a}^n, X_{2b}^n, X_{2c}^n, X_{3a}^n)$ to receiver 1, then the
total number of dimensions available to the receiver 1 is:
$$M+|G|=M+N+|X_a|=M+N+N-M=2N$$
So, $2N$ dimensions are available, and all interference can be
resolved.
\begin{eqnarray}
nR_{\Sigma} &\leq& I(W_1,W_2,W_3;\bar{Y}_1^n,\mathcal{G}_2)+n~o(\log\rho)+o(n)\\
&\leq& Mn\log\rho+h(\mathcal{G}_2|\bar{Y}_1^n)+n~o(\log\rho)+o(n)\\
&\leq& Mn\log\rho + nR_2 + h(X_{3a}^n| X_{3b}^n,X_{3c}^n)+n~o(\log\rho)+o(n)\\
\hookrightarrow 3nR&\leq& Mn\log\rho +nR+
h(X_a^n|X_b^n,X_c^n)+n~o(\log\rho)+o(n)\label{eq:g3by2second}
\end{eqnarray}

Similarly if a genie provides $\mathcal{G}_3=(W_3, X_{2c}^n)\equiv
(X_{3a}^n, X_{3b}^n, X_{3c}^n, X_{2c}^n)$ to receiver 1, we obtain
\begin{eqnarray}
3nR&\leq& Mn\log\rho +nR+
h(X_c^n|X_b^n,X_a^n)+n~o(\log\rho)+o(n)\label{eq:g3by2third}
\end{eqnarray}

Adding (\ref{eq:g3by2first}), (\ref{eq:g3by2second}),
(\ref{eq:g3by2third}) we have:
\begin{eqnarray}
9nR&\leq& 3Mn\log\rho +4nR+n~o(\log\rho)+o(n)
\end{eqnarray}
which implies that
\begin{eqnarray}
d \leq \frac{3M}{5}.
\end{eqnarray}

\subsubsection{Case: Any $M/N\in [1/2, 2/3] \Rightarrow $ DoF $\leq
\frac{2N}{5}$} In this section, consider arbitrary $M, N$ values
such that $M/N\leq 2/3$. For all such values of $M,N$, the DoF outer
bound that holds is DoF $\leq 2N/5$.
\begin{eqnarray*}
\begin{array}{r|r|c|}
\hline |X_{1a}|=(N-M)&X_{1a}&\circ\\\hline
|X_{1b}|=N-2(N-M)\leq(N-M)&X_{1b}&\circ\\\hline
|X_{1c}|=(N-M)&X_{1c}&\circ\\\hline
\end{array}
&&
\begin{array}{|c|l|l}
\hline \circ&S_{1}(X_{2a},X_{2b},X_{3b},X_{3c})&|S_{1}|=M\geq
N-M\\\hline
\end{array}
\\
&&\\
&&\\\begin{array}{r|r|c|} \hline |X_{2a}|=(N-M)&X_{2a}&\circ\\\hline
|X_{2b}|=N-2(N-M)\leq(N-M)&X_{2b}&\circ\\\hline
|X_{2c}|=(N-M)&X_{2c}&\circ\\\hline
\end{array}
&&
\begin{array}{|c|l|l}
\hline \circ&S_{2}(X_{3a},X_{3b},X_{1b},X_{1c})&|S_{2}|=M\geq
N-M\\\hline
\end{array}
\\
&&\\
&&\\
\begin{array}{r|r|c|}
\hline |X_{3a}|=(N-M)&X_{3a}&\circ\\\hline
|X_{3b}|=N-2(N-M)\leq(N-M)&X_{3b}&\circ\\\hline
|X_{3c}|=(N-M)&X_{3c}&\circ\\\hline
\end{array}
&&
\begin{array}{|c|l|l}
\hline \circ&S_{3}(X_{1a},X_{1b},X_{2b},X_{2c})&|S_{3}|=M\geq
N-M\\\hline
\end{array}
\end{eqnarray*}

\noindent Note that the size of $|X_b|\leq N-M$  as indicated above,
because $M/N\leq 2/3$.

First, a genie provides $\mathcal{G}_1=(X_{3a}^n, X_{3b}^n,
(X_{2a}^n, X_{3c}^n)_{[2N-3M]},X_{2b}^n, X_{2c}^n)$ to receiver 1.
Note that $(X_{2a}^n,X_{3c}^n)_{[\min(2N-3M,N-M)]}$ represents the
extra dimensions provided to the receiver so that the total number
of dimensions is at least $2N$ (the minimum necessary to resolve all
interference). This is verified as follows.

The receiver already has $M$ dimensions. The usual genie information
$X_{3a}^n, X_{3b}^n, X_{2b}^n, X_{2c}^n$ constitutes another
$N+|X_b^n| = N+2M-N = 2M$ dimensions, for a total of $3M$
dimensions. This is still less than $2N$ because $M/N\leq 2/3$. So
we need another $2N-3M$ dimensions. So we provide any $2N-3M$
elements from $X_{2a}^n, X_{3c}^n$, which is the \emph{extra} genie
information indicated as $ (X_{2a}^n, X_{3c}^n)_{[2N-3M]}$. Thus the
total number of dimensions available to the receiver (including
those provided by the genie) is $2N$, and all interference can be
resolved. In the outer bound, the extra dimensions are directly
accounted in the prelog factor and dropped from the conditioning
terms (which can only increase the entropy terms).
\begin{eqnarray}
nR_{\Sigma} \!\!\!\!&\leq&\!\!\!\! I(W_1,W_2,W_3;\bar{Y}_1^n,\mathcal{G}_1)+n~o(\log\rho)+o(n)\\
\!\!\!\!&\leq&\!\!\!\! Mn\log\rho+h(\mathcal{G}_1|\bar{Y}_1^n)+n~o(\log\rho)+n~o(\log\rho)+o(n)\\
\!\!\!\!&\leq&\!\!\!\! Mn\log\rho +(2N-3M)n\log\rho + h(X_{3a}^n, X_{3b}^n)+h(X_{2b}^n,X_{2c}^n)+n~o(\log\rho)+o(n)\\
\hookrightarrow 3nR \!\!\!\!&\leq&\!\!\!\! (M+2N-3M)n\log\rho +
h(X_a^n,X_b^n)+h(X_b^n,X_c^n)+n~o(\log\rho)+o(n)\label{eq:l3by2first}
\end{eqnarray}

Second, a genie provides $\mathcal{G}_2=(W_2, X_{3a}^n)\equiv
(X_{2a}^n, X_{2b}^n, X_{2c}^n, X_{3a}^n)$ to receiver 1, then the
total number of dimensions available to receiver 1 is:
$$M+|G|=M+N+|X_a|=M+N+N-M=2N$$
So, $2N$ dimensions are available, no \emph{extra} genie information
is needed, and all interference can be resolved.
\begin{eqnarray}
nR_{\Sigma} \!\!\!\!&\leq&\!\!\!\! I(W_1,W_2,W_3;\bar{Y}_1^n,\mathcal{G}_2)+n~o(\log\rho)+o(n)\\
\!\!\!\!&\leq&\!\!\!\! Mn\log\rho+h(\mathcal{G}_2|\bar{Y}_1^n)+n~o(\log\rho)+n~o(\log\rho)+o(n)\\
\!\!\!\!&\leq&\!\!\!\! Mn\log\rho + nR_2 + h(X_{3a}^n| X_{3b}^n,X_{3c}^n)+n~o(\log\rho)+o(n)\\
\hookrightarrow 3nR \!\!\!\!&\leq&\!\!\!\! M +nR+
h(X_a^n|X_b^n,X_c^n)\label{eq:l3by2second}
\end{eqnarray}

Similarly if a genie provides $\mathcal{G}_3=(W_3, X_{2c}^n)\equiv
(X_{3a}^n, X_{3b}^n, X_{3c}^n, X_{2c}^n)$ to receiver 1, we obtain
\begin{eqnarray}
3nR \leq Mn\log\rho +nR+
h(X_c^n|X_b^n,X_a^n)+n~o(\log\rho)+o(n)\label{eq:l3by2third}
\end{eqnarray}

Adding (\ref{eq:l3by2first}), (\ref{eq:l3by2second}),
(\ref{eq:l3by2third}) we have
\begin{eqnarray}
9nR \!\!\!\!&\leq&\!\!\!\! (3M+2N-3M)n\log\rho+2nR+2nR+n~o(\log\rho)+o(n)\\
\Rightarrow d \!\!\!\!&\leq&\!\!\!\! \frac{2N}{5}
\end{eqnarray}

\subsubsection{Case $(N,M)=(5,4) \Rightarrow $ DoF $\leq
\frac{20}{9}$}

\begin{eqnarray*}
\begin{array}{r|c|}
\hline X_{1a_1}&\circ\\\hline X_{1a}&\circ\\\hline
X_{1b}&\circ\\\hline X_{1c}&\circ\\\hline X_{1c_1}&\circ\\\hline
\end{array}
&&
\begin{array}{|c|l}
\hline
\circ&S_{1a_1}(X_{2a_1},X_{2a},X_{2b},X_{2c},X_{3a},X_{3b},X_{3c})\\\hline
\circ&S_{1a}(X_{2a},X_{2b},X_{3b})\\\hline
\circ&S_{1c}(X_{2b},X_{3b},X_{3c})\\\hline
\circ&S_{1c_1}(X_{2a},X_{2b},X_{2c},X_{3a},X_{3b},X_{3c},X_{3c_1})\\\hline
\end{array}
\\
&&\\
&&\\
\begin{array}{r|c|}
\hline X_{2a_1}&\circ\\\hline X_{2a}&\circ\\\hline
X_{2b}&\circ\\\hline X_{2c}&\circ\\\hline
X_{2c_1}&\circ\\\hline\end{array} &&
\begin{array}{|c|l}
\hline
\circ&S_{2a_1}(X_{3a_1},X_{3a},X_{3b},X_{3c},X_{1a},X_{1b},X_{1c})\\\hline
\circ&S_{2a}(X_{3a},X_{3b},X_{1b})\\\hline
\circ&S_{2c}(X_{3b},X_{1b},X_{1c})\\\hline
\circ&S_{2c_1}(X_{3a},X_{3b},X_{3c},X_{1a},X_{1b},X_{1c},X_{1c_1})\\\hline\end{array}
\\
&&\\
&&\\
\begin{array}{r|c|}
\hline X_{3a_1}&\circ\\\hline X_{3a}&\circ\\\hline
X_{3b}&\circ\\\hline X_{3c}&\circ\\\hline X_{3c_1}&\circ\\\hline
\end{array}
&&
\begin{array}{|c|l}
\hline
\circ&S_{3a_1}(X_{1a_1},X_{1a},X_{1b},X_{1c},X_{2a},X_{2b},X_{2c})\\\hline
\circ&S_{3a}(X_{1a},X_{1b},X_{2b})\\\hline
\circ&S_{3c}(X_{1b},X_{2b},X_{2c})\\\hline
\circ&S_{3c_1}(X_{1a},X_{1b},X_{1c},X_{2a},X_{2b},X_{2c},X_{2c_1})\\\hline
\end{array}
\end{eqnarray*}

First, a genie provides $\mathcal{G}_1=(X_{3a_1}^n,
X_{3a}^n,X_{3b}^n,X_{2b}^n,X_{2c}^n,X_{2c_1}^n)$ to receiver 1, then
we have:
\begin{eqnarray}
nR_\Sigma \!\!\!\!&\leq&\!\!\!\! I(W_1,W_2,W_3;\bar{Y}_1^n,\mathcal{G}_1)+n~o(\log\rho)+o(n)\\
\!\!\!\!&\leq&\!\!\!\! Mn\log\rho+h(\mathcal{G}_1|\bar{Y}_1^n)+n~o(\log\rho)+n~o(\log\rho)+o(n)\\
\!\!\!\!&\leq&\!\!\!\! Mn\log\rho+h(X_{3a}^n,X_{3b}^n)+h(X_{2b}^n,X_{2c}^n)+n~o(\log\rho)+o(n)\nonumber\\
&&+h(X_{3a_1}^n|X_{3a}^n,X_{3b}^n,X_{3c}^n, X_{3c_1}^n)+h(X_{2c_1}^n|X_{2a_1}^n,X_{2a}^n,X_{2b}^n,X_{2c}^n)\\
\hookrightarrow 3nR \!\!\!\!&\leq&\!\!\!\! Mn\log\rho+h(X_a^n,X_b^n)+h(X_b^n,X_c^n)+n~o(\log\rho)+o(n)\notag\\
&&+h(X_{a_1}^n|X_a^n,X_b^n,X_c^n,X_{c_1}^n)+h(X_{c_1}^n|X_{a_1}^n,X_a^n,X_b^n,X_c^n)\\
\Rightarrow 3nR \!\!\!\!&\leq&\!\!\!\!
Mn\log\rho+h(X_a^n,X_b^n)+h(X_b^n,X_c^n)+h(X_{a_1}^n,X_{c_1}^n|X_a^n,X_b^n,X_c^n)+n~o(\log\rho)+o(n)\
\ \ \ \label{eq:5by4first}
\end{eqnarray}

Second, if a genie provides $\mathcal{G}_2=(X_{3a_1}^n, X_{3a}^n,
X_{2a}^n, X_{2b}^n,X_{2c}^n,X_{2c_1}^n)$ to receiver 1, then we
have:
\begin{eqnarray}
nR_\Sigma \!\!\!\!&\leq&\!\!\!\! I(W_1,W_2,W_3;\bar{Y}_1^n,\mathcal{G}_2)+n~o(\log\rho)+o(n)\\
\!\!\!\!&\leq&\!\!\!\! Mn\log\rho+h(\mathcal{G}_2|\bar{Y}_1^n)+n~o(\log\rho)+o(n)\\
\!\!\!\!&\leq&\!\!\!\! Mn\log\rho+h(X_{2a}^n,X_{2b}^n,X_{2c}^n,X_{2c_1}^n)+h(X_{3a}^n|X_{3b}^n,X_{3c}^n)+h(X_{3a_1}^n|X_{3a}^n,X_{3b}^n,X_{3c}^n,X_{3c_1}^n)\ \ \ \ \notag\\
&&+n~o(\log\rho)+o(n)\\
\hookrightarrow 3nR \!\!\!\!&\leq&\!\!\!\! Mn\log\rho+h(X_a^n,X_b^n,X_c^n,X_{c_1}^n)+h(X_a^n|X_b^n,X_c^n)+h(X_{a_1}^n|X_a^n,X_b^n,X_c^n,X_{c_1}^n)\notag\\
&&+n~o(\log\rho)+o(n)\\
\Rightarrow 3nR \!\!\!\!&\leq&\!\!\!\!
Mn\log\rho+nR+h(X_a^n|X_b^n,X_c^n)+n~o(\log\rho)+o(n)\label{eq:5by4second}
\end{eqnarray}
Similarly if a genie provides $\mathcal{G}_3=(X_{2c_1}^n, X_{2c}^n,
X_{3c}^n, X_{3b}^n,X_{3a}^n,X_{3a_1}^n)$ to receiver 1, then we
obtain:
\begin{eqnarray}
3nR \leq
Mn\log\rho+nR+h(X_c^n|X_a^n,X_b^n)+n~o(\log\rho)+o(n)\label{eq:5by4third}
\end{eqnarray}
%Adding (\ref{eq:5by4first}), (\ref{eq:5by4second}), (\ref{eq:5by4third}),
%\begin{eqnarray}
%9R&\leq&12+
%\end{eqnarray}

Last a genie provides $\mathcal{G}_4=(X_{3a_1}^n,
\underbrace{X_{2a_1}^n, X_{2a}^n,
X_{2b}^n,X_{2c}^n,X_{2c_1}^n}_{W_2})$ to receiver 1, we have:
\begin{eqnarray}
nR_\Sigma \!\!\!\!&\leq&\!\!\!\! I(W_1,W_2,W_3;\bar{Y}_1^n,\mathcal{G}_4)+n~o(\log\rho)+o(n)\\
\!\!\!\!&\leq&\!\!\!\! Mn\log\rho+h(\mathcal{G}_4|\bar{Y}_1^n)+n~o(\log\rho)+o(n)\\
\!\!\!\!&\leq&\!\!\!\! Mn\log\rho+nR_2+h(X_{3a_1}^n|X_{3a}^n,X_{3b}^n,X_{3c}^n,X_{3c_1}^n)+n~o(\log\rho)+o(n)\\
\hookrightarrow 3nR \!\!\!\!&\leq&\!\!\!\!
Mn\log\rho+nR+h(X_{a_1}^n|X_{a}^n,X_{b}^n,X_{c}^n,X_{c_1}^n)+n~o(\log\rho)+o(n)\label{eq:5by4fourth}
\end{eqnarray}
and similarly if a genie provides $\mathcal{G}_5=(X_{2c_1}^n,
\underbrace{X_{2c_1}^n, X_{3a}^n,
X_{3b}^n,X_{3c}^n,X_{3c_1}^n}_{W_3})$ to receiver 1 then we have:
\begin{eqnarray}
3nR \leq
Mn\log\rho+nR+h(X_{c_1}^n|X_{a_1}^n,X_a^n,X_{b}^n,X_{c}^n)+n~o(\log\rho)+o(n)\label{eq:5by4fifth}
%\\ \Rightarrow 6R \!\!\!\!&\leq&\!\!\!\! 8 + 2R +
%h(X_{a_1},X_{c_1}|X_a,X_b,X_c) \label{eq:5by4sixth}
\end{eqnarray}
%where (\ref{eq:5by4sixth}) is obtained by adding (\ref{eq:5by4fourth}) and (\ref{eq:5by4fifth}).

Finally, adding (\ref{eq:5by4first}), (\ref{eq:5by4second}),
(\ref{eq:5by4third}), (\ref{eq:5by4fourth}), (\ref{eq:5by4fifth}),
we obtain
\begin{eqnarray}
15nR &\leq & 6nR + 5Mn\log\rho+n~o(\log\rho)+o(n)\\
\Rightarrow d&\leq & \frac{5M}{9}=\frac{20}{9}
\end{eqnarray}

\subsubsection{Case:  Any $M/N\geq 4/5 \Rightarrow $ DoF $\leq
\frac{5M}{9}$}
\begin{small}
\begin{eqnarray*}
\begin{array}{r|r|c|}
\hline |X_{1a_1}|=N-M&X_{1a_1}&\circ\\\hline
|X_{1a}|=N-M&X_{1a}&\circ\\\hline |X_{1b}|\geq
N-M&X_{1b}&\circ\\\hline |X_{1c}|=N-M&X_{1c}&\circ\\\hline
|X_{1c_1}|=N-M&X_{1c_1}&\circ\\\hline
\end{array}
&&
\begin{array}{|c|l|l}
\hline
\circ&S_{1a_1}(X_{2a_1},X_{2a},X_{2b},X_{2c},X_{3a},X_{3b},X_{3c})&|S_{1a_1}|=N-M\\\hline
\circ&S_{1a}(X_{2a},X_{2b},X_{3b})&|S_{1a}|=N-M\\\hline
\circ&S_{1b}(X_{2a_1},X_{2a},X_{2b},X_{2c},X_{3a},X_{3b},X_{3c},X_{3c_1})&|S_{1b}|\geq
0\\\hline \circ&S_{1c}(X_{2b},X_{3b},X_{3c})&|S_{1c}|=N-M\\\hline
\circ&S_{1c_1}(X_{2a},X_{2b},X_{2c},X_{3a},X_{3b},X_{3c},X_{3c_1})&|S_{1c_1}|=N-M\\\hline
\end{array}
\\
&&\\
&&\\
\begin{array}{r|c|}
\hline X_{2a_1}&\circ\\\hline X_{2a}&\circ\\\hline
X_{2b}&\circ\\\hline X_{2c}&\circ\\\hline
X_{2c_1}&\circ\\\hline\end{array} &&
\begin{array}{|c|l}
\hline
\circ&S_{2a_1}(X_{3a_1},X_{3a},X_{3b},X_{3c},X_{1a},X_{1b},X_{1c})\\\hline
\circ&S_{2a}(X_{3a},X_{3b},X_{1b})\\\hline \circ&S_{2b}\\\hline
\circ&S_{2c}(X_{3b},X_{1b},X_{1c})\\\hline
\circ&S_{2c_1}(X_{3a},X_{3b},X_{3c},X_{1a},X_{1b},X_{1c},X_{1c_1})\\\hline\end{array}
\\
&&\\
&&\\
\begin{array}{r|c|}
\hline X_{3a_1}&\circ\\\hline X_{3a}&\circ\\\hline
X_{3b}&\circ\\\hline X_{3c}&\circ\\\hline X_{3c_1}&\circ\\\hline
\end{array}
&&
\begin{array}{|c|l}
\hline
\circ&S_{3a_1}(X_{1a_1},X_{1a},X_{1b},X_{1c},X_{2a},X_{2b},X_{2c})\\\hline
\circ&S_{3a}(X_{1a},X_{1b},X_{2b})\\\hline \circ&S_{3b}\\\hline
\circ&S_{3c}(X_{1b},X_{2b},X_{2c})\\\hline
\circ&S_{3c_1}(X_{1a},X_{1b},X_{1c},X_{2a},X_{2b},X_{2c},X_{2c_1})\\\hline
\end{array}
\end{eqnarray*}
\end{small}
All genies work exactly as in the $(N,M)=(5,4)$ case. Re-write the
same equations, substituting $M$ instead of $4$ in each of
(\ref{eq:5by4first}), (\ref{eq:5by4second}), (\ref{eq:5by4third}),
(\ref{eq:5by4fourth}), (\ref{eq:5by4fifth}), so that we obtain the
bound:
\begin{eqnarray*}
R&\leq&\frac{5M}{9}
\end{eqnarray*}

\subsubsection{Case: Any $M/N\in [3/4, 4/5] \Rightarrow $ DoF $\leq
\frac{4N}{9}$}

\begin{eqnarray*}
\begin{array}{r|r|c|}
\hline |X_{1a_1}|=N-M&X_{1a_1}&\circ\\\hline
|X_{1a}|=N-M&X_{1a}&\circ\\\hline |X_{1b}|\leq
N-M&X_{1b}&\circ\\\hline |X_{1c}|=N-M&X_{1c}&\circ\\\hline
|X_{1c_1}|=N-M&X_{1c_1}&\circ\\\hline
\end{array}
&&
\begin{array}{|c|l|l}
\hline
\circ&S_{1a_1}(X_{2a_1},X_{2a},X_{2b},X_{2c},X_{3a},X_{3b},X_{3c})&|S_{1a_1}|=N-M\\\hline
\circ&S_{1}(X_{2a},X_{2b},X_{2c},X_{3a},X_{3b},X_{3c})&|S_{1}|\geq
N-M\\\hline
\circ&S_{1c_1}(X_{2a},X_{2b},X_{2c},X_{3a},X_{3b},X_{3c},X_{3c_1})&|S_{1c_1}|=N-M\\\hline
\end{array}
\\
&&\\
&&\\
\begin{array}{r|c|}
\hline X_{2a_1}&\circ\\\hline X_{2a}&\circ\\\hline
X_{2b}&\circ\\\hline X_{2c}&\circ\\\hline
X_{2c_1}&\circ\\\hline\end{array} &&
\begin{array}{|c|l}
\hline
\circ&S_{2a_1}(X_{3a_1},X_{3a},X_{3b},X_{3c},X_{1a},X_{1b},X_{1c})\\\hline
\circ&S_{2}\\\hline
\circ&S_{2c_1}(X_{3a},X_{3b},X_{3c},X_{1a},X_{1b},X_{1c},X_{1c_1})\\\hline\end{array}
\\
&&\\
&&\\
\begin{array}{r|c|}
\hline X_{3a_1}&\circ\\\hline X_{3a}&\circ\\\hline
X_{3b}&\circ\\\hline X_{3c}&\circ\\\hline X_{3c_1}&\circ\\\hline
\end{array}
&&
\begin{array}{|c|l}
\hline
\circ&S_{3a_1}(X_{1a_1},X_{1a},X_{1b},X_{1c},X_{2a},X_{2b},X_{2c})\\\hline
\circ&S_{3}\\\hline
\circ&S_{3c_1}(X_{1a},X_{1b},X_{1c},X_{2a},X_{2b},X_{2c},X_{2c_1})\\\hline
\end{array}
\end{eqnarray*}
Once again, only Genie 1 needs \emph{extra} dimensions. Precisely an
additional $4N-5M$ extra dimensions are needed, so that in place of
(\ref{eq:5by4first}) we get
\begin{eqnarray}
3nR&\leq&(M+4N-5M)n\log\rho+
h(X_a^n,X_b^n)+h(X_b^n,X_c^n)+h(X_{a_1}^n,X_{c_1}^n|X_a^n,X_b^n,X_c^n)\notag\\
&&+n~o(\log\rho)+o(n)
\end{eqnarray}
The remaining genie bounds remain unaffected. Adding the
corresponding bounds as before we obtain:
\begin{eqnarray}
15nR &\leq & (4N-5M+ 5M)n\log\rho+ 6nR +n~o(\log\rho)+o(n) \\
\Rightarrow d&\leq & \frac{4N}{9}
\end{eqnarray}

\subsubsection{Case: $(N,M) = (2L+3, 2L+2) \Rightarrow $ DoF $\leq
\frac{MN}{M+N}$}

\begin{eqnarray*}
\begin{array}{r|c|}
\hline X_{1a_L}&\circ\\\hline \vdots&\circ\\\hline
X_{1a_1}&\circ\\\hline X_{1a_0}&\circ\\\hline X_{1b_0}&\circ\\\hline
X_{1c_0}&\circ\\\hline X_{1c_1}&\circ\\\hline \vdots&\circ\\\hline
X_{1c_L}&\circ\\\hline
\end{array}
&&
\begin{array}{|c|l}
\hline \circ&S_{1a_L}(X_{2a_L},\cdots,X_{2c_{L-1}}, X_{3a_{L-1}},
\cdots, X_{3c_{L-1}})\\\hline \circ&\vdots\\\hline
\circ&S_{1a_1}(X_{2a_1},X_{2a_0},X_{2b_0},X_{2c_0},X_{3a_0},X_{3b_0},X_{3c_0})\\\hline
\circ&S_{1a_0}(X_{2a_0},X_{2b_0},X_{3b_0})\\\hline
\circ&S_{1c_0}(X_{2b_0},X_{3b_0},X_{3c_0})\\\hline
\circ&S_{1c_1}(X_{2a_0},X_{2b_0},X_{2c_0},X_{3a_0},X_{3b_0},X_{3c_0},X_{3c_1})\\\hline
\circ&\vdots\\\hline
\circ&S_{1c_L}(X_{2a_{L-1}},\cdots,X_{2c_{L-1}},X_{3a_{L-1}},\cdots
X_{3c_L})\\\hline
\end{array}
\\
&&\\
&&\\
\begin{array}{r|c|}
\hline X_{2a_L}&\circ\\\hline \vdots&\circ\\\hline
X_{2a_1}&\circ\\\hline X_{2a_0}&\circ\\\hline X_{2b_0}&\circ\\\hline
X_{2c_0}&\circ\\\hline X_{2c_1}&\circ\\\hline \vdots&\circ\\\hline
X_{2c_L}&\circ\\\hline
\end{array}
&&
\begin{array}{|c|l}
\hline \circ&S_{2a_L}(X_{3a_L},\cdots,X_{3c_{L-1}}, X_{1a_{L-1}},
\cdots, X_{1c_{L-1}})\\\hline \circ&\vdots\\\hline
\circ&S_{2a_1}(X_{3a_1},X_{3a_0},X_{3b_0},X_{3c_0},X_{1a_0},X_{1b_0},X_{1c_0})\\\hline
\circ&S_{2a_0}(X_{3a_0},X_{3b_0},X_{1b_0})\\\hline
\circ&S_{2c_0}(X_{3b_0},X_{1b_0},X_{1c_0})\\\hline
\circ&S_{2c_1}(X_{3a_0},X_{3b_0},X_{3c_0},X_{1a_0},X_{1b_0},X_{1c_0},X_{1c_1})\\\hline
\circ&\vdots\\\hline
\circ&S_{2c_L}(X_{3a_{L-1}},\cdots,X_{3c_{L-1}},X_{1a_{L-1}},\cdots
X_{1c_L})\\\hline

\end{array}
\\
&&\\
&&\\
\begin{array}{r|c|}
\hline X_{3a_L}&\circ\\\hline \vdots&\circ\\\hline
X_{3a_1}&\circ\\\hline X_{3a_0}&\circ\\\hline X_{3b_0}&\circ\\\hline
X_{3c_0}&\circ\\\hline X_{3c_1}&\circ\\\hline \vdots&\circ\\\hline
X_{3c_L}&\circ\\\hline
\end{array}
&&
\begin{array}{|c|l}
\hline \circ&S_{3a_L}\\\hline \circ&\vdots\\\hline
\circ&S_{3a_1}\\\hline \circ&S_{3a_0}\\\hline \circ&S_{3c_0}\\\hline
\circ&S_{3c_1}\\\hline \circ&\vdots\\\hline
\circ&S_{3c_L}\\\hline\end{array}
\end{eqnarray*}

The rigorous proof for the general case is similar to that shown in
Section 5. For brevity, we only provide the genie signals and show
the bounds that we eventually obtain, and all of them are easily to
check.

First, a genie provides $\mathcal{G}_1=(X_{3a_L}^n, \cdots,
X_{3b_0}^n, X_{2b_0}^n, \cdots, X_{2c_L}^n)$ to receiver 1, and we
obtain:
\begin{eqnarray}
3nR&\leq&Mn\log\rho+h(X_{a_0}^n,X_{b_0}^n)+h(X_{b_0}^n,
X_{c_0}^n)+\sum_{i=1}^Lh(X_{a_i}^n,X_{c_i}^n|X_{a_{i-1}^n},\cdots,X_{c_{i-1}}^n)\notag\\
&&+n~o(\log\rho)+o(n).
\end{eqnarray}

Second, if a genie provides $\mathcal{G}_2=(X_{3a_L}^n, \cdots,
X_{3a_0}^n, X_{2a_0}^n, \cdots, X_{2c_L}^n)$ and
$\mathcal{G}_3=(X_{2c_L}^n, \cdots, X_{2c_0}^n,$ $ X_{3c_0}^n,
\cdots, X_{3a_L}^n)$ to receiver 1, respectively, we eventually
obtain:
\begin{eqnarray}
6nR&\leq&2Mn\log\rho+2nR+h(X_{a_0}^n|X_{b_0}^n,X_{c_0}^n)+h(X_{c_0}^n|X_{a_0}^n,X_{b_0}^n)+n~o(\log\rho)+o(n).
\end{eqnarray}

Third, if a genie provides $\mathcal{G}_4=(X_{3a_L}^n, \cdots,
X_{3a_1}^n, X_{2a_1}^n, \cdots, X_{2c_L}^n)$ and
$\mathcal{G}_5=(X_{2c_L}^n, \cdots, X_{2c_1}^n,$ $X_{3c_1}^n,
\cdots, X_{3a_L}^n)$ to receiver 1, respectively, we eventually
obtain:
\begin{eqnarray}
6nR&\leq&2Mn\log\rho+2nR+h(X_{a_1}^n,X_{c_1}^n|X_{a_0}^n,X_{b_0}^n,X_{c_0}^n)+n~o(\log\rho)+o(n).
\end{eqnarray}

$\vdots$

Last, if a genie provides $\mathcal{G}_{2L+2}=(X_{3a_L}^n,
X_{2a_L}^n, \cdots, X_{2c_L}^n)$ and
$\mathcal{G}_{2L+3}=(X_{2c_L}^n, X_{3c_L}^n, \cdots, X_{3a_L}^n)$ to
receiver 1, respectively, we eventually obtain:
\begin{eqnarray}
6nR&\leq&2Mn\log\rho+2nR+h(X_{a_L}^n,X_{c_L}^n|X_{a_{L-1}}^n,\cdots,X_{c_{L-1}}^n)+n~o(\log\rho)+o(n).
\end{eqnarray}
Adding all the outer bounds, we have:
\begin{eqnarray}
3(2L+3)nR&\leq& (2L+3)Mn\log\rho+(2L+4)nR+n~o(\log\rho)+o(n)\\
\Rightarrow d&\leq&\frac{(2L+3)M}{4L+5}\\
\Rightarrow d&\leq&\frac{MN}{M+N}
\end{eqnarray}

And the bounds for $M/N\leq \frac{2L+2}{2L+3}$ and $M/N\geq
\frac{2L+2}{2L+3}$ follow similarly as well.

\subsection{Outerbound Proofs around $M/N$ ratios $3/4, 5/6,
\cdots$}

\subsubsection{Case $(N,M)=(4,3) \Rightarrow $ DoF $\leq
\frac{12}{7}$}

\begin{eqnarray*}
\begin{array}{r|c|}
\hline X_{1a_1}&\circ\\\hline X_{1a_0}&\circ\\\hline
X_{1c_0}&\circ\\\hline X_{1c_1}&\circ\\\hline
\end{array}
&&
\begin{array}{|c|l}
\hline \circ&S_{1a_0}(X_{2a_1},X_{2a_0},X_{2c_0}, X_{3a_0},
X_{3c_0})\\\hline \circ&S_{1b_0}(X_{2a_0}, X_{3c_0})\\\hline
\circ&S_{1c_0}(X_{2a_0},X_{2c_0},X_{3a_0}, X_{3c_0},
X_{3c_1})\\\hline
\end{array}
\\
&&\\
&&\\
\begin{array}{r|c|}
\hline X_{2a_1}&\circ\\\hline X_{2a_0}&\circ\\\hline
X_{2c_0}&\circ\\\hline X_{2c_1}&\circ\\\hline
\end{array}
&&
\begin{array}{|c|l}
\hline \circ&S_{2a_0}(X_{3a_1},X_{3a_0},X_{3c_0}, X_{1a_0},
X_{1c_0})\\\hline \circ&S_{2b_0}(X_{3a_0}, X_{1c_0})\\\hline
\circ&S_{2c_0}(X_{3a_0},X_{3c_0},X_{1a_0}, X_{1c_0},
X_{1c_1})\\\hline
\end{array}
\\
&&\\
&&\\
\begin{array}{r|c|}
\hline X_{3a_1}&\circ\\\hline X_{3a_0}&\circ\\\hline
X_{3c_0}&\circ\\\hline X_{3c_1}&\circ\\\hline
\end{array}
&&
\begin{array}{|c|l}
\hline \circ&S_{3a_0}(X_{1a_1},X_{1a_0},X_{1c_0}, X_{2a_0},
X_{2c_0})\\\hline \circ&S_{3b_0}(X_{1a_0},X_{2c_0})\\\hline
\circ&S_{3c_0}(X_{1a_0},X_{1c_0},X_{2a_0}, X_{2c_0},
X_{2c_1})\\\hline
\end{array}
\end{eqnarray*}

First, a genie provides $\mathcal{G}_1=(X_{3a_1}^n, X_{3a_0}^n,
X_{2a_0}^n, X_{2c_0}^n, X_{2c_1}^n)$ to receiver 1, then we have:
\begin{eqnarray}
nR_\Sigma&\!\!\!\!\leq\!\!\!\!& Mn\log\rho+h(X_{2a_0}^n, X_{2c_0}^n)+h(X_{3a_0}^n|X_{3c_0}^n)+n~o(\log\rho)+o(n)\nonumber\\
&&+h(X_{2c_1}^n|X_{2a_1}^n,X_{2a_0}^n, X_{2c_0}^n)+h(X_{3a_1}^n|X_{3a_0}^n, X_{3c_0}^n, X_{3c_1}^n)~~~~\\
\hookrightarrow 3nR&\!\!\!\!\leq\!\!\!\! & Mn\log\rho + h(X_{a_0}^n, X_{c_0}^n)+h(X_{a_0}^n|X_{c_0}^n)+n~o(\log\rho)+o(n)\nonumber\\
&&+h(X_{c_1}^n|X_{a_1}^n,X_{a_0}^n, X_{c_0}^n)+h(X_{a_1}^n|X_{a_0}^n, X_{c_0}^n, X_{c_1}^n)~~~~\\
&\!\!\!\!\leq\!\!\!\! & Mn\log\rho + h(X_{a_0}^n, X_{c_0}^n)+h(X_{a_0}^n|X_{c_0}^n)+h(X_{a_1}^n,X_{c_1}^n|X_{a_0}^n, X_{c_0}^n)+n~o(\log\rho)+o(n)\ \ \ \\
&\!\!\!\!\leq\!\!\!\! & Mn\log\rho +
nR+h(X_{a_0}^n|X_{c_0}^n)+n~o(\log\rho)+o(n)\label{eq:4by3a}
\end{eqnarray}
Similarly if a genie provides $\mathcal{G}_2=(X_{2c_1}^n,
X_{2c_0}^n, X_{3c_0}^n, X_{3a_0}^n, X_{3a_1}^n)$ to receiver 1, we
have:
\begin{eqnarray}
3nR&\!\!\!\!\leq\!\!\!\! & Mn\log\rho +
nR+h(X_{c_0}^n|X_{a_0}^n)+n~o(\log\rho)+o(n)\label{eq:4by3b}
\end{eqnarray}
Adding up (\ref{eq:4by3a}) and (\ref{eq:4by3b}) we have:
\begin{eqnarray}
6nR&\!\!\!\!\leq\!\!\!\! & 2Mn\log\rho +
2nR+h(X_{a_0}^n,X_{c_0}^n)+n~o(\log\rho)+o(n)\label{eq:4by3first}
\end{eqnarray}

Second, if a genie provides $\mathcal{G}_3=(X_{3a_1}^n,
\underbrace{X_{2a_1}^n, X_{2a_0}^n, X_{2c_0}^n, X_{2c_1}^n}_{W_2})$
to receiver 1, then we have:
\begin{eqnarray}
nR_\Sigma &\leq& Mn\log\rho + nR_2 + h(X_{3a_1}^n|X_{3a_0}^n, X_{3c_0}^n, X_{3c_1}^n)+n~o(\log\rho)+o(n)\\
\hookrightarrow 3nR&\leq & Mn\log\rho + nR + h(X_{a_1}^n|X_{a_0}^n,
X_{c_0}^n, X_{c_1}^n)+n~o(\log\rho)+o(n)
\end{eqnarray}
and similarly if a genie provides $\mathcal{G}_4=(X_{2c_1}^n,
\underbrace{X_{3c_1}^n, X_{3c_0}^n, X_{3a_0}^n, X_{3a_1}^n}_{W_3})$
to receiver 1, then we have:
\begin{eqnarray}
3nR&\leq & Mn\log\rho + nR + h(X_{c_1}^n|X_{a_1}^n, X_{a_0}^n,
X_{c_0}^n)+n~o(\log\rho)+o(n)
\end{eqnarray}
Thus, we have:
\begin{eqnarray}
6nR &\leq & 2Mn\log\rho + 2nR + h(X_{a_1}^n,X_{c_1}^n|X_{a_0}^n,
X_{c_0}^n)+n~o(\log\rho)+o(n)\label{eq:4by3second}
\end{eqnarray}
Adding (\ref{eq:4by3first}), (\ref{eq:4by3second}), we obtain
\begin{eqnarray}
12nR &\leq & 4Mn\log\rho + 4nR + nR+n~o(\log\rho)+o(n)\\
\Rightarrow d&\leq&\frac{4M}{7}=\frac{12}{7}
\end{eqnarray}

\subsubsection{Case $(N,M)=(6,5) \Rightarrow $ DoF $\leq
\frac{30}{11}$}

\begin{eqnarray*}
\begin{array}{r|c|}
\hline X_{1a_2}&\circ\\\hline X_{1a_1}&\circ\\\hline
X_{1a_0}&\circ\\\hline X_{1c_0}&\circ\\\hline X_{1c_1}&\circ\\\hline
X_{1c_2}&\circ\\\hline
\end{array}
&&
\begin{array}{|c|l}
\hline \circ&S_{1a_1}(X_{2a_2}, X_{2a_1},X_{2a_0},X_{2c_0},X_{2c_1},
X_{3a_1},X_{3a_0}, X_{3c_0},X_{3c_1})\\\hline
\circ&S_{1a_0}(X_{2a_1},X_{2a_0},X_{2c_0}, X_{3a_0},
X_{3c_0})\\\hline \circ&S_{1b_0}(X_{2a_0}, X_{3c_0})\\\hline
\circ&S_{1c_0}(X_{2a_0},X_{2c_0},X_{3a_0}, X_{3c_0},
X_{3c_1})\\\hline
\circ&S_{1c_1}(X_{2a_1},X_{2a_0},X_{2c_0},X_{2c_1},
X_{3a_1},X_{3a_0}, X_{3c_0},X_{3c_1},X_{3c_2} )\\\hline
\end{array}
\\
&&\\
&&\\
\begin{array}{r|c|}
\hline X_{2a_2}&\circ\\\hline X_{2a_1}&\circ\\\hline
X_{2a_0}&\circ\\\hline X_{2c_0}&\circ\\\hline X_{2c_1}&\circ\\\hline
X_{2c_2}&\circ\\\hline
\end{array}
&&
\begin{array}{|c|l}
\hline \circ&S_{2a_1}(X_{3a_2}, X_{3a_1},X_{3a_0},X_{3c_0},X_{3c_1},
X_{1a_1},X_{1a_0}, X_{1c_0},X_{1c_1})\\\hline
\circ&S_{2a_0}(X_{3a_1},X_{3a_0},X_{3c_0}, X_{1a_0},
X_{1c_0})\\\hline \circ&S_{2b_0}(X_{3a_0}, X_{1c_0})\\\hline
\circ&S_{2c_0}(X_{3a_0},X_{3c_0},X_{1a_0}, X_{1c_0},
X_{1c_1})\\\hline
\circ&S_{2c_1}(X_{3a_1},X_{3a_0},X_{3c_0},X_{3c_1},
X_{1a_1},X_{1a_0}, X_{1c_0},X_{1c_1},X_{1c_2} )\\\hline
\end{array}
\\
&&\\
&&\\
\begin{array}{r|c|}
\hline X_{3a_2}&\circ\\\hline X_{3a_1}&\circ\\\hline
X_{3a_0}&\circ\\\hline X_{3c_0}&\circ\\\hline X_{3c_1}&\circ\\\hline
X_{3c_2}&\circ\\\hline
\end{array}
&&
\begin{array}{|c|l}
\hline \circ&S_{3a_1}(X_{1a_2}, X_{1a_1},X_{1a_0},X_{1c_0},X_{1c_1},
X_{2a_1},X_{2a_0}, X_{2c_0},X_{2c_1})\\\hline
\circ&S_{3a_0}(X_{1a_1},X_{1a_0},X_{1c_0}, X_{2a_0},
X_{2c_0})\\\hline \circ&S_{3b_0}(X_{1a_0}, X_{2c_0})\\\hline
\circ&S_{3c_0}(X_{1a_0},X_{1c_0},X_{2a_0}, X_{2c_0},
X_{2c_1})\\\hline
\circ&S_{3c_1}(X_{1a_1},X_{1a_0},X_{1c_0},X_{1c_1},
X_{2a_1},X_{2a_0}, X_{2c_0},X_{2c_1},X_{2c_2} )\\\hline
\end{array}
\end{eqnarray*}

First, a genie provides $\mathcal{G}_1=(X_{3a_2}^n, X_{3a_1}^n,
X_{3a_0}^n, X_{2a_0}^n, X_{2c_0}^n, X_{2c_1}^n,X_{2c_2}^n)$ to
receiver 1, then we have:
\begin{small}
\begin{eqnarray}
nR_\Sigma&\!\!\!\!\leq\!\!\!\!& Mn\log\rho +h(X_{2a_0}^n, X_{2c_0}^n)+h(X_{3a_0}^n|X_{3c_0}^n)+n~o(\log\rho)+o(n)\nonumber\\
&\!\!\!\!\!\!\!\!&+h(X_{2c_1}^n|X_{2a_1}^n,X_{2a_0}^n,X_{2c_0}^n)+h(X_{3a_1}^n|X_{3a_0}^n, X_{3c_0}^n, X_{3c_1}^n)\nonumber\\
&\!\!\!\!\!\!\!\!&+h(X_{2c_2}^n|X_{2a_2}^n,X_{2a_1}^n,X_{2a_0}^n, X_{2c_0}^n,X_{2c_1}^n)+h(X_{3a_2}^n|X_{3a_1}^n,X_{3a_0}^n, X_{3c_0}^n, X_{3c_1}^n,X_{3c_2}^n)\\
\hookrightarrow 3nR&\!\!\!\!\leq\!\!\!\! & Mn\log\rho + h(X_{a_0}^n, X_{c_0}^n)+h(X_{a_0}^n|X_{c_0}^n)+h(X_{c_1}^n|X_{a_1}^n,X_{a_0}^n, X_{c_0}^n)+h(X_{a_1}^n|X_{a_0}^n, X_{c_0}^n, X_{c_1}^n)\nonumber\\
&\!\!\!\!\!\!\!\!&+h(X_{c_2}^n|X_{a_2}^n, X_{a_1}^n,X_{a_0}^n, X_{c_0}^n,X_{c_1}^n)+h(X_{a_2}^n|X_{a_1}^n,X_{a_0}^n, X_{c_0}^n, X_{c_1}^n,X_{c_2}^n)+n~o(\log\rho)+o(n)\\
&\!\!\!\!\leq\!\!\!\!& Mn\log\rho + h(X_{a_0}^n, X_{c_0}^n)+h(X_{a_0}^n|X_{c_0}^n)+h(X_{a_1}^n,X_{c_1}^n|X_{a_0}^n, X_{c_0}^n)+h(X_{a_2}^n,X_{c_2}^n| X_{a_1}^n,X_{a_0}^n, X_{c_0}^n,X_{c_1}^n)\nonumber\\
&\!\!\!\!\!\!\!\!&+n~o(\log\rho)+o(n)\\
&\!\!\!\!\leq\!\!\!\!& Mn\log\rho + nR +
h(X_{a_0}|X_{c_0})+n~o(\log\rho)+o(n)
\end{eqnarray}
\end{small}
Similarly if a genie provides $\mathcal{G}_2=(X_{2c_2}^n,
X_{2c_1}^n, X_{2c_0}^n, X_{3c_0}^n, X_{3a_0}^n,
X_{3a_1}^n,X_{3a_2}^n)$ to receiver 1, then we have:
\begin{eqnarray}
3nR&\!\!\!\!\leq\!\!\!\!& Mn\log\rho + nR +
h(X_{c_0}^n|X_{a_0}^n)+n~o(\log\rho)+o(n)
\end{eqnarray}
and thus we have:
\begin{eqnarray}
6nR&\!\!\!\!\leq\!\!\!\!& 2Mn\log\rho + 2nR +
h(X_{a_0}^n,X_{c_0}^n)+n~o(\log\rho)+o(n)
\end{eqnarray}

Second, a genie provides $\mathcal{G}_3=(X_{3a_2}^n, X_{3a_1}^n,
X_{2a_1}^n, X_{2a_0}^n, X_{2c_0}^n, X_{2c_1}^n,X_{2c_2}^n)$ to
receiver 1, then we have:
\begin{eqnarray}
nR_\Sigma&\leq& Mn\log\rho +h(X_{2a_1}^n, X_{2a_0}^n, X_{2c_0}^n, X_{2c_1}^n,X_{2c_2}^n)+n~o(\log\rho)+o(n)\nonumber\\
&&+h(X_{3a_1}^n| X_{3a_0}^n, X_{3c_0}^n, X_{3c_1}^n)+h(X_{3a_2}^n|X_{3a_1}^n, X_{3a_0}^n, X_{3c_0}^n, X_{3c_1}^n, X_{3c_2}^n)\\
\hookrightarrow 3nR&\leq& Mn\log\rho +h(X_{a_1}^n, X_{a_0}^n, X_{c_0}^n, X_{c_1}^n,X_{c_2}^n)+n~o(\log\rho)+o(n)\nonumber\\
&&+h(X_{a_1}^n| X_{a_0}^n, X_{c_0}^n, X_{c_1}^n)+h(X_{a_2}^n|X_{a_1}^n, X_{a_0}^n, X_{c_0}^n, X_{c_1}^n, X_{c_2}^n)\\
\Rightarrow 3nR&\leq & Mn\log\rho + nR + h(X_{a_1}^n| X_{a_0}^n,
X_{c_0}^n, X_{c_1}^n)+n~o(\log\rho)+o(n)
\end{eqnarray}
Similarly if a genie provides $\mathcal{G}_4=(X_{2c_2}^n,
X_{2c_1}^n, X_{3c_1}^n, X_{3c_0}^n, X_{3a_0}^n,
X_{3a_1}^n,X_{3a_2}^n)$ to receiver 1, then we have:
\begin{eqnarray}
3nR&\leq & Mn\log\rho + nR + h(X_{c_1}^n| X_{a_1}^n, X_{a_0}^n,
X_{c_0}^n)+n~o(\log\rho)+o(n)
\end{eqnarray}
and thus we obtain:
\begin{eqnarray}
6nR &\leq & 2Mn\log\rho + 2nR + h(X_{a_1}^n, X_{c_1}^n| X_{a_0}^n,
X_{c_0}^n)+n~o(\log\rho)+o(n)
\end{eqnarray}

Last, a genie provides $\mathcal{G}_5=(X_{3a_2}^n,
\underbrace{X_{2a_2}^n, X_{2a_1}^n, X_{2a_0}^n, X_{2c_0}^n,
X_{2c_1}^n,X_{2c_2}^n}_{W_2})$ to receiver 1, then we have:
\begin{eqnarray}
nR_\Sigma&\leq& Mn\log\rho + nR_2 + h(X_{3a_2}^n|X_{3a_1}^n, X_{3a_0}^n, X_{3c_0}^n, X_{3c_1}^n, X_{3c_2}^n)+n~o(\log\rho)+o(n)\\
\hookrightarrow 3nR&\leq & Mn\log\rho + nR + h(X_{a_2}^n|X_{a_1}^n,
X_{a_0}^n, X_{c_0}^n, X_{c_1}^n, X_{c_2}^n)+n~o(\log\rho)+o(n)
\end{eqnarray}
Similarly if a genie provides $\mathcal{G}_6=(X_{2c_2}^n,
\underbrace{X_{3c_2}^n, X_{3c_1}^n, X_{3c_0}^n, X_{3a_0}^n,
X_{3a_1}^n,X_{3a_2}^n}_{W_3})$ to receiver 1, then we have:
\begin{eqnarray}
3nR&\leq & Mn\log\rho + nR + h(X_{c_2}^n|X_{a_2}^n, X_{a_1}^n,
X_{a_0}^n, X_{c_0}^n, X_{c_1}^n)+n~o(\log\rho)+o(n)
\end{eqnarray}
and thus we have:
\begin{eqnarray}
6nR &\leq & 2Mn\log\rho + 2nR + h(X_{a_2}^n,X_{c_2}^n| X_{a_1}^n,
X_{a_0}^n, X_{c_0}^n, X_{c_1}^n)+n~o(\log\rho)+o(n)
\end{eqnarray}
Adding the final equations from each of the genies, we obtain:
\begin{eqnarray}
18nR &\leq & 6Mn\log\rho + 6nR + nR+n~o(\log\rho)+o(n)\\
d&\leq & \frac{6M}{11}=\frac{30}{11}
\end{eqnarray}

\subsubsection{Case $(N,M)=(2L+2,2L+1) \Rightarrow $ DoF $\leq
\frac{MN}{M+N}$}

\begin{eqnarray*}
\begin{array}{r|c|}
\hline X_{1a_L}&\circ\\\hline \vdots &\circ\\\hline
X_{1a_1}&\circ\\\hline X_{1a_0}&\circ\\\hline X_{1c_0}&\circ\\\hline
X_{1c_1}&\circ\\\hline \vdots&\circ\\\hline X_{1c_L}&\circ\\\hline
\end{array}
&&
\begin{array}{|c|l}
\hline \circ&S_{1a_{L-1}}(X_{2a_L}, \cdots, X_{2c_{L-1}},
X_{3a_{L-1}},\cdots, X_{3c_{L-1}})\\\hline \circ&\vdots\\\hline
\circ&S_{1a_0}(X_{2a_1},X_{2a_0},X_{2c_0}, X_{3a_0},
X_{3c_0})\\\hline \circ&S_{1b_0}(X_{2a_0}, X_{3c_0})\\\hline
\circ&S_{1c_0}(X_{2a_0},X_{2c_0},X_{3a_0}, X_{3c_0},
X_{3c_1})\\\hline \circ&\vdots\\\hline
\circ&S_{1c_{L-1}}(X_{2a_{L-1}},\cdots, X_{2c_{L-1}},
X_{3a_{L-1}},\cdots,X_{3c_L} )\\\hline
\end{array}
\\
&&\\
&&\\
\begin{array}{r|c|}
\hline X_{2a_L}&\circ\\\hline \vdots &\circ\\\hline
X_{2a_1}&\circ\\\hline X_{2a_0}&\circ\\\hline X_{2c_0}&\circ\\\hline
X_{2c_1}&\circ\\\hline \vdots&\circ\\\hline X_{2c_L}&\circ\\\hline
\end{array}
&&
\begin{array}{|c|l}
\hline \circ&S_{2a_{L-1}}(X_{3a_Ln}, \cdots, X_{3c_{L-1}},
X_{1a_{L-1}},\cdots, X_{1c_{L-1}})\\\hline \circ&\vdots\\\hline
\circ&S_{2a_0}(X_{3a_1},X_{3a_0},X_{3c_0}, X_{1a_0},
X_{1c_0})\\\hline \circ&S_{2b_0}(X_{3a_0}, X_{1c_0})\\\hline
\circ&S_{2c_0}(X_{3a_0},X_{3c_0},X_{1a_0}, X_{1c_0},
X_{1c_1})\\\hline \circ&\vdots\\\hline
\circ&S_{2c_{L-1}}(X_{3a_{L-1}},\cdots, X_{3c_{L-1}},
X_{1a_{L-1}},\cdots,X_{1c_L} )\\\hline
\end{array}
\\
&&\\
&&\\
\begin{array}{r|c|}
\hline X_{3a_L}&\circ\\\hline \vdots &\circ\\\hline
X_{3a_1}&\circ\\\hline X_{3a_0}&\circ\\\hline X_{3c_0}&\circ\\\hline
X_{3c_1}&\circ\\\hline \vdots&\circ\\\hline X_{3c_L}&\circ\\\hline
\end{array}
&&
\begin{array}{|c|l}
\hline \circ&S_{3a_{n-1}}(X_{1a_L}, \cdots, X_{1c_{L-1}},
X_{2a_{L-1}},\cdots, X_{2c_{L-1}})\\\hline \circ&\vdots\\\hline
\circ&S_{3a_0}(X_{1a_1},X_{1a_0},X_{1c_0}, X_{2a_0},
X_{2c_0})\\\hline \circ&S_{3b_0}(X_{1a_0}, X_{2c_0})\\\hline
\circ&S_{3c_0}(X_{1a_0},X_{1c_0},X_{2a_0}, X_{2c_0},
X_{2c_1})\\\hline \circ&\vdots\\\hline
\circ&S_{3c_{L-1}}(X_{1a_{L-1}},\cdots, X_{1c_{L-1}},
X_{2a_{L-1}},\cdots,X_{2c_L} )\\\hline\end{array}
\end{eqnarray*}

First, a genie provides $\mathcal{G}_1=(X_{3a_L}^n, \cdots,
X_{3a_0}^n, X_{2a_0}^n, \cdots,X_{2c_L}^n)$ and
$\mathcal{G}_2=(X_{2c_L}^n, \cdots, X_{2c_0}^n,X_{3c_0}^n,$ $
\cdots,X_{3a_L}^n)$ to receiver 1, respectively, and we eventually
have:
\begin{eqnarray}
6nR&\leq & 2Mn\log\rho+ 2nR +
h(X_{a_0}^n,X_{c_0}^n)+n~o(\log\rho)+o(n).
\end{eqnarray}

Second, a genie provides $\mathcal{G}_3=(X_{3a_L}^n, \cdots,
X_{3a_1}^n, X_{2a_1}^n, \cdots,X_{2c_L}^n)$ and
$\mathcal{G}_4=(X_{2c_L}^n, \cdots, X_{2c_1}^n, X_{3c_1}^n,$ $
\cdots,X_{3a_L}^n)$ to receiver 1, respectively, and we eventually
have:
\begin{eqnarray}
6nR &\leq & 2Mn\log\rho + 2nR + h(X_{a_1}^n, X_{c_1}^n| X_{a_0}^n,
X_{c_0}^n)+n~o(\log\rho)+o(n).
\end{eqnarray}

\vdots

Last, a genie provides $\mathcal{G}_{2L+1}=(X_{3a_L}^n,  X_{2a_L}^n,
\cdots,X_{2c_L}^n)$ and $\mathcal{G}_{2L+2}=(X_{2c_L}^n, X_{3c_L}^n,
\cdots,X_{3a_L}^n)$ to receiver 1, respectively, then we eventually
obtain:
\begin{eqnarray}
6nR &\leq & 2Mn\log\rho + 2nR + h(X_{a_L}^n,X_{c_L}^n|
X_{a_{L-1}}^n,\cdots, X_{c_{L-1}}^n)+n~o(\log\rho)+o(n).
\end{eqnarray}
Adding the final equations from each of the genies, we obtain:
\begin{eqnarray}
6(L+1)nR &\leq & 2(L+1)Mn\log\rho + 2(L+1)nR + nR+n~o(\log\rho)+o(n).\label{eq:adjust}\\
\Rightarrow d&\leq & \frac{(2L+2)M}{4L+3}=\frac{MN}{M+N}
\end{eqnarray}

\subsubsection{Case $M/N\geq (2L+1)/(2L+2) \Rightarrow$ DoF $\leq
\frac{(2L+2)M}{4L+3}$}

The proof is identical to the previous section.

\subsubsection{Case $M/N\in[2L/(2L+1), (2L+1)/(2L+2)] \Rightarrow$
DoF $\leq \frac{(2L+1)N}{4L+3}$}

This setting is also simple, but needs a small manipulation in the
size of $X_{a_0}, X_{c_0}$ (the two cannot be equal) if $N$ is not
even. As always:
\begin{eqnarray}
|X_{a_i}|=|X_{c_i}|=N-M, ~~~\forall i\in\{1,\cdots, L\}
\end{eqnarray}
In this case, if $N$ is not even, let us choose
\begin{eqnarray}
|X_{a_0}|&=&\left\lceil \frac{N -2L(N-M)}{2}\right\rceil\\
|X_{c_0}|&=&\left\lfloor \frac{N- 2L(N-M)}{2}\right\rfloor
\end{eqnarray}
so that $$ |X_{a_0}| + |X_{c_0}| = N-2L(N-M)$$

Only Genie 1 needs to provide \emph{extra} dimensions. The Genie
usually includes two symmetric cases, one that provides two terms of
the form $|X_{a_o}|$ and another that provides two terms of the form
$|X_{c_0}|$. Because the two have different sizes, the two genies
usually involved within Genie 1, are no longer symmetric.

The number of extra dimensions needed for the Genie that provides
two terms of the form $|X_{a_o}|$ is
\begin{eqnarray}
(N-M)-|X_{a_0}|
\end{eqnarray}

so that the corresponding bound instead of (\ref{eq:4by3a}) is

\begin{eqnarray}
3nR &\leq &
[M+(N-M)]n\log\rho-|X_{a_0}^n|+h(X_{a_0}^n|X_{c_0}^n)+n~o(\log\rho)+o(n)
\end{eqnarray}
Similarly, the corresponding bound instead of (\ref{eq:4by3b}) is
\begin{eqnarray}
3nR &\leq &
[M+(N-M)]n\log\rho-|X_{c_0}^n|+h(X_{c_0}^n|X_{a_0}^n)+n~o(\log\rho)+o(n)
\end{eqnarray}
and the resulting bound instead of (\ref{eq:4by3first}) is
\begin{eqnarray}
6nR &\leq &
[2M+2(N-M)-N+2n(N-M)]n\log\rho+h(X_{a_0}^n,X_{c_0}^n)+n~o(\log\rho)+o(n)
\end{eqnarray}
Thus, adding the bounds from all the genies, instead of
(\ref{eq:adjust}) we get
\begin{eqnarray}
6(L+1)nR&\!\!\!\!\leq\!\!\!\!& 2(L+1)Mn\log\rho+2(L+1)nR+nR+n~o(\log\rho)+o(n)\notag\\
&\!\!\!\!\!\!\!\!&+[2(N-M)-N+2L(N-M)]n\log\rho\\
6(L+1)nR&\!\!\!\!\leq\!\!\!\!& 2(L+1)nR+nR+(2L+1)Nn\log\rho+n~o(\log\rho)+o(n)\\
\Rightarrow d&\leq & \frac{(2L+1)N}{4L+3}.
\end{eqnarray}


\begin{thebibliography}{1}

\bibitem{Gou_Jafar_Seong_Chung}
Tiangao Gou, Syed A. Jafar, Sang-Woon Jeon, Sae-Young Chung,
``Aligned Interference Neutralization and the Degrees of Freedom of
the $2\times 2\times 2$ Interference Channel", {\em e-print
arXiv:1012.2350}, Dec. 2010.

\bibitem{Shomorony_Avestimehr}
I.~Shomorony and S.~Avestimehr, ``Two unicast wireless networks:
characterizing the degrees-of-freedom'', \emph{arXiv:1102.2498},
Mar. 2011.

%\bibitem{Wang_Gou_Jafar_layeredX}
%C. Wang, T. Gou and S. Jafar, ``Multiple Unicast Capacity of
%2-Source 2-Sink Networks", {\em e-print arXiv:1104.0954}, April 2011

%\bibitem{Lee_Jafar}
%N.~Lee, S.~Jafar, ``Aligned Interference Neutralization and the
%Degrees of Freedom of the 2 User Interference Channel with
%Instantaneous Relay", {\em e-print arXiv:1102.3833}.
%
%\bibitem{Jeon_Chung_Jafar_Alleton09}
%Sang-Woon Jeon, Sae-Young Chung, Syed A. Jafar, ``Degrees of Freedom
%of Multi-Source Relay Networks", {\em Proceedings of the 47th Annual
%Allerton Conference}, 2009
%
%\bibitem{Cai_Letaief_Fan_Feng}
%K. Cai, K. B. Letaief, P. Fan, R. Feng, ``On the Solvability of
%2-pair Unicast Networks --- A Cut-based Characterization", {\em
%arXiv:1007.0465v1}, July 2010.

%\bibitem{Abinesh}
%A. Ramakrishnan, A. Das, H. Maleki, A. Markopoulou, S. A. Jafar, S.
%Vishwanath, ``Network Coding for Three Unicast Sessions:
%Interference Alignment Approaches," {\em in Proceedings of Allerton
%Conference 2010}, Sep. 2010, Monticello, IL.

\bibitem{Jafar_Shamai}
S.~Jafar, S.~Shamai, ``Degrees of Freedom Region for the MIMO $X$
Channel'', {\em IEEE Transactions on Information Theory}, vol.~54,
No.~1, pp.~151-170, Jan. 2008.

%\bibitem{Cadambe_Jafar_X}
%V.~Cadambe and S.~Jafar, ``Interference alignment and the degrees of
%freedom of wireless $X$ networks'', {\em IEEE Trans. on Information
%Theory}, vol.~55, no. 9,  pp.~3893--3908, Sep. 2009.

\bibitem{MMK}
M.A. Maddah-Ali, A.S. Motahari , and A.K. Khandani, "Communication
Over MIMO $X$ Channels: Interference Alignment, Decomposition, and
Performance Analysis," IEEE Transaction on Information Theory, Vol.
54, No. 8, pp. 3457-3470, Aug. 2008.

\bibitem{Etkin_Ordentlich_rational}
R. Etkin and E. Ordentlich, ``The Degrees of Freedom of the K User
Gaussian Interference Channel Is Discontinuous at Rational Channel
Coefficients", {\em IEEE Trans. on Infor. Theory}, vol. 55, no. 11,
Nov. 2009.

\bibitem{Motahari_Gharan_Khandani_real}
A.S. Motahari, S. O. Gharan and A. K. Khandani, ``Real Interference
Alignment with Real Numbers," {\em arXiv:0908.1208}, Aug. 2009.

%\bibitem{Maddah-Ali_compoundbc}
%M.A. Maddah-Ali, ``On the Degrees of Freedom of the Compound MIMO
%Broadcast Channels with Finite States", {\em arXiv:0909.5006v3},
%Oct. 2009.

\bibitem{Cadambe_Jafar_int}
V.~Cadambe and S.~Jafar, ``Interference alignment and the degrees of
freedom of the $K$ user interference channel'', {\em IEEE Trans. on
Information Theory}, vol.~54, pp.~3425--3441, Aug. 2008.

\bibitem{Bresler_Tse_diversity}
G.~Bresler, D.~Tse, ``Degrees-of-freedom for the 3-user Gaussian
interference channel as a function of channel diversity," {\em
Allerton Conference on Communication, Control, and Computing},
Monticello, IL, September 2009.

%\bibitem{Suh_Tse_cellular}
%C. Suh and D. Tse, ``Interference alignment for cellular networks,"
%{\em Allerton Conference on Communication, Control, and Computing},
%Montecello, IL, Sep. 2008.
%
%\bibitem{Suh_Tse_downlink}
%C. Suh, M. Ho, D. Tse, ``Downlink Interference Alignment," {\em
%arXiv:1003.3707}, March 2010.

%\bibitem{Wang_Gou_Jafar_blindia}
%C. Wang, T. Gou, S. Jafar, ``Aiming Perfectly in the Dark - Blind
%Interference Alignment through Staggered Antenna Switching," {\em
%IEEE Transactions on Signal Processing}, June 2011.
%
%\bibitem{Gou_Jafar_Wang_compound}
%T. Gou, S. Jafar, C. Wang, ``On the Degrees of Freedom of Finite
%State Compound Wireless Networks," {\em To Appear in the IEEE
%Transactions on Information Theory}, 2010.
%
%\bibitem{Cadambe_Jafar_relay_fb}
%V. R. Cadambe and S. A. Jafar, ``Degrees of Freedom of Wireless
%Networks with Relays, Feedback, Cooperation and Full Duplex
%Operation," {\em IEEE Trans. Inform. Theory}, vol. 55, no. 5, pp.
%2334-2344, May 2009.

\bibitem{Gou_Jafar_MIMO}
T. Gou, S. Jafar, ``Degrees of Freedom of the $K$ User $M\times N$
MIMO Interference Channel," {\em IEEE Transactions on Information
Theory}, Dec. 2010, Vol. 56, Issue: 12, Page(s): 6040-6057.

\bibitem{Ghasemi_Motahari_Khandani_MIMO}
A. Ghasemi, A. Motahari, A. Khandani, ``Interference Alignment for
the $K$ User MIMO Interference Channel," {\em arXiv:0909.4604}, Sep.
2009.

\bibitem{Jafar_Fakhereddin}
S.~Jafar, M.~Fakhereddin, ``Degrees of Freedom for the MIMO
Interference Channel," {\em IEEE Transactions on Information
Theory}, July 2007, Vol. 53, No. 7, Pages: 2637-2642.

\bibitem{Cenk_Gou_Jafar_feasibility}
Cenk M. Yetis, Tiangao Gou, Syed A. Jafar, Ahmet H. Kayran, ``On
Feasibility of Interference Alignment in MIMO Interference
Networks," {\em IEEE Transactions on Signal Processing}, Sep. 2010,
Vol. 58, Issue: 9, Pages: 4771-4782.

\bibitem{Bresler_Cavendish_Tse}
Guy Bresler, Dustin Cartwright, David Tse, ``Settling the
feasibility of interference alignment for the MIMO interference
channel: the symmetric square case", {\em arXiv:1104.0888}

\bibitem{Razaviyayn_MIMO}
Meisam Razaviyayn, Gennady Lyubeznik, Zhi-Quan Luo, ``On the Degrees
of Freedom Achievable Through Interference Alignment in a MIMO
Interference Channel", {\em arXiv:1104.0992}

%\bibitem{Cadambe_Jafar_Wang_acs}
%Viveck R. Cadambe, Syed A. Jafar, Chenwei Wang, ``Interference
%Alignment with Asymmetric Complex Signaling - Settling the
%Host-Madsen-Nosratinia Conjecture", {\em IEEE Transactions on
%Information Theory}, Sep. 2010, Vol. 56, Issue: 9, Pages: 4552-4565

\bibitem{Avestimehr_Diggavi_Tse}
S. Avestimehr, S. Diggavi, and D. Tse, ``deterministic approach to
wireless relay networks", {\em Allerton Conf. Commun., Control,
Comput.}, Monticello, IL, Sep. 2007.

\bibitem{Jafar_corr}
Syed A. Jafar, ``Exploiting Channel Correlations - Simple
Interference Alignment Schemes with no CSIT", {\em e-print
arXiv:0910.0555}.

\bibitem{Maddah_Tse}
M. A. Maddah-Ali and D. Tse, ``On the degrees of freedom of miso
broadcast channels with delayed feedback", {\em EECS Department,
University of California, Berkeley, Tech. Rep. UCB/EECS-2010-122,
Sep 2010}. [Online]. Available: http://www.eecs.berkeley.edu/Pubs/
TechRpts/2010/EECS-2010-122.html

\bibitem{Maleki_Jafar_Shamai}
Hamed Maleki, Syed A. Jafar, Shlomo Shamai, ``Retrospective
Interference Alignment", {\em e-print arXiv:1009.3593}.

\bibitem{Wu_Shamai_Verdu}
Yihong Wu, Shlomo Shamai(Shitz) and Sergio Verdu, ``Degrees of
Freedom of Interference Channel: a General Formula", {\em IEEE ISIT
(International Symposium on Information Theory}, August 2011.

\bibitem{Jafar_IA}
Syed A. Jafar, ``On asymptotic interference alignment", {\em Plenary
talk, International Conference on Signal Processing and
Communications (SPCOM)}, 2010.

\bibitem{Host-Madsen_Nosratinia}
A.~Host-Madsen and A.~Nosratinia, ``The multiplexing gain of
wireless networks,'' \emph{in Proc. of IEEE ISIT} 2005.

\bibitem{Motahari_Gharan_Maddah-Ali_Khandani}
A. Motahari, S. Gharan, M. A. Maddah-Ali, A, Khandani, ``Real
Interference Alignment: Exploiting the Potential of Single Antenna
Systems", {\em arxiv.org/pdf/0908.2282}, August 2009

\bibitem{Gomadam_Cadambe_Jafar}
Krishna S. Gomadam, Viveck R. Cadambe, Syed A. Jafar, ``A
Distributed Numerical Approach to Interference Alignment and
Applications to Wireless Interference Networks", {\em IEEE
Transactions on Information Theory,} Vol. 57, No. 6, June, 2011,
Pages: 3309-3322

\bibitem{Jafar_FnT}
Syed A. Jafar, ``Interference Alignment: A New Look at Signal
Dimensions in a Communication Network", {\em Foundations and Trends
in Communications and Information Theory}, Vol. 7, No. 1, pages:
1-136.

\bibitem{Nafie_etal}
M. Amir, A. El-Keyi, M. Nafie, "A New Achievable DoF Region for the
3-user $M\times N$ Symmetric Interference Channel",
arXiv:1105.4026v1 [cs.IT].

\end{thebibliography}
\end{document}